\documentclass[12pt]{article}

\usepackage{amssymb}
\usepackage{epsf}
\usepackage{epsfig}
\advance \topmargin by -2.5cm
\advance \oddsidemargin by -1.3cm
\advance \evensidemargin by -3.3cm

\begin{document}
%\debugon

% ------------ macros to chapter 6 ---------------------
\def\be{\begin{equation}}
\def\ee{\end{equation}}
\def\bea{\begin{eqnarray}}
\def\eea{\end{eqnarray}}
\def\no{\nonumber}
\def\str{\mbox{ Str}}
\def\bG{{\bf G}}
\def\bK{{\bf K}}
\def\bA{{\bf A}}
\def\bGc{{\bf G_C}}
\def\bKc{{\bf K_C}}
\def\bAc{{\bf A_C}}
\def\bC{{\bf C}}
\def\cG{{\cal G}}
\def\cK{{\cal K}}
\def\cA{{\cal A}}
\def\cP{{\cal P}}
\def\cGc{{\cal G}_{\bf C}}
\def\cKc{{\cal K}_{\bf C}}
\def\cAc{{\cal A}_{\bf C}}
\def\cPc{{\cal P}_{\bf C}}
\def\cNc{{\cal N}_{\bf C}}
\def\tG{\tilde{\cal G}}
\def\tK{\tilde{\cal K}}
\def\tP{\tilde{\cal P}}
\def\cN{{\cal N}}
\def\bNc{{\bf N_C}}
\def\bN{{\bf N}}
\def\grad{{\rm grad}}
\def\Dr{{\cal L}^\#}
\def\bGo{{\bf G_0}}
\def\bKo{{\bf K_0}}
\def\bAo{{\bf A_0}}
\def\bNo{{\bf N_0}}
\def\cGo{{\cal G}_{\bf 0}}
\def\cKo{{\cal K}_{\bf 0}}
\def\cAo{{\cal A}_{\bf 0}}
\def\cPo{{\cal P}_{\bf 0}}
\def\cNo{{\cal N}_{\bf 0}}

% ----------------- macros to chapter 7 --------------------------------------
\def\av#1{\langle {#1} \rangle}
\def\r{{\bf{r}}}
\def\ro{{\bf{r_o}}}
%------------------------------------------------------------------

%\selectlanguage{english}

\title{Statistics of energy levels and eigenfunctions in disordered systems}
\author{Alexander D. Mirlin \thanks{Also at Petersburg Nuclear Physics
Institute, 188350 Gatchina, St. Petersburg, Russia} \\
Institut f\"ur Theorie der kondensierten Materie,\\
Universit\"at Karlsruhe, 76128 Karlsruhe, Germany}
\date{6 July 1999}
\maketitle

\begin{abstract}

The article reviews recent developments in the theory of
fluctuations and correlations of energy levels and eigenfunction
amplitudes in diffusive
mesoscopic samples. Various spatial geometries are
considered, with emphasis on low-dimensional
(quasi-1D and 2D) systems. Calculations are based on the
supermatrix $\sigma$-model approach. The method reproduces,
in so-called zero-mode approximation, the universal random
matrix theory (RMT) results for the energy-level and eigenfunction
fluctuations. Going beyond this approximation
allows us to study system-specific deviations from universality, which are
determined by the diffusive classical dynamics
in the system. These deviations are especially strong in the far
``tails'' of the distribution function of the eigenfunction
amplitudes (as well as of some related quantities, such as local density
of states, relaxation time, etc.). These asymptotic ``tails'' are
governed by anomalously localized states which are formed in rare
realizations of the random potential. The deviations of the level and
eigenfunction statistics from their RMT form strengthen  with
increasing disorder and become especially pronounced at the Anderson
metal-insulator transition. In this regime, the wave functions
are multifractal, while the level statistics acquires a
scale-independent form with distinct critical features. 
Fluctuations of the conductance and of the local intensity of a
classical wave radiated by a point-like source in the
quasi-1D geometry are also studied within the $\sigma$-model
approach. For
 a ballistic system with rough surface an appropriately
modified (``ballistic'') $\sigma$-model is used. Finally, the interplay of
the fluctuations and the electron-electron interaction in small samples is
discussed, with application to the Coulomb blockade spectra.

\end{abstract}



\newpage

\tableofcontents
\section{Introduction}

Statistical properties of energy levels and eigenfunctions of complex
quantum systems have been attracting a lot of interest of physicists
since the work of Wigner \cite{wigner}, who formulated a statistical
point of view on nuclear spectra. In order to describe
excitation spectra of complex nuclei, Wigner proposed to replace a
complicated and unknown Hamiltonian by a large $N\times N$ random
matrix. This was a beginning of the random matrix theory (RMT), which
was further developed by Dyson and Mehta in the early 1960's
\cite{dyson,mehta}. This theory predicts a universal form of the spectral
correlation functions determined solely by some global
symmetries of the system (time-reversal invariance and value of the
spin). 

Later it was realized that the random matrix theory is not
restricted to strongly interacting many-body systems, but has a much
broader range of applicability. In particular, Bohigas, Giannoni, and
Schmit \cite{bohigas} put forward a conjecture (strongly supported by
accumulated numerical evidence) that the RMT describes adequately
statistical properties of spectra of quantum systems whose classical
analogs are chaotic. 

Another class of systems to which the RMT
applies and which is of special interest to us here is that of
disordered systems. More specifically, we mean a quantum particle (an
electron) moving in a random potential created by some kind of
impurities. It was conjectured by Gor'kov and Eliashberg \cite{gorel}
that statistical properties of the energy levels in such a disordered
granule can be described by the random matrix theory. This statement
had remained in the status of conjecture until 1982, when it was proven by
Efetov \cite{efetov83}. This became possible due to development by
Efetov of a very powerful tool of treatment of the disordered systems
under consideration --- the supersymmetry method (see the review
\cite{efetov83} and the recent book \cite{ef-book}). This method allows
one to map the problem of the particle in a random potential onto a
certain deterministic field-theoretical model (supermatrix
$\sigma$-model), which generates the disorder-averaged correlation
functions of the original problem. As Efetov showed, under certain
conditions one can neglect spatial variation of the $\sigma$-model
supermatrix field (so-called zero-mode approximation), which allows
one to calculate the correlation functions. The corresponding results
for the two-level correlation function reproduced precisely the RMT
results of Dyson. 

The supersymmetry method can be also applied to the problems of the
RMT-type. In this connection, we refer the reader to the paper
\cite{vwz}, where the technical aspects of the method are 
discussed in detail.

More recently, focus of the research interest was shifted from the
proof of the applicability of RMT to the study of system-specific
deviations from the universal (RMT) behavior. For the problem of level
correlations in a disordered system, this question was addressed
for the first time by Altshuler and Shklovskii \cite{ashk} in the
framework of the diffuson-cooperon diagrammatic perturbation theory.
They showed that the diffusive motion of the particle leads to  a
high-frequency behavior of the level correlation function completely
different from its RMT form. Their perturbative treatment was however
restricted to frequencies much larger than the level spacing and was
not able to reproduce the oscillatory contribution to the level
correlation function. Inclusion of non-zero spatial modes 
(which  means going beyond universality) within the
$\sigma$-model treatment of the level correlation function
was performed in Ref.~\cite{km}. The method developed in
\cite{km} was later used for calculation of deviations from the RMT of
various statistical characteristics of a disordered system. For the
case of level statistics, the calculation of \cite{km} valid for
not too large frequencies (below the Thouless energy equal to the
inverse time of diffusion through the system) was complemented by
Andreev and Altshuler \cite{aa} whose saddle-point treatment was, in
contrast, applicable for large frequencies. Level statistics in
diffusive disordered samples is discussed in detail in Section~\ref{c2}
of the present article.

Not only the energy levels statistics but also the
statistical properties of wave functions are of considerable
interest. In the case of nuclear spectra, they determine fluctuations
of widths and heights of the resonances \cite{porter}. In the case of
disordered (or chaotic) electronic systems, eigenfunction fluctuations
govern, in particular, statistics of the tunnel conductance in the
Coulomb blockade regime \cite{jsa}. Note also that the eigenfunction
amplitude can be directly measured in microwave cavity experiments
\cite{stoeckmann,kudrolli,alt} (though in this case one considers
the intensity of a classical wave rather than of a quantum particle, all
the results are equally applicable; see also Section~\ref{c7}).
Within the random matrix theory, the distribution of eigenvector
amplitudes is simply Gaussian, leading to $\chi^2$ distribution of the
``intensities'' $|\psi_i|^2$ (Porter-Thomas distribution)
\cite{porter}.  

A theoretical study of the eigenfunction statistics in a disordered
system is again possible with use of the supersymmetry method. The
corresponding formalism, which was developed in Refs.
\cite{mf93a,fm94a,mf94a,mf94b} (see Section~\ref{s3.1}), allows one to
express various distribution functions characterizing the
eigenfunction statistics through the $\sigma$-model correlators. As in
the case of the level correlation function, the zero-mode
approximation to the $\sigma$-model reproduces the RMT results, in
particular the Porter-Thomas distribution of eigenfunction
amplitudes. However, one can go beyond this approximation. In
particular, in the case of a quasi-one-dimensional geometry,
considered in Section~\ref{s3.2}, this $\sigma$-model has been solved
exactly using the transfer-matrix method, yielding
exact analytical results for the eigenfunction statistics for
arbitrary length of the system, from weak to strong localization
regime \cite{fm92a,mf93a,fm93a,fm93b,fm94a}. The case of a quasi-1D
geometry is of great interest not only from the point of view of
condensed matter theory (as a model of a disordered wire) but also
for quantum chaos.

In Section~\ref{s3.3} we consider the case of arbitrary spatial
dimensionality of the system. Since for $d>1$ an exact solution of the
problem cannot be found, one has to use some approximate methods. In
Refs.~\cite{fm94b,fm95a} the scheme of \cite{km} was generalized to
the case of the eigenfunction statistics. This allowed us to calculate
the distribution of eigenfunction intensities and its deviation from
the universal (Porter-Thomas) form. Fluctuations of the inverse
participation ratio and
long-range correlations of the eigenfunction amplitudes, 
which are determined by the diffusive dynamics in the
corresponding classical system \cite{fm95a,bm96,bm97}, 
%%were studied in Refs.~\cite{fm95a,bm96,bm97} and 
are considered in Sec.~\ref{s3.3.3}. 

Section~\ref{c4} is devoted to the asymptotic ``tails'' of the distribution
functions of various fluctuating quantities 
(local amplitude of an eigenfunction,
relaxation time, local density of states) characterizing a
disordered system. It turns out that the
asymptotics of all these distribution functions are determined by rare
realizations of disorder leading to formation of anomalously localized
eigenstates. These states show some kind of localization while all
``normal'' states are ergodic; in the quasi-one-dimensional
case they have an effective localization length much shorter than the
``normal'' one. Existence
of such states was conjectured by Altshuler,
Kravtsov, and Lerner \cite{akl} who studied distributions of various
quantities in $2+\epsilon$ dimensions via the renormalization group
approach. More recently, Muzykantskii and Khmelnitskii \cite{mk1}
suggested a new approach to the problem. Within this method, the
asymptotic ``tails'' of the distribution functions are obtained by
finding a non-trivial saddle-point configuration of the supersymmetric
$\sigma$-model. Further development and generalization of the method
allowed one to calculate the asymptotic behavior of the distribution functions
of relaxation times \cite{mk1,m95,mk2}, eigenfunction intensities
\cite{fe1,fe2}, local density of states \cite{m96}, inverse
participation ratio \cite{mf95u,m97}, level curvatures
\cite{kravtsov97a,basu97} etc. 
The saddle-point solution describes directly the spatial
shape of the corresponding anomalously localized state \cite{mk1,m97}. 

Section~\ref{c5} deals with statistical properties of the energy
levels and wave functions at the Anderson metal-insulator transition
point. As is well known, in $d>2$ dimensions a disordered system
undergoes, with increasing strength of  disorder,
a transition from the phase of extended states to that of
localized states (see
e.g. \cite{leeram} for review). This transition changes drastically
the statistics of energy levels and eigenfunctions. While in the
delocalized phase the levels repel each other strongly and their
statistics is described by RMT (up to the deviations discussed above
and in Section~\ref{c2}), in the localized regime the level repulsion
disappears (since states nearby in energy are located far from each
other in real space). As a result, the levels form an ideal 1D gas (on
the energy axis) obeying the Poisson statistics. In particular,
the variance of the number $N$ of levels in an interval $\Delta E$
increases linearly, $\mbox{var}(N)=\langle N\rangle$, in contrast to
the slow logarithmic increase in the RMT case. What happens to the level
statistics at the transition point? This question was addressed for
the first time by Altshuler, Zharekeshev, Kotochigova, and
Shklovskii in \cite{aszk}, where a Poisson-like increase, 
$\mbox{var}(N)=\chi\langle N\rangle$, was found numerically with
a spectral compressibility
$\chi\simeq 0.3$. More recently, Shklovskii
{\it et al} \cite{merman} put forward the conjecture
 that the nearest level spacing
distribution $P(s)$ has a universal form at the critical point,
combining the RMT-like level repulsion at small $s$ with the Poisson-like
behavior at large $s$. However, these results were questioned by
Kravtsov {\it et al} \cite{klaa} who developed an analytical approach
to the problem and found, in particular, a sublinear increase of
$\mbox{var}(N)$. This controversy was resolved in
\cite{am,kraler95} where the consideration of
\cite{klaa} was critically reconsidered and the level number variance
was shown to have
generally a linear behavior at the transition point. By now, this result 
has been confirmed by numerical simulations done by several groups
\cite{sears,zhar1,zhar2,braun}. Recently, a connection between this
behavior and multifractal properties of eigenfunctions has been
conjectured \cite{ckl}. 

Multifractality is a formal way to characterize strong fluctuations of
the wave function amplitude at the mobility edge. It follows from the
renormalization group calculation of Wegner \cite{wegner80} (though
the term ``multifractality'' was not used there). Later the
multifractality of the critical wave functions was
discussed in \cite{castpel} and confirmed by numerical simulations of
the disordered tight-binding model in
\cite{schreiber,pook,huck92,janssen94,huck95}. It implies, in very
rough terms, that the eigenfunction is effectively located in a
vanishingly small portion of the system volume. A natural
question then arises: why do such extremely sparse eigenfunctions show
the same strong level repulsion as the ergodic states in the RMT? 
This problem is addressed in Sec.~\ref{s5.2}. It is shown there that
the wavefunctions of nearby-in-energy states exhibit very strong
correlations (they have essentially the same multifractal structure),
which preserves the level repulsion despite the sparsity of the wave
functions. 

In Sec.~\ref{s5.3} we consider a ``power-law random banded matrix
ensemble'' (PRBM) which describes a kind of one-dimensional system
with a long-range hopping whose amplitude decreases as $r^{-\alpha}$
with distance \cite{prbm}. Such a random matrix ensemble arises in
various contexts in the theory of  quantum chaos
\cite{jose,leval} and disordered systems
\cite{balatsky,ps,skvortsov}. The problem can again be mapped onto a
supersymmetric $\sigma$-model. It is further shown that at $\alpha=1$
the system is at a critical point of the localization-delocalization
transition. More precisely, there exists a whole family of such
critical points labeled by the coupling constant of the
$\sigma$-model (which can be in turn related to the parameters of the
microscopic PRBM ensemble). Statistics of levels and eigenfunctions in
this model are studied. At the critical point they show the critical
features discussed above (such as the multifractality of eigenfunctions
and a finite spectral compressibility $0<\chi<1$). 

The energy level and eigenfunction statistics characterize the
spectrum of an isolated sample.  For an open system
(coupled to external conducting leads), different
quantities become physically relevant. In particular, we have already
mentioned the distributions of the local density of states and 
of the relaxation times discussed
in Section~\ref{c4} in connection with anomalously localized
states. In Section~\ref{c6} we consider one of the most famous issues
in the physics of mesoscopic systems, namely that of conductance
fluctuations. We focus on the case of the quasi-one-dimensional
geometry. The underlying microscopic model describing a disordered
wire coupled to freely propagating modes in the leads 
was proposed by Iida, Weidenm\"uller, and Zuk \cite{iwz}. It can be
mapped onto a 1D $\sigma$-model with boundary terms representing
coupling to the leads. The conductance is given in this approach by the
multichannel Landauer-B\"uttiker formula.
%%\cite{landauer,fishlee,buettiker,barstone}. 
The average
conductance $\langle g \rangle$ of this system for arbitrary value of
the ratio of its length $L$ to the localization length $\xi$ was
calculated by Zirnbauer \cite{zirn1}, who developed for this purpose
the Fourier analysis on supersymmetric manifolds. The variance of the
conductance was calculated in \cite{mmz} (in the case of a
system with strong spin-orbit interaction there was a subtle error in the
papers \cite{zirn1,mmz} corrected by Brouwer and Frahm \cite{bf}). The
analytical results
which describe the whole range of $L/\xi$ from the weak localization
($L\ll\xi$) to the strong localization ($L\gg\xi$) regime were
confirmed by numerical simulations \cite{mm,wang}. 

As has been already mentioned, the $\sigma$-model formalism is not
restricted to quantum-mechanical particles, but is equally applicable
to classical waves.
Section~\ref{c7} deals with a problem of intensity distribution in
the optics of disordered media. In an optical experiment,
a source and a detector of the radiation can be placed in
the bulk of disordered media. The distribution of the detected intensity
is then described in the leading approximation by the Rayleigh law
\cite{ishimaru} which follows from the assumption of a random
superposition of independent traveling waves. This result can be also
reproduced within the diagrammatic technique \cite{shapiro}. 
Deviations from the Rayleigh distribution governed by the diffusive
dynamics were studied in \cite{mps} for the quasi-1D geometry. 
When the source and the detector are moved
toward the opposite edges of the sample, the intensity distribution
transforms into the distribution of transmission coefficients 
\cite{nvr,kk,been1}. 

Recently, it has been suggested by Muzykantskii and Khmelnitskii \cite{mk3} 
that the supersymmetric $\sigma$-model approach developed previously
for the diffusive systems is also applicable in the case of ballistic
systems. Muzykantskii and Khmelnitskii derived the ``ballistic
$\sigma$-model'' where the diffusion operator was replaced by the
Liouville operator governing the ballistic dynamics of the
corresponding classical system. This idea was further developed by
Andreev, Agam, Simons and Altshuler \cite{aasa1,aasa2} who derived
the same action via the energy averaging for a chaotic ballistic
system with no disorder. (There are some indications  that
one has to include in consideration certain amount of disorder to
justify the derivation of \cite{aasa1,aasa2}.)
Andreev {\it et al}  replaced, in this case, the Liouville
operator by its regularization known as Perron-Frobenius
operator. However, this approach has failed to provide explicit
analytical results for any particular chaotic billiard so far. This is
because the eigenvalues of the Perron-Frobenius operator are usually
not known, while its eigenfunctions are highly singular. To overcome
these difficulties and to make a further analytical progress possible,
a ballistic model with surface disorder was considered in
\cite{bmm2,DEK}. The corresponding results are reviewed in
Sec.~\ref{c8}. It is assumed that roughness of the sample
surface leads to the diffusive surface scattering, 
modelling a ballistic system with strongly chaotic classical
dynamics. Considering
the simplest (circular) shape of the system allows one to find the
spectrum of the corresponding Liouville operator and to study
statistical properties of energy levels and eigenfunctions. The results
for the level statistics show important differences as compared to the
case of a diffusive system and are in agreement with 
arguments of Berry \cite{berry1,berry2} concerning the spectral
statistics in a generic chaotic billiard.

In Section~\ref{c9} we discuss a combined effect of the level and
eigenfunction fluctuations and the electron-electron interaction on
thermodynamic properties of quantum dots. Section~\ref{s9.1} is
devoted to statistics of the  so-called 
addition spectrum of a quantum dot in the
Coulomb blockade regime. The addition spectrum, which is determined by
the positions of the Coulomb blockade conductance peaks with varying
gate voltage, corresponds to a successive addition of electrons to the
dot coupled very weakly to the outside world \cite{dotrev}. 
The two important energy scales characterizing such a dot are the
charging energy $e^2/C$ and the electron level spacing $\Delta$ (the 
former being much 
larger than the latter for a dot with large number of electrons). 
Statistical properties of the addition spectrum were experimentally
studied for the 
first time by Sivan {\it et al} \cite{sivan}. It was 
conjectured in Ref.~\cite{sivan} 
that fluctuations in the addition spectrum are of the order of $e^2/C$ and
are thus of classical origin. However, it was found in
Ref.~\cite{bmm1,berkovits97} that this is 
not the case and that the magnitude of fluctuations is set by the
level spacing $\Delta$, as in the non-interacting case. The
interaction modifies, however, the shape of the distribution
function. In particular, it is responsible for breaking the spin
degeneracy of the quantum dot spectrum. These results have been
confirmed recently by  thorough experimental studies
\cite{marcus1,marcus2}. 

%Finally, in Section~\ref{s9.2} we study
%mesoscopic fluctuations and weak localization corrections to
%polarizability  of a small metallic sample
%\cite{efetov96,noat96,bm3}. 
%For this kind of problems the electron-electron interaction is also of
%crucial importance, since it determines the screening of the
%externally applied electric field. 

The research activity in the field of disordered mesoscopic systems,
random matrix theory, and quantum chaos has been growing enormously
during the recent years, so that a review article clearly cannot 
give an account of the progress in the whole field. 
Many of the topics which are
not covered here have been extensively discussed in
the recent reviews by Beenakker \cite{been-rev} and by Guhr,
M\"uller-Groeling, and Weidenm\"uller \cite{gmw-rev}.

\section{Energy level statistics: Random matrix theory and beyond}
\label{c2}

\subsection{Supersymmetric $\sigma$-model formalism}
\label{s2.1}

The problem of energy level correlations has been attracting a lot of
research interest since the work of Wigner \cite{wigner}. The random
matrix theory (RMT) developed by Wigner, Dyson, and Mehta
\cite{dyson,mehta} was found to
describe well the level statistics of various classes of complex
systems. In particular, in 1965 Gor'kov and Eliashberg \cite{gorel}
put forward a
conjecture that the RMT is applicable to the problem of energy level
correlations of a quantum particle moving in a random potential. To
prove this hypothesis, Efetov developed the supersymmetry approach to
the problem \cite{efetov83,ef-book}. The quantity of primary interest is the
two-level correlation function\footnote{The two-level correlation
function is conventionally denoted \cite{mehta,bohigas} as
$R_2(s)$. Since we will not consider higher-order correlation
functions, we will omit the subscript ``2''.}
\begin{equation}
R(s)={1\over \langle\nu\rangle^2}
\langle\nu(E-\omega/2)\nu(E+\omega/2)\rangle\, 
\label{e2.1}
\end{equation}
where $\nu(E)=V^{-1}\mbox{Tr}\delta(E-\hat{H})$ 
is the density of states at the energy $E$,
$V$ is the system volume, $\hat{H}$ is the Hamiltonian,
$\Delta=1/\langle\nu\rangle V$ is the mean level spacing,
$s=\omega/\Delta$, and $\langle\ldots\rangle$ denote averaging over
realizations of the random 
potential. As was shown by Efetov \cite{efetov83}, the correlator
(\ref{e2.1}) can be expressed
in terms of a Green function of certain supermatrix $\sigma$--model.
Depending on whether the time reversal and spin rotation symmetries
are broken or not, one of three different $\sigma$--models is relevant, with
unitary, orthogonal or symplectic symmetry group. We will consider
first the technically simplest case
of the unitary symmetry (corresponding to the broken time reversal 
invariance); the results for two other cases will be 
presented at the end. 

We give only a brief sketch of the derivation of the expression 
for $R(s)$ in terms of the $\sigma$--model. One begins with
representing the density of states in terms of the Green's functions,
\be
\label{e2.1a}
\nu(E)={1\over 2\pi i V}\int d^d{\bf r}\left[G_A^E({\bf r},{\bf r})-
G_R^E({\bf r},{\bf r})\right]\ ,
\ee
where 
\begin{equation}
G^E_{R,A}({\bf r_1},{\bf r_2})=\langle
{\bf r_1}|(E-\hat{H}\pm i\eta)^{-1}|{\bf r_2}\rangle\ ,\qquad \eta\to
+0\ .
\label{anom16}
\end{equation}
The Hamiltonian $\hat{H}$  consists of the free part
$\hat{H}_0$ and the disorder potential $U({\bf r})$:
\begin{equation}
\hat{H}=\hat{H}_0+U({\bf r})\ ;\qquad \hat{H}_0={1\over 2m}
\hat{{\bf p}}^2\ ; \label{anom18}
\end{equation}
the latter being defined by the correlator
\begin{equation}
\langle U({\bf r})U({\bf r'})\rangle={1\over
2\pi\nu\tau}\delta({\bf r}-{\bf r'}) 
\label{anom19}
\end{equation}
A non-trivial part of the calculation is the averaging of the $G_R G_A$ terms
entering the correlation function
$\langle\nu(E+\omega/2)\nu(E-\omega/2)\rangle$. 
The following steps are:
\begin{itemize}
\item[i)] to write the  product of the Green's functions in terms
of the integral over a supervector field
$\Phi=(S_1,\chi_1,S_2,\chi_2)$: 
\begin{eqnarray}
&&G_R^{E+\omega/2}({\bf r_1},{\bf r_1}) G_A^{E-\omega/2}({\bf
r_2},{\bf r_2} )
= \int D\Phi\,
D\Phi^\dagger S_1({\bf r_1}) S_1^*({\bf r_1}) 
S_2({\bf r_2})S_2^*({\bf r_2})
\nonumber\\&&\ \ \ \times 
\exp\left\{i\int d{\bf r'} \Phi^\dagger({\bf r'}) \Lambda^{1/2}
[E+(\omega/2+i\eta)\Lambda-\hat{H}]\Lambda^{1/2} \Phi({\bf r'})\right\}\ ,
\label{e2.1b}
\end{eqnarray}
where $\Lambda=\mbox{diag}\{1,1,-1,-1\}$;
\item[ii)] to average over the disorder;
\item[iii)] to introduce a $4\times 4$ supermatrix variable
${\cal R}_{\mu\nu}({\bf r})$ conjugate to the tensor product
$\Phi_\mu({\bf r}) \Phi^\dagger_\nu({\bf r})$ ;
\item[iv)] to integrate out the $\Phi$ fields ;
\item[v)] to use the saddle-point approximation which leads to the
following equation for ${\cal R}$:
\begin{eqnarray}
&& {\cal R}({\bf r})={1\over 2\pi\nu\tau} g({\bf r},{\bf r})\ ;
\label{anom20}\\
&&g({\bf r_1},{\bf r_2})=\langle
{\bf r_1}|(E-\hat{H}_0-{\cal R})^{-1}|{\bf r_2}\rangle 
\label{anom21}
\end{eqnarray}
\end{itemize}
The relevant set of the solutions (the saddle-point manifold)  has the
form:
\begin{equation}
{\cal R}=\sigma\cdot I - {i\over 2\tau} Q
\label{anom22}
\end{equation}
where $I$ is the unity matrix, $\sigma$ is certain constant, 
and the $4\times 4$ supermatrix 
$Q=T^{-1}\Lambda T$ satisfies the condition $Q^2=1$,
with $T$ belonging to the coset space $U(1,1|2)/U(1|1)\times U(1|1)$. 
The expression for the two-level correlation function $R(s)$ then reads
\begin{eqnarray}
&&R(s)=\left({1\over 4V}\right)^2 \mbox{Re} \int DQ({\bf r}) 
\left[\int d^d{\bf r}\mbox{Str} Q\Lambda k\right]^2     \nonumber\\
&&\times\exp\left\{-{\pi\nu\over 4}\int d^d{\bf r} 
\mbox{Str}\: [-D(\nabla Q)^2 - 2i\omega\Lambda Q]\right\}   
\label{e2.2}
\end{eqnarray}
Here $k=\mbox{diag}\{1,-1,1,-1\}$,  
Str denotes the supertrace,  
and $D$ is the classical diffusion constant. We do not give here a
detailed description of the model and  mathematical entities involved,
which can be found e.g. in Refs.\cite{efetov83,ef-book,vwz,zirn86}, and restrict
ourselves to a qualitative discussion of the structure of the matrix
$Q$. The size $4$ of the matrix is due to (i) two types of the Green
functions (advanced and retarded) entering the correlation function
(\ref{e2.1}),  and (ii) necessity to introduce bosonic and fermionic
degrees of freedom to represent these Green's function in terms of a
functional integral. The matrix $Q$ consists thus of four $2\times 2$
blocks according to its advanced-retarded structure, 
each of them being a supermatrix in the boson-fermion space.  

To proceed further, Efetov \cite{efetov83} neglected spatial
variation of the supermatrix field $Q({\bf r})$ and approximated 
the functional integral in Eq.(\ref{e2.2})
by an integral over a single supermatrix $Q$ (so-called zero-mode
approximation). The resulting integral can be calculated
yielding precisely the Wigner--Dyson distribution\footnote{Strictly
speaking, the level correlation functions 
(\ref{e2.3})--(\ref{e2.5})  contain an addititonal term
$\delta(s)$ corresponding to the ``self-correlation'' of an energy
level. Furthermore, in the symplectic case all the levels are double
degenerate (Kramers degeneracy). This degeneracy is disregarded in
(\ref{e2.5}) which thus represents the correlation function of
distinct levels only, normalized to the corresponding level spacing.}:
\begin{equation}
R^{U}_{\rm WD}(s)=1-\frac{\sin^2(\pi s)}{(\pi s)^2},
\label{e2.3}
\end{equation}
the superscript $U$ standing for the unitary ensemble. The
corresponding results for the orthogonal (O) and the symplectic (Sp) 
ensemble are 
\begin{eqnarray}
&& R^{O}_{\rm WD}(s)=1-\frac{\sin^2(\pi s)}{(\pi s)^2}-
\left[{\pi\over 2}{\rm sign}(s)-\mbox{Si}(\pi s)\right]
\left[{\cos\pi s\over\pi
s}-{\sin\pi s\over (\pi s)^2}\right]\ , \label{e2.4} \\
&& R^{Sp}_{\rm WD}(s)=1-\frac{\sin^2(2\pi s)}{(2\pi s)^2}+
\mbox{Si}(2\pi s)\left[{\cos 2\pi s\over 2\pi
s}-{\sin 2\pi s\over (2\pi s)^2}\right]\ , \label{e2.5}\\
&&\mbox{Si}(x)=\int_0^x{\sin y\over y} dy\ . \nonumber
\end{eqnarray}
The aim of Sec.~\ref{s2.2} will be to study the deviations
of the level correlation function from the universal RMT results
(\ref{e2.3}), (\ref{e2.4}), (\ref{e2.5}). 

\subsection{Deviations from universality}
\label{s2.2}

The procedure
we are using in order to calculate deviations from the universality 
is as follows \cite{km}. We first decompose $Q$ into the
constant part $Q_0$ and the contribution $\tilde{Q}$ of higher modes with
non--zero momenta. Then we use the renormalization group ideas and
integrate out all fast modes. This can be done perturbatively
provided the dimensionless (measured in units of $e^2/h$)
conductance $g=2\pi E_c/\Delta=2\pi\nu DL^{d-2}\gg 1$ (here
$E_c=D/L^2$ is the Thouless energy). As a result,
we get an integral over the matrix $Q_0$ only, which has to be calculated
non-perturbatively. 

We begin with presenting  the correlator $R(s)$ in the form
\begin{eqnarray}
&& R(s)={1\over (2\pi i)^2} {\partial^2\over \partial u^2}\int DQ
\exp\{-S[Q]\}|_{u=0}\ ;  \nonumber\\
&& S[Q]=-{1\over t}\int \mbox{Str}(\nabla Q)^2
+\tilde{s}\int \mbox{Str}\Lambda Q +\tilde{u}\int\mbox{Str} Q\Lambda k
\label{e2.6}
\end{eqnarray}
where $1/t=\pi\nu D/4$, $\tilde{s}=\pi s/2iV$, $\tilde{u}=\pi u/2iV$. Now we
decompose $Q$ in the following way
\begin{equation}
Q({\bf r})=T_0^{-1}\tilde{Q}({\bf r})T_0
\label{e2.7}
\end{equation}
where $T_0$ is a spatially uniform matrix and $\tilde{Q}$ describes all modes 
with non-zero momenta. 
When $\omega\ll E_c$, the matrix $\tilde{Q}$ fluctuates only weakly near the
origin $\Lambda$ of the coset space. In the leading order, 
$\tilde{Q}=\Lambda$, thus reducing (\ref{e2.6}) to a zero-dimensional 
$\sigma$--model, which leads to the Wigner--Dyson distribution (\ref{e2.3}). 
To find the corrections, we should expand the matrix $\tilde{Q}$ around
the origin $\Lambda$:
\begin{eqnarray}
\tilde{Q}&=&\Lambda(1+W/2)(1-W/2)^{-1} \nonumber \\
&=&\Lambda\left(1+W+{W^2 \over 2}+{W^3\over 4}+\ldots\right)\ ,
\label{e2.8}
\end{eqnarray}
where $W$ is a supermatrix with the following block structure:
\begin{equation}
W=\left(\begin{array}{ll} 0 & t_{12} \\ t_{21} & 0 \end{array}\right)
\label{e2.9}
\end{equation}
Substituting this expansion into Eq.(\ref{e2.6}), we get
\begin{eqnarray}
&& S = S _0+ S _1+O(W^3)\ ;\nonumber\\
&& S _0=\int\mbox{Str}\left[{1\over t}(\nabla W)^2+\tilde{s}Q_0\Lambda+
\tilde{u}Q_0\Lambda k\right]               \nonumber\\
&& S _1={1\over 2}\int\mbox{Str}[\tilde{s}U_0\Lambda W^2 +
\tilde{u}U_{0k}\Lambda W^2]
\label{e2.10}
\end{eqnarray}
where $Q_0=T_0^{-1}\Lambda T_0$,
$U_0=T_0\Lambda T_0^{-1}$, $U_{0k}=T_0\Lambda k T_0^{-1}$.
Let us define $ S _{\rm eff}[Q_0]$ as a result of elimination of the 
fast modes:
\begin{equation}
e^{-{S}_{\rm eff}[Q_0]}=e^{-{S}_0[Q_0]}\langle e^{-
S_1[Q_0,W]+\ln J[W]} \rangle_W\ ,
\label{e2.11}
\end{equation}
where $\langle\ldots\rangle_W$ denote the integration over $W$ and 
$J[W]$ is the Jacobian of the transformation (\ref{e2.7}),
(\ref{e2.8}) from the variable $Q$ to $\{Q_0,W\}$ (the Jacobian does
not contribute to the leading order correction calculated here, but is
important for higher-order calculations \cite{fm95a,kravtsov97b}). 
Expanding
up to the order $W^4$, we get
\begin{equation} 
{S}_{\rm eff}={S}_0+\langle {S}_1\rangle -
{1\over 2}\langle{S}_1^2\rangle + {1\over 2} \langle{S}_1\rangle^2
+\ldots
\label{e2.12}
\end{equation}
The integral over the fast modes can be calculated now using the Wick theorem
and the contraction rules \cite{efetov83,akl}:
\begin{eqnarray}
&&\langle\mbox{Str} W({\bf r})PW({\bf r'})R\rangle=\Pi({\bf r},{\bf r'})
(\mbox{Str}P\mbox{Str}R-\mbox{Str}P\Lambda\mbox{Str}R\Lambda)\ ;\nonumber\\
&&\langle\mbox{Str}[W({\bf r})P]\mbox{Str}[W({\bf r'})R]\rangle=
\Pi({\bf r},{\bf r'})
\mbox{Str}(PR-P\Lambda R\Lambda),
\label{e2.13}
\end{eqnarray}
where $P$ and $R$ are arbitrary supermatrices.
The diffusion propagator $\Pi$ is the solution of the diffusion
equation
\begin{equation} \label{corrrev_diff} 
- D \nabla^2\Pi({\bf  r}_1,{\bf  r}_2) = (\pi \nu)^{-1}
[\delta({\bf  r}_1 - {\bf  r}_2)-V^{-1}]   
\end{equation}
with the Neumann boundary condition (normal derivative equal to
zero at the sample boundary) and can be presented in the form
\begin{equation}
\Pi({\bf r}, {\bf r'})={1\over \pi\nu}\sum_{\mu;\ \epsilon_\mu\ne 0}
{1\over \epsilon_\mu} \phi_{\mu}({\bf r})\phi_{\mu}({\bf r'})
\label{e2.13a}
\ee
where $\phi_{\mu}({\bf r})$ 
are the eigenfunctions of the diffusion operator $-D\nabla^2$
corresponding to the eigenvalues $\epsilon_\mu$ (equal to $D{\bf q}^2$
for a rectangular geometry). As a result, we find:
\begin{eqnarray}
&&\langle{S}_1\rangle=0\ ;\nonumber\\
&&\langle{S}_1^2\rangle=
{1\over 2}\int d{\bf r}d{\bf r'} \Pi^2({\bf r},{\bf r'})
(\tilde{s}\mbox{Str}Q_0\Lambda+\tilde{u}\mbox{Str}Q_0\Lambda k)^2
\label{e2.14}
\end{eqnarray}
Substitution of Eq.~(\ref{e2.14}) into Eq.~(\ref{e2.12}) yields
\begin{eqnarray}
&&\hspace{-1cm}
{S}_{\rm eff} [Q_0]={\pi \over 2i}s\mbox{Str}Q_0\Lambda+{\pi\over 2i} u
\mbox{Str} Q_0\Lambda k +{\pi^2 a_d\over 4 g^2} (s\mbox{Str} Q_0\Lambda +
u\mbox{Str}Q_0\Lambda k)^2  \ ; \nonumber\\
&& \hspace{-1cm}
a_d= {g^2\over 4 V^2}\int d{\bf r} d{\bf r'}\Pi^2({\bf r}, {\bf r'})
={1\over \pi^4}
\sum_{\begin{array}{c}n_1,\ldots,n_d=0\\n_1^2+\ldots +n_d^2>0 
\end{array}}^{\infty}{1\over (n_1^2+\ldots+n_d^2)^2}\ .
\label{e2.15}
\end{eqnarray}
The value of the coefficient $a_d$ depends on spatial dimensionality
$d$ and on the sample geometry; in the last line of Eq.(\ref{e2.15})
we assumed a cubic sample with hard-wall boundary conditions. Then
for $d=1,2,3$ we have $a_1=1/90\simeq 0.0111$, 
$a_2\simeq 0.0266$, and $a_3\simeq 0.0527$ respectively.
In the case of a cubic sample with periodic boundary conditions
we get instead 
\begin{equation}
a_d={1\over (2\pi)^4}
\sum_{\begin{array}{c}n_1,\ldots,n_d=-\infty\\n_1^2+\ldots +n_d^2>0 
\end{array}}^{\infty}{1\over (n_1^2+\ldots+n_d^2)^2}\ ,
\label{e2.15a}
\ee
so that $a_1=1/720\simeq 0.00139$, $a_2\simeq 0.00387$, and $a_3\simeq
0.0106$. Note that for $d<4$ the sum in Eqs.~(\ref{e2.15}),
(\ref{e2.15a}) converges, so that no ultraviolet cut-off is needed. 

Using now Eq.(\ref{e2.6}) and calculating the remaining integral over
the supermatrix $Q_0$, we finally get 
the following expression for the
correlator to the $1/g^2$ order:
\begin{equation}
R(s)=1-\frac{\sin^2(\pi s)}{(\pi s)^2}+\frac{4a_d}{ g^2}\sin^2(\pi s)
\label{e2.16}
\end{equation}
The last term in Eq.(\ref{e2.16}) just represents the correction 
of  order $1/g^2$ to the Wigner--Dyson distribution. The formula 
(\ref{e2.16}) is valid for $s\ll g$. Let us note that  the 
smooth (non-oscillating) part of this correction 
in the region $1\ll s\ll g$ can be found by using purely perturbative approach
of Altshuler and Shklovskii \cite{ashk,aszk}. 
For $s\gg 1$ the leading perturbative contribution
to $R(s)$ is given by a two--diffuson diagram:
\begin{eqnarray}
R_{\rm AS}(s)-1 &=& {\Delta^2 \over 2 \pi^2} \mbox{Re} \sum_{
\begin{array}{c}
q_i=\pi n_i/L \\ n_i=0,1,2,\ldots
\end{array} }
{1\over (D{\bf q}^2-i\omega)^2} \nonumber\\
&=& {1\over 2\pi^2}\mbox{Re}\sum_{n_i\ge 0}
{1\over [-is+(\pi/2)g{\bf n}^2]^2}
\label{e2.17}
\end{eqnarray}
At $s\ll g$ this expression is dominated by the ${\bf q}=0$ term, with
other terms giving a correction of order $1/g^2$:
\begin{equation}
R_{\rm AS}(s) = 1-{1\over 2\pi^2 s^2} + {2a_d\over  g^2}\ ,
\label{e2.18}
\end{equation}
where $a_d$ was defined in Eq.(\ref{e2.15}). This formula is obtained in the
region $1\ll s\ll g$ and is perturbative in both $1/s$ and $1/g$. It does 
not contain oscillations (which cannot be found perturbatively) and gives no
information about actual small--$s$ behavior of $R(s)$. The result
(\ref{e2.16}) is much stronger: it represents the exact
(non--perturbative in $1/s$) form of the correction in the whole region
$s\ll g$. 

The important feature of Eq.(\ref{e2.16}) is that it relates
corrections to the smooth and oscillatory parts of the level
correlation function (represented by the contributions to the last
term proportional to unity and to $\cos 2\pi s$ respectively). While
appearing naturally in the framework of the supersymmetric
$\sigma$-model, this fact is highly non-trivial from the point of view
of semiclassical theory \cite{berry1,berry2}, which represents the
level structure factor $K(\tau)$ (Fourier transform of $R(s)$) in
terms of a sum over periodic orbits. The smooth part of $R(s)$
corresponds then to the small-$\tau$ behavior of $K(\tau)$, which is
related to the properties of short periodic orbits. On the other hand,
the oscillatory part of $R(s)$ is related to the behavior of $K(\tau)$
in the vicinity of the Heisenberg time $\tau=2\pi$ ($t=2\pi/\Delta$
in dimensionful units), and thus to the properties of long periodic
orbits. 

The calculation presented above can be straightforwardly generalized to the
other symmetry classes. The result can be presented in a form valid for all
the three cases\footnote{For all the ensembles, we denote by $g$ 
 the conductance per one spin projection: $g=2\pi\nu D L^{d-2}$, without
multiplication by factor 2 due to the spin.}:
\begin{equation}
R^{(\beta)}(s)=\left(1+\frac{2a_d}{\beta g^2} \frac{d^2}{ds^2}s^2\right)
R^{(\beta)}_{\rm WD}(s)
\label{e2.19}
\end{equation}
where $\beta=1 (2, 4)$ for the orthogonal (unitary, symplectic) symmetry;
$R^{(\beta)}_{\rm WD}$ denotes the corresponding Wigner--Dyson distribution
(\ref{e2.3})--(\ref{e2.5}).  

For $s\to 0$ the Wigner--Dyson distribution has the following behavior:
\begin{eqnarray}
&& R^{(\beta)}_{WD}\simeq c_\beta s^{\beta};\qquad  s\to 0 \nonumber\\
&& c_1={\pi^2\over 6};\ 
c_2={\pi^2\over 3};\ c_4={(2\pi)^4\over 135}
\label{e2.20}
\end{eqnarray}
As is clear from Eq.(\ref{e2.19}), the found correction does not change
the exponent $\beta$, but renormalizes the prefactor $c_\beta$:
\begin{equation}
R^{(\beta)}(s)=\left(1+\frac{2(\beta+2)(\beta+1)}{\beta} \frac{a_d}{g^2}
\right) c_\beta s^\beta\ ;\quad s\to 0
\label{e2.21}
\end{equation} 
The correction to $c_\beta$ is positive, which means that 
the level repulsion becomes weaker. This is related to a tendency of
eigenfunctions to localization with decreasing $g$.

What is the behavior of the level correlation function in its
high-frequency tail $s\gg g$? The non-oscillatory part of $R(s)$ in
this region follows from the Altshuler-Shklovskii perturbative formula
(\ref{e2.17}). For $s\gg g$ summation can be replaced by integration,
yielding 
\begin{equation}
\label{e2.22}
R_{\rm AS}(s)-1\propto g^{-d/2} s^{d/2-2}
\end{equation}
(note that in 2D  the coefficient of the  term (\ref{e2.22}) vanishes, and
the result for $R_{\rm AS}$ is smaller by an additional factor $1/g$,
see \cite{kraler95}). What is the fate of the oscillations in $R(s)$
in this regime? The answer to this question was given by Andreev and
Altshuler \cite{aa} who calculated $R(s)$ using the stationary-point
method for the $\sigma$-model integral (\ref{e2.2}). Their crucial
observation was that on top of the trivial stationary point
$Q=\Lambda$ (expansion around which is just the usual perturbation
theory), there exists another one, $Q=k\Lambda$, whose vicinity
generates the oscillatory part of $R(s)$. (In the case of symplectic
symmetry there exists an additional family of stationary points, see
\cite{aa}). The saddle-point approximation
of Andreev and Altshuler is valid for $s\gg 1$; at $1\ll s\ll g$ it
reproduces the above results of Ref.~\cite{km} (we remind that the
method of \cite{km}
works for all $s\ll g$). The result of \cite{aa} has the following
form 
\begin{eqnarray}
R_{\rm osc}^U(s)&=&{\cos 2\pi s\over 2\pi^2}D(s)\ ,
\label{e2.23} \\
R_{\rm osc}^O(s)&=&{\cos 2\pi s\over 2\pi^4}D^2(s)\ ,
\label{e2.23o} \\
R_{\rm osc}^{Sp}(s)&=&{\cos 2\pi s\over 4 }D^{1/2}(s)+
{\cos 4\pi s\over 32\pi^4}D^2(s)\ ,
\label{e2.23sp}
\end{eqnarray}
where $D(s)$ is the spectral determinant
\begin{equation}
D(s)={1\over
s^2}\prod_\mu\left(1+{s^2\Delta^2\over\epsilon_\mu^2}\right)^{-1}.
\label{e2.24}
\end{equation}
The product in Eq.(\ref{e2.24}) goes over the non-zero eigenvalues
$\epsilon_\mu$ of the diffusion operator (which are equal to $D{\bf
q}^2$ for the cubic geometry). This demonstrates again the 
relation between $R_{\rm osc}(s)$ and the perturbative part (\ref{e2.17}),
which can be also expressed through $D(s)$,
\begin{equation}
R^{(\beta)}_{\rm AS}(s)-1
=-{1\over 2\beta\pi^2}{\partial^2\ln D(s)\over \partial s^2}
\label{e2.25}
\end{equation}
In the high-frequency region $s\gg g$ the spectral determinant is
found to have the following behavior 
\begin{equation}
D(s)\sim\exp\left\{-{\pi\over \Gamma(d/2)d\sin(\pi
d/4)}\left({2s\over g}\right)^{d/2}\right\}\ ,
\label{e2.26}
\end{equation}
so that the amplitude of the oscillations vanishes exponentially with
$s$ in this region. 

Taken together, the results of \cite{km} and \cite{aa}
provide complete description of the deviations of the
level correlation function from universality in the metallic regime
$g\gg 1$. They show that in the whole region of frequencies these
deviations are controlled by the classical (diffusion) operator
governing the dynamics in the corresponding classical
system.

\section{Statistics of eigenfunctions}
\label{c3}

\setcounter{equation}{0}

\subsection{Eigenfunction statistics in terms of the supersymmetric
$\sigma$-model} 
\label{s3.1}

Within the RMT, the distribution of eigenfunction amplitudes is simply
Gaussian, leading to the $\chi^2$ distribution of the ``intensities''
$y_i=N|\psi_i^2|$ (we normalized $y_i$ in such a way that $\langle
y\rangle=1$) \cite{porter}
\begin{eqnarray}
&&{\cal P}^U(y)=e^{-y}\ , \label{e3.1} \\
&&{\cal P}^O(y)={e^{-y/2}\over \sqrt{2\pi y}}\ . \label{e3.2}
\end{eqnarray}
Eq.~(\ref{e3.2}) is known as the Porter-Thomas distribution; it was
originally introduced to describe
fluctuations of widths and heights of resonances in nuclear spectra
\cite{porter}. 

Recently, interest in properties of eigenfunctions in disordered and
chaotic systems has started to grow. On the experimental side, it was
motivated by the possibility of 
fabrication of small systems (quantum dots)
with well resolved electron energy levels
\cite{meirav90,sivan94,dotrev}. Fluctuations in the tunneling
conductance of such a dot measured in recent experiments
\cite{chang96,folk96} are related to statistical properties of
wavefunction amplitudes \cite{jsa,prigefii,alhassid96,bruus96}. When
the electron-electron Coulomb interaction is taken into account, the
eigenfunction fluctuations determine the statistics of matrix
elements of the interaction, which is in turn important for
understanding the properties of excitation and addition spectra of the
dot \cite{blanter,agkl,bmm1}. Furthermore, the microwave cavity
technique allows one to observe experimentally spatial fluctuations of
the wave amplitude in chaotic and disordered cavities
\cite{stoeckmann,kudrolli,alt}. 

Theoretical study of the eigenfunction statistics in a $d$-dimensional
disordered system is again possible
with use of the supersymmetry method
\cite{mf93a,fm94a,mf94a,mf94b}. The distribution function of the
eigenfunction intensity $u=|\psi^2({\bf r_0})|$ in a point ${\bf r_0}$
is defined as 
\begin{equation}
{\cal P}(u)= {1\over\nu
V}\left\langle\sum_\alpha 
\delta(|\psi_\alpha({\bf r_0})|^2-u)\delta(E-E_\alpha)\right\rangle.
\label{anom9}
\end{equation}
The moments of ${\cal P}(u)$ can be written through the Green's
functions in the following way
\begin{equation}
\langle|\psi({\bf r_0})|^{2q}\rangle ={i^{q-2}\over 2\pi\nu
V}\lim_{\eta\to 0}(2\eta)^{q-1}\langle G_R^{q-1}({\bf r_0},{\bf r_0})
G_A({\bf r_0},{\bf r_0})\rangle.
\label{e3.3}
\end{equation}
%where 
%\begin{equation}
%G_{R,A}({\bf r_1},{\bf r_2})=\langle
%{\bf r_1}|(E-\hat{H}\pm i\eta)^{-1}|{\bf r_2}\rangle
%\label{anom16}
%\end{equation}
%Here $\hat{H}$ is the Hamiltonian which consists of the free part
%$\hat{H}_0$ and the disorder potential $U({\bf r})$:
%\begin{equation}
%\hat{H}=\hat{H}_0+U({\bf r})\ ;\qquad \hat{H}_0={1\over 2m}
%\hat{{\bf p}}^2\ ; \label{anom18}
%\end{equation}
%the latter being defined by the correlator
%\begin{equation}
%\langle U({\bf r})U({\bf r'})\rangle={1\over
%2\pi\nu_{0}\tau_s}\delta({\bf r}-{\bf r'}) 
%\label{anom19}
%\end{equation}
%The following steps are:
%\begin{itemize}
The  product of the Green's functions can be expressed in terms
of the integral over a supervector field
$\Phi=(S_1,\chi_1,S_2,\chi_2)$, 
\begin{eqnarray}
&&G_R^{q-1}({\bf r_0},{\bf r_0}) G_A({\bf r_0},{\bf r_0})
= {i^{2-q}\over (q-1)!}\int D\Phi\,
D\Phi^\dagger (S_1({\bf r_0}) S_1^*({\bf r_0}))^{q-1} 
S_2({\bf r_0})S_2^*({\bf r_0})
\nonumber\\&&\ \ \ \times 
\exp\left\{i\int d{\bf r} \Phi^\dagger({\bf r}) \Lambda^{1/2}
(E+i\eta\Lambda-\hat{H})\Lambda^{1/2} \Phi({\bf r})\right\}\ .
\label{anom17}
\end{eqnarray}
Proceeding now in the same way as in the case of the level correlation
function (Sec.~\ref{s2.1}), we represent the r.h.s. of
Eq.~(\ref{anom17}) in terms of a $\sigma$-model correlation function.
%\item[ii)] to average over the disorder;
%\item[iii)] to introduce a $4\times 4$ supermatrix variable
%$R_{\mu\nu}({\bf r})$ having the symmetry of the tensor product
%$\Phi_\mu({\bf r}) \Phi^\dagger_\nu({\bf r})$ ;
%\item[iv)] to integrate out the $\Phi$ fields ;
%\item[v)] to use the saddle-point approximation which leads to the
%following equation for $R$:
%\begin{eqnarray}
%&& R({\bf r})={1\over 2\pi\nu\tau_s} g({\bf r},{\bf r})\ ;
%\label{anom20}\\
%&&g({\bf r_1},{\bf r_2})=\langle
%{\bf r_1}|(E-\hat{H}_0-R)^{-1}|{\bf r_2}\rangle 
%\label{anom21}
%\end{eqnarray}
%\end{itemize}
%The relevant set of the solutions (the saddle-point manifold)  has the
%form:
%\begin{equation}
%R=\sigma\cdot I - {i\over 2\tau_s} Q
%\label{anom22}
%\end{equation}
%where $I$ is the unity matrix, $\sigma$ is certain constant, 
%and $Q$ belongs to the coset space
%$U(1,1|2)$ and satisfies the condition $Q^2=1$.
As a result, we find\footnote{The first two indices of $Q$ correspond
to the advanced-retarded and the last two to the boson-fermion
decomposition.} 
\begin{equation}
\langle|\psi({\bf r_0})|^{2q}\rangle=-{q\over 2V}\lim_{\eta\to
0}(2\pi\nu\eta)^{q-1} \int DQ\,
Q_{11,bb}^{q-1}({\bf r_0})Q_{22,bb}({\bf r_0})e^{-S[Q]},
\label{e3.4}
\end{equation}
where $S[Q]$ is the $\sigma$-model action,
\begin{equation}
S[Q]=-{\beta\over 2}
\int d^dr\,\mbox{Str}\left[{\pi\nu D\over 4}(\nabla
Q)^2-\pi\nu\eta\Lambda Q\right]
\label{anom_2}
\end{equation}
($\beta=2$ for the considered case of the unitary symmetry). 
Let us define now the function $Y(Q_0)$ as
\begin{equation}
Y(Q_0)=\int_{Q({\bf r_0})=Q_0} DQ({\bf r})
\exp\{-S[Q]\}\ .
\label{anom_1}
\end{equation}
Here ${\bf r_0}$ is the spatial point, in which the statistics of
eigenfunction amplitudes is studied. 
For the invariance reasons, the function $Y(Q_0)$ turns out to be
dependent in the unitary symmetry case on the two scalar variables 
$1\le \lambda_{1}<\infty $ and $ -1\le \lambda_{2} \le 1$ only, which are
the eigenvalues of the ``retarded-retarded'' block of the matrix $Q_0$.
Moreover, in the limit $\eta \to 0$ (at a fixed  value of the
system volume) only the dependence on $\lambda_1$ persists:
\begin{equation}
Y(Q_0)\equiv Y(\lambda_1,\lambda_2)\to Y_a(2\pi\nu \eta\lambda_1)
\label{anom_3}
\end{equation}
With this definition, Eq.(\ref{e3.4}) takes the form of an integral
over the single matrix $Q_0$,
\begin{equation}
\langle|\psi({\bf r_0})|^{2q}\rangle=-{q\over 2V}\lim_{\eta\to
0}(2\pi\nu\eta)^{q-1} \int DQ_0\, Q_{0;11,bb}^{q-1}Q_{0;22,bb}Y(Q_0)
\label{e3.5}
\end{equation}
Evaluating this integral, we find
\begin{equation}
\langle|\psi({\bf r_0})|^{2q}\rangle={1\over V}q(q-1)\int du\,u^{q-2}
Y_a(u).
\label{e3.6}
\end{equation}
Consequently, the distribution function of the eigenfunction intensity
is given by \cite{mf93a}
\begin{equation}
{\cal P}(u)={1\over V}{d^2 \over du^2} Y_a(u) 
%%      ={1\over V} {d^2 \over du^2} Y\left.(\lambda_1={u\over
%%     2\pi\nu_{0} \eta})\right|_{\eta\to 0} 
\qquad (U)\ ,
\label{anom_4}
\end{equation}
where $V$ is the sample volume.

In the case of the orthogonal symmetry, $Y(Q_0)\equiv
Y(\lambda_1,\lambda_2,\lambda)$, where 
$1\le \lambda_{1},\lambda_{2} <\infty $ and $ -1\le \lambda \le 1$.
In the limit $\eta\to 0$, the relevant region of values is 
$\lambda_1\gg\lambda_2,\lambda$, where
\begin{equation}
Y(Q_0)\to Y_a(\pi\nu\eta\lambda_1)
\label{anom_3o}
\end{equation}
The distribution of eigenfunction intensities is expressed
in this case through the function $Y_a$ as follows \cite{mf93a}:
\begin{eqnarray}
{\cal P}(u)&=& {1\over\pi Vu^{1/2}}\int_{u/2}^{\infty}
dz (2z-u)^{-1/2}{d^2\over dz^2} Y_a(z) \nonumber\\
&=& {2\sqrt{2}\over\pi Vu^{1/2}}{d^2\over du^2}\int_0^\infty {dz\over
z^{1/2}}  Y_a(z+u/2) \qquad (O) \label{anom_4o}
\end{eqnarray}

In the diffusive sample, typical configurations of the $Q$--field are
nearly constant in space, so that one can approximate the functional
integral  (\ref{anom_1}) by an integral over a single supermatrix $Q$. This
procedure, which makes the problem effectively zero-dimensional and is
known as zero-mode approximation (see Sec.~\ref{s2.1}), gives
\begin{equation}
Y_a(z)\approx e^{-Vz} \qquad (O,U)\ ,
\label{anom1}
\end{equation}
and consequently,
\begin{eqnarray}
&{\cal P}(u)\approx Ve^{-uV}&\qquad (U)\ , \label{anom2u}\\
&{\cal P}(u)\approx \sqrt{{V\over 2\pi u}}e^{-uV/2} & \qquad (O)\
, \label{anom2o}
\end{eqnarray}
which are just the RMT results for the Gaussian Unitary Ensemble (GUE)
and Gaussian Orthogonal Ensemble (GOE) respectively,
Eqs.~(\ref{e3.1}), (\ref{e3.2}). 
	
Therefore, like in the case of the level correlations, the zero mode
approximation yields the RMT results for the distribution of the
eigenfunction amplitudes. To calculate deviations from RMT, one has to
go beyond the zero-mode approximation and to evaluate the function
$Y_a(z)$ determined by Eqs.~(\ref{anom_1}), (\ref{anom_3}) for a
$d$-dimensional diffusive system. In the case of a quasi-1D geometry
this can be done exactly via the transfer-matrix method, see
Sec.~\ref{s3.2}. For higher $d$, the exact solution is not possible,
and one should rely on approximate methods. Corrections to the ``main
body'' of the distribution can be found by treating the non-zero modes
perturbatively, while the asymptotic ``tail'' can be found via a
saddle-point method (see Sec.~\ref{s3.3} and Sec.~\ref{c4}). 

Let us note that the formula (\ref{anom_1}), (\ref{anom_3}) can be
written in a slightly different, but completely equivalent form
\cite{fe1,fe2}. Making in (\ref{anom_1}) the transformation
$$
Q({\bf r})\ \longrightarrow\ \tilde{Q}({\bf r})=V^{-1}({\bf
r_0})Q({\bf r})V({\bf r_0}),
$$
where the matrix $V({\bf r_0})$ is defined from 
$Q({\bf r_0})=V({\bf r_0})\Lambda V^{-1}({\bf r_0})$, one gets in the
unitary case
\begin{equation}
\label{e3.7}
Y_a(u)=\int_{Q({\bf r_0})=\Lambda}DQ\,\exp\left\{
\int d^dr\,\mbox{Str}\left[{\pi\nu D\over 4}(\nabla Q)^2-{u\over
2}\Lambda Q P_{bb}\right]\right\},
\end{equation}
where $P_{bb}$ denotes the projector onto the boson-boson sector,
and a similar formula in the orthogonal case.

The above derivation can be extended to a more general correlation
function representing a product of eigenfunction amplitudes in
different points
\begin{equation}
\label{e3.8}
\Gamma^{\{q\}}({\bf r_1},\ldots,{\bf r_k})={1\over\nu V}\left\langle
\sum_\alpha|\psi_{\alpha}^{2q_1}({\bf r_1})|
|\psi_{\alpha}^{2q_2}({\bf r_2})|\ldots|\psi_{\alpha}^{2q_k}({\bf r_k})|
\delta(E-E_\alpha)\right\rangle
\end{equation}
If all the points ${\bf r_i}$ are separated by
sufficiently large distances (much larger than the mean free path
$l$), one finds for the unitary ensemble \cite{fm94a}
\begin{eqnarray}
&&\Gamma^{\{q\}}({\bf r_1},\ldots,{\bf r_k})=-{1\over 2V}
{q_1!q_2!\ldots q_k!\over (q_1+q_2+\ldots+q_k-1)!}\lim_{\eta\to 0}
(2\pi\nu\eta)^{q_1+\ldots+q_k-1} \nonumber\\
&&\qquad\times  \int DQ\, Q_{11,bb}^{q_1-1}({\bf r_1})Q_{22,bb}({\bf r_1})
Q_{11,bb}^{q_2}({\bf r_2})\ldots Q_{11,bb}^{q_k}({\bf r_k})e^{-S[Q]}
\label{e3.9}
\end{eqnarray}
In the case of the quasi-1D system one can again evaluate
Eq.~(\ref{e3.9}) via the transfer matrix method, while in higher $d$
one has to use approximate schemes. The correlation functions of the
type (\ref{e3.8}) appear, in particular, when one calculates the
distribution of the inverse participation ratio (IPR)
$P_2=\int d^dr|\psi^4_\alpha({\bf r})|$, the moments of which are given by
Eq.~(\ref{e3.8}) with $q_1=q_2=\ldots=q_k=2$. We will discuss the IPR
distribution function below in Sec.~\ref{s3.2.3} and
\ref{s3.3.3}. The case of $k=2$ in Eq.(\ref{e3.8}) corresponds
to the correlations of the amplitudes of an eigenfunction in two
different points; we will discuss such correlations in
Sec.~\ref{s3.3.3} and in Sec.~\ref{s4.1.1} (where they will describe
the shape of an anomalously localized state).

\subsection{Quasi-one-dimensional geometry}
\label{s3.2}

\subsubsection{Exact solution of the $\sigma$-model}
\label{s3.2.1}

In the case of quasi-1D geometry an exact solution of the
$\sigma$-model is possible due to the
transfer-matrix method. The idea of the method, quite general for the
one-dimensional problems, is in reducing the functional integral
(\ref{anom_1}) or (\ref{e3.9}) to solution of a differential
equation. This is completely analogous to  constructing  the
Schr\"odinger equation from the quantum-mechanical Feynman path
integral. In the present case, the role of the time is played by the
coordinate along the wire, while the role of the particle coordinate
is played by the supermatrix $Q$. In general, at finite value of the
frequency $\eta$ in Eq.~(\ref{anom_2}) (more precisely, $\eta$ plays a
role of imaginary frequency), the corresponding differential equation
is too complicated and cannot be solved analytically \cite{efetov83}.
However, a simplification appearing in the limit $\eta\to 0$, when
only the non-compact variable $\lambda_1$ survives, allows to find an
analytical solution \cite{fm94a} of the 1D $\sigma$-model\footnote{Let
us stress that we consider a sample with the hard-wall (not
periodic) boundary conditions in the logitudinal direction, i.e a wire
with two ends (not a ring).}.

There are several different microscopic models which can be
mapped onto the 1D supermatrix $\sigma$-model. First of all, this is a
model of a particle in a random potential (discussed above) in the
case of a quasi-1D sample geometry. Then one can neglect transverse
variation of the $Q$-field in the $\sigma$-model action, thus reducing
it to the 1D form \cite{eflar,efetov83}. Secondly, this is the random
banded matrix (RBM) model \cite{fm91,mf93a,fm94a} which is relevant
to various problems in the field of quantum chaos
\cite{chaos,izrailev90}. In particular, the
evolution operator of a kicked rotor (paradigmatic model of a
periodically driven quantum system) has a structure of a quasi-random
banded matrix, which makes this system to belong to the ``quasi-1D
universality class'' \cite{fm94a,alzir96}. Finally, the
Iida-Weidenm\"uller-Zuk random matrix model \cite{iwz} of the
transport in a disordered wire (see Section~\ref{c6} for more detail)
can be also mapped onto the 1D $\sigma$-model. 

The result for the function $Y_a(u)$ determining the distribution of
the eigenfunction intensity $u=|\psi^2({\bf r_0})|$ reads (for the
unitary symmetry)
\begin{equation}
Y_a(u)=
W^{(1)}(uA\xi,\tau_+)W^{(1)}(uA\xi,\tau_-)\ .
\label{anom11}
\end{equation}
Here $A$ is the wire cross-section, $\xi=2\pi\nu DA$ the localization
length, $\tau_+=L_+/\xi$, $\tau_-=L_-/\xi$, with $L_+$, $L_-$ being
the distances from the observation point ${\bf r_0}$ to the right and
left edges of the sample. For the orthogonal symmetry $\xi$ is
replaced by $\xi/2$.
The function $W^{(1)}(z,\tau)$ satisfies the equation
\begin{equation}
{\partial W^{(1)}(z,\tau)\over \partial\tau}=
\left( z^2 {\partial^2\over \partial z^2}-z\right) 
W^{(1)}(z,\tau)
\label{anom12}
\end{equation}
and the boundary condition
\begin{equation}
W^{(1)}(z,0)=1\ .
\label{anom13}
\end{equation}
The solution to Eqs.(\ref{anom12}), (\ref{anom13}) can be found in
terms of the expansion in eigenfunctions of the operator
$z^2 {\partial^2\over \partial z^2}-z$. The functions
$2z^{1/2}K_{i\mu}(2z^{1/2})$, with $K_\nu(x)$ being the modified Bessel
function (Macdonald function),  form the
proper basis for such an expansion \cite{marichev}, which is known as
the Lebedev--Kontorovich expansion; the corresponding
eigenvalues are $-(1+\mu^2)/4$.  The result is
\begin{equation}
W^{(1)}(z,\tau)=2z^{1/2}\left\{K_{1}(2z^{1/2})+\frac{2}{\pi}\int_
{0}^{\infty}d\mu\frac{\mu}{1+\mu^{2}}
\sinh{\frac{\pi\mu}{2}}K_{i\mu}(2z^{1/2})
e^{-\frac{1+\mu^{2}}{4}\tau}\right\}\ .
\label{anom14}
\end{equation}
The formulas (\ref{anom_4}), (\ref{anom_4o}), (\ref{anom11}),
(\ref{anom14}) give therefore the exact solution for the 
eigenfunction statistics for arbitrary value of the parameter
$X=L/\xi$ (ratio of the total system length $L=L_+ + L_-$ to the
localization length). The form of the distribution function ${\cal
P}(u)$ is essentially different in the metallic regime $X\ll 1$ (in
this case $X=1/g$) and in the insulating one $X\gg 1$. We will discuss
these two limiting cases below, in Sections \ref{s3.2.2} and
\ref{s3.2.3} respectively.

\subsubsection{Global statistics of eigenfunctions}
\label{s3.2.1a}

The multipoint correlation functions (\ref{e3.8}) determining the global
statistics of eigenfunctions can be also computed in a similar
way. Let us first assume that the points ${\bf r}_i$ lie sufficiently
far from each other, $|{\bf r}_i-{\bf r}_j|\gg l$. 
We order the points according to their coordinates $x_i$ 
along the wire, $0<x_1<\ldots<x_k<L$, and define $t_i=x_i/\xi$,
$\tau_i=t_{i}-t_{i-1}$, $\tau_1=t_1$. Then we find from
Eq.~(\ref{e3.9}) for the unitary symmetry 
\bea
\label{glob1}
&& V\Gamma^{\{q\}}({\bf r_1},\ldots,{\bf r_k})=
{q_1!\ldots q_k! \over (q_1+\ldots q_k-2)!(\xi A)^{q_1+\ldots + q_k-1}
} \nonumber \\
&& \qquad \times  \int _0^\infty dz\,z^{q_k-2} W^{(1)}(z;X-\sum_{s=1}^k \tau_s)
W^{(k)}(z;\tau_1,\ldots,\tau_k)\ ,
\eea
where the functions $W^{(s)}(z;\tau_1,\ldots,\tau_s)$ 
are defined by the equation
(identical to Eq.~(\ref{anom12}))
\begin{equation}
{\partial W^{(s)}(z;\tau_1,\ldots,\tau_s)\over \partial\tau_s}=
\left( z^2 {\partial^2\over \partial z^2}-z\right) 
W^{(s)}(z;\tau_1,\ldots,\tau_s)
\label{glob2}
\end{equation}
and the boundary conditions
\be
\label{glob3}
W^{(s)}(z;\tau_1,\tau_2,\ldots,\tau_{s-1},0)= z^{q_{s-1}}
W^{(s-1)}(z;\tau_1,\tau_2,\ldots,\tau_{s-1})\ .
\ee
Solving these equations consecutively via the Lebedev-Kontorovich
transformation, one can find all the correlation functions
$V\Gamma^{\{q\}}({\bf r_1},\ldots,{\bf r_k})$ (in the form of multiple
integrals over $\mu_s$). This will be in particular used in 
Section \ref{s4.2.1}, where we will study the joint distribution
function of the wave function intensities in two points ($k=2$).

We show now that the correlation functions (\ref{glob1})
allow to represent the statistics of eigenfunction envelops in a very
compact form. Making the substitution of the variable $z=e^\theta$ and
defining the functions
$\tilde{W}^{(s)}(\theta;\tau_1,\ldots,\tau_s)=z^{-1/2}
W^{(s)}(z;\tau_1,\ldots,\tau_s)$,
we can rewrite (\ref{glob2}) in the form of the imaginary time
 Schr\"odinger equation,
\be
\label{glob4}
-{\partial\tilde{W}^{(s)}\over\partial\tau_s}=\hat{H}\tilde{W}^{(s)}\
 ,\qquad\qquad
\hat{H}=-\partial_\theta^2+e^\theta+{1\over 4}
\ee
with the boundary conditions
\bea
\label{glob5}
&& \tilde{W}^{(s)}(\theta;\tau_1,\ldots,\tau_{s-1},0)=e^{q_{s-1}\theta}
\tilde{W}^{(s-1)}(\theta;\tau_1,\ldots,\tau_{s-1})\ ,\nonumber\\
&&\tilde{W}^{(1)}(\theta;0)=e^{-\theta/2}\ .
\eea
This allows to rewrite Eq.~(\ref{glob1}) as a matrix element,
\bea
\label{glob6}
&& V\Gamma^{\{q\}}({\bf r_1},\ldots,{\bf r_k})=
{q_1!\ldots q_k! \over (q_1+\ldots q_k-2)!(\xi A)^{q_1+\ldots + q_k-1}
} \nonumber \\
&& \qquad \times
\langle e^{-\theta/2}|e^{-(X-\sum_{s=1}^k\tau_s)\hat{H}}
e^{q_s\theta}e^{-\tau_s\hat{H}}e^{q_{s-1}\theta}
\ldots e^{q_1\theta}e^{-\tau_1\hat{H}}|e^{-\theta/2}\rangle
\eea
(these transformations are completely analogous to those performed
by Kolokolov in \cite{kolokol93},
where the eigenfunction statistics in the strictly 1D case was
studied). Furthermore, the matrix element can be represented as a
Feynman path integral,
\bea
\label{glob7}
&& V\Gamma^{\{q\}}({\bf r_1},\ldots,{\bf r_k})=
{q_1!\ldots q_k! \over (q_1+\ldots q_k-2)!(\xi A)^{q_1+\ldots + q_k-1}
} \int d\theta_l d\theta_r e^{-{\theta_l+\theta_r \over 2}}\nonumber \\
&& \qquad \times {\cal N}\int_{
\theta(0)=\theta_l,\,\theta(X)=\theta_r }
 D\theta(t)\exp 
\left\{-\int dt\,\left({1\over 4}\dot{\theta}^2+e^\theta
+{1\over 4}\right)\right\} \nonumber \\
&& \qquad \times
e^{q_1\theta(t_1)+q_2\theta(t_2)+
\ldots+q_k\theta(t_k) }\ ,
\eea
with ${\cal N}$ being the normalization constant, ${\cal N}^{-1}=
\int D\theta \exp \{-{1\over 4}\int dt\,\dot{\theta}^2\}$. The quantum
mechanics defined by the Hamiltonian (\ref{glob4}) (or, equivalently, by
the path integral (\ref{glob7})) is known as Liouville quantum mechanics
\cite{jackiw,shelton}; the corresponding spectral expansion is
obviously equivalent to the Lebedev-Kontorovich expansion.

Inserting here the decomposition of unity, $1=\int dw\delta(X^{-1}\int
dt\, e^\theta-w)$ and making a shift $\theta \to \theta+\ln w$, we
get
\bea
\label{glob8}
&& \Gamma^{\{q\}}({\bf r_1},\ldots,{\bf r_k})=
{q_1!\ldots q_k! \over V^{q_1+\ldots + q_k}} e^{-X/4} {\cal N} \int
D\theta(t) e^{-{\theta(0)+\theta(X)\over 2}} \nonumber \\
&& \quad \times \exp\left\{-{1\over 4}\int dt\,\dot{\theta}^2\right\}
e^{q_1\theta(t_1)+\ldots+q_k\theta(t_k) }
\delta\left(X^{-1}\int dt\,e^\theta -1\right)\ .
\eea
According to (\ref{glob8}), the eigenfunction intensity can be written
as a product
\be
\psi({\bf r})=\Phi({\bf r})\Psi(t)\ ,
\label{glob9}
\ee
where $\Phi({\bf r})$ is a quickly fluctuating (in space) function,
which has the Gaussian Ensemble statistics,
$\langle|\Phi^{2q}|\rangle = q!/V^q$, and fluctuates independently in
the points separated by a distance larger than the mean free path. The
function $\Psi(t)$ determines, in contrast, a smooth envelope of
the wave function. Its fluctuations are long-range correlated and are
described by the probability density
\be
\label{glob10}
{\cal P}\{\theta(t)=\ln\Psi^2(t)\}={\cal N} e^{-X/4} 
e^{-{\theta(0)+\theta(X)\over 2}}
\exp\left\{-{1\over 4}\int dt\,\dot{\theta}^2\right\}\ .
\ee

The above calculation can be repeated for the case, when some of the
points ${\bf r_i}$ lie closer than $l$ to each other. The result
(\ref{glob9}), (\ref{glob10}) is reproduced also in this case, with
the function $\Phi({\bf r})$ having the ideal metal statistics 
given by the zero-dimensional $\sigma$-model. This statistics
\cite{berry3,prigodin95,srednicki96} is Gaussian and is determined by
the (short-range) correlation function
\be
\label{glob11}
V\langle\Phi^*({\bf r})\Phi({\bf r'})\rangle=k_d^{1/2}(|{\bf r}-{\bf
r'}|)\ ; 
\ee
see Eq.~(\ref{corrrev_e6}) below.

The physics of these results is as follows. The short-range
fluctuations of the wave function (described by the function
$\Phi({\bf r})$)  have the same origin as in a strongly chaotic system,
where superposition of plane waves with random amplitudes and phases
leads to the Gaussian fluctuations of eigenfunctions with the
correlation function (\ref{glob11}) and, in particular, to the RMT
statistics of the local amplitude, $\langle|\Phi^{2q}|\rangle =
q!/V^q$. The second factor $\Psi(t)$ in the decomposition
(\ref{glob9}) describes the smooth envelope of the eigenfunction
(changing on a scale $\gg l$), whose statistics is given by
(\ref{glob10}) and is determined by diffusion and localization
effects.

Let us note that in the metallic regime, $X\ll 1$, the measure
(\ref{glob10}) can be approximated as
\be
\label{glob12}
{\cal P}\{\theta(x)\} = \exp \left\{-{\pi\nu AD\over 2}\int dx\,
\left({d\theta\over dx}\right)^2\right\}\ .
\ee
We will see in Section~\ref{c4}, while studying the statistics of
anomalously localized states in $d\ge 1$ dimensions, that the
probability of appearance in a metallic sample of such a rare state
with an envelope $e^\theta({\bf r})$ is given (within the
exponential accuracy) by the $d$-dimensional generalization of
(\ref{glob12}) (see, in particular, Eqs.~(\ref{e4.10}),
(\ref{e4.509})). 

Finally, we compare the eigenfunction statistics in the quasi-1D case
with that in a strictly 1D disordered system. In the latter case, the
eigenfunction can be written as
\be
\label{glob13}
\psi_{1D}(x)=\sqrt{{2\over L}}\cos(kx+\delta)\Psi(x)\ ,
\ee
where $\Psi(x)$ is a smooth envelope function. The local statistics of 
$\psi_{1D}(x)$ (i.e. the moments (\ref{e3.3})) was studied in
\cite{alprig89}, while the global statistics (the correlation
functions of the type (\ref{e3.8})) in \cite{kolokol93}. Comparing the
results for the quasi-1D and 1D systems, we find  that the
statistics of the smooth envelopes $\Psi$ is exactly the same in the
two cases, for a given value of the ratio of the system length $L$ to
the localization length 
(equal to $\beta\pi\nu A D$ in quasi-1D
and to the mean free path $l$ in 1D). In particular, the moments 
$\Gamma^{(q)}({\bf r})=\langle |\psi^{2q}({\bf r})|\rangle$ are found to
be related as
\be
\label{glob14}
A^q\Gamma^{(q)}_{Q1D} = {q!^2\over (2q-1)!!}\Gamma^{(q)}_{1D}\ ,
\ee
where the factor $q!^2/(2q-1)!!$ represents precisely the ratio of the
GUE moments, $\langle|\Phi^{2q}|\rangle=q!/V^q$, to the plane wave
moments, $\langle (2/V)^q\cos^{2q} (kx+\delta)\rangle=(2q-1)!!/q!V^q$.
For the case of the orthogonal symmetry of the quasi-1D system, this
factor is replaced by $q!$. 
Equivalence of the statistics of the eigenfunction envelopes implies,
in particular, that the distribution of the inverse participation
ratio (IPR),
\be
\label{glob15}
P_2= \int d{\bf r}\, |\psi({\bf r})|^4\ ,
\ee 
is identical in the 1D \cite{bergor80,kolokol94} and quasi-1D 
\cite{fm93a,fm94a} cases (the form of this
distribution in the localized limit $L/\xi\gg 1$ is explicitly given
in Sec.~\ref{s3.2.3} below; for arbitrary $L/\xi$ the result is
very cumbersome \cite{kolokol94}).

\subsubsection{Short wire}
\label{s3.2.2}

In the case of a short wire, $X=1/g\ll 1$, Eq.~(\ref{anom_4}),
(\ref{anom_4o}), (\ref{anom11}), (\ref{anom14}) yield
\cite{mf93a,fm94a,m97} 
\begin{eqnarray}
& {\cal P}^{(U)}(y)= e^{-y}\left[1+{\alpha X\over 6}(2-4y+y^2)+
\ldots\right]\
; &\ \    y\lesssim X^{-1/2} 
\label{anom_adm15a}\\
& {\cal P}^{(O)}(y)=\frac{e^{-y/2}}{\sqrt{2\pi y}}
\left[1+{\alpha X\over 6}\left({3\over 2}-3y+{y^2\over 2}\right)
+\ldots\right]\ ;&\ \    y\lesssim X^{-1/2} 
\label{anom_adm8a} \\
& {\cal P}^{(U)}(y)= \exp\left\{ 
-y+{\alpha \over 6}y^2X+\ldots\right\}\
;& \ \    X^{-1/2}\lesssim y\lesssim X^{-1} 
\label{anom_adm15b}\\
& {\cal P}^{(O)}(y)={1\over\sqrt{2\pi y}} \exp\left\{ {1\over 2}
\left [-y+{\alpha \over 6}y^2X+\ldots\right]\right\}
;&\ \    X^{-1/2}\lesssim y\lesssim X^{-1} 
\label{anom_adm15bo}\\
& {\cal P}(y)\sim\exp\left[-2\beta\sqrt{y/X}\right]\ ;&\ \ 
 y\gtrsim X^{-1} 
\label{anom_adm15c}
\end{eqnarray}
(a more accurate formula for the far ``tail'' (\ref{anom_adm15c})
can be found in Sec.~\ref{s4.2.1}, Eq.~(\ref{anom50})).
Here the coefficient $\alpha$ is equal to
$\alpha=2[1-3L_-L_+/L^2]$. We see that there exist three different
regimes of the behavior of the distribution function.
For not too
large amplitudes $y$, Eqs.(\ref{anom_adm15a}), (\ref{anom_adm8a})  
are just the RMT results with relatively small corrections.
In the intermediate range (\ref{anom_adm15b}), (\ref{anom_adm15bo})
the correction {\it in the
exponent} is small compared to the leading term but much larger than
unity, so that ${\cal P}(y)\gg {\cal P}_{RMT}(y)$ though
$\ln{\cal P}(y)\simeq \ln{\cal P}_{RMT}(y)$. Finally, in the large
amplitude region, 
(\ref{anom_adm15c}), the distribution function ${\cal P}(y)$ differs
completely from the RMT prediction. Note that 
Eq.~(\ref{anom_adm15c}) is not valid when
the observation point is located close to the sample boundary, in
which case the exponent of (\ref{anom_adm15c}) becomes smaller by a
factor of 2, see Sec.~\ref{s4.2.3}.

\subsubsection{Long wire}
\label{s3.2.3}

In the limit of a long sample, $X=L/\xi\gg 1$, Eqs.~(\ref{anom_4}),
(\ref{anom_4o}), (\ref{anom11}), (\ref{anom14}) reduce to
\begin{eqnarray}
&&{\cal P}^{(U)}(u)\simeq {8\xi^2 A\over L} \left[K_1^2(2\sqrt{uA\xi})+
K_0^2(2\sqrt{uA\xi})\right] \ ,
\label{anom3u}\\
&& {\cal P}^{(O)}(u)\simeq {2\xi^2 A\over L} {K_1(2\sqrt{uA\xi})\over 
\sqrt{uA\xi} } \ ,
\label{anom3o}
\end{eqnarray}
with $\xi=2\pi\nu AD$ as before.
Note that in this case the natural variable is not $y=uV$, but rather
$uA\xi$, since typical intensity of a localized wave function is
$u\sim 1/A\xi$ in contrast to $u\sim 1/V$ for a delocalized one. The
asymptotic behavior of Eqs.~(\ref{anom3u}), (\ref{anom3o}) at $u\gg
1/A\xi$ has precisely the same form, 
\be
{\cal P}(u)\sim\exp(-2\beta\sqrt{uA\xi})\ ,
\label{anom3as}
\ee
 as in the region of very large
amplitude in the metallic sample, Eq.~(\ref{anom_adm15c}). On this
basis, it was conjectured in \cite{fm94a} that the asymptotic
behavior 
(\ref{anom_adm15c}) is controlled by the probability to have a
quasi-localized eigenstate with an effective spatial extent much less
than $\xi$ (``anomalously localized state''). This conjecture was
proven rigorously in \cite{m97} where the shape of the anomalously
localized state (ALS) responsible for the large-$u$ asymptotics was
calculated via the transfer-matrix method. We will discuss this in
Section~\ref{c4} devoted to ALS and to asymptotics of different
distribution functions. 

Distribution of the inverse participation ratio (IPR) is also found to
have a simple form in the limit $L\gg\xi$ \cite{fm93a,fm95a}: 
 \begin{eqnarray}
{\cal P}(z)&=&2\pi^{2}\sum_{k=1}^{\infty}(2\pi^{2}zk^{4}-3k^{2})
e^{-\pi^{2}k^{2}z}  \nonumber\\
&\equiv &\frac{4}{\sqrt{\pi}}
\frac{\partial}{\partial z}\{z^{-3/2}\sum_{k=1}^{\infty}k^{2}
e^{-k^{2}/z}\}
\label{ijmpb_118}
\end{eqnarray}
where $z=\pi\nu DA^2P_2$ in the unitary case and $z=(\pi\nu
DA^2/3)P_2$ in the orthogonal case. 
(The second line in (\ref{ijmpb_118})
can be obtained from the first one by using
the Poisson summation formula.) Therefore, the spatial extent of a
localized eigenfunction measured by IPR fluctuates strongly (of order
of 100\%) from one eigenfunction to another. More precisely, the ratio
of the r.m.s. deviation of IPR to its mean value is equal to
$1/\sqrt{5}$ according to Eq.~(\ref{ijmpb_118}). 
The first form of  Eq.(\ref{ijmpb_118}) is more suitable for extracting
the asymptotic behavior of ${\cal P}(z)$ at $z\gg 1$, whereas
the second line gives us the leading behavior of ${\cal P}(z)$
at small $z\ll 1$:
\begin{equation} {\cal P}(z)=\left\{\begin{array}{cc}
4\pi^{4}ze^{-\pi^{2}z}\ ,&\quad z\gg 1 \\
4\pi^{-1/2}z^{-7/2}e^{-1/z}\ ,&\quad z\ll 1
\end{array}\right.
\label{ijmpb_119}
\end{equation}
Therefore, the probability to have atypically large or atypically
small IPR is exponentially suppressed.
The function ${\cal P}(z)$  is presented in Fig.~\ref{fluc_ipr}.

\begin{figure}
\centerline{\epsfxsize=120mm\epsfbox{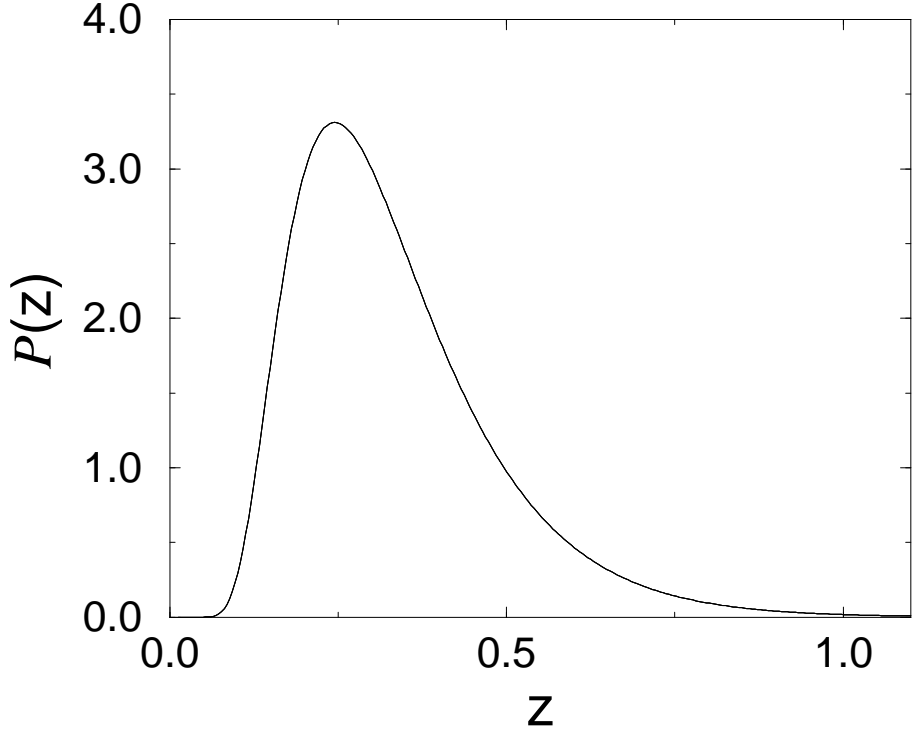}}
\vspace{3mm}
\caption{Distribution function ${\cal P}(z)$ of the normalized
(dimensionless) inverse participation ratio
$z=[\beta^2/(\beta+2)]\pi\nu DA^2 P_2$ in a long ($L\gg\xi$) quasi-1D
sample. The average value is $\langle z\rangle=1/3$. From \cite{fm94a}.} 
\label{fluc_ipr} 
\end{figure}

The above fluctuations of
IPR are due to fluctuations in the ``central bump'' of a localized
eigenfunction. 
They should be distinguished from the fluctuations in the rate of
exponential decay of eigenfunctions (Lyapunov exponent). The latter
can be extracted from another important physical quantity ---
the distribution function ${\cal P}_{x}(v)$, where 
$$
v=(2\pi\nu DA^2)^2 |\psi_\alpha^2({\bf r_1})\psi_\alpha^2({\bf r_2})|
$$
is the product of the eigenfunction intensity in the two points close
to the opposite edges of the sample $r_1\to 0$, $r_2\to L$. The result
is \cite{fm93b,fm94a}
\begin{eqnarray}
\label{ijmpb_123}
&&{\cal P}(-\ln{v})= 
F[-(\beta\ln v)/2X]
\frac{1}{2(2\pi X/\beta)^{1/2}}\exp\left\{-\frac{\beta}{8X}
(2X/\beta+\ln{v})^{2}\right\} \nonumber\\
&& F^U(u)=u{\Gamma^2[(3-u)/2]\over\Gamma(u)}\ ,\qquad
F^O(u)={u\Gamma^2[(1-u)/2]\over\pi\Gamma(u)}
\end{eqnarray}
Therefore, $\ln v$ is asymptotically distributed according to the
Gaussian law with the mean value $\langle-\ln
v\rangle=(2\beta)X=L/\beta\pi\nu AD$ and the variance
$\mbox{var}(-\ln v)=2\langle-\ln v\rangle$. The same log-normal
distribution is found for the conductance and for transmission
coefficients of a quasi-1D sample from the
Dorokhov-Mello-Pereyra-Kumar formalism \cite{smmp,been1} (see end of
Sec.~\ref{s6.2}).

Note that the formula (\ref{ijmpb_123}) is valid in the region of
$v\ll 1$ (i.e. negative $\ln v$) only, which contains almost all
normalization of the distribution function. In the region of still
higher values of $v$ the log-normal form of ${\cal P}(v)$ changes into
the much faster
 stretched-exponential fall-off $\propto\exp\{-2\sqrt{2}\beta
v^{1/4}\}$, as can be easily found from the exact solution given in
\cite{fm93b,fm94a}. The decay rate of all the moments
$\langle v^k\rangle$, $k\ge 1/2$, is four times less than $\langle-\ln
v\rangle$ and does not depend on $k$: $\langle v^k\rangle\propto
e^{-X/2\beta}$. This is because the moments $\langle v^k\rangle$,
$k\ge 1/2$ are determined by the probability to find an ``anomalously
delocalized state'' with $v\sim 1$.

\subsection{Arbitrary dimensionality: metallic regime}
\label{s3.3}

\subsubsection{Distribution of eigenfunction amplitudes}
\label{s3.3.1}

In the case of arbitrary  dimensionality $d$, deviations from the RMT
distribution ${\cal P}(y)$ for not too large $y$ can be calculated
\cite{fm94b,fm95a} via the method described in
Section~\ref{c2}. Applying this method to the moments (\ref{e3.4}), one
gets
\bea
\langle|\psi({\bf r})|^{2q}\rangle &=& {q!\over
V^{q}}\left[1+{1\over 2}\kappa q(q-1)+\ldots\right]\qquad (U)
\label{e3.10}\\ 
\langle|\psi({\bf r})|^{2q}\rangle &=& {(2q-1)!!\over
V^{q}}[1+\kappa q(q-1)+\ldots]\qquad (O)\ , 
\label{e3.11}
\eea
where $\kappa=\Pi({\bf r}, {\bf r})$. Correspondingly, the correction
to the distribution function reads
\bea
{\cal P}(y)&=&e^{-y}\left[1+{\kappa\over 2}(2-4y+y^2)+\ldots\right]
\qquad\qquad (U)
\label{e3.12}\\
{\cal P}(y)&=&{e^{-y/2}\over\sqrt{2\pi y}}
\left[1+{\kappa\over 2}\left({3\over 2}-3y+{y^2\over 2}\right)
+\ldots\right]\qquad (O)\ . 
\label{e3.13}
\eea
Deviations of the eigenfunction distribution function ${\cal P}(y)$
from its RMT form are illustrated for the orthogonal symmetry case in
Fig.~\ref{stat_wave_func}. Numerical studies of the statistics of
eigenfunction amplitudes in weak localization regime have been
performed in Ref.~\cite{mueller97} for the 2D and in
Ref.~\cite{uski98} for the 3D case. The found deviations from RMT are
well described by the above theoretical results.
Experimentally, statistical properties of the eigenfunction
intensity have been studied for microwaves in a disordered cavity
\cite{kudrolli}. For a weak disorder the found deviations are in
good agreement with (\ref{e3.13}) as well. 

\begin{figure}
\centerline{\epsfxsize=120mm\epsfbox{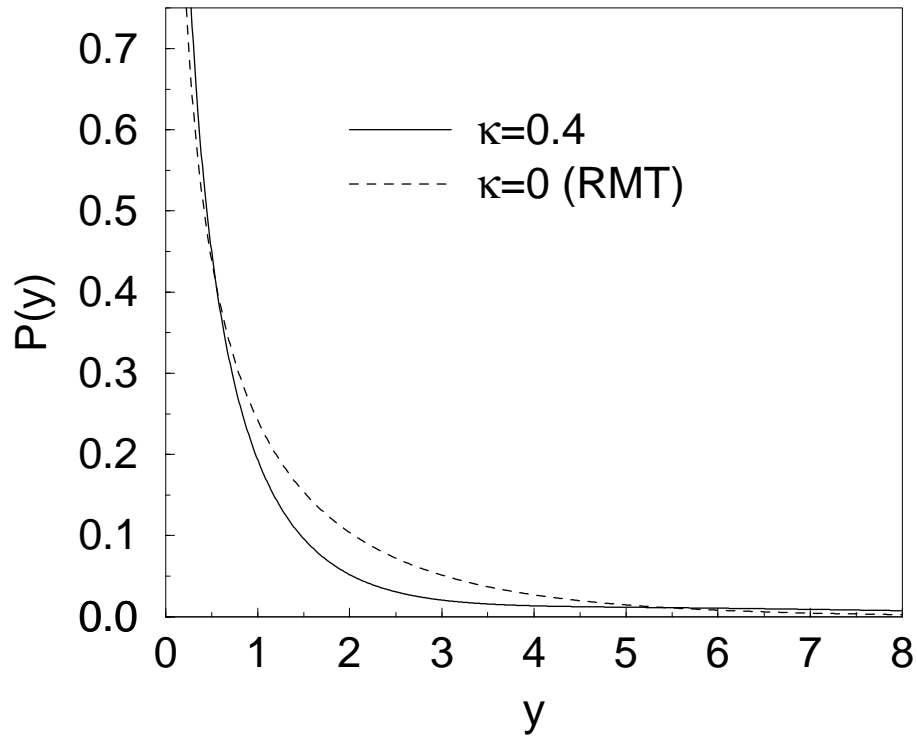}}
\vspace{3mm}
\caption{Distribution ${\cal P}(y)$ of the normalized eigenfunction
intensities $y=V|\psi^2({\bf r})|$ in the orthogonal symmetry
case. The dashed line shows the RMT result, Eq.~(\ref{e3.2}), while the
full line corresponds to Eq.~(\ref{e3.13}) with $\kappa=0.4$.} 
\label{stat_wave_func} 
\end{figure}

In the quasi-one-dimensional case (with hard wall boundary conditions
in the longitudinal direction), the one-diffuson loop $\Pi({\bf
r},{\bf r})$ is equal to
\be
\kappa\equiv\Pi({\bf r},{\bf r})={2\over g}\left[{1\over 3}-{x\over
L}\left(1-{x\over L}\right)\right]\ ,\qquad 0\le x\le L\ ,
\label{e3.14}
\ee
so that Eqs.~(\ref{e3.12}), (\ref{e3.13}) agree with the results
(\ref{anom_adm15a}), (\ref{anom_adm8a}) obtained from the exact
solution. For the periodic boundary conditions in the longitudinal
direction (a ring) we have $\kappa=1/6g$. 
In the case of 2D geometry,
\be
\Pi({\bf r},{\bf r}) = {1\over \pi g}\ln {L\over l}\ ,
\label{e3.15}
\ee
with $g=2\pi\nu D$. Finally, in the 3D case the sum over the momenta 
$\Pi({\bf r},{\bf r}) =(\pi\nu V)^{-1}\sum_{{\bf q}}(D{\bf q}^2)^{-1}$
diverges linearly at large $q$. The diffusion approximation is valid
up to $q\sim l^{-1}$; the corresponding cut-off gives
$\Pi({\bf r},{\bf r}) \sim 1/2\pi\nu Dl=g^{-1}(L/l)$. This divergency
indicates that more accurate evaluation of $\Pi({\bf r},{\bf r})$
requires taking into account also the contribution of the ballistic
region ($q>l^{-1}$) which depends on microscopic details of the
random potential. We will return to this question in
Sec.~\ref{s3.3.4}. 

The formulas (\ref{e3.12}), (\ref{e3.13}) are valid in the region of
not too large amplitudes, where the perturbative correction is smaller
than the RMT term, i.e. at $y\ll \kappa^{-1/2}$. In the region of
large amplitudes, $y>\kappa^{-1/2}$ the distribution function was
found by Fal'ko and Efetov \cite{fe1,fe2}, who applied to
Eqs.~(\ref{anom_4}), 
(\ref{anom_4o}) the saddle-point method suggested by Muzykantskii and
Khmelnitskii \cite{mk1}. We relegate the discussion of the method to
Section~\ref{c4} and only present the results here:
\bea
&& \hspace{-1.3cm} {\cal P}(y)\simeq\exp\left\{{\beta\over
2}(-y+{\kappa\over 2} y^2+\ldots)\right\}\times\left\{
\begin{array}{ll} 1 & \ (U)\\
                  {1\over\sqrt{2\pi y}} &\  (O)
\end{array}\right.
\ , \  \kappa^{-1/2}\lesssim y\lesssim
\kappa^{-1}, \label{e3.16} \\
&& \hspace{-1.3cm} {\cal P}(y)\sim\exp\left\{-{\beta\over 4\kappa}\ln^d(\kappa
y)\right\}\ ,\qquad\qquad y\gtrsim\kappa^{-1}. \label{e3.17}
\eea
Again, as in the quasi-one-dimensional case, there is an intermediate
range where a correction in the exponent is large compared to unity,
but small compared to the leading RMT term [Eq.~(\ref{e3.16})] and a
far asymptotic region (\ref{e3.17}), where the decay of ${\cal P}(y)$
is much slower 
than in RMT. In the next section we will discuss the structure of
anomalously localized eigenstates, which are responsible for the
asymptotic behavior (\ref{anom_adm15c}), (\ref{e3.17}).

\subsubsection{2D: Weak multifractality of eigenfunctions}
\label{s3.3.2}

Since $d=2$ is the lower critical dimension for the Anderson
localization problem, metallic 2D samples (with $g\gg 1$) share many
common properties with systems at the critical point of the 
metal-insulator transition. Although the localization length $\xi$ in 2D is
not infinite (as for truly critical systems), it is exponentially
large, and the criticality takes place in the very broad range
of the system size $L\ll\xi$. 

\paragraph{Multifractality: basic definitions.} The criticality of
eigenfunctions shows up via their multifractality. Multifractal
structures first introduced by Mandelbrot \cite{mandelbrot74}
are characterized by an infinite set of critical exponents describing
the scaling of the moments of a distribution of some quantity.
Since then, this feature has been observed in various objects, such
as the energy dissipating set in turbulence
\cite{frisch83,benzi84,frisch95},  
strange attractors in chaotic dynamical systems
\cite{grassberger83,hentschel83,benzi85,halsey86}, and the growth
probability distribution in diffusion-limited aggregation
\cite{halsey86a,j_lee88,blumenfeld89}; see Ref.~\cite{paladin87} for a
review.  

The fact that an eigenfunction at the mobility edge has the
multifractal structure was  noticed for the first time in
\cite{castpel}, though the underlying renormalization group
calculations were done by Wegner several years earlier
\cite{wegner80}.  For this problem, the
probability distribution is just  the eigenfunction intensity
$|\psi^2({\bf r})|$ and the corresponding moments are the 
the inverse participation ratios,  
\be
\label{multifr1}
P_q=\int d^dr|\psi^{2q}({\bf r})|\ .
\ee
The  multifractality is characterized by the anomalous
scaling of $P_q$ with the system size $L$,
\be
\label{multifr2}
P_q\propto L^{-D_q(q-1)}\equiv L^{-\tau(q)},
\ee
with $D_q$ different from the spatial dimensionality $d$ and dependent
on $q$. 
Equivalently, the eigenfunctions are characterized by the singularity
spectrum $f(\alpha)$ describing the measure $L^{f(\alpha)}$ of 
the set of those points ${\bf r}$ 
where the eigenfunction takes the value $|\psi^2({\bf
r})|\propto L^{-\alpha}$. The two sets of exponents $\tau(q)$ and
$f(\alpha)$ are  related via the Legendre transformation,
\be
\tau(q)=q\alpha-f(\alpha)\ ;\ \ f'(\alpha)=q\ ;\ \
\tau'(q)=\alpha\ .
\label{multifr3}
\ee
For a recent review on multifractality of critical eigenfunctions the
reader is referred to \cite{janssen94,janssen98}.

\paragraph{Multifractality in 2D.} We  note first that the
formulas (\ref{e3.10}), (\ref{e3.11}) for the IPR's with $q\lesssim
\kappa^{-1/2}$ can be rewritten in the 2D case (with (\ref{e3.15})
taken into account) as
\be
\label{multifr4}
{\langle P_q\rangle \over P_q^{\rm RMT}}\simeq\left({L\over
l}\right)^{{1\over\beta\pi g}q(q-1)}\ ,
\ee
where $P_q^{\rm RMT}$ is the RMT value of $P_q$ equal to
$q!L^{-2(q-1)}$ for GUE and $(2q-1)!!L^{-2(q-1)}$ for GOE. We see that
(\ref{multifr4}) has precisely the form (\ref{multifr2}) with 
\be
D_q=2-{q\over\beta\pi g}
\label{multifr5}
\ee
As was found in \cite{fe1,fe2}, the eigenfunction amplitude
distribution (\ref{e3.16}), (\ref{e3.17}) leads to the same result
(\ref{multifr5}) for all $q\ll 2\beta\pi g$. Since deviation of $D_q$
from the normal dimension 2 is proportional to the small parameter
$1/\pi g$, it can be termed ``weak multifractality'' (in analogy with
weak localization). The result (\ref{multifr5}) was in fact obtained
for the first time by Wegner \cite{wegner80} via the renormalization
group calculations. 

The limits of validity of Eq.~(\ref{multifr5}) are not unambiguous and
should be commented here. The singularity spectrum $f(\alpha)$
corresponding to (\ref{multifr5}) has the form
\be
\label{multifr6}
f(\alpha)=2-{\beta\pi g\over 4}\left(2+{1\over\beta\pi
g}-\alpha\right)^2\ ,
\ee
so that $f(\alpha_\pm=0)$ for
\be
\label{multifr7}
\alpha_\pm=2\left[1\pm{1\over(2\beta\pi g)^{1/2}}\right]^2\ .
\ee
If $\alpha$ lies outside the interval $(\alpha_-,\,\alpha_+)$, the
corresponding $f(\alpha)<0$, which means that the most likely the
singularity $\alpha$ will not be found for a given
eigenfunction. However, if one considers the {\it average} $\langle
P_q\rangle$ over a sufficiently large ensemble of
eigenfunctions (corresponding to different realizations of disorder),
a negative value of $f(\alpha)$ makes sense (see a related discussion
in \cite{mandelbrot90,mandelbrot91}). This is the definiton which
was assumed  in \cite{fe1,fe2} where
Eq.~(\ref{multifr5}) was obtained for all positive $q \ll 2\beta\pi g$. 

In contrast, if one studies
a {\it typical} value of $P_q$, the regions $\alpha>\alpha_+$ and
$\alpha<\alpha_-$ will not contribute. In this case, Eq.~(\ref{multifr5})
is valid only within the interval $q_-\le q\le q_+$ with
$q_\pm=\pm(2\beta\pi g)^{1/2}$; outside this region one finds
\cite{chamon96,castillo97} 
\be
\tau(q)\equiv D_q(q-1)=\left\{  \begin{array}{ll}
q\alpha_-\ , &\ \ q>q_+\\
q\alpha_+\ , &\ \ q<q_-\ .
\end{array}
\right.
\label{multifr8}
\ee
Therefore, within this definition the multifractal dimensions $D_q$
saturate at the values $\alpha_+$ and $\alpha_-$ for $q\to+\infty$ and
$q\to-\infty$ respectively. This is in agreement with results of
numerical simulations \cite{schreiber,pook,huck92,janssen94,huck95}.

\subsubsection{Correlations of eigenfunction amplitudes and
fluctuations of the inverse participation ratio}
\label{s3.3.3}

In this subsection, we study correlations of eigenfunctions in the
regime of a good conductor \cite{fm95a,bm96,bm97,m98,prigal98}.
The correlation function of amplitudes of one and the same
eigenfunction with energy $E$ can be formally defined as follows:
\begin{equation}
\label{corrrev_e1}
\alpha({\bf  r}_1, {\bf  r}_2, E) = \left\langle
\vert \psi_k({\bf  r}_1) \psi_k({\bf  r}_2) \vert^2
\right\rangle_{E}\equiv
\Delta \left\langle \sum_k \vert \psi_k({\bf 
r}_1) \psi_k({\bf  
r}_2) \vert^2 \delta (E- \epsilon_k) \right\rangle.
\end{equation}
An analogous correlation function for two
different eigenfunctions is defined as
\begin{eqnarray}
\label{corrrev_e2}
&& \sigma({\bf  r}_1, {\bf  r}_2, E, \omega) =
\left\langle \vert \psi_k({\bf  r}_1) \psi_l ({\bf  r}_2)
\vert^2 \right\rangle_{E,\omega} \nonumber\\
&&\qquad \equiv  \Delta^2 R^{-1}(\omega)
\left\langle \sum_{k\ne l} \vert \psi_k({\bf 
r}_1) \psi_l({\bf   
r}_2) \vert^2 \delta (E - \epsilon_k) \delta(E + \omega
- \epsilon_l) \right\rangle,
\end{eqnarray}
where $R(\omega)$ is the two-level correlation function (\ref{e2.1}).
To evaluate $\alpha({\bf  r}_1, {\bf  r}_2, E)$ and
$\sigma({\bf  r}_1, {\bf  r}_2, E, \omega)$, we employ an
identity    
\begin{eqnarray}
&& 2\pi^2\left[\Delta^{-1}\alpha({\bf r}_1, {\bf r}_2, E)
\delta(\omega)+\Delta^{-2}
\tilde{R}(\omega)\sigma({\bf  r}_1, {\bf  r}_2, E,\omega)\right]
 \label{corrrev_e4}\\
&&= \mbox{Re}\left[\left\langle
G^R({\bf  r}_1, {\bf  r}_1, E)G^A({\bf  r}_2, {\bf  r}_2, E+\omega)  -
 G^R({\bf  r}_1, {\bf  r}_1, E)
G^R({\bf  r}_2, {\bf  r}_2, E+\omega)\right\rangle \right] \nonumber,
\end{eqnarray}
where $G^{R,A}({\bf  r}, {\bf  r'}, E)$ are retarded and advanced
Green's functions and 
$\tilde{R}(\omega)$ is the non-singular part of the level-level
correlation function:
$R(\omega)=\tilde{R}(\omega)+\delta(\omega/\Delta)$.
A natural question, which arises at this point, is whether the
r.h.s. of Eq.(\ref{corrrev_e4}) cannot be simply found within the
diffuson-Cooperon perturbation theory \cite{ashk}. Such a
calculation would, however, be justified only for $\omega\gg\Delta$
(more precisely, one has to introduce an imaginary part of frequency:
$\omega\to\omega +i\Gamma$, and require that
$\Gamma\gg\Delta$). Therefore, it would only allow to find 
a smooth in $\omega$ part of
$\sigma({\bf  r}_1, {\bf  r}_2, E, \omega)$ for $\omega\gg \Delta$. 
Evaluation of $\alpha({\bf  r}_1, {\bf  r}_2, E)$, as well as of
$\sigma({\bf  r}_1, {\bf  r}_2, E, \omega)$ at $\omega\sim\Delta$,
cannot be done within such a calculation. For this reason the
non-perturbative supersymmetry approach is to be used.

The r.h.s. of Eq.(\ref{corrrev_e4}) can be expressed in terms of the
supermatrix $\sigma$-model, yielding:
\begin{eqnarray} \label{corrrev_e5}
& & 2\pi^2 \left[\Delta^{-1} {\alpha({\bf r}_1, {\bf r}_2,
E)} \delta(\omega) + \Delta^{-2}
{\sigma({\bf r}_1, {\bf r}_2,E, \omega)}  
\tilde{R} (\omega) \right]  \\
& & = (\pi\nu)^2 \left[1-\mbox{Re} \langle  Q_{bb}^{11}
({\bf  r}_1) Q_{bb}^{22} ({\bf  r}_2) \rangle_S 
 - k_d({\bf r}_1-{\bf r}_2) \mbox{Re} \langle Q^{12}_{bb}
({\bf  r}_1) Q^{21}_{bb}({\bf  r}_1) \rangle_S \right], \nonumber
\end{eqnarray}
where $\langle \dots \rangle_S$ denotes the averaging with the 
sigma-model action and\\
$k_d({\bf r}) = (\pi\nu)^{-2} \langle\mbox{Im}G^R
({\bf r})\rangle^2$ is a short-range function  explicitly given by 
\begin{equation}
 k_d({\bf r}) =  \exp (-r/l) \left\{
\begin{array}{ll} 
J_0^2(p_F r), & \ \ 2D\\
(p_Fr)^{-2} \sin^2 p_Fr, & \ \ 3D 
\end{array}
\right..
\label{corrrev_e6}
\end{equation}
We consider the unitary ensemble first; results for the orthogonal
symmetry will be presented in the end. 
Evaluating the $\sigma$-model correlation functions in the
r.h.s. of Eq.(\ref{corrrev_e5}) and separating the result into the singular 
the singular (proportional to $\delta(\omega)$) 
and regular at $\omega = 0$ parts, one can obtain the correlation functions
$\alpha({\bf r}_1,{\bf r}_2,E)$ and
$\sigma({\bf r}_1,{\bf r}_2,E,\omega)$.
The two-level correlation function $R (\omega)$ entering
Eq.(\ref{corrrev_e5}) was studied in  Section~\ref{c2}.
We employ again the method of \cite{km} described in Section~\ref{c2}
to calculate
the sigma-model correlation functions $\langle Q_{bb}^{11}({\bf 
r}_1) Q_{bb}^{22} ({\bf  r}_2) \rangle_S$ and $\langle
Q_{bb}^{12}({\bf  r}_1) Q_{bb}^{21} ({\bf  r}_2) \rangle_S$ 
for relatively low frequencies $\omega \ll E_c$. First, we
restrict ourselves to the terms of order $g^{-1}$. Then, the result
for the first correlation function reads as  
\begin{equation} \label{corrrev_1122a}
 \langle Q_{bb}^{11}({\bf  r}_1) Q_{bb}^{22} ({\bf  r}_2)
\rangle_S 
= -1 - 2i {\exp (i\pi s) \sin \pi s\over (\pi s)^{2}} - 
\frac{2i}{\pi s}
\Pi({\bf  r}_1, {\bf  r}_2)\ ,
\end{equation}
where $s=\omega/\Delta+i0$.
The first two terms in Eq. (\ref{corrrev_1122a}) represent the result of
the zero-mode approximation; the last term is the correction of order
$g^{-1}$. An analogous calculation for the second correlator yields:
\begin{equation} \label{corrrev_Q3}
\langle Q_{bb}^{12} ({\bf  r}_1) Q_{bb}^{21} ({\bf  r}_2)
\rangle_S =  -2\left\{ {i\over \pi s }
+ \left[ 1 +
 i { \exp (i\pi s) \sin \pi s\over (\pi s)^{2}} 
\right] \Pi({\bf  r}_1, {\bf  r}_2) \right\}.
\end{equation}
Now, separating regular and singular parts in r.h.s. of Eq. (\ref{corrrev_e5}),
we obtain the following result for the autocorrelations of the same 
eigenfunction:
\begin{equation} \label{corrrev_fin1}
 V^2\langle \vert \psi_k({\bf  r}_1) \psi_k({\bf  r}_2) \vert^2
\rangle_{E} -1  
 =   k_d(r) 
 [1 + \Pi({\bf  r}_1,{\bf  r}_1)] + \Pi({\bf  r}_1,{\bf  r}_2),
\end{equation}
and for the correlation of amplitudes of two different eigenfunctions
\begin{equation} \label{corrrev_fin2}
V^2\langle \vert \psi_k({\bf  r}_1) \psi_l({\bf  r}_2) \vert^2
\rangle_{E, \omega} -1  =  k_d(r) 
 \Pi({\bf  r}_1,{\bf  r}_1), \ k \ne l 
\end{equation}
In particular, for ${\bf  r}_1
= {\bf  r}_2$ we have 
\begin{equation} \label{corrrev_fin3}
V^2 \langle \vert \psi_k ({\bf  r}) \psi_l {\bf  r}) \vert^2
\rangle_{E, \omega} - 1 = \delta_{kl} + (1 + \delta_{kl})
\Pi({\bf  r},{\bf  r}). 
\end{equation}
Note that the result (\ref{corrrev_fin1}) for
${\bf  r}_1 = {\bf  r}_2$ is the inverse participation ratio
calculated above (section \ref{s3.3.1}); on the other hand, neglecting
the terms with the diffusion propagator (i.e. making the zero-mode
approximation), we reproduce the result of
Refs.~\cite{berry3,prigodin95,srednicki96}.  

Eqs. (\ref{corrrev_fin2}), (\ref{corrrev_fin3})  show
that the correlations between different eigenfunctions are relatively
small in the weak disorder regime. Indeed, they are proportional to
the small parameter $\Pi({\bf r},{\bf r})$.
The correlations are enhanced by disorder; when the system approaches
the strong localization regime, the
relative magnitude of correlations, $\Pi({\bf  r}, {\bf  r})$
ceases to be small. The correlations near the Anderson localization
transition will be discussed in  Section~\ref{c5}.

Another correlation function, generally used for the calculation of
the linear response of the system,
\begin{eqnarray} 
\label{corrrev_gamma}
&& \hspace{-0.5cm} \gamma({\bf  r}_1, {\bf  r}_2, E, \omega) =
\left\langle \psi^*_k({\bf r}_1) \psi_l({\bf r}_1) 
\psi_k({\bf r}_2) \psi^*_l({\bf  r}_2) \right\rangle_{E,
\omega} \nonumber \\
&& \hspace{-0.5cm} \equiv  \Delta^2 R ^{-1}(\omega)
\left\langle \sum_{k\ne l}  
\psi^*_k({\bf r}_1) \psi_l({\bf r}_1) 
\psi_k({\bf r}_2) \psi^*_l({\bf  r}_2)
\delta (E - \epsilon_k) \delta(E + \omega
- \epsilon_l) \right\rangle
\end{eqnarray}
can be calculated in a similar way. The result reads
\begin{equation} \label{corrrev_states}
V^2 \langle \psi^*_k({\bf r}_1) \psi_l({\bf r}_1) 
\psi_k({\bf r}_2) \psi^*_l({\bf  r}_2) \rangle_{E, \omega}
 = k_d(r) + \Pi({\bf 
r}_1,{\bf  r}_2),\ \ \ k \ne l. 
\end{equation}

As is seen from Eqs. (\ref{corrrev_fin1}), (\ref{corrrev_fin2}),
 (\ref{corrrev_states}), 
 in the $1/g$ order the correlation functions 
$\alpha({\bf  r}_1, {\bf  r}_2, E)$
and $\gamma({\bf  r}_1, {\bf  r}_2, E, \omega)$
survive for the large separation between the points, $r \gg l$, 
while $\sigma({\bf  r}_1, {\bf  r}_2, E, \omega)$ decays
exponentially for the distances larger than the mean free path
$l$. This is, however, an artifact of the $g^{-1}$ approximation, and
the investigation of the corresponding tails requires the
extension of the above calculation to the terms proportional to
$g^{-2}$. The correlator $\langle Q_{bb}^{11}({\bf
 r_1})Q_{bb}^{22}({\bf r_2})\rangle_S$ gets the following correction
 in the $g^{-2}$ order: 
\begin{eqnarray} \label{corrrev_correct}
 \delta \langle Q_{bb}^{11}({\bf r_1})Q_{bb}^{22}({\bf r_2}) \rangle_S
&=& -f_1 + 2f_4 +\exp(2i\pi s) f_3 
 -2i {\exp(2i\pi s)\over \pi s} (f_2 - f_3) \nonumber \\
&-& {\exp(2i\pi s)-1\over 2(\pi s)^{2}} (f_1 - 4f_2 + 3f_3 - 2f_4)\ , 
\end{eqnarray}
where we defined the functions
\begin{eqnarray} \label{fun1}
f_1({\bf  r}_1, {\bf  r}_2) & = & \Pi^2 ({\bf  r}_1,
{\bf  r}_2), \nonumber \\
f_2({\bf  r}_1, {\bf  r}_2) & = & (2V)^{-1} \int d{\bf  r}
\left[\Pi^2 ({\bf  r}, {\bf  r}_1) + \Pi^2 ({\bf  r},
{\bf  r}_2) \right], \nonumber\\
f_3 & = & V^{-2} \int d{\bf  r}
d{\bf  r}' \Pi^2 ({\bf  r}, {\bf  r}'), \nonumber \\
f_4({\bf  r}_1, {\bf  r}_2) & = & V^{-1} \int d{\bf  r}
\Pi ({\bf  r}, {\bf  r}_1) \Pi ({\bf  r},
{\bf  r}_2). 
\end{eqnarray}
Consequently, we obtain the following results for the correlations of
different ($k\ne l$) eigenfunctions at $r>l$:
\begin{eqnarray} 
&&V^2\langle \vert \psi_k({\bf  r}_1) \psi_l({\bf  r}_2) \vert^2
\rangle_{E, \omega} -1   =  
{1\over 2}(f_1-f_3-2f_4) \nonumber\\
&&+2(f_2-f_3)\left({\sin^2 \pi s\over (\pi s)^2}-
{\sin 2\pi s\over 2\pi s}\right)
\left (1 - { \sin^2 \pi s\over (\pi s)^2}\right)^{-1}.
\label{corrrev_corr2}
\end{eqnarray}
As it should be expected, the double integral over the both coordinates of 
this correlation function is equal to zero. This property is just the
normalization condition and should hold in arbitrary order of
expansion in $g^{-1}$.  

The quantities $f_2$, $f_3$, and $f_4$ are proportional to $g^{-2}$,
with some (geometry-dependent) prefactors of order unity. 
On the other hand, $f_1$ in $2D$ and $3D$ geometry depends
essentially on the distance $r=|{\bf r}_1-{\bf r}_2|$. In
particular, for $l\ll r\ll L$ 
\begin{eqnarray*}
& & f_1({\bf r}_1,{\bf r}_2)=\Pi^2({\bf r}_1,{\bf r}_2)\approx
\left\{
\begin{array}{ll}
\displaystyle{{1\over (\pi g)^2}\ln^2{L\over r}\ }, & \qquad 2D\\
\displaystyle{{1\over (4\pi^2\nu Dr)^2}}        \ , & \qquad 3D\ .
\end{array}
\right.
\end{eqnarray*}
Thus, for $l<r\ll L$, the contributions proportional to $f_1$ dominate
in Eq.(\ref{corrrev_corr2}), yielding
\begin{equation}
V^2\langle \vert \psi_k({\bf  r}_1) \psi_l({\bf  r}_2) \vert^2
\rangle_{E, \omega} -1 = {1\over
2}\Pi^2({\bf r}_1,{\bf r}_2)\ ,\ \ k \ne l. 
\label{corrrev_adm3}
\end{equation}
On the other hand, for the case of the quasi-1D geometry (as well as in 2D
and 3D for $r\sim L$), all quantities $f_1$, $f_2$, $f_3$, and $f_4$
are of order of $1/g^2$. Thus, the correlator 
$\sigma({\bf r_1},{\bf r_2},E,\omega)$ acquires a
non-trivial (oscillatory) frequency dependence on a scale $\omega\sim\Delta$
described by the second term in the r.h.s. of Eq.(\ref{corrrev_corr2}). In
particular, in the quasi-1D case the function $f_2-f_3$ determining
the spatial dependence of this term has the form
\begin{equation}
f_2-f_3=-{2\over 3g^2}\left[B_4\left({r_1\over L}\right)+
B_4\left({r_2\over L}\right)\right]\ ,
\label{f2f3}
\end{equation}
where $B_4(x)=x^4-2x^3+x^2-1/30$ is the Bernoulli polynomial. 

Let us remind the reader that the above derivation is valid for $\omega \ll
E_c$. In the range $\omega \gtrsim E_c$ 
the $\sigma$-model correlation functions entering
Eqs. (\ref{corrrev_e5}) can be calculated by means of the perturbation
theory \cite{ashk}, yielding 
\begin{eqnarray} 
\label{corrrev_pert}
V^2\langle \vert \psi_k({\bf  r}_1) \psi_l({\bf  r}_2) \vert^2
\rangle_{E, \omega} &=& 1 + \mbox{Re} \left\{ k_d(r)
\Pi_{\omega}({\bf r}_1, 
{\bf r}_2) \phantom{\frac{1}{V^2} \int} \right. \nonumber \\
 &+&\left.  \frac{1}{2} \left[ \Pi^2_{\omega} ({\bf r}_1,{\bf r}_2)
- \frac{1}{V^2} \int d{\bf r} d{\bf r}' \Pi^2_{\omega}
({\bf r},{\bf r}') \right] \right\}, \nonumber \\
V^2 \langle \psi^*_k({\bf r}_1) \psi_l({\bf r}_1) 
\psi_k({\bf r}_2) \psi^*_l({\bf  r}_2) \rangle_{E, \omega} 
&=& k_d(r) + \mbox{Re}\, 
\Pi_{\omega} ({\bf r}_1,{\bf r}_2), 
\end{eqnarray}
where $\Pi_{\omega} ({\bf r}_1,{\bf r}_2)$ is the finite-frequency
diffusion propagator
\begin{equation} \label{corrrev_Ask}
\Pi_{\omega} ({\bf r}_1,{\bf r}_2) = (\pi\nu)^{-1} \sum_{{\bf q}}
\frac{ \phi_q ({\bf r}_1) \phi_q ({\bf r}_2)}{Dq^2 - i\omega}, 
\end{equation}
and the summation in Eq. (\ref{corrrev_Ask}) now includes ${\bf q} = 0$.
As was mentioned, the perturbation theory should give correctly the
non-oscillatory (in $\omega$) part of the correlation functions at
$\omega\gg\Delta$. Indeed, it can be checked that
Eqs.(\ref{corrrev_pert}) 
match the results (\ref{corrrev_fin2}), (\ref{corrrev_states}) of the
non-perturbative calculation in this
regime. Furthermore, in the $1/g$ order [which means keeping only linear
in $\Pi_\omega$ terms in (\ref{corrrev_pert}) and neglecting $-i\omega$ in
denominator of Eq.(\ref{corrrev_Ask})] Eqs.(\ref{corrrev_pert}),
(\ref{corrrev_Ask}) reproduce 
the exact results (\ref{corrrev_fin2}), (\ref{corrrev_states}) even at small
frequencies $\omega\sim\Delta$. We stress, however, that the perturbative 
calculation is not justified in this region and only the supersymmetry
method provides a rigorous derivation of these results. 

 Generalization to a system with unbroken time 
reversal symmetry (orthogonal ensemble) is straightforward \cite{bm3}; in
the $1/g$-order Eqs.(\ref{corrrev_fin1}), (\ref{corrrev_fin2}), and
(\ref{corrrev_states}) are modified as follows:
\begin{equation} \label{corrrev_fin1o}
V^2\langle \vert \psi_k({\bf r_1}) \psi_k({\bf r_2}) \vert^2
\rangle_{E} 
= \left[ 1 + 2k_d (r) \right] \left[ 1 + 2 \Pi ({\bf r_1},
{\bf r_2}) \right]\ ,
\end{equation}
\begin{equation} \label{corrrev_fin2o}
V^2\langle \vert \psi_k({\bf r_1}) \psi_l({\bf r_2}) \vert^2
\rangle_{E, \omega} - 1  =  2 k_d(r) \Pi ({\bf r_1},
{\bf r_2})\ ,
\end{equation}
\begin{equation} \label{corrrev_stateso}
 V^2 \langle \psi^*_k({\bf r_1}) \psi_l({\bf r_1}) 
\psi_k({\bf r_2}) \psi^*_l({\bf r_2}) \rangle_{E, \omega}
 = k_d(r) + \left[ 1 + k_d(r) \right]
\Pi({\bf r_1},{\bf r_2}),\ \ \ k \ne l.  
\end{equation}

\paragraph{IPR fluctuations.}
Using the supersymmetry method, one can calculate also higher order
correlation functions of the eigenfunction amplitudes. In particular, the
correlation function $\langle|\psi_k^4({\bf r_1})||\psi_k^4({\bf
r_2})|\rangle_E$ determines fluctuations of the inverse participation
ratio (IPR) $P_2=\int d{\bf r}|\psi^4({\bf r})|$. Details of the
corresponding calculation can be found in Ref.\cite{fm95a}; the result
for the relative variance of IPR, $\delta(P_2)=\mbox{var}(P_2)/\langle
P_2\rangle^2$ reads
\begin{equation}
\delta(P_2)={8\over\beta^2}\int {d{\bf r}d{\bf r'}\over V^2}
\Pi^2({\bf r}, {\bf r'})={32 a_d\over\beta^2 g^2},
\label{corrrev_varipr}
\end{equation}
with a numerical coefficient $a_d$ defined in Section~\ref{c2}
(see Eqs.~(\ref{e2.15}), (\ref{e2.15a})). 
The fluctuations (\ref{corrrev_varipr}) have the same relative
magnitude ($\sim 1/g$)
as the famous universal conductance fluctuations. Note also
that extrapolating Eq.(\ref{corrrev_varipr}) to the Anderson transition point,
where $g\sim 1$, we find $\delta(P_2)\sim 1$, so that the magnitude of
IPR fluctuations is of the order of its mean value (which is, in turn,
much larger than in the metallic regime; see Section~\ref{c5}).

Eq.~(\ref{corrrev_varipr}) can be generalized onto higher IPR's $P_q$
with $q>2$,
\be
{\mbox{var}(P_q)\over\langle P_q\rangle^2}\simeq {2\over
\beta^2}q^2(q-1)^2 
\int {d{\bf r}d{\bf r'}\over V^2}
\Pi^2({\bf r}, {\bf r'})={8 q^2(q-1)^2 a_d\over\beta^2 g^2},
\label{ipr41}
\end{equation}
so that the relative magnitude of fluctuations of $P_q$ is $\sim
q(q-1)/g$. Furthermore, the higher irreducible moments (cumulants)
$\langle\langle P_q^n \rangle\rangle$, $n=2,3,\ldots$, have the form
\begin{eqnarray}
{\langle\langle P_q^n \rangle\rangle\over \langle P_q\rangle^n}&=&
{(n-1)!\over 2} \left[{2\over\beta}q(q-1)\right]^n\int
{d{\bf r_1}\ldots d{\bf r_n}\over V^n}\Pi({\bf r_1},{\bf r_2})\ldots
\Pi({\bf r_n},{\bf r_1}) \nonumber \\
&=&{(n-1)!\over 2} \mbox{Tr}  \left[{2\over\beta}q(q-1)\Pi\right]^n\ ,
\label{ipr42}
\end{eqnarray}
where $\Pi$ is the integral operator with the kernel $\Pi({\bf r},{\bf
r'})/V$. This is valid provided $q^2 n\ll 2\beta\pi
g$. Prigodin and Altshuler  \cite{prigal98} obtained
Eq.~(\ref{ipr42}) starting from the assumption that the eigenfunction
statistics is described by the Liouville theory. 
According to (\ref{ipr42}), the distribution function ${\cal P}(P_q)$
of the IPR $P_q$ (with $q^2/\beta\pi g\ll 1$) decays exponentially in the
region $q(q-1)/g\ll P_q/\langle P_q\rangle -1\ll 1$,
\be
\label{ipr43}
{\cal P}(P_q)\sim\exp\left\{-{\pi\beta\over 2}{\epsilon_1\over\Delta}
{ P_q/\langle P_q\rangle -1\over q(q-1)}\right\}\ ,
\ee
where $\epsilon_1$ is the lowest non-zero eigenvalue of the diffusion
operator $-D{\bf\nabla}^2$.

The perturbative calculations show that the cumulants of the IPR's are
correctly reproduced 
(in the leading order in $1/g$) if one assumes  \cite{prigal98}
that the statistics of the eigenfunction envelopes 
$|\psi^2({\bf r})|_{\rm smooth}=e^{\theta({\bf r})}$ is governed by
the Liouville theory (see e.g. \cite{zz,i_kogan96})
 defined by the functional integral
\be
\label{ipr44}
\int D\theta\,\delta\left(\int {d^d r\over V} e^\theta-1\right)\exp\left\{
-{\beta\pi\nu D\over 4}\int d^d{\bf r}({\bf
\nabla}\theta)^2\right\}\ldots
\ee
We will return to
this issue in Sec.~\ref{c4} where  the asymptotics of
the IPR distribution function will be discussed. 
We will see that these ``tails'' 
governed by rare realizations of disorder
are described by saddle-point solutions
which can be also obtained from the Liouville theory description
(\ref{ipr44}). The multifractal dimensions (\ref{multifr5}) can 
be found from the Liouville theory as well \cite{zz,i_kogan96}. It should
be stressed, however, that this agreement between the supermatrix
$\sigma$-model governing the eigenfunctions statistics and the
Liouville theory is not exact, but only holds 
in the leading order in $1/g$.

Let us note that
the correlations of eigenfunction amplitudes determine also
fluctuations of matrix elements of an operator of some (say, Coulomb) 
interaction computed on
eigenfunctions $\psi_k$ of the one-particle Hamiltonian in a random
potential. Such a problem naturally arises, when one wishes to study
the effect of interaction onto statistical properties of excitations in
a mesoscopic sample (see Section~\ref{c9}).

\subsubsection{Ballistic effects}
\label{s3.3.4}

\paragraph{Ballistic systems.}
The above consideration can be generalized to a ballistic chaotic
system, by applying a recently developed ballistic generalization of
the $\sigma$-model \cite{mk3,aasa1,aasa2}. The results are then
expressed \cite{bmm2}
in terms of the (averaged over the direction of velocity)
kernel $g({\bf r_1}, {\bf n_1}; {\bf r_2}, {\bf n_2})$ of the
Liouville operator $\hat{K}=v_F{\bf n}\nabla$ governing the classical
dynamics in the system,
\begin{eqnarray}
\label{corrrev_ball}
& & \Pi_B ({\bf r_1}, {\bf r_2})
= \int d{\bf n_1} d{\bf n_2}\, g({\bf r_1}, {\bf n_1}; {\bf r_2},
{\bf n_2}); \nonumber\\
& & \hat K g({\bf r_1}, {\bf n_1}; {\bf r_2}, {\bf n_2}) =
 \left( \pi \nu \right)^{-1} \left[ \delta({\bf r_1} - {\bf r_2})
\delta({\bf n_1} - {\bf n_2}) - V^{-1} \right].
\end{eqnarray} 
Here ${\bf n}$ is a unit vector determining the direction of momentum,
and normalization $\int d{\bf n}=1$ is used. 
Equivalently,  the function $\Pi_B({\bf r_1},{\bf r_2})$ can be defined
as
\begin{equation}
\label{corrrev_ball2}
\Pi_B({\bf r_1},{\bf r_2})=\int_0^\infty dt\int d{\bf n_1}
\,\tilde{g}({\bf r_1},{\bf n_1},t;{\bf r_2})\ ,
\end{equation}
where $\tilde{g}$ is determined by the evolution equation
\begin{equation}
\label{corrrev_ball3}
\left({\partial\over\partial t}+v_F{\bf n_1 \nabla_1}\right)
\tilde{g}({\bf r_1},{\bf n_1},t;{\bf r_2})=0\ ,\qquad t>0
\end{equation}
with the boundary condition
\begin{equation}
\label{ball4}
\tilde{g}|_{t=0}=(\pi\nu)^{-1}[\delta({\bf r_1}-{\bf r_2})-V^{-1}].
\end{equation}
Eq.(\ref{corrrev_ball}) is a
natural ``ballistic'' counterpart of Eq.(\ref{corrrev_diff}). 
Note, however, that $\Pi_B({\bf r_1},{\bf r_2})$ contains a
contribution $\Pi_B^{(0)}({\bf r_1},{\bf r_2})$
of the straight line motion from ${\bf r_2}$ to
$\bf{r_1}$ (equal to $1/(\pi p_F |\bf r_1-\bf r_2|)$ in 2D and to
$1/2(p_F |\bf r_1-\bf r_2|)^2$ in 3D), which is 
nothing else but the smoothed version of the function 
$k_d(|{\bf r_1}-{\bf r_2}|)$. For this reason, $\Pi({\bf r_1},{\bf
r_2})$ in Eqs.~(\ref{corrrev_fin1o})--(\ref{corrrev_stateso}) should
be replaced in the ballistic case by 
$\Pi({\bf r_1},{\bf r_2})=\Pi_B({\bf r_1},{\bf r_2})-\Pi_B^{(0)}
({\bf r_1},{\bf r_2})$. At large distances $|{\bf r_1}-{\bf r_2}|\gg
\lambda_F$ the (smoothed) correlation function takes in the leading
approximation the form
\begin{equation}
\label{corrrev_ball1}
V^2\alpha({\bf r_1}, {\bf r_2}, E) = 1 + {2\over \beta} \Pi_B({\bf r_1},
{\bf r_2}).
\end{equation}
 A formula
for the variance of matrix elements closely related to
Eq.(\ref{corrrev_ball1})
was obtained in the semiclassical approach in Ref.~\cite{eckhardt95}.
In a  recent paper \cite{srednicki98} a similar
generalization of the Berry formula for $\langle\psi_k^*({\bf
r_1})\psi_k({\bf r_2})\rangle$ was proposed.

Eq.(\ref{corrrev_ball1}) shows that correlations in eigenfunction amplitudes
in remote points are determined by the classical dynamics in the
system. It is closely related to the phenomenon of scarring of
eigenfunctions by the classical orbits
\cite{heller91,agfish93}. 
Indeed, if $\bf {r_1}$ and ${\bf r_2}$ belong to a short periodic
orbit, the function $\Pi_B({\bf r_1},{\bf r_2})$ is positive, so that
the amplitudes $|\psi_k({\bf r_1})|^2$ and $|\psi_k({\bf r_2})|^2$
are positively correlated. This is a reflection of the ``scars''
associated with this periodic orbits and a quantitative
characterization of their strength in the coordinate space. 
Note that this effect gets smaller
with increasing energy $E$ of eigenfunctions. Indeed, for a strongly
chaotic system and for $|{\bf r_1}-{\bf r_2}|\sim L$ ($L$ being the
system size), we have in the 2D case
$\Pi_B({\bf r_1},{\bf r_2})\sim\lambda_F/L$,
so that the magnitude of correlations decreases as $E^{-1/2}$. 
The function $\Pi_B({\bf r_1},{\bf r_2})$ was explicitly calculated 
in Ref.~\cite{bmm2}
for a circular billiard with  diffuse surface scattering (see
Section~\ref{c8}).

\paragraph{Ballistic effects in diffusive systems.}
We return now to the question of deviations of the eigenfunction
amplitude distribution from the RMT in a diffusive 3D sample.
As was shown in Sec.~\ref{s3.3.1}, such 
deviations are 
controlled by the parameter $\kappa=\Pi({\bf r},{\bf r})$, see
Eqs. (\ref{e3.12})--(\ref{e3.15}). 
The physical meaning of the parameter $\kappa$ is the time-integrated
return probability, see Eq. (\ref{corrrev_ball2}) generalizing
definition of $\Pi({\bf r_1},{\bf r_2})$ to the ballistic case. The
contribution to this return probability from the times larger than the
momentum relaxation time, $t>\tau$, is given by
$$\Pi^{\rm diff}({\bf r},{\bf r}) =(\pi\nu V)^{-1}\sum_{|{\bf q}|\lesssim 1/l}
(D{\bf q}^2)^{-1}\ .$$ 
The sum over the momenta diverges on the ultraviolet bound in $d\ge
2$, so that the cut-off at $q\sim 1/l$ is required. This results in
Eq.~(\ref{e3.15}) in 2D and in $\Pi^{\rm diff}({\bf r},{\bf r})\sim
1/(k_F l)^2$ in the 3D case.  There exists, however, an additional --
ballistic -- contribution to $\Pi({\bf r},{\bf r})$, which comes from
the times $t$ shorter than the mean free time
$\tau$. Diagrammatically, it is determined by the first term of the
diffuson ladder contributing to $\Pi({\bf r},{\bf r})$ (that with one
impurity line), i.e. by the probability to hit an impurity and to be
rejected back after a time $t\ll\tau$. Contrary to the diffusive
contribution, which has a universal form and is determined by the
value of the diffusion constant $D$ only, the ballistic one is strongly
dependent on the microscopic structure of the disorder. In particular,
in the case of the white-noise disorder we find
\be 
\label{3d.7}
\Pi^{\rm ball}({\bf r},{\bf r})=\left\{
\begin{array}{ll}
\displaystyle{
{1\over \pi\nu v_F^2\tau}\int{(dq)\over q^2}={1\over 2\pi
g}\ln{l\over\lambda_F}  }\ ,& \ \ 2D \\
\displaystyle{
{\pi\over 4\nu v_F^2\tau}\int{(dq)\over q^2} \sim {1\over k_Fl}   }
\ ,&\ \ 3D
\end{array}
\right.
\ee
Note that the integrals over the momenta are again divergent at large
$q$ -- precisely in the same way as in the diffusive region, but with
different numerical coefficients -- and are now cut-off at $q\sim
k_F$. 

The total return probability is given by the sum of the short-scale
(ballistic) and long-scale (diffusive) contributions. It is important
to notice, however, that the single-scattering contribution
(\ref{3d.7}) should be divided by 2 in the orthogonal symmetry case,
because the corresponding trajectory is identical to its time reversal.
Thus, $\kappa=\Pi^{\rm diff}({\bf r},{\bf r})+
\Pi^{\rm ball}({\bf r},{\bf r})$ for $\beta=2$ and 
$\kappa=\Pi^{\rm diff}({\bf r},{\bf r})+
(1/2)\Pi^{\rm ball}({\bf r},{\bf r})$ for $\beta=1$.
We see that for the white-noise random potential in 3D the
return probability is dominated by the ballistic
contribution, yielding $\kappa\sim 1/k_Fl$. 
In the 2D case, taking into account of the
ballistic contribution modifies only the argument of the logarithm in
(\ref{e3.15}). Furthermore, even in the quasi-1D geometry the
non-universal short-scale effects can be important. Indeed, if we
consider a 3D sample of the quasi-1D geometry ($L_x,\,L_y\ll L_z$),
the diffusion contribution will be given by Eq.~(\ref{e3.14}),
$\Pi^{\rm diff}({\bf r},{\bf r})\sim 1/g$, 
while the ballistic one will be
$\Pi^{\rm ball}({\bf r},{\bf r})\sim 1/k_f l$. Therefore, the
diffusion contribution is dominant only provided $g\ll k_F l$.

On the other hand, let us consider the opposite case of a smooth
random potential with correlation length $d\gg\lambda_F$. Then the
scattering 
is of small-angle nature and the probability for a particle to return
back in a time $t\ll\tau$ is exponentially small, so that 
$\Pi^{\rm ball}({\bf r},{\bf r})$ can be neglected. Therefore, the
return probability $\kappa$ in Eqs.~(\ref{e3.10})--(\ref{e3.13}) is
correctly given by the diffusion contribution, see Eq.~(\ref{e3.15})
for 2D and the estimate below it for 3D. Thus the corrections to the
``body'' of the distribution function are properly given by the
$\sigma$-model in this case.

\section{Asymptotic behavior of distribution functions and
anomalously localized states}
\label{c4}

\setcounter{equation}{0}

In this section, we  discuss  asymptotics of distribution
functions of various quantities characterizing wavefunctions in a
disordered system. Asymptotic behavior of these distribution functions
is determined by rare realizations of the disorder producing the
states, which show much stronger localization features than typical
states in the system. We call such states ``anomalously localized
states'' (ALS). 

It was found
by Altshuler, Kravtsov and Lerner (AKL) \cite{akl} that distribution
functions of conductance, density of states, local density of
states, and relaxation times have slowly decaying
logarithmically--normal (LN) asymptotics at large values of the
arguments. These results were obtained within the renormalization
group treatment of the  $\sigma$--model. The validity of
this RG approach is restricted to 2D and $2+\epsilon$--dimensional
systems, with $\epsilon\ll 1$. 
On the other hand, the conductance, LDOS and relaxation times
fluctuations in strictly 1D disordered chains, where all states are
strongly localized, were studied with the use of Berezinski and
Abrikosov--Ryzhkin techniques
\cite{melnikov80a,melnikov80b,abrikosov81,alprig88,alprig89}. The
corresponding 
distributions were found to be of the LN form, too. It was conjectured on
the basis of this similarity \cite{akl,alprig89,krav}
that even in a metallic sample there is a
finite probability to find ``almost localized'' eigenstates, and that  
these states govern the slow asymptotic decay of the distribution
functions. Similar conclusion \cite{fm95a} is implied by the exact
results for the statistics of the eigenfunction amplitude in the
quasi-one-dimensional case, which shows the identical asymptotic
behavior in the
localized and metallic regimes, see Section 3,
Eqs.~(\ref{anom_adm15c}),  (\ref{anom3as}). 

A new boost to the activity in this direction was given by the paper of
Muzykantskii and Khmelnitskii \cite{mk1}, who proposed to use the
saddle-point method for the supersymmetric $\sigma$-model in order to
calculate the long-time dispersion of the average conductance
$G(t)$. Their idea was to reproduce the AKL result by means of a more
direct calculation. However, they found a different, power-law decay
of $G(t)$ in an intermediate range of times $t$ in 2D. As was shown by
the author \cite{m95} (and then reproduced in \cite{mk2} within the
ballistic $\sigma$-model approach), the far asymptotic behavior is of
log-normal form and is thus in agreement with AKL. The saddle-point
method of Muzykantskii and Khmelnitskii allowed also to study the
asymptotic behavior of distribution functions of other quantities:
relaxation times \cite{mk1,m95,mk2}, eigenfunction intensities
\cite{fe1,fe2}, local density of states \cite{m96}, inverse
participation ratio \cite{mf95u,m97}, level curvatures
\cite{kravtsov97a} etc. 
The form of the saddle-point solution describes directly the spatial
shape of the corresponding anomalously localized state \cite{mk1,m97}. 

We will consider the unitary symmetry ($\beta=2$) throughout this
section; in the general case, 
the conductance $g$ in the exponent  of the distribution
functions is replaced
by $(\beta/2)g$ (we will sometimes do it explicitly in the
end of the calculation).

\subsection{Long-time relaxation}
\label{s4.1}

In this subsection we study (mainly following Refs.~\cite{mk1,m95})
the asymptotic (long-time) behavior of
the relaxation processes in an open disordered conductor. One possible
formulation of the problem is to consider the time-dependence of the
average conductance $G(t)$ defined by the non-local (in time)
current-voltage relation
\be
I(t)=\int_{-\infty}^t\,dt'G(t-t')V(t')
\label{e4.1}
\ee
Alternatively, one can study the decay low, i.e. the survival
probability $P_s(t)$ 
for a particle injected into the sample at $t=0$  to be found 
there after a time $t$. Classically, $P_s(t)$  decays
according to the exponential law, $P_s(t)\sim e^{-t/t_D}$, where
$t_D^{-1}$ is the lowest eigenvalue of the diffuson operator
$-D\nabla^2$ with the proper boundary conditions. The time $t_D$ has
the meaning of the time of diffusion through the sample, and
$t_D^{-1}$ is the Thouless energy (see Section~\ref{c2}). The same
exponential decay holds for the conductance $G(t)$, where it is induced by
the weak-localization correction. The quantities
of interest can be expressed in the form of the $\sigma$-model
correlation function
\be
G(t),\ P_s(t)\sim \int {d\omega\over 2\pi}e^{-i\omega t}\int DQ({\bf
r})A\{Q\} e^{-S[Q]}\ ,
\label{e4.2}
\ee
where $S[Q]$ is given by Eq.~(\ref{anom_2}) with $\eta\to -2i\omega$.
The preexponential factor $A\{Q\}$ depends on specific
formulation of the problem but is not important for the leading
exponential behavior studied here.

Varying the exponent in
Eq.~(\ref{e4.2}) with respect to $Q$ and $\omega$, one gets
the equations \cite{mk1}
\bea
&& 2D\nabla(Q\nabla Q)+i\omega[\Lambda,Q]=0 
\label{e4.3}\\
&& {\pi\nu\over 2} \int d{\bf r}\,\mbox{Str}(\Lambda Q)=t 
\label{e4.4}
\eea
[We assume unitary symmetry ($\beta=2$); in the orthogonal symmetry
case the calculation is applicable with minor modifications and we
will present the result in the end.]
Note that in fact $\omega$ plays in (\ref{e4.2}) a role of the Lagrange
multiplier corresponding to the condition (\ref{e4.4}). Therefore, it
remains 
\begin{enumerate}
\item[i)] to find a solution $Q_\omega$ of Eq.(\ref{e4.3}) (which will
depend on $\omega$);
\item[ii)] to substitute it into the self-consistency equation
(\ref{e4.4}) and thus to fix $\omega$ as a function of $t$;
\item[iii)] to substitute the found solution $Q_t$ into
Eq.(\ref{e4.2}), which will yield
\end{enumerate}
\be
P_s(t)\sim\exp\left\{ {\pi\nu D\over 4}\mbox{Str}
\int (\nabla Q_t)^2\right\} 
\label{e4.5}
\ee
Note that Eq.~(\ref{e4.3}) is to be supplemented by the boundary
conditions
\be
Q|_{\mbox{leads}}=\Lambda
\label{e4.6}
\ee
at the open part of the boundary, and
\be
\nabla_{\bf n}Q|_{\mbox{insulator}}=0
\label{e4.7}
\ee
at the insulating part of the boundary (if it exists); $\nabla_{\bf
n}$ denotes here the normal derivative. 

It is not difficult to show \cite{mk1} that the solution of
Eq.~(\ref{e4.3}) has in the standard parametrization the only
non-trivial variable -- bosonic ``non-compact
angle''\footnote{$\theta_1$ is related to the eigenvalue $\lambda_1$
used in Sec.~\ref{s3.1} as $\lambda_1=\cosh\theta_1$.}
$0\le\theta_1<\infty$; all 
other coordinates being equal to zero. As a result, Eq.~(\ref{e4.3}) 
reduces to an equation for $\theta_1({\bf r})$ (we drop the subscript
``1'' below)
\be
\nabla^2\theta+{i\omega\over D}\sinh\theta=0\ ,
\label{e4.8}
\ee
the self-consistency condition (\ref{e4.4}) takes the form
\be
\pi\nu\int d^dr(\cosh\theta-1)=t\ ,
\label{e4.9}
\ee
and Eq.~(\ref{e4.5}) can be rewritten as
\be
\ln P_s(t)=-{\pi \nu D\over 2}\int d^dr(\nabla\theta)^2\ .
\label{e4.10}
\ee
For sufficiently small times, $\theta$ is small according to
(\ref{e4.9}), so that Eqs.~(\ref{e4.8}), (\ref{e4.9}) can be linearized
\bea
&&\nabla^2\theta+2\gamma\theta=0\ ;\qquad 2\gamma=i\omega/D\ ;
\label{e4.11} \\
&&{\pi\nu\over 2}\int d^dr \theta^2=t\ .
\label{e4.12}
\eea
This yields
\be
\theta=\left({2t\over\pi\nu}\right)^{1/2}\phi_1({\bf r})\ ,
\label{e4.13}
\ee
where $\phi_1$ is the  eigenfunction of the Laplace operator
corresponding to the lowest eigenvalue $2\gamma_1=1/Dt_D$. The
survival probability (\ref{e4.10}) reduces thus to
\be
\ln P_s(t) = {\pi\nu D\over 2}\int d^dr\theta\nabla^2\theta=-\pi\nu
D\gamma_1\int d^dr\theta^2=-t/t_D\ ,
\label{e4.14}
\ee
as expected. Eq.~(\ref{e4.14}) is valid (up to relatively small
corrections) as long as $\theta\ll 1$, i.e. for $t\Delta\ll 1$
($\Delta=1/\nu V$ being the mean level spacing). To find the behavior
at $t\gtrsim\Delta^{-1}$, as well corrections at $t<\Delta^{-1}$, one
should consider the exact (non-linear) equation (\ref{e4.8}), solution
of which depends on the sample geometry.

\subsubsection{Quasi-1D geometry}
\label{s4.1.1}

We consider a wire of a length $L$ and a cross-section $A$
  with open boundary conditions at
both edges, $\theta(-L/2)=\theta(L/2)=0$. Eqs.~(\ref{e4.8}),
(\ref{e4.9}) take the form
\bea
&& \theta''+2\gamma\sinh\theta=0 
\label{e4.15} \\
&&\int_{-L/2}^{L/2}dx(\cosh\theta-1)=t/\pi\nu A
\label{e4.16}
\eea
From the symmetry consideration $\theta(x)=\theta(-x)$ and
$\theta'(0)=0$, so that it is sufficient to consider the region
$x>0$. The solution of Eq.~(\ref{e4.15}) reads
\be
x=\int_{\theta(x)}^{\theta_0}
{d\vartheta\over 2\sqrt{\gamma(\cosh\theta_0-\cosh\vartheta)}} ,
\label{e4.17}
\ee
where $\theta_0$ is determined by the condition $\theta(L/2)=0$ yielding
\be
L=\int_0^{\theta_0}
{d\vartheta\over \sqrt{\gamma(\cosh\theta_0-\cosh\vartheta)}}
\label{e4.18}
\ee
In the large-$t$ region ($t\Delta\gg 1$) we will have $\theta_0\gg 1$,
and Eqs.~(\ref{e4.17}), (\ref{e4.18}) can be simplified to give
\bea
&& \theta(x)\simeq\theta_0 \left(1-{2x\over L}\right)\ ,
\label{e4.19} \\
&&\theta_0\simeq\ln {2\theta_0^2\over \gamma L^2}
\label{e4.20}
\eea
Substituting this into the condition (\ref{e4.16}) allows to relate
$\theta_0$ to $t$,
\be
e^{\theta_0}={2\over\pi}t\Delta\theta_0
\label{e4.21}
\ee
Finally, substitution of Eqs.~(\ref{e4.19}) and
(\ref{e4.21}) into (\ref{e4.10}) yields the log-normal asymptotic
behavior of $P_s(t)$ \cite{mk1}:
\be
\ln P_s(t)\simeq -g\ln^2 {t\Delta\over\ln(t\Delta)}
\ ; \qquad t\Delta\gg 1\ ,
\label{e4.22}
\ee
with $g=2\pi\nu AD/L$ being the dimensionless conductance. We remind
that Eq.~(\ref{e4.22}) has been derived for the unitary ensemble
($\beta=2$); in the general case, its r.h.s. should be multiplied by
$\beta/2$. 

Equation (\ref{e4.22}) has essentially the same form as the asymptotic
formula for $G(t)$ found by Altshuler and Prigodin \cite{alprig88} for the
{\it strictly} 1D sample with a length much {\it exceeding} 
the localization length:
\begin{equation}
G(t)\sim\exp\left\{-{l\over L}\ln^2(t/\tau)\right\}
\label{relax_13}
\end{equation}
If we replace in Eq.~(\ref{relax_13}) the 1D localization length
$\xi=l$ by the 
quasi-1D localization length $\xi=\beta\pi\nu AD$, we  reproduce 
the asymptotics (\ref{e4.22}) (up to a normalization of $t$ in the
argument of
$\ln^2$, which does not affect the leading term in the exponent for
$t\to\infty$). This is one more manifestation of the equivalence of
statistical properties of smooth envelopes of the
wave functions in 1D and quasi-1D samples \cite{fm94a} (see
Sec.~\ref{c3}). Furthermore, agreement of the results for the
metallic and the insulating samples demonstrates clearly that the
asymptotic ``tail'' (\ref{e4.22}) in the metallic 
sample is indeed due to anomalously localized eigenstates.

As another manifestation of this fact, Eq.~(\ref{e4.22}) can be
represented 
as a superposition of the simple relaxation processes with mesoscopically
distributed relaxation times \cite{akl}:
\begin{equation}
P_s(t)\sim\int dt_\phi e^{-t/t_\phi} {\cal P}(t_\phi)
\label{anom106a}
\end{equation}
The distribution function ${\cal P}(t_\phi)$ then behaves as follows:
\begin{equation}
{\cal P}(t_\phi)\sim\exp\{-g\ln^2(g\Delta t_\phi)\}\ ;\qquad t_\phi\gg
{1\over g\Delta}\equiv t_D
\label{anom107}
\end{equation}
This can be easily checked by substituting Eq.~(\ref{anom107}) into
Eq.~(\ref{anom106a}) and calculating the integral via the stationary
point method; the stationary point equation being
\begin{equation}
2gt_\phi\ln(g\Delta t_\phi)=t
\label{anom108}
\end{equation}
Note that Thouless energy $t_D^{-1}$ determines
the typical width of a level of an open system. Therefore, 
the formula (\ref{anom107}) concerns  indeed 
the states with anomalously small widths $t_\phi^{-1}$
in the energy space. 

The saddle-point solution $\theta({\bf r})$ provides a direct
information on the spatial shape of the corresponding ALS. This was
conjectured by Muzykantskii and Khmelnitskii \cite{mk1} and was
explicitly proven in \cite{m97} for the states
determining the distribution of eigenfunction amplitudes, see
Sec.~\ref{s4.2.1} below. Specifically, the smoothed (over a scale
larger than the Fermi wave length) intensity of the ALS is
$|\psi^2({\bf r})|={\cal N}^{-1} e^\theta({\bf r})$, where ${\cal N}$
is the normalization factor determined by the requirement
$\int d^dr |\psi^2({\bf r})|=1$. We get thus from Eq.~(\ref{e4.19})
\begin{equation}
|\psi^2({\bf r})|_{\mbox{smooth}}={\theta_0\over AL}e^{-2\theta_0|x|/L}
\label{anom112}
\end{equation}
with
\begin{equation}
\theta_0\simeq\ln\left({2\over\pi}t\Delta\ln (t\Delta)\right)\simeq
\ln\left({4\over\pi}g\Delta t_\phi \ln^2(g\Delta t_\phi)\right).
\label{anom111}
\end{equation}
Thus, the ALS, which gives a minimum to the level width $t_\phi^{-1}$,
has an exponential shape (\ref{anom112}), (\ref{anom111}). 

The saddle-point method allows also to find the corrections to
Eq.~(\ref{e4.14}) in the intermediate region $t_D\ll t\ll
\Delta^{-1}$, where $\theta_0\ll 1$ \cite{mirmuz}.  For this purpose, we
expand $\cosh\theta_0$ and $\cosh\vartheta$ in Eq.~(\ref{e4.18}) up to
the 4th order terms, which leads to the following relation between
$\gamma$ and $\theta_0$:
\be
\gamma={\pi^2\over 2L^2}\left(1-{1\over 8}\theta_0^2+\ldots\right)
\label{e4.24}
\ee
Further, we substitute Eq.~(\ref{e4.17}) into (\ref{e4.16}),
\be
\int_0^{\theta_0}{d\vartheta(\cosh\vartheta-1)\over
2\sqrt{\gamma(\cosh\theta_0-\cosh\vartheta)} } ={t\over 2\pi\nu A}
\label{e4.25}
\ee
and expand $\cosh\vartheta$ in (\ref{e4.25}) up to the 4th order
terms. This gives the relation
\be
{t\over 2\pi\nu A}=\theta_0^2{\pi\over 8\sqrt{2\gamma}}\left(1-{1\over
96}\theta_0^2\right) 
\label{e4.26}
\ee
Using Eqs.~(\ref{e4.15}), (\ref{e4.16}), we can rewrite the action in
the form
\be
S\equiv{\pi\nu AD\over 2}\int dx (\theta')^2=
2\pi\nu ADL\gamma(\cosh\theta_0-1)-2Dt\gamma
\label{e4.27}
\ee
Expressing now $\theta_0$ and $\gamma$ through $t$ according to
Eqs.~(\ref{e4.25}), (\ref{e4.26}), we find
\be
-\ln P_s(t) = S = {t\over t_D}\left(1-{1\over 2\pi^2 g}{t\over
t_D}+\ldots\right)\ ,
\label{e4.28}
\ee
with $t_D=L^2/\pi^2D$. In the general case, $g$ is
replaced by $(\beta/2)g$ here. Eq.~(\ref{e4.28}) is completely
analogous to the formula (\ref{anom_adm15b}),
(\ref{anom_adm15bo}) for the statistics of
eigenfunction amplitudes. It shows that the correction to the leading
term $-t/t_D$ in $\ln P_S$ becomes large compared to unity at
$t\gtrsim\sqrt{g}t_D$, though it remains small compared to the leading
term up to $t\sim gt_D\sim\Delta^{-1}$. 

The result (\ref{e4.28}) was also obtained
by Frahm \cite{frahm97} from rather involved calculations based
on the equivalence between the 1D $\sigma$-model and the Fokker-Plank
approach and employing the approximate solution of the
Do\-ro\-khov-Mello-Pe\-rey\-ra-Kumar 
equations in the metallic limit. The fact that the logarithm of the
quantum decay probability, $\ln P_S(t)$, starts to deviate strongly
(compared to unity) from the classical law, 
$\ln P_s^{\rm cl}(t)=-t/t_D$
at $t\sim t_D\sqrt{g}$ was observed in numerical simulations by
Casati, Maspero, and Shepelyansky \cite{casati97}. For related results
in the framework of the random matrix model see Sec.~\ref{s4.1.3}.

\subsubsection{2D geometry}
\label{s4.1.2}

We consider now a 2D disk-shaped sample of a radius $R$ with an open
boundary. If the problem is formulated in terms of
the conductance,
we can assume the two leads attached to the disk boundary
to be of almost semicircular shape, with relatively narrow insulating
intervals between them. Then we can approximate the boundary
conditions by using Eq.~(\ref{e4.6}) for all the boundary. 
In fact, in view of the logarithmic dependence of the saddle
point action on $R$ (see below), the result does not depend to the
leading approximation on the specific shape of the sample and  the
leads attached. With the rotationally invariant form of the boundary
condition, the minimal action corresponds to the function $\theta$
depending on the radial coordinate $r$ only. We get therefore the
radial equation 
\begin{equation}
\theta'' + \theta'/r+2\gamma\sinh\theta=0\ ;\qquad 0\le r\le R
\label{relax_1}
\end{equation}
(the prime denotes the derivative $d/dr$)
with the boundary conditions:
\begin{eqnarray}
&&\theta(R)=0  \label{relax_6a}\ ,\\
&& \theta'(0)=0 \label{relax_7a}
\end{eqnarray}
The condition (\ref{relax_7a}) follows from the requirement of analyticity
of the field in the disk center.

Assuming that characteristic values of $\theta$ satisfy the condition
$\theta\gg 1$ (which corresponds to $t\Delta\gg 1$),
one can replace
$\sinh\theta$ by $e^\theta/2$. Eq.(\ref{relax_1}) can be then easily
integrated, and its general solution reads:
\begin{equation}
e^{\theta(r)}={2C_1^2\over\gamma} 
{C_2r^{C_1-2}\over (C_2 r^{C_1}+1)^2}\ ,
\label{relax_g}
\end{equation}
with two integration constants $C_1$ and $C_2$.  
 To satisfy the boundary condition (\ref{relax_7a}), we have to choose
$C_1=2$. Furthermore, the above assumption $\theta(0)\gg1$ implies
that $2C_2/\gamma\gg1$. Therefore, the second boundary condition
(\ref{relax_6a}) is satisfied if $C_2\simeq 8/\gamma R^4$, and the
solution can be written in the form
\begin{equation}
e^{\theta(r)}\simeq[(r/R)^2 + \gamma R^2/8]^{-2}
\label{relax_3}
\end{equation}
Using now the self-consistency equation (\ref{e4.9})
one finds $\gamma=4\pi^2\nu/t$. Finally, calculating the action 
on the saddle point (\ref{relax_3}), we find \cite{mk1}
\begin{equation}
-\ln P_s(t)=S\simeq 8\pi^2\nu D\ln(t\Delta)\ ,
\label{relax_5b}
\end{equation} 
The above treatment is valid provided $\theta'(r)<l^{-1}$ on the saddle
point solution, which is the condition of the applicability of the
diffusion approximation (here $l$ is the mean free path). In
combination with the assumption $\theta(0)\gg 1$ this means that $1\ll
t\Delta\ll (R/l)^2$. 

Now let us consider the region of still longer times, $t\gg \Delta^{-1}
(R/l)^2$. In order to support the applicability of the diffusion
approximation, we should search for a function $\theta(r)$ 
minimizing the action with an additional restriction $\theta'\le
Al^{-1}$. Here $A$ is a parameter of the order of unity, which can not
be fixed within the diffusion approximation. We will see however that
the saddle--point action depends on $l$ through $\ln(R/l)$ only, and
thus does not depend on $A$ in the leading order, so that we can set
$A=1$. 
Since the derivative has a tendency to increase in the vicinity of $r=0$,
the restriction can be implemented \cite{m95}
via replacing the boundary conditions
(\ref{relax_7a}) by $\theta'(r_*)=0$, where the parameter
$r_*$ will be specified below. The solution reads now:
\begin{equation}
e^{\theta(r)}=\frac { (r/R)^{C-2}} {[(r/R)^C + {C+2\over C-2}
(r_*/R)^C]^2}\ ;\ \ r_*\le r\le R
\label{relax_6}
\end{equation}
The function $\theta(r)$ is meant as being constant within the vicinity
$|r|\le r_*$ of the disk center. 
The condition $\theta'\le l^{-1}$ yields $r_*\sim lC$. 
It is important to note that the result does not depend on details
of the cut-off procedure. For example, one gets the same results if
one chooses the boundary condition in the form $\theta'(r_*)=1/l$.
The crucial point  is that the maximum derivative $\theta'$ should not
exceed $1/l$. The constant $C$ is to be found from the
self-consistency  equation
(\ref{e4.9}) which can be reduced to the following form:
\begin{equation}
\left({R\over r_*}\right)^C= {2t\over\pi^2\nu R^2} {C^2\over C-2}
\label{relax_8a}
\end{equation}
Neglecting corrections of the $\ln(\ln\cdot)$ form, we find
\begin{equation}
C\simeq {\ln (t\Delta) \over \ln (R/r_*)}\simeq
{\ln (t\Delta) \over \ln (R/l)}
\label{relax_7}
\end{equation}
The action (\ref{e4.10}) is then equal to
\begin{equation}
S\simeq\pi^2\nu D (C+2)^2\ln (R/r_*)
 \simeq\pi^2\nu D \frac {\ln^2[t\Delta(R/l)^2]} {\ln (R/l)}
\label{relax_8}
\end{equation}

Combining Eqs.(\ref{relax_5b}) and (\ref{relax_8}) and introducing the
factor $\beta/2$ for generality, we get thus the following 
long-time asymptotics of $P_s(t)$ (or $G(t)$) \cite{m95}:
\begin{eqnarray}
&  P_s(t)\sim
(t\Delta)^{-2\pi\beta g}\ , & \quad 1\ll t\Delta\ll (R/l)^2 
\label{relax_9d}\\
& P_s(t)\sim \exp\left\{-{\pi\beta g\over 4} {\ln^2(t/g\tau)\over
\ln(R/l)}\right\}\ ,  &\quad 
t\Delta\gg (R/l)^2
\label{relax_9}
\end{eqnarray}
where $g=2\pi\nu D$ is the dimensionless conductance per square in 2D
and $\tau$ is the mean free time. 

The far asymptotic behavior (\ref{relax_9}) is of the
log-normal form and very similar to that found by AKL (see Eq.~(7.8) in
Ref.~\cite{akl}). It differs only by the factor $1/g$ in the argument
of $\ln^2$. It is easy to see however that this difference disappears if
one does the last step of the AKL calculation with a better accuracy.
Let us consider for this purpose the intermediate expression of AKL
(Ref.~\cite{akl}, Eq.~(7.11)):
\begin{equation}
G(t)\propto -{\sigma\over\tau}\int_0^\infty e^{-t/t_\phi}\exp\left[
-{1\over 4u}\ln^2{t_\phi\over\tau}\right]{dt_\phi\over t_\phi}
\label{relax_10}
\end{equation}
where $u\simeq{1\over 2\pi^2\nu D}\ln{R\over l}$ in the weak
localization region in 2D, which we are considering. Evaluating the
integral (\ref{relax_10}) by the saddle point method, we find
\begin{eqnarray}
G(t)&\sim& \exp\left\{-{1\over 4u}\ln^2{2ut\over\tau}\right\}
\nonumber\\
&\sim& \exp\left\{-{\pi g\over 4} {\ln^2(t/g\tau)\over
\ln(R/l)}\right\} \ ,
\label{relax_11}
\end{eqnarray}
where we have kept only the leading term in the
exponent. Eq.~(\ref{relax_11}) is in {\it exact} agreement with  
Eq.~(\ref{relax_9}) for $\beta=1$ (AKL assumed the orthogonal
symmetry of the ensemble). 
Therefore, the supersymmetric treatment confirms the AKL result and
also establishes the region of its validity. It is instructive to
represent the obtained results in terms of the superposition of simple
relaxation processes with mesoscopically distributed relaxation times
$t_\phi$:
\begin{equation}
G(t), P_s(t)\sim\int {dt_\phi\over t_\phi} e^{-t/t_\phi} 
{\cal P}(t_\phi)\ .
\label{relax_11s}
\end{equation}
Equations (\ref{relax_9d}), (\ref{relax_9}) lead then to the following
result for the distribution function ${\cal P}(t_\phi)$ \cite{m95}:
\begin{equation}
{\cal P}(t_\phi)\sim\left\{
\begin{array}{ll}
(t_\phi/t_D)^{-2\pi\beta g}\ , &\ \ 
t_D\ll t_\phi\ll t_D \left({R\over l}\right)^2 \\
\exp\left\{-{\pi\beta g\over 4} {\ln^2(t_\phi/\tau)\over \ln
(R/l)}\right\} \ ,&\ \ t_\phi\gg t_D \left({R\over l}\right)^2\ , 
\end{array}
\right.
\label{relax_11t}
\end{equation}
where $t_D\simeq R^2/D$ is the time of diffusion through the sample. 

The smooth envelope of the ALS corresponding to the intermediate region
$t_D\ll t_\phi\ll t_D(R/l)^2$ has according to Eq.~(\ref{relax_g}) the
following spatial structure
\begin{equation}
|\psi^2(r)|_{\mbox{smooth}}={\cal N}^{-1}e^{\theta(r)}=
{1\over 8\beta\pi D t_\phi}{1\over [(r/R)^2+R^2/8\beta Dt_\phi]^2}\ ,
\label{anom125}
\end{equation}
Thus, this ALS has an effective localization length
$\xi_{ef}\sim R(t_D/t_\phi)^{1/2}$, with the intensity decreasing
 as $1/r^4$ outside
the region of the extent $\xi_{ef}$. As to the ultra-long-time region,
Eqs.~(\ref{relax_6}), (\ref{relax_7}) indicate that now the effective
localization length is given by $l_*^{(t)}$ defined as
\begin{equation}
l_*^{(t)}=\gamma_t l\ ;\qquad
\gamma_t\simeq{\ln(t_\phi/t_D)\over\ln(R/l)}\ . 
\label{anom125b}
\end{equation}
The ALS intensity decays in a power-law manner for $r>l_*^{(t)}$, with
an exponent depending on $t_\phi$:
\begin{equation}
|\psi^2(r)|\sim{1\over l_*^{(t)2}}\left({r\over
l_*^{(t)}}\right)^{-\gamma_t-2} \ , \qquad l_*^{(t)}\le r\le R.
\label{anom125c}
\end{equation}

\subsubsection{Random matrix model}
\label{s4.1.3}

Here we mention briefly the results on the quantum decay law obtained
by Savin and Sokolov \cite{savinsok} within the RMT model. 
This will allow us to see the similarities and the differences between
the diffusive systems and the random matrix model.
The model describes 
a Hamiltonian of an open chaotic system by a Gaussian random matrix
coupled to $M$ external (decay) channels. The found decay law has the
form
\be
P_s(t)\sim (1+\Gamma_Wt/M)^{-M}\ ,
\label{e4.500}
\ee 
where $\Gamma_W=MT\Delta/2\pi$ is a typical width of the eigenstate,
with $T$ characterizing the channel coupling ($T=1$ for ideal
coupling, see also Sec.~\ref{c6}). In this case, the product $MT$
plays a role of the dimensionless conductance $g$ (in contrast to the
diffusive case where $g$ is governed by the bulk of the system,
here it is determined by the number of decay channels and the strength
of their coupling). For not too large
$t$ ($t\Delta T\ll 1$), Eq.~(\ref{e4.500}) yields the classical decay
law, $P_s(t)\sim e^{-t\Gamma_W}$, with the corrections of the form
\be
\ln P_s(t)=-t\Gamma_W(1-\Gamma_Wt/2M+\ldots)\ ,
\label{e4.501}
\ee
which is similar to the results found for the diffusive systems (see
e.g. Eq.~(\ref{e4.28})). 
At large $t\gg (\Delta T)^{-1}$, the decay take the power-law
asymptotic form \cite{lewenkopf91}
\be
\ln P_s(t)\simeq -M\ln(\Gamma_Wt/M)\ . 
\label{e4.502}
\ee

\subsubsection{Distribution of total density of states}
\label{s4.1.4}

Here we discuss the contribution of ALS to the asymptotic
behavior of the distribution function ${\cal P}({\nu})$ of the
total density of states (DOS), 
\begin{equation}
\nu(E)={-1\over \pi V}\,\mbox{Im} \int d^dr\, G_R({\bf r}, {\bf r}; E)
\label{anom121}
\end{equation}
(in the present subsection we denote the average DOS as $\nu_0$ to
distinguish it from the fluctuating quantity $\nu(E)$).
A resonance state with an energy $E$ and width $t_\phi^{-1}$ gives a
following contribution to ${\nu}(E)$:
\begin{equation}
{\nu}_{ALS}(E)={2\over \pi}{t_\phi\over
V}={2\over\pi}t_\phi\Delta\nu_{0}
\label{anom122}
\end{equation}
We expect that the asymptotic behavior of ${\cal
P}({\nu})$ at sufficiently large $\delta \nu=\nu-\nu_0$ is 
determined by a single anomalously localized state. This yields
\cite{m97} 
\begin{equation}
{\cal P}({\nu})\sim {\cal P}\left(t_\phi={\pi\,{\delta\nu}\over
2\Delta\nu_{0}}\right)\sim
\exp\left\{-g\ln^2(g\,{\delta\nu}/\nu_{0})\right\}. 
\label{anom123}
\end{equation}
in the quasi-1D case and
\begin{equation}
{\cal P}({\nu})\sim {\cal P}\left(t_\phi={\pi\,{\delta\nu}\over
2\Delta\nu_{0}}\right)\sim 
\left\{
\begin{array}{ll}
(g\,{\delta\nu}/\nu_{0})^{-4\pi g}\ 
,&\qquad {{\delta\nu}\over\nu_{0}}\ll{1\over
g}\left({L\over l}\right)^2 \\
\displaystyle{ \exp\left\{-{\pi g\over
2}{\ln^2({\delta\nu}/\nu_{0}\Delta\tau)\over\ln(L/l)}\right\}\ , }  
&\qquad {{\delta\nu}\over\nu_{0}}\gg{1\over
g}\left({L\over l}\right)^2
\end{array}
\right.
\label{anom132}
\end{equation}
in the 2D geometry. 
The far LN asymptotic tail  in Eq.~(\ref{anom132}) is in full agreement
with the RG calculation by Altshuler, Kravtsov, and Lerner
\cite{akl}. We find also an intermediate power-law behavior, which
could not be obtained from the study of cumulants  in Ref.~\cite{akl}. We
note, however, that this power-law form is fully consistent with the
change of the behavior of cumulants 
$\langle\!\langle(\delta\nu)^n\rangle\!\rangle$ at $n\sim\pi g$ 
discovered in \cite{akl}.

\subsection{Distribution of eigenfunction amplitudes}
\label{s4.2}

\subsubsection{Quasi-1D geometry}
\label{s4.2.1}

The spatial shape of the ALS determining the asymptotics of the
distribution function of eigenfunction intensities can be found
\cite{m97} via  the exact solution of the $\sigma$-model.
We define
\begin{equation}
\langle|\psi^2({\bf r})|\rangle_u = \frac {{\cal Q}(u,{\bf r})}
{{\cal P}(u)}\ ,
\label{anom8}
\end{equation}
where
\begin{eqnarray}
{\cal Q}(u,{\bf r})&=& {1\over\nu
V}\left\langle\sum_\alpha |\psi_\alpha({\bf r})|^2
\delta(|\psi_\alpha(0)|^2-u)\delta(E-E_\alpha)\right\rangle\ , 
\label{anom10}
\end{eqnarray}
and ${\cal P}(u)$ is the distribution function of $u=|\psi^2(0)|$
defined formally by Eq.~(\ref{anom9}). According to Eq.~(\ref{anom8}),
$\langle|\psi^2({\bf r})|\rangle_u$ is the average intensity of an
eigenstate, which has in the point ${\bf r}=0$ the intensity $u$
(which will be assumed to be atypically large). The exact result for 
${\cal P}(u)$ is given in the form of the Lebedev-Kontorovich
expansion by Eq.~(\ref{anom_4}), (\ref{anom11}),
(\ref{anom14}). Calculating the moments
$\langle|\psi(0)|^2 |\psi({\bf r})|^{2q}\rangle$, and restoring the
function ${\cal Q}(u,{\bf r})$, we find for $r>l$
\begin{equation}
{\cal Q}(u,{\bf r}) = -{1\over V\xi A}{\partial\over\partial u}
\left[{W^{(2)}(u\xi A;\tau_1,\tau_2)W^{(1)}(u\xi A,\tau_-)\over
u}\right]\ ,
\label{anom35}
\end{equation}
where the function $W^{(2)}(z;\tau_1,\tau_2)$ satisfies the same
equation, as $W^{(1)}$,
\begin{equation}
{\partial W^{(2)}(z;\tau_1,\tau_2)\over \partial\tau_1}=
\left( z^2 {\partial^2\over \partial z^2}-z\right) 
W^{(2)}(z;\tau_1,\tau_2)
\label{anom31}
\end{equation}
with the boundary condition
\begin{equation}
W^{(2)}(z;0,\tau_2)=zW^{(1)}(z,\tau_2)
\label{anom32}
\end{equation}
The solution of Eqs.~(\ref{anom31}), (\ref{anom32}) is
\begin{eqnarray}
&&W^{(2)}(z,\tau_1,\tau_2)=2\sqrt{z}\int_0^\infty d\mu
b(\mu,\tau_2)K_{i\mu}(2\sqrt{z}) e^{-{1+\mu^2\over 4}\tau_1}\ ;
\nonumber\\
&& b(\mu,\tau_2)={\mu\sinh(\pi\mu)\over 2 \pi^2}\int_0^\infty dt\, 
K_{i\mu}(t) W^{(1)}(t^2/4,\tau_2)
\label{anom33}
\end{eqnarray}
Substituting here the formula (\ref{anom14}) for $W^{(1)}(z,\tau_2)$
and evaluating the integral over $z$, we can reduce Eq.~(\ref{anom33}) for
$b(\mu,\tau_2)$ to the form
\begin{eqnarray}
&& b(\mu,\tau_2)={\mu\sinh(\pi\mu)\over 16\pi^2}
\left|\Gamma\left({1+i\mu\over 2}\right)\right|^4(1+\mu^2)
+{\mu\sinh(\pi\mu)\over 2\pi^3}
\int {d\mu_1\,\mu_1\over 1+\mu_1^2} \nonumber \\
&& \ \ \times 
\sinh{\pi\mu_1\over 2}
\left|\Gamma\left(1+i{\mu+\mu_1\over 2}\right)\right|^2
\left|\Gamma\left(1+i{\mu-\mu_1\over 2}\right)\right|^2
e^{-{1+\mu_1^2\over 4}\tau_2}
\label{anom34}
\end{eqnarray}

In the opposite case $r<l$ we find
\begin{equation}
{\cal Q}(u,{\bf r})={1\over V}\left\{k_d(r)\left(u{d^2\over du^2}+
{d\over du}\right)-{d\over du}\right\} Y_a(u)
\ , \label{anom36}
\end{equation}
where the function $Y_a(u)$ was defined in Eq.~(\ref{anom_3}). This formula
is valid for any sample, which is locally $d$--dimensional. In the
case of the quasi--1D geometry we get
\begin{equation}
{\cal Q}(u,{\bf r})={1\over V}\left\{k_d(r)\left(u{d^2\over du^2}+
{d\over du}\right)-{d\over du}\right\} 
\left[W^{(1)}(u\xi A,\tau_-)W^{(1)}(u\xi A,\tau_+)\right]\ ,\qquad
r\ll l
\label{anom37}
\end{equation}

\paragraph{Insulating sample ($L\gg\xi$)}
%\label{s4.2.1.1}

The distribution ${\cal P}(u)$ is given by (\ref{anom3u}),
(\ref{anom3o}). The ``tail'', Eq.~(\ref{anom3as}), at $u\gg1/A\xi$
corresponds to atypically large local amplitudes. Analyzing the
general formula for $\langle|\psi^2({\bf r})|\rangle$ in this case, we
find \cite{m97} the following spatial structure of the ALS with
$|\psi^2(0)|=u$: 
\begin{eqnarray}
&\langle|\psi^2(r)|\rangle_u=
{\pi^{3/2}\over 16}u^{-1/2}A^{-3/2}r^{-3/2}e^{-r/4\xi}\ ,&\qquad
r\gg\xi \label{anom48a}\\
&\langle|\psi^2(r)|\rangle_u= 
\displaystyle{
{1\over 2}\left({u\over\xi A}\right)^{1/2} {1\over
\left(1+r\sqrt{uA\over\xi}\right)^2 }    }
\ ,&\qquad l<r\ll\xi
\label{anom48b}\\
&\langle|\psi^2(r)|\rangle_u=
{1\over 2}\left({u\over\xi
A}\right)^{1/2}\left[1+2\sqrt{uA\xi}k_d(r)\right]\ ,&\qquad r<l
\label{anom48c}
\end{eqnarray}
We see from
Eqs.~(\ref{anom48a}), (\ref{anom48b}), (\ref{anom48c}) 
that the eigenfunction normalization is dominated
by the region $r\sim\xi_{ef}$, where $\xi_{ef}\sim\sqrt{\xi/uA}\ll\xi$
plays the role of an effective localization length. In the region
$\xi_{ef}\ll r\ll\xi$ the wave intensity falls down as $1/r^2$, and
crosses over to the conventional localization behavior at $r\gg\xi$. 
Therefore, the appearance of an anomalously high amplitude
$|\psi^2(0)|=u\gg 1/A\xi$ is  not just a local fluctuation, but
rather a kind of a cooperative phenomenon corresponding to existence
of a whole region $r\lesssim\xi_{ef}$ with an unusually large amplitude 
$|\psi^2(r)|={1\over 2}\sqrt{u/\xi A}\sim 1/A\xi_{ef}$.  

Eq.~(\ref{anom48c}) describes a sharp drop of the amplitude from
$|\psi^2(0)|=u$ to 
$\langle|\psi^2(r)|\rangle={1\over 2}\sqrt{u/\xi A}$ at $r\sim l$.
This ``quasi-jump'' happens on a short scale $r_0\sim
k_F^{-1}(uA\xi)^{1/[2(d-1)]}$. To understand the reason for it, let us
recall that the above formulas represent  the {\it
average} intensity 
$|\psi^2({\bf r})|$ (under the condition $|\psi^2(0)|=u$). One can
also study the {\it fluctuations} of the intensity. It turns out
\cite{m97} that
in the region $r_0\ll r\ll\xi$ the fluctuations are of usual GUE type
superimposed on the envelope (\ref{anom48b}). 
It is not difficult to understand that
the quasi-jump  has the same origin as the GUE-like fluctuations at
$r\gg r_0$. One can ask, of course, why this  short-scale
fluctuation happens exactly in the center of the smooth ALS ``bump''
with a probability close to unity. The answer is as follows. We are
studying the states with an anomalously large local intensity $u$,
which is an exponentially rare event. There are two sources which 
may favor the formation of such a high intensity: i) formation of an
ALS with a spatially non-uniform smooth envelope, and ii) short-scale
GUE-like fluctuations. Both these mechanisms have exponentially small
probabilities to produce an enhancement of the intensity by a large
factor. The found configuration of $\langle|\psi^2(r)|\rangle_u$ (short-scale
quasi-jump (\ref{anom48c}) on top of the smooth configuration
(\ref{anom48b})) represents just the {\it optimal combination} of the two
mechanisms:
\be
u = |\psi^2(0)|_{\mbox{smooth}}\cdot J(u) \equiv {1\over
2}\left({u\over\xi A}\right)^{1/2} \cdot 2 (uA\xi)^{1/2}
\label{e4.505}
\ee
For arbitrary geometry of the sample,
the magnitude of the ``quasi-jump'' $J(u)$ is given 
according to Eq.~(\ref{anom36}) by
\be
J(u)\simeq -u{d\over du}\ln Y_a(u) \simeq -u{d\over du}\ln {\cal P}(u)
\label{e4.506}
\ee
(in the quasi-1D case this reduces to $J(u)= 2 (uA\xi)^{1/2}$ as
stated above). The formula (\ref{e4.506}) can be reproduced (within
the saddle-point approximation) by writing the quantity $u$ as a
product $u=u_sJ$ of the smooth part $u_s$ and the local fluctuating
quantity $J$, with the latter distributed according to ${\cal
P}(J)=e^{-J}$.

\paragraph{Metallic sample ($L\ll\xi$)}

The asymptotic behavior of the intensity distribution function has the
same stretched-exponential form, as in the localized regime, see
Eq.~(\ref{anom_adm15c}). More
accurately (with the subleading factors included), this formula reads
\cite{m97} 
\begin{eqnarray}
{\cal P}(u) &=&{16\over \pi^2}\sqrt{{A\xi\over u}}{\sqrt{L_+ L_-}\over
L} \exp \left\{-4\sqrt{u\xi A}
+{\pi^2\xi\over 4L_+}\left(1-{\sqrt{\xi/ uA}\over L_+}+\ldots\right)
\right.\nonumber\\&& \left. 
+{\pi^2\xi\over 4L_-}\left(1-{\sqrt{\xi/ uA}\over L_-}+\ldots\right)   
\right\}\ .
\label{anom50}
\end{eqnarray}
Calculation of ${\cal Q}(u,{\bf r})$ shows \cite{m97} 
that the ALS intensity has
for $l<r\ll L$ the same form (\ref{anom48b}), provided the 
condition $V|\psi^2(0)|\gg g$ is fulfilled. This condition, under which
the asymptotic behavior (\ref{anom_adm15c}), (\ref{anom50}) is valid, acquires
now a very
transparent meaning. This is just the condition that the effective
localization length of an ALS, $\xi_{ef}=\sqrt{\xi/uA}$ is much less
than the sample size $L$. Indeed,
$\xi_{ef}/L=\sqrt{\xi/uAL^2}=\sqrt{g/uV}$. 

Near the sample edges, $r\sim
L\gg \xi_{ef}$, the form of the ALS intensity is slightly modified by
the boundary of the sample, see \cite{m97}. Finally, the
``quasi-jump''  of $\langle|\psi^2(r)|\rangle_u$
at $r\ll l$ has the same form (\ref{anom48c}) as in
the insulating regime. 

\paragraph{Saddle-point method}

The saddle-point method of Muzykantskii and Khmelnitskii can be also
applied to the problem of the statistics of eigenfunction amplitudes,
as was done by Fal'ko and Efetov \cite{fe1,fe2}. In this case, one
should look for the saddle-point of the functional integral
(\ref{anom_1}) determining the function $Y_a(u)$ (which in turn
determines the eigenfunction statistics, see Eqs.~(\ref{anom_4}),
(\ref{anom_4o})). The saddle-point is  again parametrized by the
bosonic non-compact angle $\theta({\bf r})$ only, and the
corresponding saddle-point equation has the form
\be
\pi\nu D\nabla^2\theta-ue^\theta=0\ .
\label{e4.507}
\ee
It is similar to the equation (\ref{e4.8}) of the long-time relaxation
problem, but with different sign of the second term. Also, the
boundary conditions have now a different form:
\be
\theta(0)=0
\label{e4.508}
\ee
and the condition (\ref{e4.7}) at the boundary (since we consider now
a closed sample).  Alternatively, one can write Eqs.~(\ref{e4.507}),
(\ref{e4.508}) in a slightly different form by shifting the variable
$\theta+\ln(uV)\to\theta$. Then $u$ is removed from the saddle-point
equation and from the action, but appears in the boundary condition.

The action determining the distribution function ${\cal P}(u)$ is
given by
\be
-\ln{\cal P}(u)=S=\int d^dr\left[{\pi\nu D\over
2}(\nabla\theta)^2+ue^\theta\right] 
\label{e4.509}
\ee
The formula (\ref{e4.509}) acquires a very transparent meaning if we take
into account what was written in Section~\ref{s4.2.1} concerning the
two factors contributing to the large amplitude
$|\psi^2(0)|=u$. Firstly, this is the non-uniform smooth envelope 
$\propto e^{\theta({\bf r})}$ yielding
$$
|\psi^2(0)|_{\mbox{smooth}}={e^{\theta(0)}\over \int d^dr
 e^{\theta({\bf r})}} ={1\over \int d^dr e^{\theta({\bf r})}}  ;
$$
the corresponding weight is represented by the first term in the
action (\ref{e4.509}). Secondly, these are the local Gaussian
(GUE-like) fluctuations of the wavefunction amplitude, which should
provide the remaining factor (``quasi-jump'')
$$
J={u\over |\psi^2(0)|_{\mbox{smooth}}} = u \int d^dr
 e^{\theta({\bf r})}\ ;
$$
the corresponding probability ${\cal P}(J)=e^{-J}$ reproduces
the second term in the action (\ref{e4.509}). 

In the quasi-1D case and under the condition $u\gg gV^{-1}$, the
solution of Eq.~(\ref{e4.507}) reads \cite{fe2,m96}
\begin{equation}
e^{\theta(r)}=
{1\over \left(1+r\sqrt{{u\over 2\pi\nu D}}\right)^2}\ ;\qquad 0<r\ll L_+
\label{anom86}
\end{equation}
Comparing Eq.~(\ref{anom86})
with Eqs.~(\ref{anom48b}), we see that the saddle-point
solution nicely reproduces the average intensity of the ALS,
$\langle|\psi^2(r)|\rangle_u$ for $r>l$, up to an overall normalization
factor. Also, the form of ${\cal P}(u)$ found in the quasi-1D case by
the saddle-point method \cite{fe2} is in a very good agreement with
the exact results presented above. This agreement does credit to the
saddle-point method and allows to use the saddle-point configuration
for characterizing the shape of  ALS in higher dimensions and for
other distribution functions, where the exact solution is not
available.

\subsubsection{2D geometry}
\label{s4.2.2}

For a 2D disk-shaped sample of a radius $L$
with the high amplitude point
${\bf r}=0$ in the center of the disk, the saddle-point solution 
of Eqs.~(\ref{e4.507}), (\ref{e4.508}) is
found to have the form \cite{fe1,fe2}
\begin{eqnarray}
e^{\theta(r)}&=&\left({r\over l_*}\right)^{-2\mu}
\left\{1-{l_*^2 u\over 8(1-\mu)^2\pi\nu  D}
\left({r\over l_*}\right)^{2-2\mu}\right\}^{-2}\ ,\qquad r\ge l_* 
\no \\
&\approx& \left({r\over l_*}\right)^{-2\mu}\qquad \mbox{for}\  l_*\le r\ll
L\ ,
\label{anom87}
\end{eqnarray}
where the exponent $0<\mu<1$ depends on $u$ and  satisfies the equation
\begin{equation}
\left({L\over l_*}\right)^{2\mu}={2-\mu\over 8\mu(1-\mu)^2} {L^2
u\over\pi\nu  D} 
\label{anom88}
\end{equation}
We are interested in the asymptotic region $uL^2\gg \pi\nu 
D\ln^{-1}(L/l)$, where the distribution of the eigenfunction intensity
is given by Eq.~(\ref{e3.17}), and an ALS is formed. Then the exponent
$\mu$ can be approximated as
\begin{equation}
\mu\simeq{\ln\left({L^2 u\over 2\pi\nu  D}\ln{L\over
l_*}\right)\over 2\ln(L/ l_*) }
\label{anom88a}
\end{equation}
The lower cut-off scale $l_*$ appears in Eq.~(\ref{anom87}) 
for the same reason as in the long-time relaxation problem
(Sec.~\ref{s4.1.2}), i.e. because of the
restriction of the diffusion approximation on the momenta $q$ of the
$\sigma$-model field: $q<l^{-1}$. It is determined by
the condition 
$\theta('r)|_{r=l_*}\sim l^{-1}$,
which yields $l_*\sim\mu l$.
The corresponding asymptotic behavior of $Y_a(u)$ (and consequently of
the intensity distribution function), which was already quoted in
Sec.~\ref{s3.3.1}, is
\begin{equation}
Y_a(u), {\cal P}(u) \sim\exp\left\{-\pi^2\nu  D
{\ln^2\left({V u\over 2\pi^2\nu  D}\ln{L\over
l_*}\right)\over\ln(L/l_*) } \right\}
\label{anom90}
\end{equation}
 
Normalizing the expression (\ref{anom87}), we find that the
average ALS density for $r>l_*$ is equal to
\begin{equation}
\langle|\psi^2(r)|\rangle_u={u\over 4\pi^2\nu  D\mu}\left({r\over
l_*}\right)^{-2\mu} 
\left\{1-{l_*^2 u\over 8(1-\mu)^2\pi\nu  D}
\left({r\over l_*}\right)^{2-2\mu}\right\}^{-2}
\ ,\qquad r\ge l_*
\label{anom89}
\end{equation}
The saddle-point calculation  
assumes that $\theta(r)$ is constant for
$r<l_*$, so that Eq.(\ref{anom89}) gives
$\langle|\psi^2(r)|\rangle_u\simeq{u\over 4\pi^2\nu D\mu}$ in this
region.  However, for very small $r<l$ the average intensity 
$\langle|\psi^2(r)|\rangle_u$ changes
sharply, as we have seen in the quasi--1D case. 
Using  Eq.~(\ref{anom36}) for ${\cal Q}(u,r)$ and Eq.~(\ref{anom_4}) for
${\cal P}(u)$ , we get
\begin{equation}
\langle|\psi^2(r)|\rangle_u\equiv {{\cal Q}(u,r)\over {\cal P}(u)}=
[1-k_2(r)+J(u)k_2(r)]\langle|\psi^2(r=l_*)|\rangle_u\ ,\qquad r<l_*
\label{anom91}
\end{equation}
According to (\ref{e4.506}), (\ref{anom90}),
 the height of the quasi-jump is given by 
\begin{equation}
J(u)\simeq -u{d\over du}\ln Y_a(u)\simeq {2\pi^2\nu  D\over
\ln(L/l_*)}
\ln\left({Vu\over 2\pi^2\nu  D}\ln{L\over l_*}\right)\simeq 4\pi^2\nu 
D\mu\ ,
\label{anom92}
\end{equation}
which is precisely the factor by which the value of
$\langle|\psi^2(r=l_*)|\rangle_u$ found above differs from
$\langle|\psi^2(0)|\rangle_u\equiv u$.    
Combining Eqs.~(\ref{anom89}), (\ref{anom91}), and  (\ref{anom92}), we
get
\begin{equation}
\langle|\psi^2(r)|\rangle_u={u\over J(u)}[1-k_2(r)+k_2(r)J(u)]\ ,\qquad
r<l_*
\label{anom93}
\end{equation}
Therefore, in the 2D case the ALS determining the asymptotics of the
amplitude distribution function has the power-law shape (\ref{anom89})
with the short-scale bump (\ref{anom93}).

\subsubsection{States localized near the boundary.}
\label{s4.2.3}
We  assumed in the above calculations that the center of an ALS is located
far enough from the sample edge. For a quasi-1D sample, this means
that $\xi_{ef}\ll L_+,L_-$. In the 2D case this implies that the
distance from the observation point to the boundary is of the same
order of magnitude in all directions, so that $\ln(L/l)$ is defined
without ambiguity. Here, we will consider briefly the role of ALS
situated close to the boundary, when these conditions are violated
\cite{m97}. 

We first consider the quasi-1D geometry. Let us calculate the distribution
function ${\cal P}(u)$ in a point located very close to one of the
sample edges. Formally, this means that $L_-\ll\xi_{ef}$. Then the
function $W^{(1)}(uA\xi,\tau_-)$ in Eq.~(\ref{anom11}) can be
approximated by unity, and we get
\begin{equation}
{\cal P}(u)={2\over \pi}\xi^{3/4} A^{1/4}L^{-1/2} u^{-3/4}
 \exp \left\{-2\sqrt{u\xi A}
+{\pi^2\xi\over 4L_+}\left(1-{\sqrt{\xi/ uA}\over L_+}+\ldots\right)
\right\}
\label{anom135}
\end{equation}
We see therefore that close to the boundary the distribution ${\cal
P}(u)$ has the asymptotic decay ${\cal P}(u)\sim\exp\{-2\sqrt{uA\xi}\}$,
which is slower than in the bulk of the sample, 
${\cal P}(u)\sim\exp\{-4\sqrt{uA\xi}\}$. This means that if we consider
the distribution ${\cal P}(u)$ averaged over the position of the
observation point, its asymptotic tail will be always dominated by
contribution of the points located close to the boundary, 
${\cal P}(u)\sim\exp\{-2\sqrt{uA\xi}\}$. This could be already
anticipated 
from Eq.~(\ref{anom50}), where the factor $\exp\left\{{\pi^2\over
4}\left({\xi\over L_+} + {\xi\over L_-}\right)\right\}$ increases
strongly with approaching one of the sample edges. The same tendency,
but in a weaker form, is observed in Eqs.~(\ref{anom_adm15a}),
(\ref{anom_adm8a}), 
(\ref{anom_adm15b}), (\ref{anom_adm15bo}). 
Calculating the average intensity
$\langle|\psi^2(r)|\rangle_u$ of the corresponding ALS, we find that
at $r>l$ the ALS spatial shape retain the form (\ref{anom48b}),
with an additional overall factor of 2. At small $r$,
Eq.~(\ref{anom48c}) is slightly modified:
\begin{equation}
 \langle|\psi^2(r)|\rangle_u=
\left({u\over\xi
A}\right)^{1/2}\left[1+\sqrt{uA\xi}k_d(r)\right]
\label{anom136}
\end{equation}

In 2D, we can consider a sample of the semicircular shape, with the
observation point located in the center of the diameter serving as a
boundary. The saddle-point solution then has exactly the same form
(\ref{anom87}), and the ALS intensity is still given by
Eq.~(\ref{anom89}), with an additional factor 2. The asymptotic form
of the distribution function ${\cal P}(u)$ contains an extra factor $1/2$ in
the exponent:
\begin{equation}
{\cal P}(u)\sim\exp\left\{-{\pi^2\nu  D\over 2}
{\ln^2\left({V u\over 2\pi^2\nu  D}\ln{L\over
l_*}\right)\over\ln(L/ l_*) } \right\}
\label{anom137}
\end{equation}
This result is expected to be applicable to any 2D sample of a
characteristic size $L$, with a smooth boundary and the observation
point taken in the vicinity of the boundary.

We see therefore, that, very generally, the probability of formation
of an ALS with the center in a given point is strongly enhanced (via
an extra factor $1/2$ in the exponent), if this point lies close to
the sample edge. This leads to the additional factor $1/2$ in the
exponent in the asymptotic form of the distribution 
${\cal P}(u)$ near the boundary.

\subsection{Distribution of local density of states}
\label{s4.3}

We again assume the sample to be open, as in the problem of the
distribution of relaxation times, Sec.~\ref{s4.1}. Then it is
meaningful to speak about the statistics of the local density of
states (LDOS) $\rho(E,{\bf r})=(-1/\pi)\mbox{Im}\,G_R({\bf r}, {\bf
r};E)$. In a metallic sample, the LDOS is a weakly fluctuating
quantity, whose distribution ${\cal P}(\rho)$
is mostly concentrated in a narrow
Gaussian peak \cite{akl,lerner88}
with mean value $\langle\rho\rangle=\nu$ and the
variance $\mbox{var}(\rho/\nu)\sim\kappa\ll 1$, where $\kappa$ is the
usual parameter of the perturbation theory $\kappa=\Pi({\bf r}, {\bf
r}) =\sum_q 1/\pi\nu VDq^2$, which already appeared in
Sec.~\ref{s3.3.1}. This
close-to-Gaussian shape of ${\cal P}(\rho)$ holds however in the region 
$|\rho/\nu-1|\ll 1$ only. We will consider in contrast the ``tails'' of
the distribution, $\rho/\nu\ll 1$ and $\rho/\nu\gg 1$, where a much slower 
decay of ${\cal P}(\rho)$ will be found.  

We begin by discussing on a qualitative level a relation of the
asymptotics of ${\cal P}(\rho)$ to the behavior of the distribution
functions considered above. 
Typically, in an open metallic sample the LDOS
$\rho(E,{\bf r})$ is given by a superposition of $\sim 1/t_D\Delta=g$
adjacent levels, since their widths are of order of $1/t_D$. However,
we can expect that for $\rho$ much larger than its average value
$\nu $, the asymptotic form of ${\cal P}(\rho)$ is determined by a
probability to have a single narrow resonance, which gives this value
 of LDOS $\rho(E,{\bf r})$. The most favorable situation happens when the
resonance is located around the point ${\bf r}$ in the real space and
around the energy $E$ in energy space. The LDOS provided by such a
resonance is:
\begin{equation}
\rho_{ALS}=|\psi^2({\bf r})|{2 t_\phi\over\pi}\ ,
\label{anom113}
\end{equation}
where $t_\phi^{-1}$ is the resonance width. 
Thus, the optimal fluctuation should provide now a maximum to the
product of the local amplitude $u=|\psi^2({\bf r})|$ and the inverse
level width $t_\phi$, and the asymptotics of the distribution ${\cal
P}(\rho)$ should be related to that of ${\cal P}(u)$ and ${\cal
P}(t_\phi)$. In particular, in the quasi-1D case, where
 the distribution ${\cal P}(t_\phi)$,
Eq.~(\ref{anom107}), decays much more slowly than ${\cal P}(u)$,
Eq.~(\ref{anom3as}), one should expect the asymptotic behavior of ${\cal
P}(\rho)$ to be mainly determined by ${\cal P}(t_\phi)$. We will see
below that this is indeed the case.

Now we turn to a formal calculation.
The distribution function $P_\rho(\rho)$ of LDOS can be expressed
through the function $Y(\lambda_1,\lambda_2)$ introduced in
Sec.~\ref{s3.1} as follows \cite{mf94a,mf94b,efprig93,mf94c} 
\begin{equation}
P(\rho)
=\delta(\rho-1)
+\frac{1}{4\pi}\frac{\partial^{2}}{\partial \rho^{2}}\left\{
\int_{\frac{\rho^{2}+1}{2\rho}}^{\infty}
d\lambda_{1}\overline{Y}(\lambda_1)
\left(\frac{2\rho}{\lambda_{1}-
\frac{\rho^{2}+1}{2\rho}}\right)^{1/2}\right\}\ ,
\label{ldosas_5}
\end{equation}
where
\begin{equation}
\overline{Y}(\lambda_1)=\int_{-1}^{1}d\lambda_2
{Y(\lambda_1,\lambda_2)\over \lambda_1-\lambda_2}
\label{ldosas_33}
\end{equation}
and $\rho$ is normalized by its mean value: $\rho/\nu\to \rho$.
Let us note the symmetry relation found in
\cite{mf94a,mf94b} 
\begin{equation}
P(\rho^{-1})=\rho^3 P(\rho)\ .
\label{ldosas_5a}
\end{equation}
It follows from Eq.~(\ref{ldosas_5}) and is completely independent of a
particular form of the function $Y(\lambda_1,\lambda_2)$. Obviously,
Eq.~(\ref{ldosas_5a}) relates the small--$\rho$ asymptotic behavior of the
distribution ${\cal P}(\rho)$ to its large-$\rho$ asymptotics. 

Asymptotic behavior of ${\cal P}(\rho)$ was studied in
\cite{m96} via the saddle-point method supplemented
in the quasi-1D case by the exact solution. The saddle-point equation
has for this problem a very simple form, $$\nabla^2\theta=0\ ,$$ with the
boundary conditions 
$$\theta|_{\mbox{leads}}=0\ ,\qquad \theta(0)=\rho/2\ .$$

\paragraph{Quasi-1D geometry.}

In the quasi-1D case, the solution reads
\begin{equation}
e^{\theta(r)}\simeq\left\{
\begin{array}{ll}
(\rho/\nu )^{1-r/L_+}\ ,&\qquad r>0\\
(\rho/\nu )^{1-|r|/L_-}\ ,&\qquad r<0\ ,
\end{array}
\right.
\label{anom115}
\end{equation}
where, as before, $L_+$ and $L_-$ are the distances from the
observation point $r=0$ to the sample edges.
This yields the  asymptotics of the distribution function
\begin{equation}
{\cal P}(\rho)\sim\exp\left\{-{\xi\over 4}\left({1\over L_+}+{1\over
L_-}\right)\ln^2(\rho/\nu )\right\}\ .
\label{anom114}
\end{equation}

Let us note that this asymptotic behavior of ${\cal P}(\rho)$ in an
open sample is strongly different from the asymptotics of ${\cal
P}(u)$ in a closed sample, Eq.~(\ref{anom3as}). As was explained above,
the difference originates from the fact that ${\cal P}(\rho)$ is
essentially determined by ${\cal P}(t_\phi)$. To
demonstrate this explicitly, 
we put the observation point in the middle of the sample,
$L_+=L_-=L/2$. The configuration (\ref{anom115}) acquires then
precisely the same
form as the optimal configuration (\ref{e4.19}) for the relaxation
time $t_\phi$, and the asymptotics (\ref{anom114}) of ${\cal P}(\rho)$
is identical to that of ${\cal P}(t_\phi)$, Eq.~(\ref{anom107}). 

The corresponding values of $t_\phi$ and $\rho$ are
related as follows:
\begin{equation}
{4\over \pi}g\Delta t_\phi\ln^2(g\Delta t_\phi)=\rho/\nu \ .
\label{anom116}
\end{equation}
Now we calculate the value of the local amplitude $|\psi^2(0)|$ 
for an ALS corresponding to the configuration (\ref{anom115}). First,
its smoothed intensity is given by
\begin{equation}
|\psi^2(r)|_{\mbox{smooth}}={\cal N}^{-1}
e^{\theta(r)}={\ln(\rho/\nu )\over V}
\left({\rho\over\nu }\right)^{-2|r|/L}. 
\label{anom118}
\end{equation}
Second, the quasi-jump induced by the GUE-type fluctuations gives an
additional factor, which is found according to 
Eq.~(\ref{e4.506}) to be
\begin{equation}
J(\rho)=-\rho{\partial\over\partial\rho}\ln {\cal
P}(\rho)=2g\ln(\rho/\nu )\ .
\label{anom119}
\end{equation}
Combining Eqs.(\ref{anom116}), (\ref{anom118}), and (\ref{anom119}), 
we can compute the LDOS (\ref{anom113})
determined by this resonance state:
\begin{equation}
\rho_{ALS}(E,0)=|\psi^2(0)|_{\mbox{smooth}}\cdot J(\rho) \cdot
{2t_\phi\over\pi}=
{\ln(\rho/\nu )\over V}\cdot 2g\ln(\rho/\nu )\cdot {\rho V\over
2g\ln^2(\rho/\nu )}=\rho
\label{anom120}
\end{equation}
We have explicitly checked therefore that the LDOS $\rho$ is indeed
determined by a single ALS, smoothed intensity of which is given by
Eq.~(\ref{anom118}). There are three sources of the enhancement of
LDOS: i) amplitude of the smooth envelope of the wave function, ii) the 
short-scale GUE ``bump'', and iii) the inverse resonance width. They
are represented by the three factors in Eq.~(\ref{anom120}),
respectively. 

The result (\ref{anom114}) can be also obtained \cite{m96}  from the exact
solution of the $\sigma$-model. It was also shown in \cite{m96}  that
in the long wire limit ($L\gg\xi$) the whole distribution function
takes a 
log-normal form analogous to that found in \cite{alprig89} for the
case of a strictly 1D sample by the Berezinskii technique.

\paragraph{2D geometry.}

In the 2D case, we introduce again (as in Sec.~\ref{s4.1.2},
\ref{s4.2.2}) the small-$r$ cutoff, $l^{(\rho)}_*$.
The saddle-point solution reads
\begin{equation}
e^{\theta(r)}\simeq{\rho\over\nu }\left({l^{(\rho)}_*\over
r}\right)^{\gamma_\rho}\ , 
\label{anom127}
\end{equation}
where $l^{(\rho)}_*=\gamma_\rho l$, and
$$
\gamma_\rho={\ln(\rho/\nu )\over \ln (L/l^{(\rho)}_*) }.
$$
The distribution ${\cal P}(\rho)$ has the following asymptotics \cite{m96}
\begin{equation}
{\cal P}(\rho)\sim\exp\left\{-\frac{\pi^2\nu  D \ln^2\rho}
{\ln (L/l^{(\rho)}_*)}\right\}.
\label{anom126}
\end{equation}
So, the LDOS distribution has (as in the quasi-1D case) the log-normal
form, which is now similar to the distribution (\ref{anom90})
of the eigenfunction intensities (while the distribution of relaxation
times has an intermediate power-law regime). This result is 
in perfect agreement with the asymptotic behavior of ${\cal P}(\rho)$
found by the renormalization group method in \cite{akl}. 

As for all the other distribution functions studied, the relevant ALS
decay in a power-law fashion,
$|\psi^2(r)|\propto(r/l^{(\rho)}_*)^{-\gamma_\rho}$. Like in the
quasi-1D case, one can explicitly verify \cite{m97} that a single
state with the spatial shape determined by (\ref{anom127}) indeed
provides, by virtue of Eq.~(\ref{anom113}), the value of LDOS equal to
$\rho$.

\subsection{Distribution of inverse participation ratio}
\label{s4.3a}

In this subsection, we study  the asymptotics of the
distribution function of the IPR $P_2$, Eq.(\ref{glob15}). 
We have already considered the fluctuations of $P_2$ in
Sec.~\ref{s3.3.3}. As was explained there, the relative magnitude of
the fluctuations is $[{\rm r.m.s.}(P_2)]/\langle P_2\rangle\sim
1/g$. At $1/g\ll P_2/\langle P_2\rangle-1\ll 1$ the distribution
function is of the exponential form,
\be
\label{ipr45}
{\cal P}(P_2)\sim\exp\left\{-{\pi\beta\over 4}{\epsilon_1\over\Delta}
\left({ P_2\over\langle P_2\rangle} -1\right)\right\}\ .
\ee
Note that for negative deviations $ P_2/\langle P_2\rangle-1$
with $| P_2/\langle P_2\rangle-1|\gg 1/g$ the distribution function
decays much faster \cite{prigal98}, so that the distribution is
strongly asymmetric. As was mentioned in Sec.~\ref{s3.3.2}, 
the ``body'' of the distribution ${\cal P}(P_2)$ is  described
properly (in the leading order in 
$1/g$) by the Liouville theory. We will see below that this is also
true for the asymptotic ``tail'' of ${\cal P}(P_2)$.

Our consideration of  asymptotics of the IPR distribution
 is based on unpublished results \cite{mf95u}
(partially announced in \cite{m97}). We
 derive first a relation between ${\cal P}(P_2)$ and the distribution
 of level 
velocities ${\cal P}_v(v)$ \cite{fm95a}. To this end, we consider
a Hamiltonian ${\cal H}+\alpha{\cal W}$, where 
 ${\cal W}$ is a random perturbation. Specifically, the matrix elements
${\cal W}_{r_1r_2}$ are supposed to be independent Gaussian-distributed
random variables with
the mean value equal to zero and the
variance 
$$\langle {\cal W}^{*}_{r_1r_2}{\cal
W}_{r_2'r_1'}\rangle={\cal W}_{0}(|{\bf r_1}-{\bf r_2}|) 
\delta({\bf r_1}-{\bf r_1'})\delta({\bf r_2}-{\bf r_2'}).$$
We will assume that  ${\cal W}_{0}(r)$ is a short-ranged function with
some characteristic scale $\zeta$. 
The level velocity $v_{n}$ corresponding to an energy level $E_n$ 
is defined as $v_{n}=dE_n(\alpha)/d\alpha$ (where $E_n(\alpha)$ is the
level of the perturbed Hamiltonian ${\cal H}+\alpha{\cal W}$) and can
be found within the conventional perturbation theory as
\begin{equation}
\label{vel_eq10}
v_{n}=\int d^d r\int d^d r' {\cal W}_{rr'}\psi_{n}^*({\bf r})
\psi_{n}({\bf r'})\ .
\end{equation}
Using (\ref{vel_eq10}), we find 
\begin{equation}
\langle \overline{v_n^2}\rangle = w_{0}P_2\ ;\qquad
 w_{0}=c\int d^d r {\cal W}_0 ({\bf r})\ ,
\label{vel_c}
\end{equation}
where $c=1\ (1/2)$ if $k_f\zeta\ll 1$ (resp. $k_f\zeta\gg 1$).

This consideration can be extended to higher moments of the level
velocity 
$\langle\overline{v_n^{2q}}\rangle$
as well. This leads to the following 
relation between the two distributions:
\begin{eqnarray}
{\cal P}_v(v)&\equiv& \langle\overline{\delta( v-v_n )}\rangle
\nonumber \\ &=& \int_{0}^{\infty}\frac{dP_2}{\left[2\pi
{w}_{0}P_2\right]^{1/2}}\exp{\left[-\frac{v^2}{2{w}_0 P_2}\right]}
{\cal P}_I(P_2)\ .
\label{vel_eq12}
\end{eqnarray}
On the other hand, the level velocity distribution can be expressed
through the $\sigma$-model correlation function in the following way
\cite{f94}. According to the definition,
\bea
\label{ipr1}
{\cal P}_v(v) & = & {1\over\nu V}\left.\left\langle\sum_n\delta(E-E_n)
\delta(v-\partial E_n/\partial\alpha)\right\rangle
\right|_{\alpha\to 0}
\nonumber \\
& = & \lim_{\alpha\to 0}{\alpha\over\nu
V}\left\langle\sum_n\delta(E-E_n(0)) 
\delta(E+\alpha v-E_n(\alpha))\right\rangle \ ,
\eea
so that ${\cal P}_v(v)$ is determined by the $\alpha\to 0$ limit of the
parametric level correlation function. The latter can be represented
[as a generalization of Eq.~(\ref{e2.2})] in terms of a correlation
function of the $\sigma$-model
\cite{szafer93,simons93a,simons93b}. This yields
\bea
\label{ipr2}
&& {\cal P}_v(v)=\lim_{\eta\to 0}{-1\over 8V^2}{\eta\over\nu Vv}\int
DQ \left(\int d^dr \mbox{Str} Q_{11}k\right) 
\left(\int d^dr \mbox{Str} Q_{22}k\right)
e^{-S_v[Q]}\ ; 
\nonumber \\
&& S_v[Q]=\int d^dr \mbox{Str} \left[-{\pi\nu D\over 4}(\nabla Q)^2
-{i\pi\nu\eta \over 2} Q\Lambda + {w_0(\pi\nu)^2\eta^2 \over 4v^2}
Q\Lambda Q\Lambda \right]\ .
\eea
Combining Eqs.~(\ref{vel_eq12}) and (\ref{ipr2}), we find the expression
for the IPR distribution function in terms of the $\sigma$-model,
\bea
\label{ipr3}
{\cal P}_I(P_2) &=&{-1\over 8V^2}{1\over 2i\sqrt{2\pi}P_2^{3/2}}
\lim_{\eta\to 0} \eta\nu V \int_{c-i\infty}^{c+i\infty}
{du\over \sqrt{u}}
\nonumber \\
&\times& \int DQ
\left(\int d^dr \mbox{Str} Q_{11}k\right) 
\left(\int d^dr \mbox{Str} Q_{22}k\right)
e^{-S_u[Q]}\ ;
\eea
where $S_u[Q]=-u/2P_2+S_v[Q]|_{v^2/w_0=u}$. 
The saddle point configuration is
again parametrized by the bosonic ``angle'' $\theta({\bf r})$ only;
the action on such a configuration is
\be
\label{ipr4}
S_u[\theta]=-{u\over 2P_2}+\int d^dr\left[{\pi\nu D\over
2}(\nabla\theta)^2 - {i\pi\nu\eta\over 2}e^\theta + {(\pi\nu\eta)^2\over
4u}e^{2\theta}\right] \ .
\ee
The corresponding saddle-point equations
can be readily obtained by varying $S_u[\theta]$ with respect to
$\theta({\bf r})$ and $u$:
\bea
\label{ipr5}
&& -D \nabla^2\theta - {i\eta\over 2} e^\theta + {\pi\nu\eta^2\over
2u}e^{2\theta}=0\ ; \nonumber \\
&& {(\pi\nu\eta)^2\over 4 u^2} \int d^dr e^{2\theta} + {1\over 2P_2} =0\ .
\eea
Making a shift $\theta=\tilde{\theta}+\ln (iu/\pi\nu\eta)$ and dropping
the tilde, we can reduce them to the form
\bea
\nabla^2\theta-\gamma(e^\theta-e^{2\theta})=0\ ; \label{ipr6} \\
P_2 =2\left/\int d^dr e^{2\theta}\right.\ , \label{ipr7}
\eea
where $\gamma=u/2\pi\nu D$. Integration of Eq.~(\ref{ipr6}) with the
Neumann boundary condition yields $\int d^dr
(e^\theta-e^{2\theta})=0$, so that Eq.~(\ref{ipr7}) can be rewritten
in the following form (invariant with respect to a shift of the
variable $\theta$):
\be
\label{ipr8}
P_2=2{\int d^dr\, e^{2\theta} \over \left(\int d^dr\,
e^{\theta}\right)^2}\ .
\ee
The meaning of Eq.~(\ref{ipr8}) is completely transparent if we recall
that $|\psi^2({\bf r})|_{\mbox{smooth}}
\propto e^{\theta({\bf r})}$. The factor $2$
comes from the GUE-like short-scale fluctuations.

Taking into account Eqs.~(\ref{ipr6}), (\ref{ipr7}), we find
the action (\ref{ipr4}) on the saddle-point configuration to be equal to
\be
\label{ipr9}
S_u[\theta]={\pi\nu D\over 2}\int d^dr\,(\nabla\theta)^2\ .
\ee
The problem that we are solving is easily seen to be equivalent to
searching for the minimum of (\ref{ipr9}) under the conditions $\int
e^\theta d^dr=1$, $\int e^{2\theta}d^dr=P_2$ (or, equivalently, under
the condition (\ref{ipr8}) invariant with respect to normalization of
$e^\theta$). This is nothing else but the optimum fluctuation problem
for ${\cal P}(P_2)$ within the Liouville  theory (\ref{ipr44}). 
Therefore, the Liouville theory (\ref{ipr44}) describes properly the
asymptotics of the IPR distribution.

\paragraph{Quasi-1D geometry.}

Equation
\be
\label{ipr10}
\theta''-\gamma e^\theta+\gamma e^{2\theta}=0
\ee
has the following general solution
\be
\label{ipr11}
e^{-\theta}={\gamma\over C_1}\left[1+\sqrt{1-C_1/\gamma}
\sin(\sqrt{C_1}x-C_2)\right]\ .
\ee
The constants $C_1$ and $C_2$ should be found from the boundary
conditions
$$
\theta'(-L/2) = \theta'(L/2) = 0\ ,
$$ 
yielding 
$$
\cos\left(\sqrt{C_1}{L\over 2}\pm C_2\right)=0\ .
$$
The solution providing the minimum to the action corresponds to
$C_2=0$, $\sqrt{C_1}=\pi/L$ and gives the wave function intensity
\be
\label{ipr12}
A|\psi^2({\bf r})|_{\mbox{smooth}}= {e^{\theta(x)}\over \int
 e^{\theta(x)}dx} = {\pi\over \gamma^{1/2}L^2}
{1\over 1+\sqrt{1-{\pi^2/ L^2\gamma}}\sin(\pi x/L)}
\ee
(with $A$ being the sample transverse 
cross-section) and the IPR value
\be
\label{ipr13}
P_2={2\gamma^{1/2}\over A\pi}= P_2^{\rm GUE}\cdot{\gamma^{1/2}L\over
\pi}\ .
\ee
Calculating the action (\ref{ipr9}) we find the following asymptotic
behavior of the IPR distribution function
\be
\label{ipr14}
-\ln {\cal P}(P_2)\simeq S_u= {\pi^3\nu D A^2\over
 4}\left(P_2-{2\over LA}\right)={\pi^2\over 4}g
\left({P_2\over P_2^{\rm GUE}}-1\right)\ .
\ee
Eq.~(\ref{ipr14}) is valid for $1/g\ll P_2/P_2^{\rm GUE}-1\ll L/l$. 
Therefore, in the quasi-1D case the exponential behavior (\ref{ipr45})
is not restricted to the region of small deviations from the average
value; there is no change of the behavior of ${\cal P}(P_2)$ at
$P_2/P_2^{\rm GUE}-1\sim 1$ (we will find such a change below in the 2D
geometry case).

The far asymptotics at $P_2\gg P_2^{\rm GUE}$ is determined by the
states, which have an effective localization length $\xi_{ef}$ much
smaller than the sample length $L$. The shape of these states is
according to (\ref{ipr12})
\be
\label{ipr15}
S_1|\psi^2({\bf r})|_{\mbox{smooth}}={2\over\pi}{\xi_{ef}\over
(x+L/2)^2+\xi_{ef}^2} \ ; \qquad\xi_{ef}={2\over \pi A P_2}\ .
\ee
Let us note that the density of such a state is concentrated near the
edge ($x=-L/2$) of the sample (of course, there exists an equivalent
solution located near the opposite edge, $x=L/2$). Since the shape of
a state with $\xi_{ef}\ll L$ and the corresponding action 
$S_u\simeq \pi^3\nu DA^2P_2/4$ are only weakly affected by the sample
length $L$, we may expect such states to determine the asymptotic
behavior of ${\cal P}_I(P_2)$ also in the localized regime,
$L\gg\xi$. However, if we compare the action found above with the
asymptotics of ${\cal P}_I(P_2)$ at $P_2\gg 1/\xi A$ in an infinitely
long sample [first line of Eq.(\ref{ijmpb_119})], 
we find that the absolute value of
the exponent in the latter case is larger by factor of $4$. The
explanation is as follows. The states determining the large-$P_2$ asymptotics
in the limit $L\to\infty$ can be obtained from Eq.~(\ref{ipr11}) by
taking a solution which has a maximum in the middle of the sample
($C_2=\pi/2$, $\sqrt{C_1}=2\pi/L$); the result is
\be
\label{ipr16}
A|\psi^2({\bf r})|_{\mbox{smooth}}={1\over\pi}{\xi_{ef}\over
x^2+\xi_{ef}^2} \ ; \qquad \xi_{ef}={1\over \pi A P_2}\ .
\ee
The only difference between (\ref{ipr15}) and (\ref{ipr16}) is that
ALS is now located in the bulk of the sample. This leads to an extra
factor 4 in the action (see similar discussion in Sec.~\ref{s4.2.3}
for the case of the statistics of eigenfunction amplitudes). However,
in the limit of a long sample, $L\xi\gg 1$ the contribution of the
states located near the boundary is additionally suppressed by a
factor $\sim 1/L$ as compared to that of the bulk ALS (which may be
located everywhere in the sample). Therefore, if one fixes $P_2$ and
considers ${\cal P}_I(P_2)$ in the limit $L\to\infty$, only the
contribution of the bulk ALS survives, despite the fact that it has
the exponent 4 times larger than that of the ALS located near the
boundary. In other words, the contribution of the states located near
the boundary to the first line of Eq.(\ref{ijmpb_119})
(large-$P_2$ asymptotics of ${\cal P}_I(P_2)$ at
$X=L/\xi\gg 1$) is $\propto {1\over X}e^{-\pi^2z/4}$. 

On the other hand, if a sample with periodic boundary conditions in
the longitudinal direction (a ring) is considered, only the bulk
solution ($C_2=\pi/2$, $\sqrt{C_1}=2\pi/L$) will
survive. Consequently, the asymptotic form of $\ln {\cal P}(P_2)$
will be different from (\ref{ipr14}) by an extra factor of 4.

\paragraph{2D geometry.}

Now we calculate the far asymptotics of ${\cal P}(P_2)$ at 
$P_2 \gg P_2^{\rm GUE}$.
We assume first the periodic boundary conditions (so that the sample
has no boundary)
and search for a rotationally-invariant solution, which should satisfy
the differential equation
\be
\label{ipr21}
\theta''+{1\over r}\theta'-\gamma(e^\theta-e^{2\theta})=0\ .
\ee
(In fact, for the hard-wall boundary conditions
the asymptotics is determined by the states located near
the boundary; however, such states can be obtained from the symmetric
solution by putting the center at the boundary and restricting the
solution to the interior of the sample; see below.) From our 
experience in the quasi-1D case, we expect the solution to have a form
of a bump concentrated in a region $r\lesssim l_p$ and decreasing with
$r$ outside this region. For $r\gg l_p$ the term $e^{2\theta}$ is
thus expected to become irrelevant, so that the equation takes the
form (\ref{e4.507}). Its solution is given by (\ref{anom87}) and can
be approximated at $l_p\ll r\ll L$ as $e^{\theta(r)}\simeq A/r^\alpha$
with some coefficient $A$ and exponent $\alpha$. In order to have large
$P_2$, we require $\alpha>1$. Using the condition (\ref{ipr8}), we
find scale $l_p$ to be given by
\be
\label{ipr22}
l_p\sim \left\{
\begin{array}{ll}
P_2^{-1/2}\ ,\qquad &  \alpha> 2\ ; \\
(L^{4-2\alpha}P_2)^{-1/[2(\alpha-1)]}\ ,\qquad & \alpha< 2\ , \\
\end{array}
\right.
\ee
up to a numerical coefficient of order unity. Therefore, the action
(\ref{ipr9}) is equal to 
\be
\label{ipr23}
S_u=\alpha^2\pi^2\nu D\ln (L/l_p) = {\cal F}(\alpha)\pi^2\nu D\ln (L^2P_2)\ ,
\ee
where
\be
\label{ipr24}
{\cal F}(\alpha)=\left\{
\begin{array}{ll}
\alpha^2/2\ ,\qquad & \alpha\ge 2\ ; \\
{\alpha^2\over 2(\alpha-1)}\ ,\qquad & 1<\alpha\le 2\ .
\end{array} \right.
\ee
Thus, the minimum of $S_u$ corresponds to $\alpha=2$, yielding
\be
\label{ipr27}
{\cal P}(P_2)\sim \left( {P_2\over \langle P_2
\rangle}\right)^{-\beta\pi g/2}\ .
\ee
The upper border of validity of (\ref{ipr27}) is $P_2\sim P_2^{\rm
RMT}(L/l)^2\sim 1/l^2$.

For hard wall boundary conditions,
the asymptotic behavior of ${\cal P}(P_2)$ will be, however,
determined by configurations with a maximum located near the sample
boundary. Assuming that the sample has a smooth boundary with the
single characteristic scale $L$ (for example, it is of the circular
form), we can get such a state from the rotationally invariant bulk
state by putting its center on the boundary and removing that half of
the state which is outside of  the sample.  Such a
truncated state will have the twice
larger IPR and twice smaller action compared to its parent bulk
state. Consequently, the asymptotics of the distribution function
for a sample with a boundary will be different from (\ref{ipr27}) by
an extra factor $1/2$ in the exponent.

Using the Liouville theory description (\ref{ipr44}), one can
generalize the above consideration to the distribution ${\cal P}(P_q)$
of higher IPR's $P_q$ (\ref{ipr41}) with $q>2$ \cite{mf98}. We will assume
that $q$ is not too large, $q^2<2\beta\pi g$,
so that the average value $\langle
P_q\rangle$ is at the same time the typical value of $P_q$ (see
Sec.~\ref{s3.3.2}). Then in the region $q^2/\beta\pi g \lesssim
P_q/\langle P_q\rangle-1\lesssim 1$ the distribution has the exponential form
(\ref{ipr43}). At larger $P_q$ the optimal configuration is again of
the form $|\psi^2({\bf r})|_{\rm smooth}\equiv e^{\theta({\bf
r})}=A/r^\alpha$ for $r>l_p$; minimizing the action, we find
$\alpha=q/4$ and the distribution function
\be
\label{ipr46}
{\cal P}(P_q)\sim {1\over P_q} \left({P_q\over \langle
P_q\rangle}\right)^{-2\beta\pi g/q^2}\ .
\ee
This is valid for $P_q<P_q^{\rm RMT}(L/l)^2$; for still larger $P_q$
the corresponding optimal fluctuation would violate the condition of
the applicability of the diffusion approximation $\theta'\lesssim
1/l$. Incorporating this restriction (cf. similar situation for the
distribution of relaxation times, Sec.~\ref{s4.1.2}) leads to
\be
\label{ipr47}
\alpha={1 \over q}\left[{\ln(P_q/P_q^{\rm RMT})\over
\ln(L/l)}-2\right]
\ee
and to the log-normal far asymptotics
\be
\label{ipr48}
{\cal P}(P_q)\sim {1\over P_q} \exp\left\{-{\beta\pi g\over 4 q^2}\,
{\ln^2[(P_q/P_q^{\rm RMT})(L/l)^2]\over\ln (L/l)}\right\}
\ee
for   $(L/l)^2\lesssim P_q/P_q^{\rm RMT}\lesssim (L/l)^{2q-2}$.

\subsection{3D systems}
\label{s4.4}

As we will see below, in the 3D case the states determining the
asymptotics of the distribution functions have just a local
short-scale spike on top of a homogeneous background. In this sense,
no ALS is formed, in contrast to the quasi-1D and 2D situations. As a
closely related feature, we find that the results in 3D are strongly
dependent on microscopic details of the random potential. We start
from discussing what  the $\sigma$-model calculation gives when
applied to the 3D geometry. Then we compare this with the results of
the direct optimal fluctuation method \cite{smolyarenko97}. 

Let us consider e.g. the long-time relaxation problem. The solution of
the $\sigma$-model saddle-point equation has the form \cite{m95}
\be
\label{3d.1}
\theta(r)\simeq C\left({l_*\over r}-{l_*\over R}\right)\ , \qquad
l_*\le r\le R\ ,
\ee
where $R$ is the system size (which is, in fact, irrelevant here) and
\be
\label{3d.2}
l_*\sim l \ln\left[{t\over \tau(k_Fl)^2}\right]\ ; \qquad C\sim 
\ln\left[{t\over \tau(k_Fl)^2}\right]\ .
\ee
The reason for introduction of the short-scale cut-off length $l_*$ is
the same as in 2D (see Sections \ref{s4.1.2}, \ref{s4.2.2}): the
gradient $\theta'$ should not exceed $\sim 1/l$. 
Calculating the action, we find
\be
\label{3d.3}
-\ln P_s(t) \sim (k_Fl)^2 \ln^3\left[{t\over \tau(k_Fl)^2}\right]
\ee
with an uncertainty in numerical prefactor. This uncertainty
originates from the fact that the action is dominated by the
ultraviolet (short-distance) region $r\sim l_*$. Similar result is
obtained for the eigenfunction amplitude statistics \cite{fe2,m97}
\be
\label{3d.4}
-\ln {\cal P}(u) \sim  (k_Fl)^2 \ln^3{Vu\over (k_Fl)^2}
\ee
and for the LDOS distribution function \cite{m96}
\be
\label{3d.5}
-\ln {\cal P}(\rho) \sim  (k_Fl)^2 \ln^3{\rho\over\nu}\ .
\ee
As we see, all these distribution functions are found to have the
exponential-log-cube asymptotic form. The corresponding eigenstates
have the shape
$$
|\psi^2({\bf r})|_{\rm smooth} \simeq {1\over V} \exp \left\{
C_i {l\over r} \ln^2 Z_i \right\}\ ,
$$
where $C_i\sim 1$, $Z_t=t/[\tau(k_F l)^2]$, $Z_u=Vu/(k_Fl)^2$, and
$Z_\rho=\rho/\nu$, i.e. they consist of the usual homogeneous
background $|\psi^2({\bf r})|_{\rm smooth}=1/V$ supplemented by a
short-scale bump containing only a minor part of the eigenfunction
normalization. 

Muzykantskii and Khmelnitskii \cite{mk2} applied the ballistic
$\sigma$-model approach to the problem of the long-time
relaxation. They reproduced the result (\ref{3d.3}) and found the
numerical prefactor there to be equal to $\pi/9\sqrt{3}$. On the other
hand, Smolyarenko and Altshuler \cite{smolyarenko97} employed the
direct optimal fluctuation method to study the asymptotics of the
distribution function ${\cal P}(u)$. This approach is similar to that
used many years ago  \cite{halperin66,zittartz66} to calculate the
tails of the density of states of a disordered system. The authors of 
\cite{smolyarenko97} did not use the $\sigma$-model description, but
rather searched directly for the optimal realization of the random
potential (hence their term ``direct optimal fluctuation
method''). While having confirmed the exponential-log-cube form of the
asymptotics in the 3D case, they found a prefactor smaller by $\sim
k_Fl$ compared to the $\sigma$-model result, i.e. they obtained
\be
\label{3d.6}  
{\cal P}(u)\sim k_Fl \ln^3(uV)\ .
\ee

The physical reason for this difference lies in the ballistic effects
which have been already discussed in Sec.~\ref{s3.3.3} in connection
with perturbative corrections to the eigenfunction amplitude
distribution. It was shown there that in the 3D case and for the
white-noise random potential the parameter $\kappa$ governing these
corrections is dominated by a non-universal (depending on the type of
the disorder) ballistic contribution yielding $\kappa\sim 1/k_F l$,
while the diffusive contribution is $\sim 1/(k_F l)^2$. This is in
direct correspondence with Eqs.~(\ref{3d.4}), (\ref{3d.6}), which show
precisely the same difference. Eq.~(\ref{3d.6}) is again
non-universal; its derivation by Smolyarenko and Altshuler 
\cite{smolyarenko97}  relies on
the white-noise disorder assumption.  The corresponding optimal
configuration of the potential found in \cite{smolyarenko97} is
nothing else but a potential wall surrounding the observation point
with the height several times larger than $E_F$ and the thickness
$\sim\lambda_F\ln(uV)$. Such configurations are not included in the
$\sigma$-model consideration, which assumes that the absolute value
of the particle velocity does not change appreciably in space. If one
would consider a smooth random potential, whose magnitude is limited
from above by some value $U_{\rm max}\ll E_F$, such configurations
would not be allowed. Whether in this case the $\sigma$-model result
would hold remains to be seen.

\subsection{Discussion}
\label{s4.5}

In Section \ref{c4}, we have studied the asymptotic behavior of various
distribution functions characterizing the eigenfunction statistics in
a disordered sample. For this purpose, we used two methods of treatment
of the $\sigma$-model:  exact solution (in the quasi-1D case) and
the saddle-point method. Physically, the saddle-point solution
describes the relevant optimum fluctuations of the wave function
envelope; probability of formation of such a fluctuation is found to
be governed by the Liouville theory (\ref{ipr44}).  
In the quasi-1D case, the results of the saddle-point method are in
agreement with those of the exact solution of the $\sigma$-model.

The 2D geometry is of special interest, since the eigenfunctions show
the features of criticality.
In this case, a full  agreement between the saddle-point calculation
and the renormalization-group (RG) treatment of Altshuler, Kravtsov and
Lerner \cite{akl} was found for all the distributions, where such a
comparison was possible, namely for ${\cal P}(t_\phi)$, ${\cal
P}(\rho)$ and ${\cal P}(\nu)$. 
This agreement is highly non-trivial, for the following
reason. The RG treatment is based on a resummation of the perturbation
theory expansion and can be equally well performed within the replica
(bosonic or fermionic) or supersymmetric formalism. At the same time,
the present approach based on the supersymmetric formalism relies
heavily on the topology of the saddle-point manifold combining
non-compact ($\lambda_1$) and compact ($\lambda_2$) degrees of
freedom. The asymptotic behavior of the distribution functions
considered  is determined by the region $\lambda_1\gg 1$ which is very
far from the ``perturbative'' region of the manifold $Q\simeq\Lambda$
(i.e. $\lambda_1,\lambda_2\simeq 1$). It is well known \cite{verbzirn}
that for the problem
of energy level correlation, the replica approach 
fails, since it does not reflect properly 
the topology of the $\sigma$-model manifold.
The success of the  RG treatment of \cite{akl} seems to be determined
by the fact that for the present problem (in contrast to that of level
correlation) only the non-compact sector
of the supersymmetric $\sigma$-model is essential, with
compact one playing an auxiliary role. Let us note that the same
situation appears in the vicinity of the Anderson transition 
\cite{zirn86,mf94a,mf94b}
where the function $Y(\lambda_1,\lambda_2)$ 
acquires a role of the order parameter function and depends on the
non-compact variable $\lambda_1$ only. The above agreement found
for the ``tails'' of the  distributions in the metallic region
 provides therefore  support to the
results concerning the Anderson transition
obtained with making use of the renormalization group approach
and $2+\epsilon$ expansion.

We have
found that the spatial structure of ALS relevant to the asymptotic
behavior of different distributions may be different. This is because
an ALS constitutes an optimal fluctuation for one of the above
quantities, and the form of this fluctuation depends on the specific
characteristic, which is to be optimized. 
Finally, we have discussed
interrelations between asymptotics of various distribution functions.
In the quasi-1D and 2D cases, we thus presented a
comprehensive picture which explains all the studied 
asymptotics as governed
by exponentially rare events of formation of ALS.

\section{Statistics of energy levels and eigenfunctions at the
Anderson transition}
\label{c5}

\setcounter{equation}{0}

In $d>2$ dimensions a disordered system
undergoes, with increasing strength of the disorder, 
 a transition from the phase of extended states to that of
localized states  (see
e.g. \cite{leeram} for review). This transition changes drastically the
statistics of energy levels and eigenfunctions. In particular, at the
mobility edge these statistics acquire distinct features reflecting
criticality of the theory. This is the subject of the present
section. Sec.~\ref{s5.2} is devoted to the level statistics and
Sec.~\ref{s5.3} to the eigenfunctions correlations at the mobility
edge. In Sec.~\ref{s5.4} we study the level and eigenfunction
statistics in a quasi-one-dimensional model with long-range
(power-law) hopping which undergoes the Anderson transition and shows
at criticality all the features characteristic for a conventional 
metal-insulator transition point in $d>2$.

\subsection{Level statistics. Level number variance}
\label{s5.2}

The problem of the energy level statistics at the mobility edge was
addressed for the first time by Altshuler {\it et al},
Ref. \cite{aszk}, who considered the variance
$\langle\delta N^2(E)\rangle=\langle N^2(E)\rangle-\langle 
N(E)\rangle^2$ of the number of levels within a band of a width $E$. 
This quantity is related to the 
two--level correlator (\ref{e2.1}) via
\begin{equation}
\langle\delta N(E)^2\rangle=\int_{-\langle N(E)\rangle}^{\langle N(E)\rangle}
(\langle N(E)\rangle-|s|) R^{(c)}(s) ds\ ,
\label{crit_eq2}
\end{equation}
or, equivalently,
\begin{equation}
{d \over d\langle N(E)\rangle}
\langle\delta N(E)^2\rangle=\int_{-\langle N(E)\rangle}^{\langle N(E)\rangle}
R^{(c)}(s) ds\ ,
\label{crit_eq2b}
\end{equation}
where $\langle N(E)\rangle=E/\Delta$ and $R^{(c)}(s)=R(s)-1$ 
is the connected part of the two-level correlation function.
In RMT, the $1/s^2$ behavior of $R^{(c)}(s)$ leads to the 
logarithmic behavior of the variance $\langle\delta N^2\rangle_{WD}
\simeq (2/\pi^2\beta) \ln \langle N\rangle$ for $\langle N\rangle\gg 1$. 
In the opposite situation characteristic for the phase of localized
states, when all 
energy levels are completely uncorrelated (known as the Poisson statistics),
one gets $\langle\delta N^2\rangle_P=\langle N\rangle$. Supported by
their numerical simulations, Altshuler et al.
\cite{aszk} put forward a conjecture that at the critical point
\begin{equation} 
\langle\delta N^2\rangle\simeq\chi\langle N\rangle\ ,
\label{crit_eq2a}
\end{equation}
where $0<\chi<1$ is a numerical coefficient (which is
conventionally called now ``spectral compressibility''). 
More recently, Shklovskii et al. \cite{merman} introduced the
concept of new universal statistics at the mobility edge (see also
Ref.~\cite{hofstetter93}). In Ref.~\cite{klaa} 
the correlator $R(s)$ at the mobility edge was studied by means 
of perturbation theory combined with scaling assumptions about a form
of the diffusion propagator. It was found that for $s\gg 1$,
\begin{equation}
R^{(c)}(s)\propto s^{-2+\gamma}
\label{crit_eq3}
\end{equation}
where $\gamma<1$ is certain critical index. The consideration of
Ref. \cite{klaa} led to the conclusion that $\gamma=1-1/\nu d$ (where
$\nu$ is the critical exponent of the localization length), which was
however questioned later \cite{kravtsov96} in view of the
oversimplified treatment of the diffusion propagator at the transition
point in \cite{klaa}. At any rate, the behavior (\ref{crit_eq3}) 
with some $\gamma$ is what one expects to hold at the mobility edge;
the condition $\gamma<1$ follows from the requirement of convergence
of $\int R(s)ds$.

Using Eqs. (\ref{crit_eq2}), (\ref{crit_eq3}) and the sum rule
\begin{equation}
\int R(s)ds=0
\label{crit_eq4}
\end{equation}
(implied by the conservation of the number of energy levels), 
the authors of \cite{klaa} concluded that
\begin{equation}
\langle \delta N^2\rangle\propto N^\gamma
\label{crit_eq5}
\end{equation}
with $\gamma<1$, in contradiction with Ref.~\cite{aszk}. This
conclusion was critically reexamined in Refs.~\cite{am,kraler95}, where
it was shown that  the asymptotic behavior (\ref{crit_eq3}) of the
correlator $R(s)$ at $s\gg 1$ does not imply the absence of the
linear term (\ref{crit_eq2a}). The flaw in reasoning of
Ref.~\cite{klaa} was in the assumption that the universal part of the 
correlator $R(s)$ (which is the one surviving in the limit
$E/\Delta=\langle N\rangle={\rm const}$, $L\to\infty$) satisfies the sum rule
(\ref{crit_eq4}). It turns out, however, that the sum rule
is fulfilled only if all contributions are taken into account, including
the non-universal contribution of the ``ballistic'' 
region $\omega\sim 1/\tau$, where
$\tau$ is the elastic mean free time. To demonstrate this, we estimate
below (following Ref.~\cite{am})  the
contributions to the sum rule from all regions of the variable $s$.

In fact, for the conventional model of a particle in a random potential
the definition (\ref{e2.1}) of the correlator and the sum rule relation
(\ref{crit_eq3}) should be modified, when the vicinity of the critical point is 
considered. The reason is that the Anderson transition point corresponds 
to a strong disorder regime $E_F\tau\sim 1$, so that the condition
$\omega\sim 1/\tau$ implies $\omega\sim E_F$. On the other hand,
the density of states $\nu(E_F)$ can be considered as a constant 
only for small variations of energy $\omega\ll E_F$. This means that
variation of $\nu(E_F)$ should be taken into account in (\ref{e2.1}),
(\ref{crit_eq4}). Besides, the condition $E_F\tau\sim 1$ leads to a
breakdown of the perturbation theory, that complicates the analysis of the
``ballistic'' region contribution. To get rid of these problems, we 
consider a different microscopic model which has exactly the same
{\it universal} part of $R(s)$, but whose density of states does not
change within the range of $\omega\sim 1/\tau$. This is so-called
$n$--orbital Wegner model \cite{wegner80a}, 
which can be considered as a system
of metallic granules forming a $d$--dimensional lattice, each granule
being coupled to its nearest neighbors. In the limit $n\gg 1$ this model 
can be mapped onto
a supersymmetric $\sigma$--model defined on a lattice. The action of this
$\sigma$--model reads \cite{zirn86} 
(we consider the unitary symmetry for definiteness)
\begin{equation}
S\{Q\}={\Gamma\over 2}\sum_{\langle ij\rangle}\mbox{Str} Q_i Q_j +
\varepsilon \sum_i \mbox{Str} Q_i\Lambda\ ,
\label{crit_eq4a}
\end{equation}
where the supermatrices $Q_i$ are defined
on sites $i$ of a $d$--dimensional lattice with a lattice spacing
$a$. Summation in the first term of 
Eq.(\ref{crit_eq4a}) goes over the pairs $\langle ij\rangle$ of
nearest neighbors. The parameters $\Gamma$ and $\varepsilon$
are related to the classical diffusion constant $D$, the density of
states $\nu$, and the frequency $\omega$ as follows:
\begin{equation}
\Gamma=\pi\nu D a^{d-2}\ ,\qquad \varepsilon=-i\omega\pi\nu a^d/2\ .
\label{crit_eq5a}
\end{equation}

The two--level correlator can be expressed through a correlation function
of this $\sigma$--model via the discretized version of Eq.~(\ref{e2.2}):
\begin{equation}
R(s)=\mbox{Re}\int \prod_j DQ_j\left\{\left[{a^d\over 4V}
\sum_i \mbox{Str} Q_i k\Lambda\right]^2-1\right\} \exp(-S\{Q\})\ ,
\label{crit_eq6}
\end{equation}
where $V$ is the system volume. 
It is not difficult to prove explicitly \cite{am}
that the two-level correlation
function of this lattice
$\sigma$--model, Eq.~(\ref{crit_eq6}), satisfies exactly the 
sum rule (\ref{crit_eq4}). 
As will become clear below, the continuum version of the
$\sigma$--model does not possess this property: there is a deficiency
of the sum rule for it which is related with the contribution of the
range of $\omega$ close to the ultraviolet cut-off. 
%This phenomenon is
%similar to what is known as quantum anomalies in the quantum field theory
%(see e.g. \cite{anom}). 

Let us stress that in the region $\omega\ll D/a^2\equiv E_c(a)$ the 
correlator $R(s)$ is universal, i.e. does not depend on microscopical 
details of the model. The region $\omega\sim D/a^2$ plays a role
analogous to that of the  ballistic region, $\omega\sim 1/\tau$, in
the case of 
the usual model of particle in a random potential. Despite the 
non-universality of the correlation function $R_{nu}(s)$ in this
``ballistic'' domain, the corresponding integral contribution $I_{nu}$
to the sum
rule is universal, because it determines, according
to Eq.~(\ref{crit_eq4}), the sum rule deficiency for the universal part
$R_{u}(s)$:
\begin{eqnarray}
&&I_{u}+I_{nu}=0\ ;  \nonumber\\
&&I_u=\int R^{(c)}_u(s) ds\ ;\qquad I_{nu}=\int R^{(c)}_{nu}(s) ds
\label{crit_eq8}
\end{eqnarray}
As will be seen below, 
when the system is close to the Anderson transition 
the two regions of the variable $s$ dominating the
integrals $I_u$ and $I_{nu}$, respectively,  are separated
by a parametrically broad range of $s$ 
giving a negligible contribution to the sum rule. For energy band width $E$
lying in this range, we get from Eqs.(\ref{crit_eq2b}), (\ref{crit_eq8})
\begin{equation}
\langle\delta N(E)^2\rangle\simeq I_u\langle N(E)\rangle=
-I_{nu}\langle N(E)\rangle\ , 
\label{crit_eq9}
\end{equation}
i.e. just the linear term (\ref{crit_eq2a}) with $\chi=I_u=-I_{nu}$.

We turn now to the analysis of the correlator $R(s)$ and of the sum
rule in various situation. For completeness, 
we start from the case of a good metal,
then we consider the critical point and the critical region cases.

(1) Good metal. Here the following three regions with different behavior
of $R(s)$ can be found:

(A) ``Wigner--Dyson'' (WD) region: $\omega\ll E_c$. The correlator
$R(s)$ in this region was studied in Sec.~\ref{s2.2}, see
Eqs.~(\ref{e2.16}), (\ref{e2.19}).  The corresponding contribution
to the sum rule can be estimated as (we omit
numerical factors of order unity)
\begin{equation}
I_A\simeq\int_0^g R(s)\propto +1/g\ .
\label{crit_eq9a}
\end{equation}

(B) ``Altshuler--Shklovskii'' (AS) region:
$E_c\ll\omega\ll E_c(a)$. Here $E_c(a)=D/a^2$ is the
Thouless energy at the scale of lattice spacing $a$, which plays a role
of the ultraviolet cut-off for the diffusion theory. The level
correlation function is given by Eq.~(\ref{e2.22}), yielding
\begin{equation}
I_B\simeq\int_g^{D/a^2 \Delta } R(s) ds\propto
g^{-d/2}\left({D\over a^2\Delta}\right)^{{d\over 2}-1}\propto +1/g(a)\ ,
\label{crit_eq11}
\end{equation}
where $g(a)=g(a/L)^{d-2}$ is the conductance at the scale $a$. Note that
for the case of a particle in a random potential, the following substitutions
should be done: $a\to l$; $E_c(a)\to E_c(l)=1/\tau$; 
$1/g(a)\to 1/g(l)\propto (\epsilon\tau)^{d-1}$. 

(C) ``Ballistic'' region: $\omega\gtrsim E_c(a)$.  To find the
correlator $R(s)$ in this range, we can neglect in the leading
approximation 
the first term in Eq.~(\ref{crit_eq4a}), that gives
\begin{eqnarray}
&&R^{(c)}(s)\propto - \left({L\over a}\right)^d {1\over s^2}\ , 
\label{crit_eq12}\\
&&I_C\simeq \int_{D/a^2\Delta}^\infty R^{(c)}(s)ds\propto -1/g(a)\ .
\label{crit_eq13}
\end{eqnarray}
The contribution (\ref{crit_eq9a}) to the sum rule is dominated by the
region $\omega\sim\Delta$, whereas the contributions (\ref{crit_eq11}),
(\ref{crit_eq13}) are dominated by a vicinity of the ultraviolet cut-off
$\omega\sim D/a^2$. In $d=3$ $g\gg g(a)$, and these non-universal
contributions $I_B$, $I_C$ are much larger (by absolute value) than $I_A$,
but they should cancel each other according to the sum rule. The situation
is more interesting in $d=2$, where the contribution (\ref{e2.22}), 
(\ref{crit_eq11}) of the AS region is absent in view of special analytical 
properties of the diffusion propagator \cite{algef93,algef95,kraler95}. 
In this case $g(a)=g$,
so that the contributions $I_A$ and $I_C$ are of the same order,
in agreement with the sum rule which prescribes their sum to be zero. 
We can identify then $I_A$ with $I_u$ and $I_C$ with $I_{nu}$
in Eq.~(\ref{crit_eq8}), so that $I_u=-I_{nu}\propto 1/g$.
According to Eq.~(\ref{crit_eq9}), this leads to a linear behavior of
the variance,
 \begin{equation}
\langle \delta N^2\rangle\propto (1/g) \langle N\rangle\ ;\qquad
E_c\ll E\ll D/a^2\ ,
\label{crit_eq14}
\end{equation}
but the corresponding coefficient is of order $1/g\ll 1$. The
numerical coefficient in (\ref{crit_eq14}) can be calculated using the
explicit form of $R^{(c)}(s)$, yielding
\be
\label{crit1}
\chi\equiv I_u \simeq {1\over 2\beta\pi g}\ , \qquad g\gg 1\ .
\ee

With an increase in  disorder strength, the coupling constant $\Gamma$
in Eq.~(\ref{crit_eq4a}) decreases. 
When it approaches a critical value $\Gamma_c$,
corresponding to the metal--insulator transition, the correlation
length $\xi$ becomes large: $\xi\gg a$. Depending on the relation between
$\xi$ and the system size $L$, the following two situations are to be
distinguished.

(2) Metal in the critical region: $a\ll\xi\ll L$. In this case $g(a)\simeq g_*$,
but $g\equiv g(L)\gg g_*$, where $g_*$ is the critical value of the 
conductance\footnote{In 3D the value of $g_*$ is, of course, of
order unity. 
However, if one consider formally $d=2+\epsilon$ with $\epsilon\ll 1$, then
$g_*\sim 1/\epsilon$ is parametrically large. It is thus instructive
to keep $g_*$ as a parameter in all estimates.}. 
We find then as much as four different regions of
$s$ for the correlator $R(s)$:

(A)  WD region: $\omega\ll E_c$. The correlator $R(s)$ and the contribution 
to the sum rule for this region are given by the same
Eqs.~(\ref{e2.16}), (\ref{e2.19}) and (\ref{crit_eq9a}), as for a good metal. 

(B)  AS region: $E_c\ll \omega\ll E_c(\xi)=g_*\Delta_\xi$, where $E_c(\xi)$
and $\Delta_\xi$ are Thouless energy and level spacing for a piece of the
sample with linear size $\xi$. The correlator $R^{(c)}(s)$ is given by 
Eq.~(\ref{e2.22}), and we find
\begin{equation}
I_B\simeq\int_g^{g_*\Delta_\xi/\Delta} R^{(c)}(s)ds
\propto g^{-d/2}g_*\left({\Delta_\xi\over\Delta}\right)^{{d\over 2}-1}
\propto +1/g_*\ ,
\label{crit_eq15}
\end{equation}
where we have used the relations $\Delta_\xi/\Delta=(L/\xi)^d$
and $g/g_*\propto(L/\xi)^{d-2}$. 

(C) ``Kravtsov--Lerner--Altshuler--Aronov'' (KLAA) region:
$E_c(\xi)\equiv g_*\Delta_\xi\ll\omega\ll g_*\Delta_a\equiv E_c(a)$.
The correlator $R^{(c)}(s)$ in this range was studied in Ref.~\cite{akl95}; 
the result reads:
\begin{equation}
R^{(c)}(s)\simeq -g_*^{-\gamma}\left({\Delta_\xi/\Delta}\right)^{1-\gamma}
s^{-2+\gamma}
\label{crit_eq16}
\end{equation}
and, consequently,
\begin{equation}
I_C\simeq\int_{g_*\Delta_\xi/\Delta}^{g_*\Delta_a/\Delta}R^{(c)}(s)ds
\propto -1/g_*
\label{crit_eq17}
\end{equation}

(D) ``Ballistic'' region: $\omega\gtrsim E_c(a)$. Here the equations
(\ref{crit_eq12}), (\ref{crit_eq13}) hold, yielding
\begin{equation}
I_D\propto -1/g_*
\label{crit_eq18}
\end{equation}

The contribution $I_A$ is dominated by  $\omega\sim\Delta$,
the contributions $I_B,\: I_C$ by $\omega\sim\Delta_\xi$, and finally,
the contribution $I_D$ comes from the ``non-universal'' region
$\omega\sim E_c(a)$. Therefore, this last contribution determines
$I_{nu}$ in Eq.~(\ref{crit_eq8}), leading according to 
Eq.~(\ref{crit_eq9}) to the linear behavior of the variance
\begin{equation}
\langle \delta N^2\rangle\propto (1/g_*)\langle N\rangle\ ;
\qquad E_c(\xi)\ll E\ll E_c(a)
\label{crit_eq19}
\end{equation}
with a coefficient $\sim 1/g_*$ (which is of order unity in 3D). 

(3) Finally, a system is at the critical point, when $\xi\gg L$. In this case,
$g(L)=g(a)=g_*$. The AS region disappears, and we have, 
in full analogy with the previous estimates:

(A) WD region: $\omega\lesssim E_c=g_*\Delta$.
\begin{equation}
I_A\propto +1/g_*
\label{crit_eq20}
\end{equation}

 (B) KLAA region: $g_*\Delta\ll\omega\ll \Delta_a g_*$.
\begin{equation}
 R(s)\propto-g_*^{-\gamma}s^{-2+\gamma}\ ;\qquad
 I_B\propto -1/g_*
\label{crit_eq21}
\end{equation}

(C) ``Ballistic'' region: $\omega\gtrsim\Delta_a g_*$.
\begin{equation}
I_C\propto -1/g_*
\label{crit_eq22}
\end{equation}

Again the same conclusion, as for the critical region case, can be drawn:
the ``ballistic'' contribution (\ref{crit_eq22}) can be identified 
as $I_{nu}$ in Eq.~(\ref{crit_eq8}), yielding:
\begin{equation}
\langle \delta N^2\rangle\simeq\chi\langle N\rangle\ ,\qquad
E_c\ll E\ll E_c(a)\ ,
\label{crit_eq23}
\end{equation}
with $\chi\sim 1/g_*$.

Recently, a relation between the spectral compressibility $\chi$ and
the multifractal dimension $D_2$ (for discussion of multifractality
and corresponding bibliographic references, see Sec.~\ref{s3.3.2} and
\ref{s5.3}) was proposed \cite{ckl,kravtsov96}:
\be
\label{crit2}
\chi = {d-D_2\over 2d}\ .
\ee
The central idea of the derivation \cite{ckl} is to consider the
motion of energy levels, when the system is subject to a random
perturbation. This allows to link the spectral statistics with the
wave function correlations.
In 2D, one can check, by comparing Eqs.~(\ref{multifr5}) and (\ref{crit2}),
that Eq.~(\ref{crit2}) is indeed satisfied in
leading order in $1/g$. The general derivation of Eq.~(\ref{crit2})
is, however, based on certain approximate decoupling of a higher order
correlation function \cite{chalker96a,chalker96b,ckl}, so that it is
not completely clear whether this is indeed an exact formula as argued
in \cite{ckl} or only an approximation valid for $g_*\gg 1$. 

The linear behavior (\ref{crit_eq2a})
of the level number variance at the Anderson transition
 has been confirmed by now in numerical simulations
by several groups \cite{sears,zhar1,zhar2,braun}. The spectral
compressibility $\chi$ is an important universal parameter
characterizing the critical point of Anderson transition. Its
universality is of the same sort as that of the critical indices,
i.e. $\chi$ depends only on the spatial dimensionality and on the
symmetry (universality) class. Let us note that the whole
level correlation function $R^{(c)}(s)$ is not as universal,
since it depends also on the shape of the sample and on the boundary
conditions \cite{braun,potempa98a,potempa98b,kravtsov99}. This can be
expected 
already from the perturbative $1/g^2$ correction, Eqs.~(\ref{e2.16}),
(\ref{e2.19}), where the coefficient $a_d$ does depend on the shape
and on the boundary conditions. Also, the shape dependence becomes
evident if one considers the limit of an elongated sample with a
length considerably larger than the transverse sample size. Indeed,
let us consider (in 3d) a rectangular sample with
$L_y=L_z=L_x/\alpha$, where $\alpha$ is a numerical factor
\cite{potempa98a}. 
If we fix $\alpha$ and consider the limit $L\to\infty$, the level
correlation function has a limiting form, which, depends, however on
$\alpha$. In particular, at $\alpha\gg 1$ the sample is of quasi-1D
geometry with a ratio of the sample length to the localization length
$\sim\alpha$, so that the level statistics will be close to
Poissonnian. 

Asymptotic behavior of the nearest-neighbor level spacing distribution
function $P(s)$ for $s\gg 1$ at the mobility edge has been also a
controversial issue. We remind that for the Poissonnian statistics
$P(s)=e^{-s}$, while in the RMT $P(s)\sim e^{-{\rm const}\,s^2}$.
While Refs.~\cite{aszk,merman} conjectured that 
$P(s)\sim e^{-{\rm const}\,s}$, the authors of Ref.~\cite{akl94} found
$P(s)\sim e^{-{\rm const}\,s^{2-\gamma}}$ with the same index $\gamma$
as in Eq.~(\ref{crit_eq3}). Recent numerical
studies  \cite{zhar97,zhar98}
support the former result, $P(s)\sim e^{-{\rm const}\,s}$.

\subsection{Strong correlations of eigenfunctions near the Anderson
transition}
\label{s5.3}

In this section, we discuss correlations between the amplitudes of
different, but close in energy, eigenstates at and near the critical
point of the Anderson transition. Let us recall that in the metallic
phase far from the transition point a typical eigenfunction covers
essentially uniformly the sample volume. This is reflected in the
inverse participation ratio $P_2$, as well as in the higher moments $P_q=\int
d^dr |\psi^{2q}({\bf r})|$, which differ only weakly from their RMT
values, see Eqs.~(\ref{e3.10}), (\ref{e3.11}). When the
system approaches the point of the
Anderson transition $E_c$, these extended eigenfunctions become less and
less homogeneous in space, showing regions with larger and smaller
amplitudes and eventually forming a multifractal structure in the
vicinity of $E_c$, see Sec.~\ref{s3.3.2}. 
This multifractal behavior is characterized by the
following behavior of the moments $P_q$ at the critical point
\be
\label{e5.1}
P_q\propto L^{-D_q(q-1)}\ ,\qquad L<\xi\ ,
\ee
as well as in the conducting phase in the vicinity of the critical
point:
\be
\label{e5.2}
P_q\propto \xi^{(d-D_q)(q-1)}L^{-d(q-1)}\ ,\qquad L>\xi\ .
\ee
Here $\xi\propto |E-E_c|^{-\nu}$ is the correlation length and $D_q$ is
the set of multifractality exponents, $d>D_2>D_3>\ldots$.  As
Eq.~(\ref{e5.2}) indicates, the eigenfunctions become more and more
sparse when the system approaches the critical point of the Anderson
transition from the metallic phase (i.e. when $\xi$ increases). Just at the
mobility edge the scaling of IPR with the system size $L$ becomes
different, see Eq.~(\ref{e5.1}), so that the eigenfunction effectively
occupies a vanishing fraction of the system volume.
At last, in the insulating phase any eigenstate
is localized in a domain of finite extension $\xi$, and IPR remains
finite in the limit of infinite system size $L\to \infty$.

This transparent picture serves as a basis for qualitative
understanding of spectral properties of disordered conductors.
Indeed, as long as eigenstates are well extended and cover the whole
sample, they overlap substantially, and corresponding energy levels
repel each other almost
in the same way as in RMT. As a result, the Wigner-Dyson
(WD) statistics describes 
well energy levels in a good metal, see Section~\ref{c2}.
In contrast, in the insulating phase different eigenfunctions
corresponding to levels close in energy are localized far apart from
one another and their overlap is negligible. This is the reason for
absence of correlations of energy levels in this regime, i.e.
the Poisson statistics.

A naive extrapolation of this argument to the vicinity of
the transition point would lead to a wrong  conclusion. Indeed, one might
expect that sparse (multifractal in the
critical point) eigenstates fail to overlap, that would result in  
essential weakening of level correlations close to the mobility edge and
vanishing level repulsion at $E=E_c$. However, numerical
simulations show \cite{aszk,merman,evangelou94,braun95,zhar95} that even
at the mobility edge levels
repel each other strongly (although the level statistics is different
from RMT). As we are going to explain now, 
this apparent contradiction is resolved in the following way:
The critical eigenstates for
nearby levels are so strongly correlated that they
overlap well in spite of their sparse structure.

To demonstrate the strong correlations, we will consider the relation
between the overlap function $\sigma({\bf r},{\bf r},E,\omega)$,
Eq.~(\ref{corrrev_e2}), and the ``self-overlap'' $\alpha({\bf r},{\bf
r},E)$, Eq.~(\ref{corrrev_e1}) (the latter determining the IPR $P_2$
according to $P_2=\int d{\bf r}\,\alpha({\bf r},{\bf r},E)$.
As was shown\footnote{Eq.~(\ref{corrrev_9a}) was
explicitly derived in \cite{fm97} for the case of the sparse random
matrix model \cite{mf91a,fm91a} corresponding to the limit of infinite
dimensionality, $d=\infty$. So, even at $d=\infty$, where the wave
function sparsity (multifractality) takes its extreme form, nearby in
energy eigenfunctions are fully correlated.} in \cite{fm97}, everywhere in the
metallic phase (including the vicinity of the critical point)
\begin{equation}
\sigma(r,r,E,\omega)=\frac{\beta}{\beta+2}\alpha(r,r,E)
\label{corrrev_9a}
\end{equation}
for $\omega\ll\Delta_\xi$, where $\Delta_\xi\sim 1/\nu\xi^d$ is the
level spacing in the correlation volume. Eq.~(\ref{corrrev_9a}) 
implies the following structure of eigenfunctions within
an energy interval $\omega<\Delta_\xi$.
Each eigenstate can be represented as a product
$\Psi_i({\bf r})=\psi_i({\bf r})\Phi_E({\bf r})$.
Here the function $\Phi_E({\bf r})$ is an eigenfunction envelope 
of ``humps and dips''.   It is the  same for
all eigenstates around  energy E, reflects the underlying gross  
(multifractal)
spatial structure, and governs the divergence of the inverse
participation ratios $P_q$
at the critical point. In contrast, $\psi_i({\bf r})$ shows
RMT-like fluctuations on the scale of the wave length.
It fills the envelope function $\Phi_E({\bf r})$ in an individual way for
each eigenfunction, but is not critical, i.e. is not sensitive to  
the vicinity
of the Anderson transition. These Gaussian fluctuations  are
responsible for the factor $\beta/(\beta+2)$ (which is the same as in
the corresponding Gaussian Ensemble) in Eq.~(\ref{corrrev_9a}).

As was already mentioned, this picture is valid in the energy window
$\omega\sim \Delta_\xi$ around the energy $E$; the number of
levels in this window being large as $\Delta_\xi/\Delta\sim
(L/\xi)^d\gg 1$ in the  limit $L\gg\xi$. These states
form a kind of Gaussian Ensemble on a spatially non-uniform
(multifractal for $E\to E_c$) background $\Phi_E({\bf r})$. Since the
eigenfunction correlations are described by the formula (\ref{corrrev_9a}),
which has exactly the same form as in the Gaussian Ensemble, it is not
surprising that the level statistics has the WD form
everywhere in the extended phase \cite{mf91a,fm91a}. For larger
$\omega$, $\sigma({\bf r},{\bf r},E,\omega)$ is expected to decrease as
$\omega^{-\eta/d}$, where $\eta=d-D_2$, according to the scaling arguments
\cite{chalker88,chalker90,huck94,brandes96,pracz96,janssen98} (see
below), so that  
\be
\label{critcorr_add1}
\sigma({\bf r},{\bf r},E,\omega)/\alpha({\bf r},{\bf r},E)\sim
(\omega/\Delta_\xi)^{-\eta/d}\ ;\qquad \omega > \Delta_\xi\ ,
\ee
up to a numerical coefficient of order unity. 

The above formulas are valid in the metallic regime, i.e. for
$L\gg\xi$. Exactly at the critical point ($\xi\gg L$) they take the form
\begin{equation}
\label{corrrev_10}
\sigma({\bf r},{\bf r},E,\omega)/\alpha({\bf r},{\bf r},E)\sim 1\
 ,\qquad \omega<\Delta 
 \end{equation}
and 
\be
\label{critcorr_add2}
\sigma({\bf r},{\bf r},E,\omega)/\alpha({\bf r},{\bf r},E)\sim
 (\omega/\Delta)^{-\eta/d}\ ,\qquad \omega>\Delta\ .
\ee
Of course, Eq.(\ref{corrrev_10}) is not sufficient to ensure the WD statistics
at the critical point, since there is only of order of one level
 within its validity range $\omega\sim\Delta$. Indeed, the
numerical simulations show that the level
 statistics on the mobility edge is different from the WD one
\cite{aszk,merman,evangelou94,braun95,zhar95}.

However, Eq.~(\ref{corrrev_10}) allows us to make an important conclusion
concerning the behavior of $R(s)$ (or,
which is essentially the same, of the nearest neighbor
spacing distribution $P(s)$)
at small $s=\omega/\Delta$. For this
purpose, it is sufficient to consider only two neighboring levels. Let
their energy difference be $\omega_0\sim\Delta$. Let us now perturb
the system by a random potential $V({\bf r})$ with
$\langle V({\bf r})\rangle=0$,
$\langle
V({\bf r})V({\bf r'})\rangle=\Gamma\delta({\bf r}-{\bf r'})$.
For the two-level system it reduces to a $2\times 2$ matrix
$\{V_{ij}\},\ i,j=1,2$, with  elements $V_{ij}=\int d^d{\bf r}\,  
V({\bf r})
\Psi_i^*({\bf r})\Psi_j({\bf r})$. The crucial point is that the
variances of the diagonal and off-diagonal matrix elements are
according to Eq.~(\ref{corrrev_10}) equal
to each other up to a factor of order of unity:
\begin{equation}
\langle V_{11}^2\rangle/\langle|V_{12}^2|\rangle =
\sigma({\bf r},{\bf r},E,\omega)/\alpha({\bf r},{\bf r},E)\sim 1
\label{corrrev_11}
\end{equation}
The distance between the perturbed levels is given by
$\omega=[(V_{11}-V_{22}+\omega_0)^2+|V_{12}|^2]^{1/2}$. Choosing the
amplitude of the  potential in such a way that the typical energy
shift $V_{11}\sim\Delta$ and using Eq.~(\ref{corrrev_11}), we find 
$\langle|V_{12}|^2\rangle\sim\Delta$.
As a result, the probability density for the level separation 
$\omega$ is for $\omega\ll\Delta$ of the form
$dP\sim(\omega/\Delta)^\beta d\omega/\Delta$, with some prefactor of
order of unity. We thus conclude that in the critical
point 
\be
\label{critcorr_add3}
P(s)\simeq c_\beta s^\beta\ ,\qquad s\ll 1\ ,
\ee
 with a coefficient
$c_\beta$ of order of unity, in agreement with the numerical 
findings \cite{merman,evangelou94,zhar95}. 

\paragraph{Spatial correlations at the mobility edge.} Another
question that one can ask concerning the wave function correlations at
the mobility edge is how the correlations decay with distance in real
space. Let us consider correlations $\alpha({\bf r}, {\bf r'}, E)$ 
of the eigenfunction amplitudes in two different
points. Matching the behavior of $\alpha({\bf r}, {\bf r'}, E)$ at
${\bf r}={\bf r'}$ with that at $|{\bf r}-{\bf r'}|\sim {\rm
min}\{L,\xi\}$ (where no strong correlations is expected), we find in
the metallic phase ($L\gg\xi\gg l$)
\be
\label{spatcorr1}
\alpha({\bf r}, {\bf r'}, E)\sim L^{-2d}\xi^\eta|{\bf r}-{\bf r'}|^{-\eta}\
;\qquad l\lesssim |{\bf r}-{\bf r'}| \lesssim \xi
\ee
and at the critical point ($\xi\gg L$)
\be
\label{spatcorr2}
\alpha({\bf r}, {\bf r'}, E)\sim L^{-2d+\eta}|{\bf r}-{\bf r'}|^{-\eta}\
;\qquad l\lesssim |{\bf r}-{\bf r'}| \lesssim L\ .
\ee
The same consideration can be straightforwardly applied to the higher
correlation functions $\langle |\psi^{2q_1}({\bf r})||\psi^{2q_2}({\bf
r'})|\rangle$, leading to the conclusion that they decay as
$$ 
\langle |\psi^{2q_1}({\bf r})||\psi^{2q_2}({\bf r'})|\rangle
\propto|{\bf r}-{\bf r'}|^{-[d+\tau(q_1)+\tau(q_2)-\tau(q_1+q_2)]}\ ,
$$
where $\tau(q)\equiv (q-1)D_q $ has been already introduced in
Sec.~\ref{s3.3.2}. These results were obtained for the first time by
Wegner \cite{wegner85} from renormalization-group calculations.
The above formulas for the spatial correlations at the mobility edge can be
generalized to the correlator  $\sigma({\bf r}, {\bf r'}, E,\omega)$
of two different eigenfunctions by using the scaling assumptions that
this is the length
$L_\omega=L(\Delta/\omega)^{1/d}=(1/\omega\nu)^{1/d}$ which will 
set the correlation range at finite $\omega$. This give for the
correlations at criticality ($l<|{\bf r}-{\bf r'}|<L_\omega<L,\,\xi$) 
\be
\label{spatcorr3}
\sigma({\bf r}, {\bf r'}, E,\omega)\sim L^{-2d}(\omega\nu)^{-\eta/d}
|{\bf r}-{\bf r'}|^{-\eta}\ .
\ee
Chalker and Daniell \cite{chalker88,chalker90} conjectured that the
diffusion propagator $\gamma({\bf r}, {\bf r'}, E,\omega)$,
Eq.~(\ref{corrrev_gamma}), has the same behavior (\ref{spatcorr3}) at
the mobility edge. Extensive numerical simulations
\cite{chalker88,huck94,brandes96,pracz96} fully confirm this
conjecture, which determines the form of the anomalous diffusion at
the critical point of the Anderson transition.

\subsection{Power-law random banded matrix ensemble: Anderson transition
in 1D.}
\label{s5.4}

The ensemble of random banded matrices (RBM)
 is defined as the set of matrices with elements
\begin{equation}
\label{prbm_1a}
H_{ij}=G_{ij}a(|i-j|)\ ,
\end{equation}
 where the matrix $G$ runs over the Gaussian ensemble,
and $a(r)$ is some function satisfying the condition $\lim_{r\to
\infty}a(r)=0$
and determining the shape of the band. In the most frequently
considered case
of RBM the function $a(r)$ is considered to be 
(exponentially) fast
decaying when $r$ exceeds some typical value $b$ called the bandwidth.
Matrices of this sort were first introduced as an
 attempt to describe an intermediate level statistics
 for Hamiltonian systems in a transitional regime between complete
integrability and fully developed chaos \cite{seligman85} and then appeared
in various contexts ranging from atomic physics (see \cite{flambaum94}
and references therein) to solid state physics \cite{fm94a} and
especially
in the course of investigations of the quantum behavior of
periodically driven
Hamiltonian systems \cite{izrailev95,casati91}. The most studied
system of the latter type 
is the  quantum kicked rotor  \cite{izrailev90} characterized
by the Hamiltonian
\begin{equation}
\label{prbm_2a}
\hat{H}=\frac{\hat{l}^2}{2I}+
V(\theta)\sum_{m=-\infty}^{\infty}\delta(t-mT)\;,
\end{equation}
where $\hat{l}=-i\hbar\,\partial/\partial\theta$
is the angular momentum operator
conjugated to the angle $\theta$. The constants $T$ and $I$
are the period of kicks and the moment of inertia, correspondingly,
and $V(\theta)$ is usually taken to be $V(\theta)=k\cos{\theta}$.
Classically, the kicked rotor
 exhibits an unbound diffusion in the angular momentum
space when the strength of kicks $k$ exceeds some critical value.
It was found, however, that the
 quantum effects suppress the classical diffusion 
 in close analogy with the localization of a
 quantum particle by a random potential \cite{fishman82,izrailev90}. 

It is natural to consider
the evolution (Floquet) operator $\hat{U}$ that relates values
 of the wavefunction over one period of perturbation,
$\psi(\theta,t+T)=\hat{U}\psi(\theta,t)$,
in the ``unperturbed'' basis of eigenfunctions of the operator
$\hat{l}$: $|l\rangle=\frac{1}{(2\pi)^{1/2}}\exp(i n \theta)$, $n=\pm 0,
\pm 1,\ldots$. The matrix elements $\langle m|U|n\rangle$ tend to
zero when
$|m-n|\to \infty$. In the case $V(\theta)=k\cos{\theta}$ this decay
is faster
than exponential when $|m-n|$ exceeds $b\approx k/\hbar$, whereas
within the band of the effective width $b$ matrix
 elements prove to be pseudorandom \cite{izrailev90}. 

Let us note, however, that the exponentially rapid
 decay of $\langle m|U|n\rangle$
in the above mentioned situation is due to the infinite
 differentiability of $V(\theta)=\cos{\theta}$. If we take a function
$V(\theta)$ having a discontinuity in a derivative of some order, the
corresponding matrix elements of the evolution operator
 would decay in a power-law fashion in the limit $|n-m|\to
\infty$. In fact, there is an interesting example of a
 periodically driven system where the matrix
elements of the evolution operator decay in a power-law way,
 namely the so-called Fermi accelerator \cite{jose}.
The power-law (pseudo-)random banded matrices appear also in other
models of the quantum chaos, such as a close-to-circular Bunimovich
stadium \cite{borgonovoi,casati98,borgonovi98} or a Coulomb center
inside an integrable billiard \cite{leval}.

One may also consider the random matrix (\ref{prbm_1a}) as  the Hamiltonian
of a one-dimensional tight-binding model with long-ranged off-diagonal
disorder (random hopping). A closely related problem with non-random
long-range hopping and diagonal disorder was studied numerically
in Ref. \cite{yeung}. Effect of weak long-range
 hopping on the localized states in a
3D Anderson insulator was discussed by Levitov
\cite{levitov89,levitov90,levitov99}. 
 Similar models with a power-law hopping appear
also in other physical contexts \cite{balatsky,ps,skvortsov,cizeau94}. 

As was shown in \cite{fm91,fm94a},
the conventional RBM model can be mapped
onto a 1D supermatrix non-linear $\sigma-$model, which allows for
 an exact analytical solution. The same $\sigma-$model was derived
initially \cite{eflar,efetov83}
for a particle moving in a quasi-1D system (a wire) and being subject to
a random potential. All states are found to be asymptotically localized,
with the localization length
\begin{equation}
\label{prbm_1}
\xi=\frac{B_2(2B_0-E^2)}{8B_0^2}\propto b^2;
\quad B_k=\sum_{r=-\infty}^{\infty}
a^2{(r)}r^k    \;.
\end{equation}
However, for the case of a power-like shape of the band,
\begin{equation}
\label{prbm_2}
a(r)\propto r^{-\alpha}\quad \mbox{for large}\  r \ ,
\end{equation}
this derivation should be reconsidered.
As explained below, this leads \cite{prbm} to a more general 1D
 $\sigma$-model with long-range interaction, which is much richer than
 the conventional short-range one.  In particular,
 it exhibits the Anderson localization transition at $\alpha=1$.
We mainly follow Ref.~\cite{prbm} in our presentation.

\subsubsection{Mapping onto the effective $\sigma$-model}
\label{s5.4.1}

Let us consider the ensemble of large $N\times N$ matrices
($N\to\infty$) defined by Eq. ({\ref{prbm_1a}) with a
function $a(r)$ having the form
\begin{equation}
a(r)\simeq\left\{\begin{array}{ll} 
   1,\qquad              & r< b  \\
   (r/b)^{-\alpha},\qquad & r>b
\end{array}\right.  \;.
\label{prbm_3}
\end{equation}
The parameter $b$ will serve to label the critical models with
$\alpha=1$.
We will consider $b$ to be large: $b\gg 1$, in order to justify formally
the derivation of the $\sigma-$model. We will argue later on that
 our conclusions are qualitatively valid for arbitrary $b$ as well.
We will call the ensemble (\ref{prbm_1a}), (\ref{prbm_3}) 
the power-law random banded matrix (PRBM) model.
While considering localization properties, we will restrict ourselves
by one-loop calculation, and for this reason, by the orthogonal
ensemble. 

The PRBM model can be mapped onto the $\sigma$-model on a lattice,
\begin{equation}
\label{prbm_6}
{\cal S}\{Q\}=-\frac{1}{4}(\pi\nu A_0)^2\mbox{Str}
\sum_{ij}\left[(A^{-1})_{ij}-A_0^{-1}\delta_{ij}\right]Q(i)Q(j)-
\frac{i\pi\nu\omega}{4}\sum_{i}\mbox{Str}{Q(i)\Lambda}    \;.
\end{equation}

Here $Q(i)=T_i^{-1}\Lambda T_i$ satisfies the constraint $Q^2(i)=1$,
$A$ is the matrix with elements $A_{ij}=a^2(|i-j|)$,
$A_0$ is given by\footnote{The expression for $A_0$ is valid for
$\alpha>1/2$
and in the limit $N\to \infty$. When $\alpha<1/2$,
$A_0$ starts to have a dependence on $N$
which can be removed by a proper rescaling of the matrix elements
$H_{ij}$
in Eq. (\ref{prbm_1a}). Then, the properties of the model with
 $\alpha<1/2$ turn out to be equivalent to those of the GOE, so we will not
 consider this case any longer.}
$A_0=\sum_{l}A_{kl}\approx\sum_{r=-\infty}^{\infty}a^2(r)$,
and $\nu$ is the density of states:
\begin{equation}
\label{prbm_7}
\nu=\frac{1}{2\pi A_0}\left(4A_0-E^2\right)^{1/2}  \;.
\end{equation}

The standard next step is to restrict oneself to the long wavelength
fluctuations of the $Q-$field. For usual RBM characterized by a
function
$a(r)$ decreasing faster than any power of $r$ as $r\to \infty$,
 this is achieved by the momentum expansion of the first term in
the action
 (\ref{prbm_6}):
\begin{eqnarray}\nonumber
\sum_{ij}\left[(A^{-1})_{ij}-A_{0}^{-1}\delta_{ij}\right]Q(i)Q(j)\equiv
N^{-1}\sum_q\left[A_q^{-1}-A_0^{-1}\right]Q_qQ_{-q}\\
\approx \frac{B_2}{2A_0^2}
N^{-1}\sum_q q^2Q_qQ_{-q}=\frac{B_2}{2A_0^2}\int dx\left(\partial_x
Q\right)^2   \;,
\label{prbm_8}
\end{eqnarray}
where $Q_q$ is the Fourier transform of $Q(i)$, and 
$B_2=\sum_lA_{kl}(k-l)^2$, as defined in Eq. (\ref{prbm_1}).
This immediately leads to the standard continuous version of
 the nonlinear $\sigma-$model:
\begin{equation}
\label{prbm_9}
{\cal S}\{Q\}=-\frac{\pi\nu}{4}\mbox{Str}\int dx \left[\frac{1}{2}D_{0}
\left(\partial_xQ\right)^2+i\omega Q\Lambda\right]
\end{equation}
with the classical diffusion constant $D_0=\pi\nu B_2$, which implies
the exponential localization of eigenstates with the localization length
$\xi=\pi\nu D_0\propto b^2$.

Let us try to implement the same procedure for the present case of
power-like bandshape (\ref{prbm_3}). Restricting ourselves to the
lowest  order
term in the momentum expansion, we arrive again at Eq. (\ref{prbm_9}) as
long as $\alpha\ge 3/2$. This suggests that
for $\alpha\ge 3/2$ the eigenstates
of the present model should be localized in the spatial domain of the
extension $\xi\propto \nu D_0$. However, in contrast to the usual RBM
model 
this localization is power-like rather than exponential:
$|\psi(r)|^2\propto r^{-2\alpha}$ at $r\gg\xi$.
This is quite evident due to the possibility of direct hopping with the
power-law amplitude. 
On a more formal level the appearance of power-law tails of
wavefunctions is a consequence of the breakdown of the momentum expansion for
the function $A_q^{-1}-A_0^{-1}$ in higher orders in $q^2$.
The presence of power-law ``tails'' of the wave functions,  with an
exponent
$\alpha$ determined by the decay of hopping elements, was found in
numerical simulations in Refs. \cite{yeung,casati98,borgonovi98}.

The most interesting region $1/2<\alpha<3/2$ requires a separate
consideration.
In this case, Eq. (\ref{prbm_8}) loses its validity in view of the
divergence
 of the coefficient $B_2$. We find instead
\begin{eqnarray}\nonumber
A_0^2(A_q^{-1}-A_0^{-1})&\approx&A_0-A_q=
2\int_0^{\infty}dr a^2(r)(1-\cos{qr})\\
&=&\frac{2}{|q|}\left\{\int_0^{b|q|}dx (1-\cos{x})+(b|q|)^{2\alpha}
\int_{b|q|}^{\infty}\frac{dx}{x^{2\alpha}}(1-\cos{x})\right\}
\nonumber\\
&\approx& c_{\alpha}b^{2\alpha}|q|^{2\alpha-1}\ \ \mbox{for}\
1/2<\alpha<3/2
\ \mbox{and}\  |q|\ll 1/b \;,
\label{prbm_10}
\end{eqnarray}
where $c_{\alpha}=2\int_{0}^{\infty}\frac{dx}{x^{2\alpha}}(1-\cos{x})$
is a numerical constant,
$c_\alpha=[2\Gamma(2-2\alpha)/(2\alpha-1)]\sin\pi\alpha$ for $\alpha\ne
1$ and $c_1=\pi$ for $\alpha=1$.

The corresponding long-wavelength part of the action,
\begin{equation}
\label{prbm_11}
{\cal S}_0\{Q\}=-\frac{1}{t}\mbox{Str}\int (dq)\,|q|^{2\alpha
-1}Q_qQ_{-q} \;,
\end{equation}
where $\int(dq)\equiv\int {dq\over 2\pi}\equiv N^{-1}\sum_q$, 
can not be reduced to the local-in-space form in the
coordinate representation any longer. Here, $1/t=\frac{1}{4}
(\pi\nu)^2c_{\alpha}b^{2\alpha}\propto b^{2\alpha-1}\gg 1$
plays the role of coupling constant, justifying the
 perturbative and renormalization group treatment of the model described
below.

Let us mention that if we consider the RBM model as a tight-binding
Hamiltonian,
 the corresponding classical motion described by the master equation
on the 1D lattice is superdiffusive for $1/2<\alpha<3/2$ with a
typical displacement in a time $t$ being 
$r\propto t^{1/(2\alpha-1)}$ (Levy law of index $2\alpha-1$; see
Ref.~\cite{bouchaud90} for a review). As will be discussed in
Sec.~\ref{s5.4.3},
this influences the asymptotic behavior of the spectral
correlation function
for the corresponding quantum system.

\paragraph{Perturbative treatment of the non-linear $\sigma-$model:
General formulas.}

Here we derive one-loop perturbative corrections to the
density--density correlation function and inverse participation ratios.
Analysis of these expressions for various values of the power-law
parameter
$\alpha$ will be presented in Sec.~\ref{s5.4.2}.

The  density--density correlation function (diffusion propagator)
$$K(r_1,r_2;\omega)=G_R^{E+\omega/2}(r_1,r_2)G_A^{E-\omega/2}(r_2,r_1)$$ 
can be expressed in terms of the $\sigma-$model as follows \cite{efetov83}:
\begin{equation}
K(r_1,r_2;\omega)=-(\pi\nu)^2\int DQ \:Q_{12,\alpha\beta}(r_1) k_{\beta\beta}
Q_{21,\beta\alpha}(r_2)e^{-S\{Q\}}  \;.
\label{prbm_12}
\end{equation}
Here the indices $p,p'$ of the matrix $Q_{pp',\alpha\beta}$ correspond to
its advanced--retarded block structure, whereas $\alpha,\beta$
discriminate
between bosonic and fermionic degrees of freedom. The matrix
$k_{\beta\beta}$
is equal to 1 for bosons and (-1) for fermions. To calculate the
correlation function (\ref{prbm_12}) perturbatively, we
use here the following parametrization \cite{efetov83} of the matrix $Q$:
\begin{equation}
Q=\Lambda(W+\sqrt{1+W^2})=\Lambda\left(1+W+{W^2\over 2}-{W^4\over
8}+\ldots\right)\ ,
\label{prbm_13}
\end{equation}
where $W$ is block-off-diagonal in the advanced--retarded representation.
To get the perturbative expansion for $K(r_1,r_2;\omega)$, one has to
substitute Eq.~(\ref{prbm_13}) into (\ref{prbm_12}), to separate the part
quadratic in $W$
 from the rest in the exponent and to apply the Wick theorem.
 In the usual case, when
the action is given by Eq. (\ref{prbm_9}), the leading order (tree level)
result reads in the momentum space as follows:
\begin{equation}
K_0(q,\omega)={2\pi\nu\over D_0 q^2-i\omega} \;.
\label{prbm_14}
\end{equation}
The perturbative quantum corrections do not modify the general
form (\ref{prbm_14}), but change the value of the diffusion constant. In
particular, in one-loop order one gets Eq. (\ref{prbm_14}) with $D_0$
replaced by
\cite{glk79}
\begin{equation}
D=D_0\left\{1-{1\over\pi\nu V}\sum_{q_i=\pi n_i/L_i}{1\over D_0
q^2-i\omega}
\right\}                       \;.
\label{prbm_15}
\end{equation}
This induces the standard weak-localization correction to the
conductivity.

Now we implement an analogous procedure for the non-local $\sigma-$model
of the type of Eqs.~(\ref{prbm_6}), (\ref{prbm_11}):
\begin{equation}
\label{prbm_16}
{\cal S}\{Q\}=\frac{1}{t}\mbox{Str}\sum_{r.r'}U(r-r')Q(r)Q(r')-
i\frac{\pi\nu\omega}{4}\sum_{r}\mbox{Str}{\Lambda Q(r)}\ ,
\end{equation}
with $U(r)\propto r^{-2\alpha}$ as $r\to\infty$, 
so that the Fourier-transform of $U(r)$ behaves at small momenta as
\begin{equation}
\label{prbm_17}
\tilde{U}(q)=-|q|^{\sigma};\quad 1/2<\sigma<2   \;.
\end{equation}
The exponent $\sigma$ is related to the parameter $\alpha$ of the RBM
model by $\sigma=2\alpha-1$. In  leading order,
we keep in the action the terms quadratic in $W$ only, which yields:
\begin{equation}
\label{prbm_18}
K_0(q,\omega)=\frac{2\pi\nu}{8(\pi\nu t)^{-1}|q|^{\sigma}-i\omega} \;,
\end{equation}
corresponding to a superdiffusive behavior.

To calculate the one-loop correction to $K_0(q)$ (we set $\omega=0$
for simplicity) we expand the kinetic term in ${\cal S}\{Q\}$ up to
 fourth order in $W$:
\begin{equation}
\label{prbm_19}
\sum_{r,r'}\mbox{Str}\,U(r-r')Q(r)Q(r')\mid_{\mbox{4th order}}
=\sum_{r,r'}\frac{1}{4}
\mbox{Str}\,W^2(r)W^2(r')     \;.
\end{equation}
The contraction rules are given by Eqs. (8), (16) of
Ref.~\cite{fm95a} (reproduced below as Eq.~(\ref{e2.13}) for the unitary
symmetry),
with the propagator $\Pi(q)$ replaced by
\begin{equation}
\label{prbm_20}
\Pi(q)=\frac{t}{8|q|^{\sigma}}   \;.
\end{equation}
%
%There is only one  one-loop diagram contributing to the self-energy part
%in the present parametrization, see Fig. 1. Evaluating it, we find:
%\begin{equation}
%\label{prbm_21}
%\delta
%\Gamma_1^{(2)}=\frac{1}{2}\int(dk)\frac{|q+k|^{\sigma}}{|k|^{\sigma}} \;.
%\end{equation}
% There is also
%a contribution to $ \Gamma^{(2)}$
%from the Jacobian of transformation (\ref{prbm_13}),
%which is equal to \cite{zinnjustin}
%\begin{equation}
%\label{prbm_22}
%\delta\Gamma_2^{(2)}=-\frac{1}{2}\delta(0)\equiv -\frac{1}{2}\int (dk) \;.
%\end{equation}
%\begin{equation}
%\label{prbm_23}
%\delta\Gamma^{(2)}=\frac{1}{2}\int(dk)\frac{|q+k|^{\sigma}-|k|^{\sigma}}
%{|k|^{\sigma}} \;.
%\end{equation}
%
%
Evaluating the one-loop correction, we get the following expression
for the density-density correlation
function up to the one-loop order: 
\begin{equation}
K^{-1}(q)=K_0^{-1}(q)-\frac{(\pi\nu)^2}{2}
\int(dk)\frac{|q+k|^{\sigma}-|k|^{\sigma}}
{|k|^{\sigma}}    \;.
\label{prbm_24}
\end{equation}

Now we calculate the perturbative correction to the inverse
participation ratios $P_q$. 
The  results of Ref.~\cite{fm95a} presented in Sec.~\ref{s3.3.1} are
 straightforwardly applicable to the present case of power-law RBM,
provided the appropriate modification of the diffusion propagator
entering the contraction rules is made,
see the text preceding Eq. (\ref{prbm_20}). One finds
\begin{equation}
\langle P_q\rangle
=\left\{1+{1\over N}q(q-1)\sum_r\Pi(r,r)\right\}{(2q-1)!!\over
N^{q-1}}\ ,
\label{prbm_25}
\end{equation}
 where $\Pi(r,r')=(1/N)\sum_q\Pi(q)\exp[iq(r-r')]$ and $\Pi(q)$ is
given by Eq.~(\ref{prbm_20}).

\paragraph{Renormalization group treatment.}

The effective $\sigma-$model, Eq. (\ref{prbm_16}), is actually of
one-dimensional
nature. However, for the sake of generality, we find it convenient
to consider it here to be defined in $d$-dimensional space
with arbitrary $d$. The form (\ref{prbm_18}) of the generalized diffusion
propagator implies that $d=\sigma$  plays the role of the logarithmic
dimension for the problem. In the vicinity of this critical value we
can carry out a renormalization group (RG) treatment of the
model,
following the procedure developed for general non-linear
$\sigma-$models
in \cite{brezin76a,brezin76b,brezin80,zinnjustin}. 
We begin by expressing the action in terms of the
renormalized coupling constant $t=Z_1^{-1}t_B\mu^{d-\sigma}$, where $t_B$
is the bare coupling constant, $Z_1$ is the renormalization constant, 
and $\mu^{-1}$ is the length scale governing
the renormalization \footnote{Note that the $W$-field renormalization
is absent due to the supersymmetric character of the problem, which is
physically related to the particle number conservation
\protect\cite{efetov83}.}:
\begin{eqnarray}
S&=&{\mu^{d-\sigma}\over 2t Z_1}\sum_{rr'}U(|r-r'|)\:\mbox{Str}\,
\left[-W(r)W(r')+\sqrt{1+W^2(r)}\sqrt{1+W^2(r')}\right]
\nonumber \\    &-&{i\pi\nu\omega\over 4}
\sum_r \mbox{Str}\sqrt{1+W^2(r)}         \;.
\label{prbm_26}
\end{eqnarray}
Expanding the action in powers of $W(r)$ and keeping  terms up to
4-th order, we get
\begin{eqnarray}
S&=&S_0+S_1+O(W^6) \;, \nonumber\\
S_0&=&{\mu^{d-\sigma}\over 4t Z_1}\sum_{rr'}U(|r-r'|)\:\mbox{Str}\,
(W(r)-W(r'))^2-{i\pi\nu\omega\over 8}\sum_r \mbox{Str}\,W^2(r) \;,
\nonumber\\
S_1&=&{\mu^{d-\sigma}\over 8t Z_1}\sum_{rr'}U(|r-r'|)\mbox{Str}\,
W^2(r)W^2(r')+{i\pi\nu\omega\over 32}\sum_r \mbox{Str}\,W^4(r) \;.
\label{prbm_27}
\end{eqnarray}
We have restricted ourselves to 4-th order terms, since they are
sufficient for obtaining the renormalized quadratic part of the action
in one-loop order. The calculation yields, after the cancelation of an
$\int(dk)\propto \delta(0)$ term with the contribution of the Jacobian:
\begin{equation}
S_{{\rm quad}}=S_0+\langle S_1\rangle={1\over 2}\int(dq)\mbox{Str}W_q W_{-q}
\left[-{\mu^{d-\sigma}\over t Z_1}\tilde{U}(q)-{1\over 2}
\int(dk){\tilde{U}(k)-\tilde{U}(k+q)\over
-{\mu^{d-\sigma}\over t Z_1}\tilde{U}(k)-{i\pi\nu\omega\over 4}}\right]
\;.
\label{prbm_28}
\end{equation}
According to the renormalization group idea, one has to chose the
constant
$Z_1(t)=1+at+\ldots$ so as to cancel the divergence in the
coefficient in front of the leading $|q|^\sigma$ term. 

\subsubsection{Localization and eigenfunction statistics.}
\label{s5.4.2}

In this subsection we analyze the model in different regions of the
exponent $\sigma=2\alpha-1$ using the general formulas derived in
Sec.~\ref{s5.4.2}. 

\paragraph{Localized regime: $1<\sigma<2$ ($1<\alpha<3/2)$.}

To evaluate the one-loop correction (\ref{prbm_24}) to the diffusion
propagator,
we use the expansion
\begin{equation}
\label{prbm_29}
{|{\bf q}+{\bf k}|^\sigma\over|{\bf k}|^\sigma}-1\simeq\left\{
\begin{array}{ll}
\displaystyle{
\sigma{{\bf qk}\over k^2}+{\sigma\over 2}{q^2\over k^2}+
\sigma\left({\sigma\over 2}-1\right)\left({{\bf qk}\over k^2}\right)^2
+\ldots   }
\ ,&\qquad q\ll k\\
\displaystyle{
{|q|^\sigma\over |k|^\sigma}     }
\ ,&\qquad q\gg k
\end{array}\right.  \;.
\end{equation}
Thus, the integral in Eq. (\ref{prbm_24}) can be estimated as
\begin{equation}
I\equiv\int(dk)\left[
{|{\bf q}+{\bf k}|^\sigma\over|{\bf k}|^\sigma}-1\right]
\approx\int_{k>q}(dk)\sigma\left(1+{\sigma-2\over
d}\right){q^2\over 2k^2}
+\int_{k<q} (dk) {|q|^\sigma\over |k|^\sigma}  \;,
\label{prbm_30}
\end{equation}
and we are interested in the particular case of the dimension $d=1$. 
For $\sigma>d=1$ the integral diverges at low $k$, and  the second
term in Eq. (\ref{prbm_30})  dominates. This gives
\begin{equation}
I\sim L^{\sigma-1}|q|^{\sigma}\ ,
\label{prbm_31}
\end{equation}
where $L$ is the system size determining the infrared cut-off (in the
original RBM formulation it is just the matrix size $N$). This leads
to the following one-loop expression for the diffusion correlator
\begin{equation}
K(q)={(\pi\nu)^2\tilde{t}\over 4|q|^\sigma}\ ; \qquad
\tilde{t}^{-1}=t^{-1}-\mbox{const} \, L^{\sigma-1}  \;.
\label{prbm_32}
\end{equation}

Now we turn to the renormalization group analysis, as described in
Sec. \ref{s5.4.1}. In the one-loop order, the expression for the
renormalization
constant $Z_1$ is determined essentially by the same integral $I$,
Eq. (\ref{prbm_30}), with the RG scale $\mu$ playing the
role of the infrared cut-off (analogous to that of the system size $L$
in Eq. (\ref{prbm_32})). This yields (in the minimal subtraction scheme):
\begin{equation}
Z_1(t)=1-{1\over 2\pi}{t\over\sigma-1}+O(t^2)\ ,
\label{prbm_33}
\end{equation}
leading to the following relation between the bare and
the renormalized coupling constant analogous to Eq. (\ref{prbm_32}):
\begin{equation}
{1\over t}\mu^{1-\sigma}\equiv{Z_1\over t_B}={1\over t_B}-
{1\over 2}{1\over\sigma-1}\mu^{1-\sigma}  \;.
\label{prbm_34}
\end{equation}
~From Eq. (\ref{prbm_33}), we get the expression for the $\beta$-function:
\begin{equation}
\beta(t)=\left.{\partial t\over\partial\ln\mu}\right|_{t_B}=
{(1-\sigma)t\over 1+t\partial_t\ln Z_1(t)}=-(\sigma-1)t-{t^2\over 2\pi}+
O(t^3)  \;.
\label{prbm_35}
\end{equation}
Both Eqs. (\ref{prbm_32}) and (\ref{prbm_35}) 
show that the coupling constant $t$
increases with the system size $L$ (resp. scale $\mu$), which is
analogous
to the behavior found in the conventional scaling theory of localization
in $d<2$ dimensions \cite{aalr,wegner80a}. The RG flow reaches the strong
coupling regime $t\sim 1$ at the scale $\mu\sim t_B^{1/(\sigma-1)}$.
Remembering the relation of the bare coupling constant $t_B$ and the
index $\sigma$ to the parameters of the original PRBM model:
$t_B^{-1}\propto b^{2\alpha-1}$; $\sigma\propto 2\alpha-1$, we conclude
that the length scale
\begin{equation} \label{prbm_36}
\xi\sim t_B^{-1/(\sigma-1)}\sim b^{2\alpha-1\over 2\alpha-2}
\end{equation}
plays the role of the
localization length for the PRBM model.

This conclusion is also supported by an inspection of the
expression for the
IPR, Eq. (\ref{prbm_25}). Evaluation of the one-loop perturbative
correction in Eq. (\ref{prbm_25}) yields
\begin{equation}
\langle P_q\rangle
={(2q-1)!!\over N^{q-1}}\left\{1+q(q-1){t\over 8\pi^\sigma}\zeta(\sigma)
N^{\sigma-1}\right\}\ ,
\label{prbm_37}
\end{equation}
where $\zeta(\sigma)$ is Riemann's zeta-function;
$\zeta(\sigma)\simeq
1/(\sigma-1)$ for $\sigma$ close to unity. The correction term becomes
comparable to the leading (GOE) contribution for the system size
$N\sim t^{1/(\sigma-1)}$,
parametrically coinciding with the localization length $\xi$. For larger
$N$ the perturbative expression (\ref{prbm_37}) loses its validity, and the IPR
is expected to saturate at a constant value
$\langle P_q\rangle\sim\xi^{1-q}$ for
$N\gg\xi$.

In concluding this subsection, let us stress once more that the
localized
eigenstates in the present model are expected to have integrable
power-law
tails: $|\psi^2(r)|\propto r^{-2\alpha} = r^{-\sigma-1}$ at $r\gg\xi$.

\paragraph{Delocalized regime: $0<\sigma<1\ \ (1/2<\alpha<1)$.}

We begin again by considering the perturbative corrections
to the diffusion propagator (\ref{prbm_24}). The first term in the
r.h.s. of Eq.(\ref{prbm_30}) proportional to $\int dk/k^2$ is determined by
the vicinity of its lower cut-off (i.e. $k\sim q$), whereas the second
one is proportional to $\int dk/k^\sigma$ and thus determined by the
vicinity of its upper cut-off (i.e. again $k\sim q$). Therefore,  
the integral (\ref{prbm_30}) is now dominated by the region $k\sim q$, 
and is proportional to $|q|$. We get, therefore:
\begin{equation}\label{prbm_39}
(\pi\nu)^2 K^{-1}(q)=\frac{4|q|^{\sigma}}{t}-C_{\sigma}|q|
\end{equation}
with a numerical constant $C_{\sigma}$.

We see that the correction term is of higher order in
$|q|$ as compared to the leading one.
Thus, it does not lead to a renormalization of the coupling constant $t$.
This is readily seen also in the framework of the RG scheme,
where the one-loop integral in Eq. (\ref{prbm_28}) does not give
rise to  terms of the form $|q|^{\sigma}$.
One can check that this feature is not specific to the
one-loop RG calculation, but holds in higher orders as well.
For the case of a vector model with long-ranged interaction 
this conclusion was reached by Brezin {\it et al} \cite{brezin76c}.
Thus, the renormalization constant $Z_1$
is equal to unity and the $\beta-$function is trivial:
\begin{equation}\label{prbm_40}
\beta(t)=(1-\sigma)t    \;.
\end{equation}
This means that the model does not possess a critical
 point, and, for $\sigma<1$, all states  are delocalized, for any value
of the bare coupling constant $t$. This property should be contrasted
with the behavior of a $d-$dimensional conductor
 described by the conventional local non-linear $\sigma-$
model and undergoing an Anderson transition at some critical coupling
$t=t_c$ \cite{wegner80a}.

Though all states of the model are delocalized, 
their statistical properties are different from GOE.
In particular, calculating the variance
of the inverse participation ratio $P_2$ 
(see Sec.~\ref{s3.3.3}, Eq.~(\ref{corrrev_varipr})) we get:
\begin{equation}
\label{prbm_41}
\delta(P_2)=\frac{\langle P_2^2\rangle-\langle P_2\rangle ^2}
{\langle P_2\rangle ^2}=
\frac{8}{N^2}\sum_{rr'}\Pi^2(r,r')=\frac{8}{N^2}\sum_{q=\pi n/N;\:
n=1,2,..}
\Pi^2(q) \;.
\end{equation}
At $\sigma>1/2$ the sum over $q$ is convergent yielding
\begin{equation}\label{prbm_42}
\delta(P_2)=\frac{t^2}{8\pi^{2\sigma}}\frac{\zeta(2\sigma)}{N^{2-2\sigma}}
\;.
\end{equation}
Thus, in this regime the fluctuations of the IPR are much stronger
than for the GOE where $\delta(P_2)\propto 1/N$. Only for $\sigma < 1/2$
 ($\alpha < 3/4$)
the IPR fluctuations acquire the GOE character.
Considering higher irreducible moments (cumulants) of the IPR, 
$\langle\langle P_2^n\rangle\rangle$,
one finds that the GOE behavior is restored at
$\sigma<\sigma_c^{(n)}\equiv 1/n$. In this sense, the model is
analogous to a
 $d$-dimensional conductor at $d=2/\sigma$.
 Therefore, only when $\sigma\to 0$
 (correspondingly, $\alpha\to 1/2$ in the original PRBM formulation)
all statistical properties become equivalent to those characteristic
of GOE. 

\paragraph{Critical regime: $\sigma=1$ ($\alpha=1$).}

As we have seen,  $\sigma=1$ separates the regions of
localized ($\sigma>1$) and extended ($\sigma<1$) states. It is then
natural
to expect some critical properties showing up just at $\sigma=1$.
Let us again  consider the generalized diffusion propagator,
 Eq. (\ref{prbm_24}). At $\sigma=1$ the one-loop correction yields
\begin{equation}\label{prbm_43}
\frac{1}{(\pi\nu)^2}K^{-1}(q)=4|q|\left[t^{-1}-\frac{1}{8\pi}\ln(|q|L)\right]
\;.
\end{equation}
As expected at the critical point, the correction to the
coupling constant is of logarithmic nature. However, Eq. (\ref{prbm_43})
differs essentially from that typical for a 2D disordered conductor:
\begin{equation}
\label{prbm_44}
t^{-1}=t_B^{-1}-\ln(L/l)\ ,
\end{equation}
where the bare coupling constant $t_B$ corresponds to  scale $l$.
Comparing the two formulas, we see that in Eq. (\ref{prbm_43}) the mean
free path $l$ is replaced by the inverse momentum $q^{-1}$. Therefore,
the correction to the bare coupling constant is small for low momenta
$q\sim 1/L$, and the correlator $K(q)$ is not renormalized.
This implies the absence of eigenstate localization, in contrast
 to the 2D diffusive conductor case, where Eq. (\ref{prbm_44}) results in an
exponentially large localization length $\xi\propto \exp{t_B^{-1}}$.
On a more formal level, the absence of essential corrections to
the low-$q$ behavior of $K(q)$ is due to the fact that the region $k>q$
does not give a logarithmic contribution.
This is intimately connected with the absence of $t$
 renormalization at $\sigma<1$.

To study in more details the structure of critical eigenfunctions,
let us consider the set of IPR $P_q$. The perturbative correction,
 Eq. (\ref{prbm_25}), is evaluated at $\sigma=1$ as
\begin{equation}\label{prbm_45}
\langle P_q\rangle
=\left\{1+q(q-1)\frac{t}{8\pi}\ln{(N/b)}\right\}\frac{(2q-1)!!}
{N^{q-1}}    \;,
\end{equation}
where the microscopic
scale $b$, Eq. (\ref{prbm_3}), enters as the ultraviolet cut-off
for the $\sigma-$model, the role usually played by the mean free
path $l$.
This formula is valid as long as the correction is small: $q\ll
\left[\frac{t}{8\pi}\ln{(N/b)}\right]^{-1/2}$. For larger $q$
the perturbation theory breaks down, and one has to
 use the renormalization group approach. This 
requires \cite{wegner80,akl} 
introduction of higher vertices of the type $z_q\int\mbox{Str}^q
(Qk\Lambda)dr$ into the action of the non-linear $\sigma-$model and their
subsequent renormalization. The resulting RG equations
for the charges $z_q$ read, in the one-loop order, 
\begin{equation}
\label{prbm_46}
\frac{dz_q}{d\ln{\mu^{-1}}}=q(q-1)\frac{t}{8\pi}z_q  \;,
\end{equation}
where $\mu^{-1}$ is the renormalization scale. Integrating 
Eq. (\ref{prbm_46}),
we find:
\begin{equation}
\label{prbm_47}
\langle P_q\rangle
=\frac{(2q-1)!!}{N^{q-1}}{\left(N\over b\right)}^{q(q-1)\frac{t}{8\pi}} \;.
\end{equation}
Note, that this formula is reduced to the perturbative expression,
Eq. (\ref{prbm_45}), in the regime
$q\ll\left[\frac{t}{8\pi}\ln{(N/b)}\right]^{-1/2}$.

The behavior described by Eq. (\ref{prbm_47}) is characteristic of
a multifractal structure of wave functions 
(see Sec.~\ref{s3.3.2}, \ref{s5.3}).
Comparing (\ref{prbm_47}) with (\ref{multifr2}), we find the
multifractality dimensions $D_q=1-qt/8\pi$. The general formula valid
for both orthogonal ($\beta=1$) and unitary ($\beta=2$) universality
classes is  
\begin{equation}\label{prbm_49}
D_q=1-q\frac{t}{8\pi\beta} \;.
\end{equation}
This form of the fractal dimensions is reminiscent of that found in two
and $2+\epsilon$ dimensions for the usual diffusive conductor, see
Eq.~(\ref{multifr5}). 
The one-loop result (\ref{prbm_49}) holds for $q\ll 8\pi/t$.

Now we can understand the reason for the $q$-dependent logarithmic
correction to the diffusion propagator $K(q)$,
Eq.~(\ref{prbm_43}). As was mentioned in the end of Sec.~\ref{s5.3},
the multifractality of eigenfunctions determines
the momentum dependence of the diffusion propagator at high $q$ in the
critical point \cite{chalker88,chalker90}
\be
\label{prbm_501}
K^{-1}(q)\propto |q|(|q|L)^{-\eta}\ ;\qquad \eta=d-D_2
\ee
(at finite frequency $\omega$ the system size $L$ is replaced by
$L_\omega\sim(D/\omega)^{1/2}$). On the other hand, the logarithmic
correction in Eq.~(\ref{prbm_43}) is the first term of the expansion
\bea
\label{prbm_502}
\frac{1}{(\pi\nu)^2}K^{-1}(q)&=&{4\over
t}|q|\left\{1-\frac{t}{8\pi}\ln(|q|L)
+C_2 \left[\frac{t}{8\pi}\ln(|q|L)\right]^2+\ldots\right\}
\nonumber \\
& = & {|q|\over t} F\left({t\over 8\pi}\ln(|q|L)\right)\ ,
\eea
where $F(x)$ is some parameterless function. Since $\eta=1-D_2\simeq
t/4\pi$, Eq.~(\ref{prbm_502}) has precisely the form expected from
(\ref{prbm_501}), assuming that $F(x)\sim e^{-2x}$ at $x\gg 1$.

The set of fractal dimensions $D_q$ (as well as spectral properties at
 $\sigma=1$, see Sec.~\ref{s5.4.3}) is parametrized by the coupling
constant $t$.
Strictly speaking, the above $\sigma-$model derivation is justified for
$t\ll 1$
 (i. e. $ b\gg 1$). However, the opposite limiting case can be also
studied, following  Levitov \cite{levitov89,levitov90,levitov99}. 
It corresponds to
a $d-$dimensional Anderson insulator, perturbed by a weak long-range
 hopping with an amplitude decreasing with distance as $r^{-\sigma}$.
The arguments of Levitov imply that the states
delocalize at $\sigma\le d$, carrying  fractal properties at
$\sigma=d$.
The PRBM model in the limit $b\ll 1$ is just the 1D version of
this problem.
This shows that the conclusion about localization (delocalization)
of eigenstates for $\sigma>1$ (resp. $\sigma<1$), with $\sigma=1$
being a critical point holds irrespective of the particular value of the
parameter $b$. The $b\ll 1$ limit of the PRBM model was also studied
 in Ref.~\cite{ps} where the IPR $P_2$ was calculated, yielding the
 fractal dimension $D_2 \propto b$.
Alternatively, the regime of the Anderson insulator
with weak
power-law hopping can be described in the framework of the non-linear
 $\sigma-$model, eq. (\ref{prbm_16}), by considering the limit $t\gg 1$.
Formally, the non-linear $\sigma-$model for arbitrary $t$ can be
derived from a microscopic tight-binding model by allowing $n\gg 1$
``orbitals'' per site \cite{wegner80a}.

To summarize, the PRBM model (\ref{prbm_1a}), (\ref{prbm_3}) with
$0<b<\infty$ or the $\sigma-$model (\ref{prbm_16}) with
a coupling constant $0<t<\infty$ represent at $\sigma=1$ a
continuous family of critical theories
parametrized by the value of $b$ (respectively, $t$).

\begin{figure}
%\centerline{\epsfxsize=70mm\epsfbox{prbm_fig2a.ps}
%\epsfxsize=70mm\epsfbox{prbm_fig2b.ps}}  
\centerline{\epsfig{file=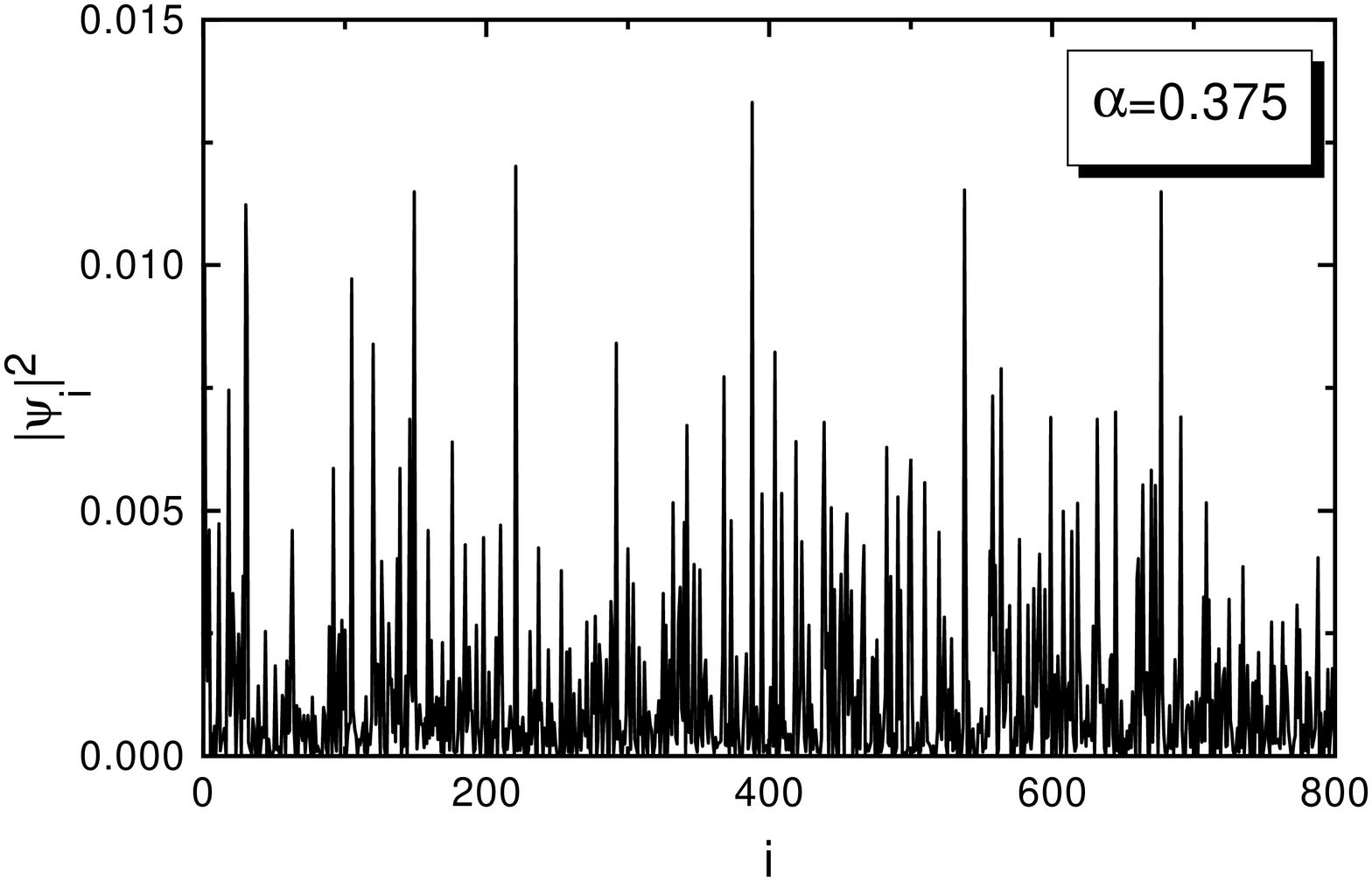,width=55mm,angle=270}
\epsfig{file=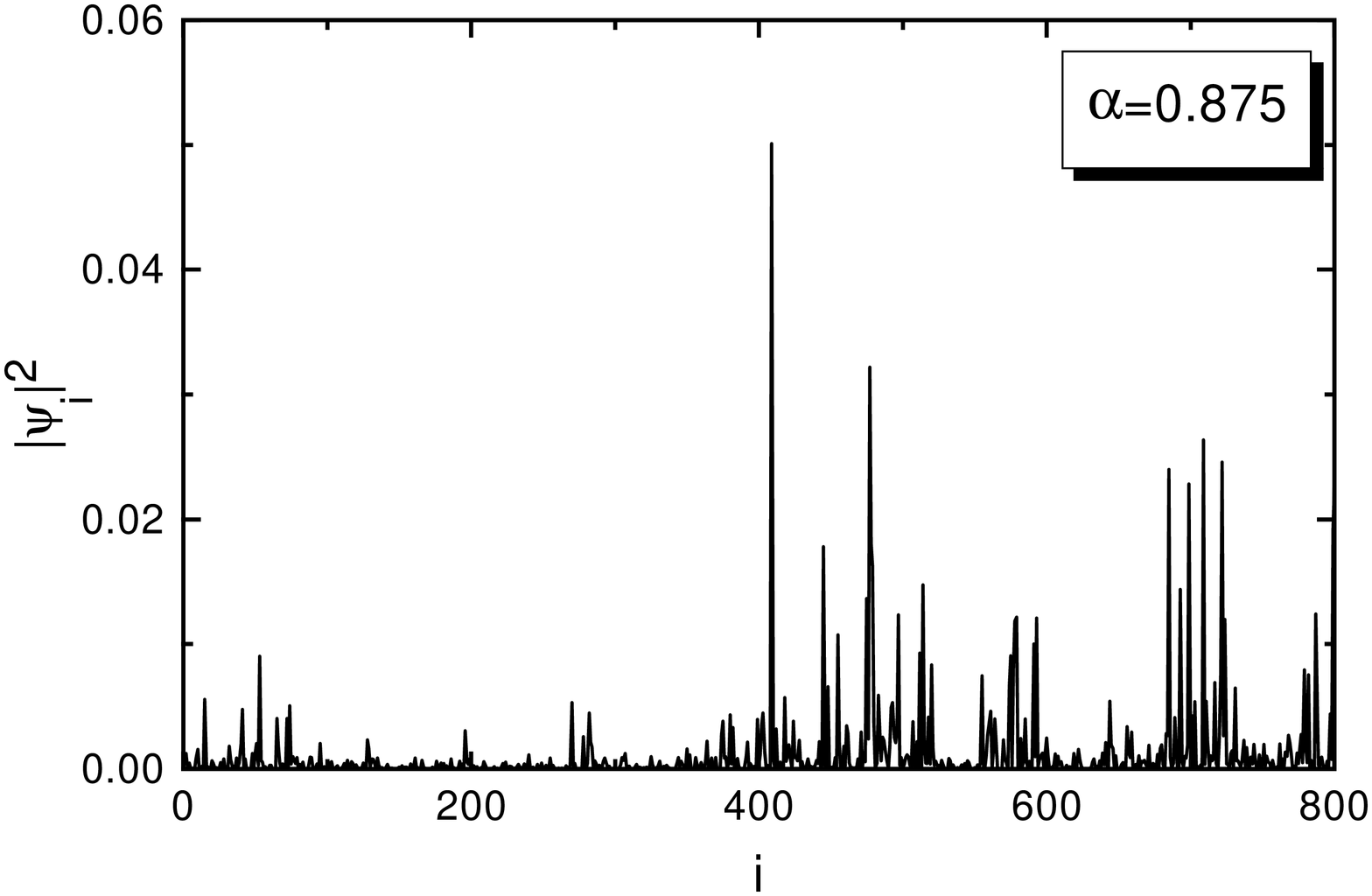,width=55mm,angle=270}}
\vspace{5mm}
%\centerline{\epsfxsize=70mm\epsfbox{prbm_fig2c.ps}
%\epsfxsize=70mm\epsfbox{prbm_fig2d.ps}}  
\centerline{\epsfig{file=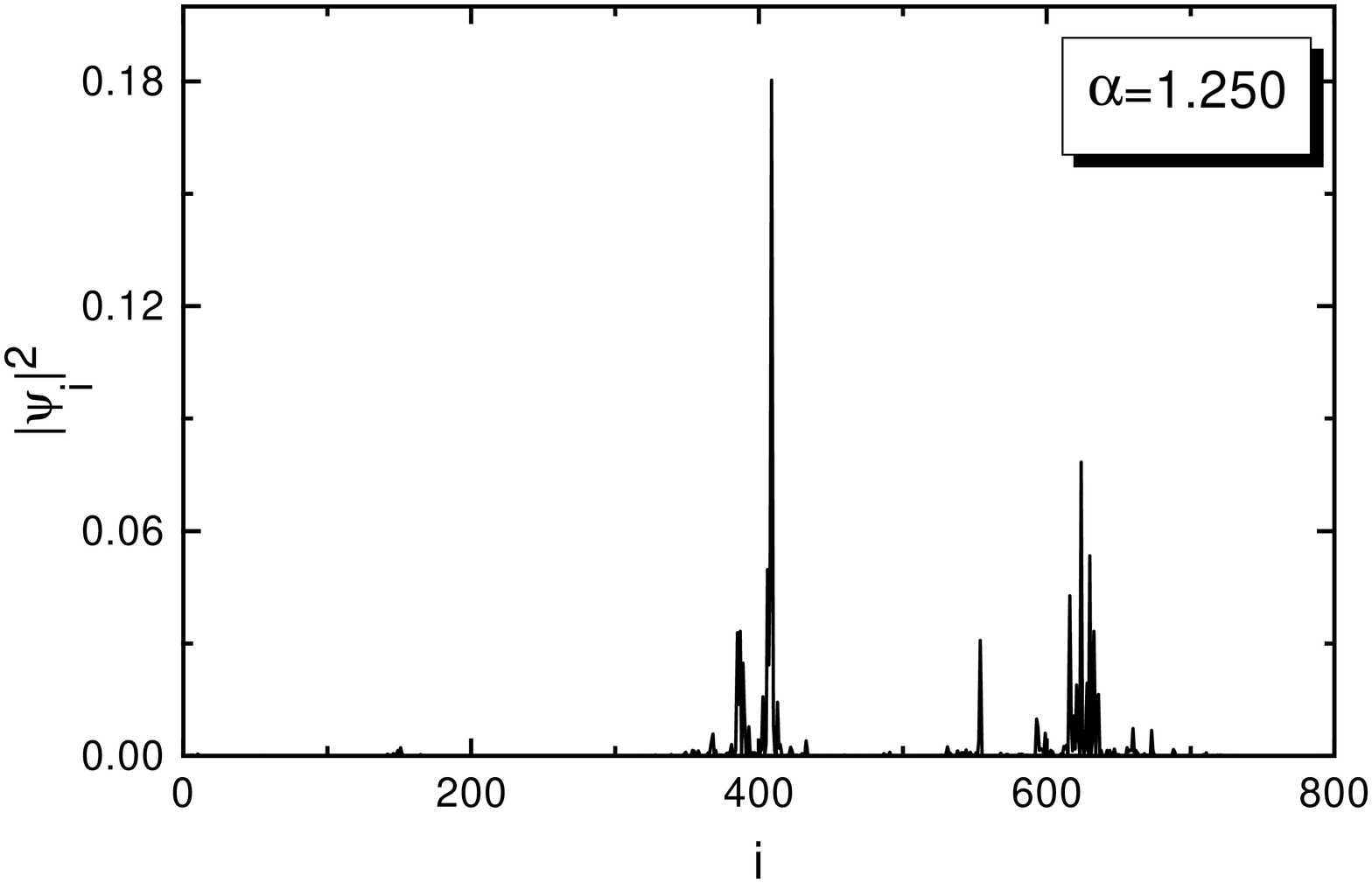,width=55mm,angle=270}
\epsfig{file=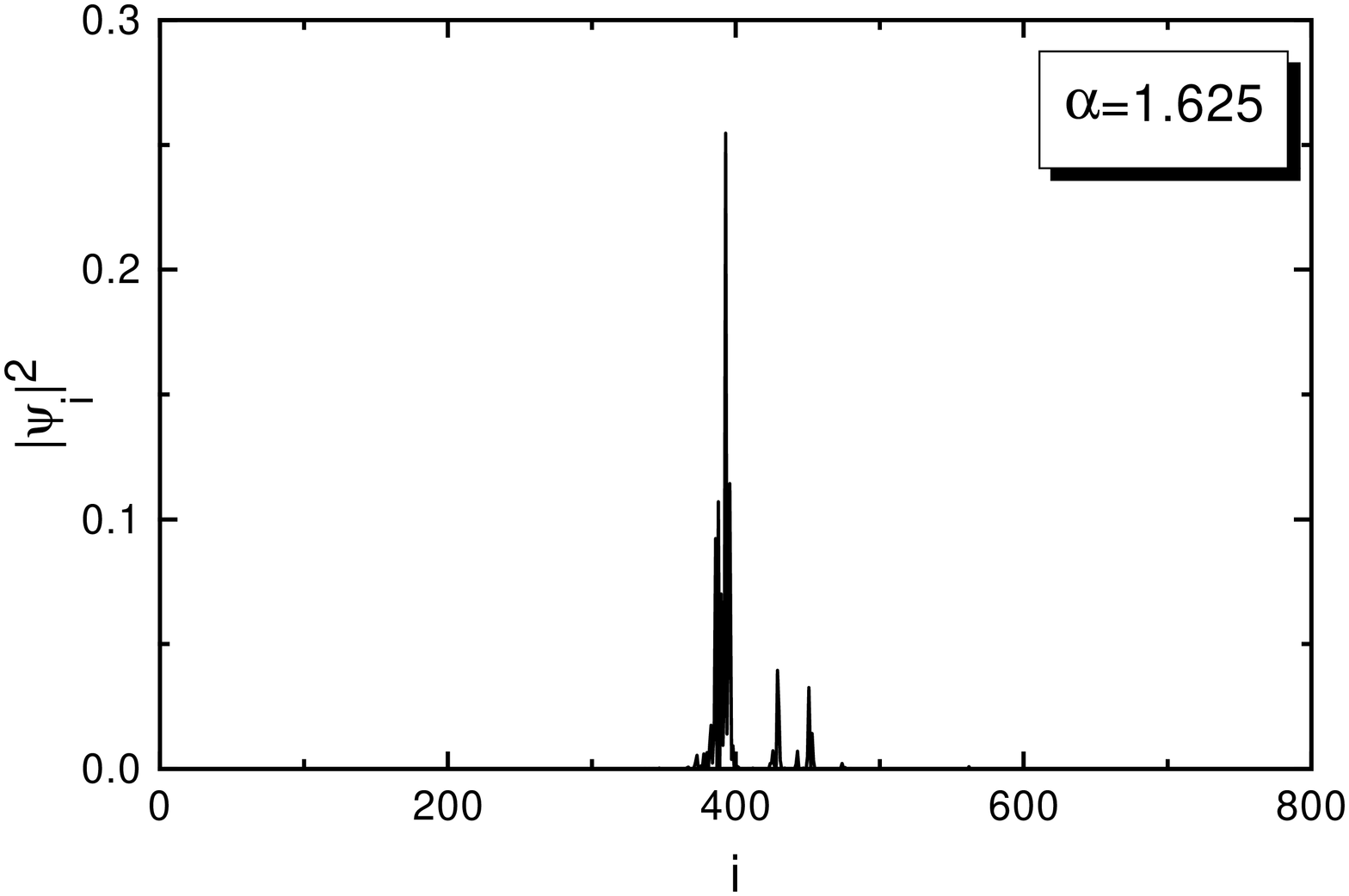,width=55mm,angle=270}}
\vspace{3mm}
\caption{Typical eigenfunctions for the matrix size $N=800$ and four
different values of $\alpha$: 0.375, 0.875, 1.250, and 1.625. From
\cite{prbm}.} 
\label{prbm_fig2} 
\end{figure}

\paragraph{Numerical simulations.}

Numerical simulations of the PRBM model were performed in
  Ref.~\cite{prbm}  for values of
$\alpha\in [0,2]$ and $b=1$. In Fig.~\ref{prbm_fig2} we present typical
eigenfunctions for
 four different regions of $\alpha$. In agreement with the
theoretical picture
presented above, the eigenstates corresponding to $\alpha=0.375$ and
$\alpha=0.875$ are extended, whereas those corresponding to $\alpha=1.25$
and $\alpha=1.625$ are localized. At the same time, one can notice
that the
states with $\alpha=0.875$ and $\alpha=1.25$ exhibit a quite sparse
structure,
as opposed to the other two cases. This can be explained
by the proximity of the former two values of $\alpha$ to the
critical value
$\alpha=1.0$ where eigenstates should show the multifractal
behavior.
To get a  quantitative information about  the properties of the
eigenstates, the mean value
of the IPR, $\langle P_2\rangle$, and its relative variance,
$\delta=(\langle P_2^2\rangle - \langle P_2\rangle^2)/\langle
P_2\rangle^2$, were calculated in \cite{prbm}.
At any given $\alpha$ the dependences of the quantities
$\langle P_2\rangle$ and $\delta$ on the matrix size $N$ were 
approximated by the power-laws $\langle P_2\rangle\propto
1/N^\nu$,
$\delta\propto 1/N^\mu$ for $N$ ranging from $100$ to $2400$.
In Figs.~\ref{prbm_fig3} and \ref{prbm_fig4} 
the values of the exponents $\nu$ and $\mu$ obtained
in this way are plotted 
versus the PRBM parameter $\alpha$. The expected theoretical
curves  are presented as well.
We see from Fig.~\ref{prbm_fig3} 
that the data show a crossover  from the behavior
typical for extended states ($\nu=1$) to that typical for localized
states
($\nu=0$), centered approximately at the critical point $\alpha=1$. 
The deviations from the sharp step-like theoretical curve
$\nu(\alpha)$ can be presumably attributed
to the finite-size effects which are unusually pronounced
in the PRBM model due to the long-range nature of the off-diagonal
coupling. The data for the exponent $\mu$ (Fig.~\ref{prbm_fig4})
also show a reasonable agreement with the
expected linear crossover, $\mu=4(1-\alpha)$ for $3/4<\alpha<1$, see
Eq. (\ref{prbm_42}).

\begin{figure}
\centerline{\epsfig{file=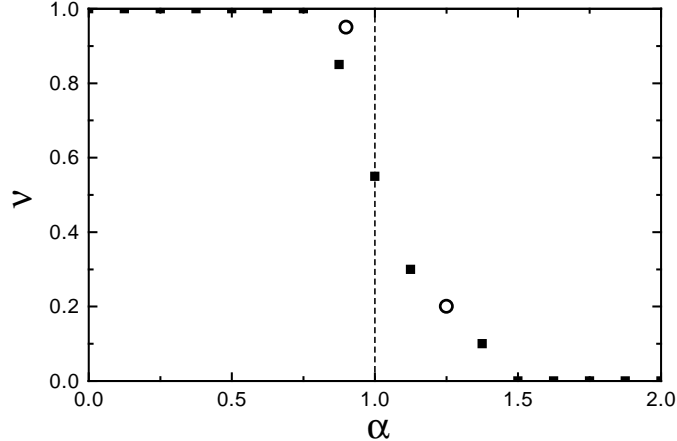,width=70mm,angle=270}}
%\centerline{\epsfxsize=120mm\epsfbox{prbm_fig3.ps}}  
\vspace{3mm}
\caption{Index $\nu $ characterizing the dependence of the inverse participation
ratio $\langle P_2\rangle$ on the matrix size $N$ via
$\langle P_2\rangle\propto 1/N^\nu$, as a function of $\alpha $. Points
refer to the best-fit values obtained from matrix sizes between $N=100$
and $N=1000$ (squares) or $N=2400$ (circles). 
The dashed line is the theoretical prediction for the
transition from $\nu  = 1$, at small $\alpha $, to $\nu  = 0$, at large
$\alpha $. From \cite{prbm}.} 
\label{prbm_fig3} 
\end{figure}

\begin{figure}
\centerline{\epsfig{file=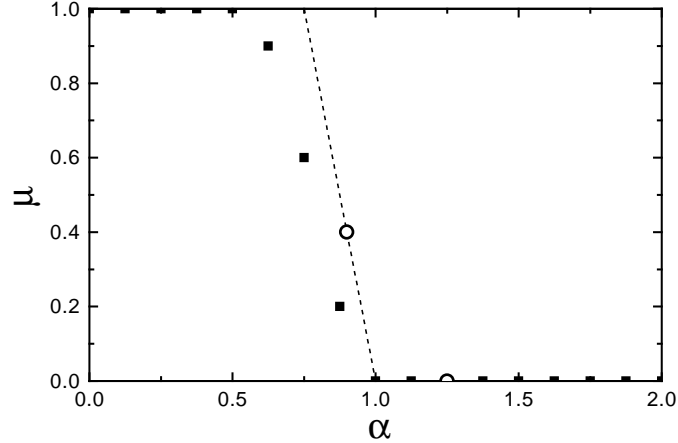,width=70mm,angle=270}}
%\centerline{\epsfxsize=120mm\epsfbox{prbm_fig4.ps}}  
\vspace{3mm}
\caption{The same as Fig.~\ref{prbm_fig3}, 
but for the index $\mu $, derived from the $N$ dependence 
of the variance $\delta $ of the inverse participation ratio:
$\delta \equiv (\langle P_2^2\rangle - \langle P_2\rangle^2)/\langle
P_2\rangle^2 \propto 1/N^\mu $. The dashed line corresponds to
the predicted linear crossover from $\mu =1$ at $\alpha < 3/4$ to $\mu =0$ at
$\alpha > 1$. From \cite{prbm}.}
\label{prbm_fig4} 
\end{figure}

While the presented
data are in good agreement with the above theoretical picture,
a more detailed numerical investigation of the structure of
eigenstates and of spectral statistics is certainly
desirable. In particular, it would
be especially interesting to study the critical manifold, $\alpha=1$, where
the multifractal properties of eigenstates and intermediate level
statistics are predicted by the theory.

\subsubsection{Spectral properties.}
\label{s5.4.3}

Let us consider now the spectral statistics of the PRBM
model. In the ``metallic'' regime, the leading correction to 
the Wigner--Dyson form $R_{WD}(s)$ of the level correlation function
is given by the properly modified Eq.~(\ref{e2.19}), i.e.
\begin{equation}
R(s)=\left[1+{1\over 2\beta}{\cal C}{d^2\over ds^2}s^2\right]R_{WD}(s)\ ,
\label{prbm_51}
\end{equation}
where
\begin{eqnarray}
{\cal C}&=&{1\over N^2}\sum_{rr'}\Pi^2(r,r')={1\over N^2}
\sum_{q=\pi n/N;\:n=1,2,\ldots} \Pi^2(q)= \nonumber\\
&=& \left\{
\begin{array}{ll}
\displaystyle{t^2\over 64\pi^{2\sigma}}N^{2\sigma-2}\ ,& \qquad
\sigma>1/2\\
\mbox{const}
\displaystyle{t^2\over 64\pi^{2\sigma}b^{1-2\sigma}}N^{-1}\ ,& \qquad
\sigma<1/2\ .
\end{array}
\right.
\label{prbm_52}
\end{eqnarray}
At $\sigma<1/2$ the sum divergent at high momenta  is cut off at
$q\sim\pi/b$;
the procedure leaving undetermined a constant of order of unity.
The correlation function $R(s)$ is close to its RMT value if $\sigma<1$
(the region of delocalized states), or else if $\sigma>1$ and the system
size $N$ is much less than the localization length $\xi$,
Eq. (\ref{prbm_36}). 
Under these conditions, Eq. (\ref{prbm_51}) holds as long as the
correction term
is small compared to the leading one. This requirement produces the
following restriction on the frequency $s=\omega/\Delta$:
\begin{equation}
s<s_c\sim\left\{
\begin{array}{ll}
t^{-1}N^{1-\sigma}\ , &\qquad \sigma>1/2\\
t^{-1}b^{{1\over 2}-\sigma}N^{1/2}\propto (Nb)^{1/2}\ , &\qquad
\sigma<1/2
\end{array}
\right. \;.
\label{prbm_53}
\end{equation}

At larger frequencies ($s>s_c$), the form of the level correlation
function
changes from the $1/s^2$ behavior typical for RMT to a 
different
one, in full similarity with the Altshuler-Shklovskii regime
(\ref{e2.22}) in the case of a conventional $d$-dimensional
conductor. Extending Eq.~(\ref{e2.17}) to the present case, we find
\begin{eqnarray}
R^{(c)}(s)&=&{\Delta^2\over \beta\pi^2}\mbox{Re}\sum_{n=0,1,...}
{1\over \left[{8\over\pi\nu t}\left({\pi n\over
N}\right)^\sigma-i\omega\right]
^2}\nonumber\\ \label{prbm_54} &\propto&
\left\{
\begin{array}{ll}
N^{1-1/\sigma}t^{1/\sigma}s^{1/\sigma-2}\ ,&\qquad \sigma>1/2\ \
(\sigma\ne 1)\\
t^2N^{-1}b^{2\sigma-1}\propto (Nb)^{-1}\ ,&\qquad \sigma<1/2
\end{array}
\right.     \;.
\end{eqnarray}

At last, let us consider the level statistics at criticality,
$\sigma=1$. In this case the coefficient of proportionality  in the
asymptotic
expression (\ref{prbm_54}) vanishes in view of  analyticity:
\begin{equation}
R^{(c)}(s)\sim {\Delta t\over
16\pi^2\beta}\int_{-\infty}^{\infty}{dx\over(x-i\omega)^2}
=0                  \;.
\label{prbm_55}
\end{equation}
This is similar to what happens in the case of a 2D diffusive
conductor (see
Sec.~\ref{s5.2}). A more accurate consideration requires
taking into account the high-momentum cut-off at $q\sim b^{-1}$.
In full analogy with the  2D situation mentioned
we find then a linear term in the level number variance:
\begin{equation} 
\langle\delta N^2(E)\rangle\simeq\chi\langle N(E)\rangle\ ;\qquad
\chi=\int R^{(c)}(s)ds={t\over 8\pi\beta} \ .
\label{prbm_56}
\end{equation}
The presence of the linear term (\ref{prbm_56}) (as well as the
multifractality
of eigenfunctions, Sec. \ref{s5.4.2}) makes the case $\sigma=1$  similar
to the
situation on the mobility edge of a disordered conductor in $d>2$
discussed in Sec.~\ref{s5.2}, \ref{s5.3}.
Let us finally mention that the value of $\chi$,
Eq. (\ref{prbm_56}), is in  agreement with the formula
(\ref{crit2}), where $D_2$ is given by Eq. (\ref{prbm_49}) with $q=2$,
and $d=1$ in the present case.
We stress, however, that both Eqs.~(\ref{prbm_49}) and (\ref{prbm_56})
have been derived in the leading order in $t/8\pi\ll 1$. As we discuss
below, at $t/8\pi\gtrsim 1$ Eq.~(\ref{crit2}) appears to be violated.

In fact, one can calculate the whole two-level correlation function
using the results of Refs.~\cite{km,aa} presented in
Sec.~\ref{s2.2}. We consider the unitary symmetry ($\beta=2$) here
for simplicity. In this case, the level correlation function is
described by a single formula $R(s)=R_{pert}(s)+R_{osc}(s)$ [with
$R_{pert}(s)$ and $R_{osc}(s)$ given by Eqs.~(\ref{e2.25}),
(\ref{e2.23}) respectively] in the whole range of $s$. The spectral
determinant $D(s)$, Eq.(\ref{e2.24}) can be easily calculated
\cite{kravtsov97c}, yielding
\be
\label{prbm_503}
D(s)={1\over s^2}{(\pi s t/16)^2\over \sinh^2(\pi s t/16)}
\ee
in the case of periodic boundary conditions and
\be
\label{prbm_504}
D(s)={1\over s^2}{\pi s t/8\over \sinh(\pi s t/8)}
\ee
in the case of hard-wall boundary conditions.
The resulting expressions for $R(s)$ are
\be
\label{prbm_505}
R^{(c)}(s)=-{\sin^2\pi s\over(\pi s)^2}{(\pi s t/16)^2\over
\sinh^2(\pi s t/16)} 
\ee
for periodic boundary conditions and
\be
\label{prbm_506}
R^{(c)}(s)=-{1\over 4\pi^2s^2}\left[1+{(\pi s t/8)^2\over \sinh^2(\pi s
t/8)}-2\cos(\pi s){\pi s t/8\over \sinh(\pi st/8)}\right]
\ee
for the hard-wall boundary conditions. Calculating the spectral
compressibility $\chi=\int R^{(c)}(s)ds$, we reproduce
Eq.~(\ref{prbm_56}).

There exists a deep connection between the PRBM ensemble and the
random matrix ensemble introduced by Moshe, Neuberger and Shapiro
in Ref.~\cite{moshe94}. This latter ensemble is determined by the
probability density
\be
\label{prbm_507}
{\cal P}(H)\propto \int
dU\exp\{-\mbox{Tr}H^2-h^2N^2\mbox{Tr}([U,H][U,H]^\dagger)\} \ ,
\ee
where $\int dU$ denotes the integration over the group of unitary
matrices with the Haar measure. The parameter $h$ allows to interpolate
between the limits of the Wigner-Dyson ($h=0$) and the Poisson
($h\to\infty$) statistics. The relation between the 
 ensemble (\ref{prbm_507}) and the PRBM ensemble is
established as follows \cite{kravtsov97c}.
For any given unitary matrix $U$ the level statistics in the ensemble
determined by the density (\ref{prbm_507}) (without integration over
$U$) is determined by the eigenvalues $e^{i\theta_k}$ ($k=1,\ldots,N$)
of $U$. For $N\gg 1$ a typical matrix $U$ has an essentially uniform
density of eigenvalues, so that we can consider as a typical
representative the matrix with equidistant eigenvalues, $\theta_k=2\pi
k/N$. Indeed, weak fluctuations of $\theta_k$ around
these values do not affect essentially the behavior of $a(r)$ in
Eq.~(\ref{prbm_509}) below. On the other hand,
the matrices with strongly nonuniform density of eigenvalues represent
a vanishingly small fraction of the whole group volume and can be
neglected. Making a transformation to the eigenbasis of $U$, we can
transform ${\cal P}(H)$ to
\be
\label{prbm_508}
{\cal P}(H)\propto
\exp\left\{-\sum_{ij}|H_{ij}|^2
\left[1+(2Nh)^2\sin^2{\pi\over N}(i-j)\right]\right\} \ ,
\ee
so that the matrix $H$ has a form of the PRBM in the critical regime
$\alpha=1$ with
\bea
\label{prbm_509}
a^2(r)&=&{\beta \over 4}{1\over 1+
{1\over b^2}{\sin^2(\pi r/N)\over (\pi/N)^2}} \nonumber \\
& \simeq & {\beta\over 4}{1\over 1+(r/b)^2} \qquad \mbox{for } r\ll
N\ ,
\eea
and $b=1/2\pi h$. Calculating (for $h\ll 1$)
the coupling constant in the center of
the band ($E=0$) according to Sec.~\ref{s5.4.1}, we get $t=4/b=8\pi h$. 

A nice feature of the ensemble
(\ref{prbm_507}) is that at $\beta=2$ the level correlation function
can be calculated exactly for arbitrary value of $h$ \cite{moshe94}
(we remind that the above $\sigma$-model
study of the PRBM ensemble was restricted to
the regime $1/2\pi b= h\ll 1$). After the integration over
the unitary group with making use of the Itzykson-Zuber formula, the
joint probability distribution of the eigenvalues of $H$ is found to
be equal to the probability density of coordinates of a system of
non-interacting 1D fermions in a harmonic confinement potential  
at finite temperature. The parameter $h$ of the model (\ref{prbm_507})
determines, in this formulation, the ratio of the temperature to the
Fermi energy, $h=T/E_F$, or, in other words, the degree of degeneracy
of the Fermi gas. Calculating the
two-particle correlation function for non-interacting fermions, one
finds the sought level correlation function \cite{moshe94}. 
In the center of the band the result reads
\be
\label{prbm_510}
\langle\nu(-\omega/2)\nu(\omega/2)\rangle-\langle\nu(0)\rangle^2=
-\left[{Nh\over \pi}\int_{-\infty}^\infty{dp\over
1+Ce^{p^2}}e^{2iphN\omega}\right]^2\ ,
\ee
where $C=[e^{1/h}-1]^{-1}$ and the mean level spacing $\Delta$ is
given by 
\be
\label{prbm_511}
\Delta^{-1}=\langle\nu(0)\rangle={Nh\over\pi}\int{dp\over 1+Ce^{p^2}}\ .
\ee
The spectral compressibility $\chi$ is found from (\ref{prbm_510}),
(\ref{prbm_511}) to be
\be
\label{prbm_512}
\chi=1-{\int dp (1+Ce^{p^2})^{-2}\over \int dp (1+Ce^{p^2})^{-1}} 
\ee
The formulas (\ref{prbm_510}), (\ref{prbm_511}) can be simplified in
the limits $h\ll 1$ and $h\gg 1$. For $h\ll 1$, when the gas is
strongly degenerate, one gets (as usual, we introduce
$s=\omega/\Delta$) 
\be
\label{prbm_513}
R^{(c)}(s)=-\left({\sin\pi s\over\pi s}\right)^2
\left(\pi^2 hs/2\over \sinh
\pi^2hs/2\right)^2 \ ,
\ee
in precise agreement with the result (\ref{prbm_505}) obtained
directly for the PRBM ensemble (with the above
identification of the parameters of the two models, $t=8\pi h$). 

In the opposite limit, $h\gg 1$, the gas is almost classical and the
correlations are weak,
\be
\label{prbm_514}
R^{(c)}(s)=-e^{-2\pi h^2s^2}\ .
\ee
Note that the corresponding spectral compressibility 
$\chi\simeq 1-1/\sqrt{2}h$ approaches unity in the limt $h\to\infty$,
whereas the formula (\ref{crit2}) would imply $\chi\le
1/2$. Therefore, at least for the PRBM model, Eq.~(\ref{crit2}) is not
an exact relation, but rather an approximation valid in the
close-to-RMT regime $1/2\pi b= h\ll 1$.

\section{Conductance fluctuations in quasi-one-dimensional wires}
\label{c6}

\setcounter{equation}{0}

This section is devoted to a study of the conductance fluctuations of
a quasi-one-dimensional disordered system \cite{zirn1,mmz}. 
As was already mentioned in
Section~\ref{c3}, there exist different microscopic models which can
be mapped onto the same 1D supersymmetric $\sigma$-model and thus
belong to the same ``quasi-one-dimensional universality class''. Our
treatment below will be based on the Iida-Weidenm\"uller-Zuk (IWZ)
model 
\cite{iwz} representing a wire as a sequence of coupled $N\times N$
random matrices, the first and the last of which are coupled to the
states propagating in the leads. Using the multi-channel
B\"uttiker-Landauer formula
\cite{landauer,fishlee,buettiker,barstone},  
the mean conductance and its variance can be expressed in terms of end-to-end
correlation functions of the $\sigma$-model. Similarly to what was done
in Sec.~\ref{s3.2}, where the eigenfunction statistics was studied,
these correlation functions can be calculated exactly via the
transfer-matrix method. This means that calculation of the functional
integral can be reduced to solution of a ``Schr\"odinger equation'',
which can be found in terms of an expansion in corresponding
eigenfunctions. The results depend on a single parameter $L/\xi$
(where $L$ is the sample length and $\xi$ the localization length) and
do not depend on details of the underlying microscopic model. On top
of the IWZ-model and two other models from the quasi-1D class already
discussed in Sec.~\ref{s3.2} (the thick wire model of Efetov and Larkin
and the random banded matrix model), we mention here a system of weakly
coupled 2d layers in strong magnetic field
\cite{chalkdohm,balfish,kim,mathur,bfz,gruzberg1,gruzberg2,%
druist,wang}. In the quantum Hall regime
the transport in each layer is due to edge states. Tunneling between
the layers leads to appearance of the transport in the transverse
direction. The coupled edge states propagating on a surface of the
cylinder form a 2d chiral metal, which can be described by a
directed network model introduced by Chalker and Dohmen
\cite{chalkdohm}.  Mapping of this problem onto the 
supersymmetric spin chain was done in \cite{balfish,bfz}; further
mapping onto the supersymmetric
$\sigma$-model was presented in \cite{gruzberg1}. 
If the number of layers is sufficiently large, the system is of
quasi-one-dimensional nature. Very
recent numerical study of the directed network model \cite{wang}
 showed perfect agreement with the analytical results presented
below. First 
experimental realization of the multilayer quantum Hall system
has been also reported recently \cite{druist}.

\subsection{Modeling a disordered wire and mapping onto 1D $\sigma$-model.} 
\label{s6.1}

The presentation below is based on Ref.~\cite{mmz}.
We consider a quasi one--dimensional disordered wire of length $L$, 
decomposed into $K$ boxes with linear dimension $\ell$.
With each box $i$ ($i=1,\ldots,K$) we associate $N$ electronic 
states $|i\mu\rangle$ ($\mu = 1,\ldots,N$). The boxes $1$ and $K$ 
are coupled to ideal leads $\kappa$ on the left ($\kappa=L$) and right
($\kappa=R$) side of the disordered region. In each of these leads we have
a number of modes $|E,a,\kappa\rangle$ ($a=1,\ldots$) with transverse
energy $\epsilon_a$ whose longitudinal momentum $k$ is defined by their
total energy $E$ being equal to $\epsilon_a + \hbar^2k^2/(2m_e^*)$. Here 
$m_e^*$ is the effective mass of the electrons. We work at zero 
temperature. The Hamiltonian of the system reads 
\begin{eqnarray}
{\cal H} &=& \sum_{\kappa,a}\int_{\epsilon_a} dE \
|E,a,\kappa\rangle E \langle E,a,\kappa| + 
\sum_{i,j,\mu,\nu} |i\mu\rangle H^{ij}_{\mu\nu} \langle j\nu| \nonumber\\
& & \quad + \sum_{i=1,K}\sum_{\mu,a,\kappa}\int_{\epsilon_a} dE
\left( |E,a,\kappa\rangle W^i_{a\mu}(E,\kappa) \langle i\mu| +
{\rm c.c.} \right).
\end{eqnarray}
The $N\times N$-matrices $(H^{ii}_{\mu\nu})$ are taken to be members of
the Gaussian unitary (GUE), orthogonal (GOE) or symplectic (GSE)
ensemble. For definiteness, we will consider the orthogonal symmetry
case in the present section; the results for all the three symmetry
classes will be presented in Sec.~\ref{s6.2}. In the GOE case, 
the elements $H^{ii}_{\mu\nu}$ are independent real Gaussian
random variables with zero mean value and 
\begin{equation}
\langle H^{ii}_{\mu\nu} H^{ii}_{\mu'\nu'}\rangle =
\frac{\lambda^2}{N} ( \delta_{\mu\nu'}\delta_{\nu\mu'} +
                      \delta_{\mu\mu'}\delta_{\nu\nu'} ). 
\label{e6.1}
\end{equation}
States in adjacent boxes are coupled by another set of Gaussian 
random variables with zero mean values and 
\begin{equation}
\langle{H^{ii+1}_{\mu\nu} H^{i+1i}_{\nu\mu}}\rangle
 = \frac{v^2}{N^2}. 
\label{e6.2}
\end{equation}
All other matrix elements $H^{ij}_{\mu\nu}$ vanish. The coupling 
between channel and box states is effected by the matrix elements 
$W^i_{a\mu}(E,\kappa)$.
We assume that they do not depend on $E$ and $\kappa$, that they obey the
symmetry $W^i_{\mu a} = W^{i*}_{a\mu} = W^{i}_{a\mu}$ 
 and that they fulfill the orthogonality relation
\begin{equation}
\pi\sum_\mu W^i_{a\mu} W^i_{\mu b} = \delta_{ab} x \hspace{3cm} (i=1,K)
\label{mmz_e1.7}
\end{equation}
with $x$ a normalization constant. Eq.~(\ref{mmz_e1.7}) is convenient but
does not result in any 
loss of generality \cite{vwz}. We note that $W^i_{a\mu}$ vanishes unless
$i=1,K$.

The conductance of the system is given by the many--channel 
Landauer--B\"uttiker formula \cite{landauer,fishlee,buettiker,barstone}, 
\be
g = \sum_{ab} \left( |S_{ab}^{RL}|^2 + |S_{ba}^{LR}|^2 \right)
= 2\sum_{ab}  |S_{ab}^{RL}|^2 
\label{mmz_e1.8}
\ee
The $S$--matrix $S^{\kappa\kappa'}_{ab}$ of the IWZ--model reads
\cite{mahaux}  
\be
S_{ab}^{\kappa\kappa'} = \delta_{ab}\delta^{\kappa\kappa'} -
2\pi i \sum_{i,j,\mu,\nu} W^i_{a\mu}(\kappa) 
(D^{-1})_{\mu\nu}^{ij} W^j_{\nu b}(\kappa'),
\label{mmz_e1.9}
\ee
where 
\be
D^{ij}_{\mu\nu} = E\delta^{ij}\delta_{\mu\nu} - H^{ij}_{\mu\nu} + 
i\Omega^1_{\mu\nu}\delta^{i1}\delta^{j1} + i\Omega^{K}_{\mu\nu}
\delta^{iK}\delta^{jK},\
\label{mmz_e1.10}
\ee
\be
\qquad \Omega^i_{\mu\nu}=\pi\sum_{a=1}^{N_{\rm ch}}W_{\mu a}^i W_{a\nu}^i
\hspace{2cm} (i=1,K).
\label{mmz_e1.10a} 
\ee
The $S$-matrix in Eq.~(\ref{mmz_e1.8}) is taken at the Fermi energy $E = E_F$, 
and $N_{\rm ch}$ denotes the number of open channels at this energy.
It follows that in order to obtain $\langle{g}\rangle$ and 
$\langle{g^2}\rangle$ 
we have to calculate the ensemble--average of a product of two resp.four
propagators $D^{-1}$. Further manipulations are completely analogous
to those outlined in Sec.~\ref{s2.1} for the case of the level
correlation function. We define a supervector
$\psi = (S_1^1,S_1^2,\chi_1,\chi^*_1,S_2^1,S_2^2,\chi_2,\chi_2^*)$  of
real commuting ($S$) and complex anticommuting ($\chi$) variables,
with the first four components corresponding to the retarded and the
last four to the advanced sector. 
Then
\be
(D^{-1})^{ij}_{\mu\nu} =
\int d[\psi] \ (S^1_1)^j_\nu (S^1_1)^i_\mu \
\exp\left( 
  \frac{i}{2} \psi^\dagger \Lambda^{1/2} \hat{D} \Lambda^{1/2} \psi
    \right) 
\label{mmz_e1.11}
\ee
where 
$\hat{D} = {\rm diag}(D,D,D,D,D^\dagger,D^\dagger,D^\dagger,D^\dagger)$,
$\Lambda = {\rm diag}(1,1,1,1,-1,-1,-1,-1)$. 
Products of two and four propagators can be expressed in similar fashion
(see Eq.(\ref{mmz_e1.18}) below). After averaging of
Eq.(\ref{mmz_e1.11})  over the Gaussian distribution
of random variables entering $\hat{D}$, we perform
the Hubbard--Stratonovitch transformation 
introducing $8\times 8$ supermatrices ${\cal R}_i\ (i=1\ldots K)$ conjugate to
the dyadic product  $\psi_i\psi_i^\dagger$ and then take
the large--$N$ limit. As a result, the integration over ${\cal R}_i$
is restricted to solutions of the saddle--point equation
${\cal R}_i = \lambda^2 (E-{\cal R}_i)^{-1}$,
which have the form 
\be
{\cal R} = \frac{E}{2} - i\sqrt{\lambda^2 - \frac{E^2}{4}}T\Lambda T^{-1}
 \equiv \sigma\cdot I - i\delta\cdot Q\ ,
\ee
with $Q=T\Lambda T^{-1}$, $Q^2=1$. 
As a result,  Eq.~(\ref{mmz_e1.11}) takes the form:
\bea
\langle{ (D^{-1})^{ij}_{\mu\nu} }\rangle &=&
\int D[Q] \ \exp\left(
\frac{-\delta^2v^2}{4\lambda^4} \sum_i \str\, Q_i Q_{i+1} 
                \right)                                \no\\ 
& &\quad \times \int d[\psi] \ (S^1_1)^j_\nu (S^1_1)^i_\mu \ \exp \left(
  \frac{i}{2} \psi^\dagger \Lambda^{1/2} 
                (\hat{E} + i\hat{\Omega}\Lambda +i\delta \hat{Q})
              \Lambda^{1/2} \psi  \right) \no \\
& \equiv & \langle (S^1_1)^j_\nu (S^1_1)^i_\mu \rangle_Q .
\label{mmz_e1.12}
\eea
Here 
$(\hat{E})^{ij}_{\mu\nu}= E\delta_{\mu\nu}\delta^{ij}$,
$(\hat{\Omega})^{ij}_{\mu\nu} = (\Omega^1)_{\mu\nu}\delta^{i1}\delta^{j1} +
 (\Omega^K)_{\mu\nu}\delta^{iK}\delta^{jK} $
and
$(\hat{Q})^{ij}_{\mu\nu} = \delta_{\mu\nu} Q_i \delta^{ij}$. The last
line of Eq.~(\ref{mmz_e1.12}) introduces a short-hand notation
$\langle\ldots\rangle_Q$. 

According to Eqs.~(\ref{mmz_e1.8}), (\ref{mmz_e1.9}), the first two
moments of the conductance are given by
\be
\langle{g}\rangle = 8\sum_{\mu\nu\mu'\nu'}\Omega^1_{\nu\nu'}
\Omega^K_{\mu'\mu}
\langle{(D^{-1})^{K1}_{\mu\nu}
(D^{-1\,\dagger})^{1K}_{\nu'\mu'}}\rangle\ , 
\label{mmz_e1.17a}
\ee
\be
\langle{g^2}\rangle = 64\sum_{\mu\nu\mu'\nu'\rho\tau\rho'\tau'}
\Omega^1_{\nu\nu'}\Omega^K_{\mu'\mu}\Omega^1_{\tau\tau'}\Omega^K_{\rho'\rho}
\langle{
(D^{-1})^{K1}_{\mu\nu} (D^{-1\,\dagger})^{1K}_{\nu'\mu'}
(D^{-1})^{K1}_{\rho\tau}
(D^{-1\,\dagger})^{1K}_{\tau'\rho'}}\rangle. 
\label{mmz_e1.17b}
\ee
The averages of products of the Green's functions
relevant for $\langle g\rangle$ and $\langle g^2\rangle$ 
can be expressed similarly to
Eq.~(\ref{mmz_e1.12}) as follows:
\bea
\langle{(D^{-1})^{K1}_{\mu\nu}
(D^{-1\,\dagger})^{1K}_{\nu'\mu'}}\rangle 
&=& 
\langle (S_1^1)^1_\nu (S^1_1)^K_\mu (S_2^1)_{\mu'}^K (S_2^1)^1_{\nu'}
\rangle_Q\ ,                      \no\\
\langle{
(D^{-1})^{K1}_{\mu\nu} (D^{-1\,\dagger})^{1K}_{\nu'\mu'}
(D^{-1})^{K1}_{\rho\tau}
(D^{-1\,\dagger})^{1K}_{\tau'\rho'}}\rangle 
&=&    \no\\
& & \hspace{-6cm}\langle
(S_1^1)^1_\nu (S^1_1)^K_\mu (S_2^1)_{\mu'}^K (S_2^1)^1_{\nu'}
(S_1^2)^1_\tau (S^2_1)^K_\rho (S_2^2)_{\rho'}^K (S_2^2)^1_{\tau'}
\rangle_Q.
\label{mmz_e1.18}
\eea
For simplicity, we set $E=0$ below. Evaluating the contractions in
(\ref{mmz_e1.18}), performing the convolution with the projectors
$\Omega^1$, $\Omega^K$ in (\ref{mmz_e1.17a}), (\ref{mmz_e1.17b}) and
using $N_{\rm ch}\gg 1$, one comes after some algebraic manipulations
to the following $\sigma$-model representation\footnote{We use here a
brief notation for the matrix elements of the $Q$-matrices. Each of
the two upper indices runs from 1 to 8 according to ordering of
components of a supervector introduced before Eq.~(\ref{mmz_e1.11})} 
 of $\langle g\rangle$ and $\langle g^2\rangle$ \cite{zirn1,mmz}:
\be
\langle g\rangle = \frac{\tilde{N}_{\rm ch}^2}{2} \int D[Q] \
Q^{51}_1 Q^{51}_K \:  \exp\{-S[Q]\}\ , 
\label{mmz_e1.26a}
\ee
\be
\langle g^2\rangle = \frac{\tilde{N}_{\rm ch}^4}{4} \int D[Q] \
Q^{51}_1 Q^{62}_1 Q^{51}_K Q^{62}_K \
\exp \{-S[Q]\}\ ,
\label{mmz_e1.26}
\ee
\be
\label{mmz_e1.26n}
S[Q]=\frac{\xi}{8\ell} \sum_i \str\, Q_i Q_{i+1}+
\frac{\tilde{N}_{\rm ch}}{8}\str \Lambda(Q_1+Q_K)\ . 
\ee
Here $\tilde{N}_{\rm ch}=N_{\rm ch} T_0$, with 
$T_0=4\delta x/(\delta+x)^2$ being the so--called transmission 
coefficient, and $\xi=(4v^2/\lambda^2)\ell$.
In the weak-disorder limit, $\xi\gg\ell$, the discrete sum 
$-(\xi/8\ell)\sum_i \str Q_i Q_{i+1}$ can be replaced by an integral
\be
\frac{\xi}{16}\int_0^L dz\:\str\left(\frac{\partial Q}{\partial z}\right)^2 ,
\label{mmz_e1.26i}
\ee
which is precisely the continuous version of the 1D $\sigma$-model (see
Sections~\ref{c2},~\ref{c3}) with $\xi$ being the
localization length equal to $\xi=2\pi\nu A D$ for a thick wire.
The second term in Eq.~(\ref{mmz_e1.26n}) containing $Q_1$ and $Q_K$
describes the coupling of the wire to the leads.

Let us note that the duplication of the number 
of field components in the supervector $\psi$ (which is forced by the
supersymmetric formulation) gave us enough flexibility to write down 
the expressions Eq.~(\ref{mmz_e1.18}) for products of two as well as of  
four propagators. As a result, we were able to express both
$\langle{g}\rangle$ 
and $\langle{g^2}\rangle$ in terms of correlation functions of the usual 
``minimal'' $\sigma$--model. We did not have to introduce $Q$--matrices
of larger dimension. This fact no longer holds true when higher moments 
of the conductance are to be calculated. For that purpose we would have to 
introduce supervectors of larger size to express higher-order 
products of propagators in a form analogous to (\ref{mmz_e1.18}). 
The increase in the number of supervector components would force us
to deal with $Q$--matrices of larger dimension. This enlargement, while
posing no difficulty to perturbative calculations \cite{akl}, gives rise 
to substantial complications in the case of the exact treatment.

To introduce the transfer operator
for the ``partition sum'' in Eq.~(\ref{mmz_e1.26}), we define the
function 
\be
W(Q_K,Q_1;L/2\xi) = \int D[Q_2] \ldots D[Q_{K-1}] \
\exp\left( \frac{\xi}{32\ell} \sum_{i=1}^{K-1} 
\str (Q_{i+1}-Q_i)^2 \right),
\label{mmz_e1.27}
\ee
which has the property
\bea
\lefteqn{
W(Q_{K+1},Q_1;(L+\ell)/2\xi) =
\int D[Q_K] ~ W(Q_{K+1},Q_K;\ell/2\xi) W(Q_K,Q_1;L/2\xi) }\no\\
& &=
\int D[Q_K] \ 
\exp\left(\frac{\xi}{32\ell} \str(Q_{K+1}-Q_K)^2 \right)
W(Q_K,Q_1;L/2\xi). 
\label{mmz_e1.28} 
\eea
In the continuum limit $\xi\gg \ell$, when $Q_i$ and $Q_{i+1}$ are
close to each other, one can reduce the integral equation
(\ref{mmz_e1.28}) to a differential one: 
\be
\partial_t W = \Delta_Q W ,
\label{mmz_e1.30}
\ee
\be 
\lim_{t \to 0} W(Q,Q';t) = \delta(Q,Q') ,
\ee
where $ \Delta_Q$ is the Laplace operator on the nonlinear space of
$Q$-matrices. 
Eq.(\ref{mmz_e1.30}) has a form of the heat (or diffusion)
equation on this space, with $W$ being the corresponding heat kernel.

\subsection{Conductance fluctuations}
\label{s6.2}

Expressions (\ref{mmz_e1.26a}), (\ref{mmz_e1.26}) contain the
effective number of open channels,
$\tilde{N}_{\rm ch}$, as a parameter. 
As was shown in \cite{iwz}, the existence of this 
parameter sets an additional length scale $L_0=\xi/\tilde{N}_{\rm ch}$. 
For a very short
system ($L\ll L_0$), the conductance $\langle g\rangle$ is controlled by
the sample--leads coupling and is of order of $\tilde{N}_{\rm ch}$. In
this case, 
the zero--mode (i.e, spatially constant $Q$) approximation can be used.
We are interested here in the opposite limit $L\gg L_0$,
where the resistance of the sample is dominated by the bulk of the
system. In this case, $\tilde{N}_{\rm ch}$ 
becomes an irrelevant parameter and the heat kernel itself already
contains all information about $\langle g\rangle$ and 
$\langle g^2\rangle$. 
Since we assume that $\tilde{N}_{\rm ch}\gg 1$, the crossover length scale
$L_0$ is much less than $\xi$, so that the ratio of the sample length
to the localization length $L/\xi$ in the bulk-dominated regime can be
arbitrary. 

Using the generalized Fourier expansion of the heat kernel in the
eigenfunctions of the Laplace operator $\Delta_Q$, one can find 
after very lengthy calculations \cite{zirn1,mmz} (there was a subtle error
in the GSE case in these papers corrected in \cite{bf}) the following
exact
results for $\langle g\rangle$ and $\langle g^2\rangle$ for an
arbitrary value of the parameter $L/\xi$:
\bea
{\rm GOE}: & & \no\\
\langle g^n\rangle(L) &=& \frac{\pi}{2} \int_0^\infty d\lambda
\tanh^2(\pi\lambda/2) (\lambda^2+1)^{-1} p_n(1,\lambda,\lambda)
\exp\left[-{L\over 2\xi}(1+\lambda^2)\right] \no\\
&+ &2^4\sum_{l\in 2\bN+1} \int_0^\infty d\lambda_1 d\lambda_2
l(l^2-1) \lambda_1 \tanh(\pi\lambda_1/2) \lambda_2 \tanh(\pi\lambda_2/2)
                                                            \no\\
& \times & p_n(l,\lambda_1,\lambda_2)\!\!\!
\prod_{\sigma,\sigma_1,\sigma_2=\pm 1}\!\!\!\!
(-1+\sigma l+i\sigma_1\lambda_1+i\sigma_2\lambda_2)^{-1} 
\exp\left[-{L\over 4\xi}(l^2+\lambda_1^2+\lambda_2^2+1)\right],   \no\\
{\rm GUE}: & & \no \\
\langle g^n\rangle(L) &=& 2^2 \sum_{l\in2\bN-1} \int_0^\infty
d\lambda \lambda \tanh(\pi\lambda/2)l(\lambda^2+l^2)^{-2}
p_n(l,\lambda) \exp\left[-{L\over 4\xi}(l^2+\lambda^2)\right],     \no\\
{\rm GSE}: & & \no\\
\langle g^n\rangle(L) &=& 2^5 \sum_{\begin{array}{c}l_1,l_2\in 2\bN-1\\
l_1+l_2\in 4\bN-2\end{array}} \int_0^\infty
d\lambda \lambda(\lambda^2+1) \tanh(\pi\lambda/2) l_1l_2 
p_n(\lambda,l_2,l_2)                  \no\\
& \times & \prod_{\sigma,\sigma_1,\sigma_2=\pm 1}
(-1+i\sigma \lambda+\sigma_1 l_1+\sigma_2 l_2)^{-1} 
\exp\left[-{L\over 8\xi}(l_1^2+l_2^2+\lambda^2-1)\right].
\label{mmz_e2.87}
\eea
 The polynomials $p_n$ in the above expressions are given by
\bea
{\rm GOE}: \quad & &
p_1(l,\lambda_1,\lambda_2) = 
l^2+\lambda_1^2+\lambda_2^2+1,                   \no\\
& & p_2(l,\lambda_1,\lambda_2) =
\frac{1}{2}(\lambda_1^4+\lambda_2^4+2l^4+3l^2(\lambda_1^2+\lambda_2^2) +
             2l^2 - \lambda_1^2 - \lambda_2^2 -2),   \no\\
{\rm GUE}: \quad & &
p_1(l,\lambda) =  l^2+\lambda^2, \no\\
& &p_2(l,\lambda) = \frac{1}{2}(l^2+\lambda^2)^2, \no\\
{\rm GSE}: \quad & &  
p_1(\lambda,l_1,l_2)  =
\lambda^2+l_1^2+l_2^2-1,                   \no\\
& & p_2(\lambda,l_1,l_2) =
\frac{1}{4}(l_1^4+l_2^4+2\lambda^4+3\lambda^2(l_1^2+l_2^2) -
             2\lambda^2 + l_1^2 + l_2^2 -2). \no\\
\label{mmz_e2.88}
\eea

\begin{figure}
\centerline{\epsfig{file=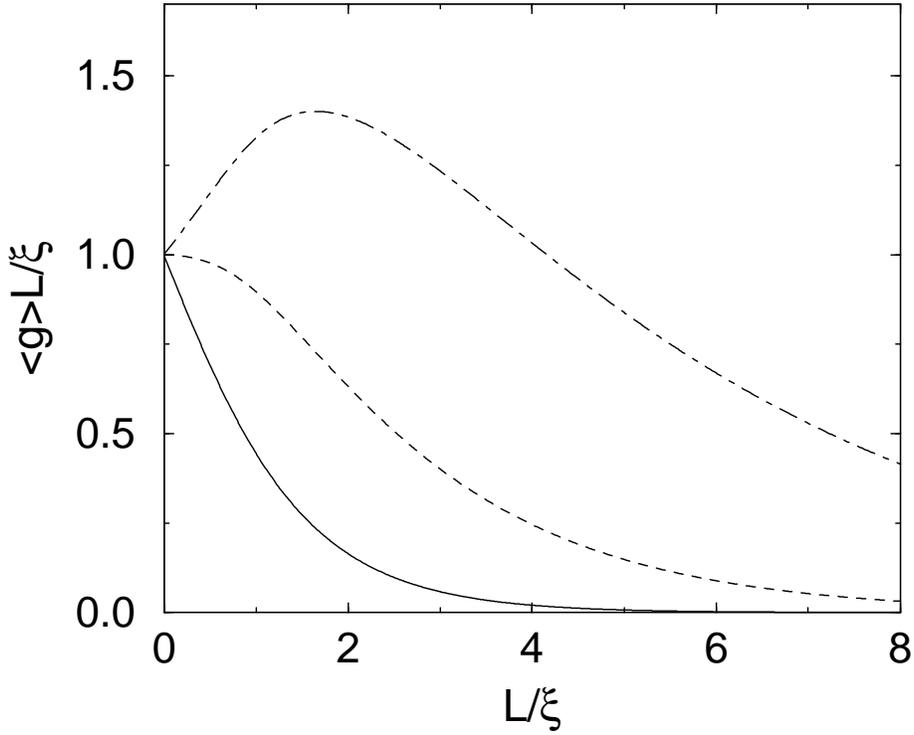,width=120mm,angle=0}}
%\centerline{\epsfxsize=120mm\epsfbox{tube.eps}}  
\vspace{3mm}
\caption{Average conductance $\langle g \rangle$ normalized to its
Ohm's law value as a function of the sample length (measured in units
of $\xi=2\pi\nu AD$). The full, dashed
and dot-dashed lines correspond to the orthogonal, unitary, and
symplectic symmetry, respectively.}
\label{cond_av} 
\end{figure}

\begin{figure}
\centerline{\epsfig{file=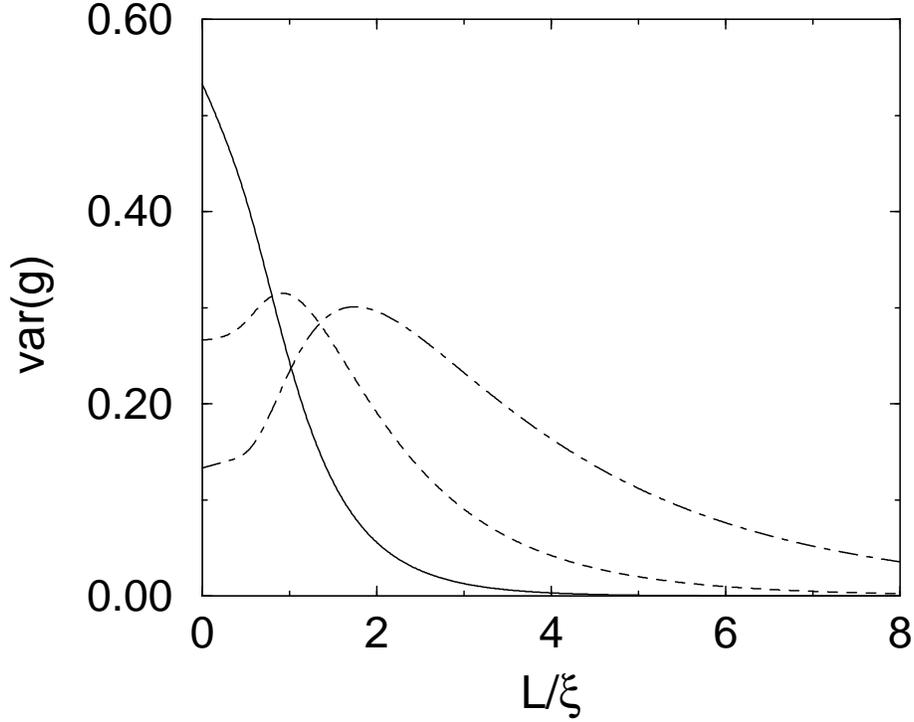,width=120mm,angle=0}}
%\centerline{\epsfxsize=120mm\epsfbox{tube.eps}}  
\vspace{3mm}
\caption{The same as Fig.~\ref{cond_av}, but for the conductance
variance ${\rm var}(g)$.}
\label{cond_var} 
\end{figure}

The results (\ref{mmz_e2.87}) are presented in Figs.~\ref{cond_av},
\ref{cond_var} for all the three symmetry classes.
The above GUE formula is written for the case of broken time-reversal
symmetry, but preserved spin-rotation invariance.
The result for systems with broken time-reversal symmetry and with
strong spin interactions (e.g. systems with magnetic impurities, or systems
with strong spin-orbit interaction in magnetic field; to be denoted as
GUE$'$ below) is 
\be
\label{e6.3}
\langle g^n\rangle^{\rm GUE'}(L) = 
(1/2)^n \langle g^n\rangle^{\rm GUE}(L/2).
\ee
The results (\ref{mmz_e2.87}) have been confirmed by numerical
simulations of the IWZ-model \cite{mm}  and of the directed network model
\cite{wang}. In particular, we present in Figs.~\ref{cond_gse_num} and
\ref{var_gse_num} the numerical results of \cite{mm} for the symplectic
symmetry class.

\begin{figure}
\centerline{\epsfig{file=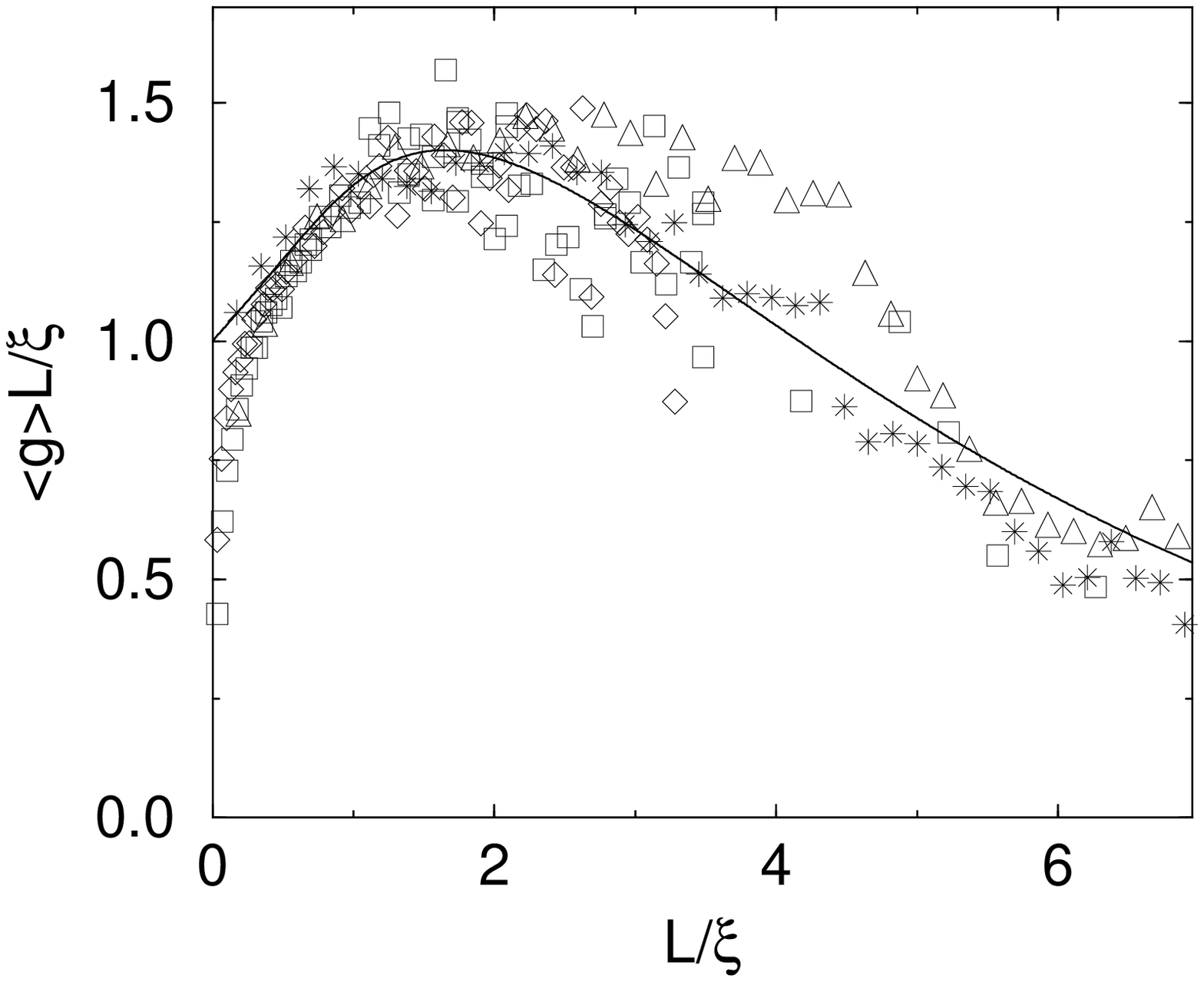,width=120mm,angle=0}}
%\centerline{\epsfxsize=120mm\epsfbox{tube.eps}}  
\vspace{3mm}
\caption{Numerical data \cite{mm} for the average conductance in the
case of symplectic symmetry. The box size $N$ and the number of
channels $N_{\rm ch}$ are equal to 10 (squares), 20 (diamonds), 60
(triangles), and 100 (stars). Each data point corresponds to an average
over 100 realizations of disorder. The full line is the theoretical
prediction.}
\label{cond_gse_num} 
\end{figure}

\begin{figure}
\centerline{\epsfig{file=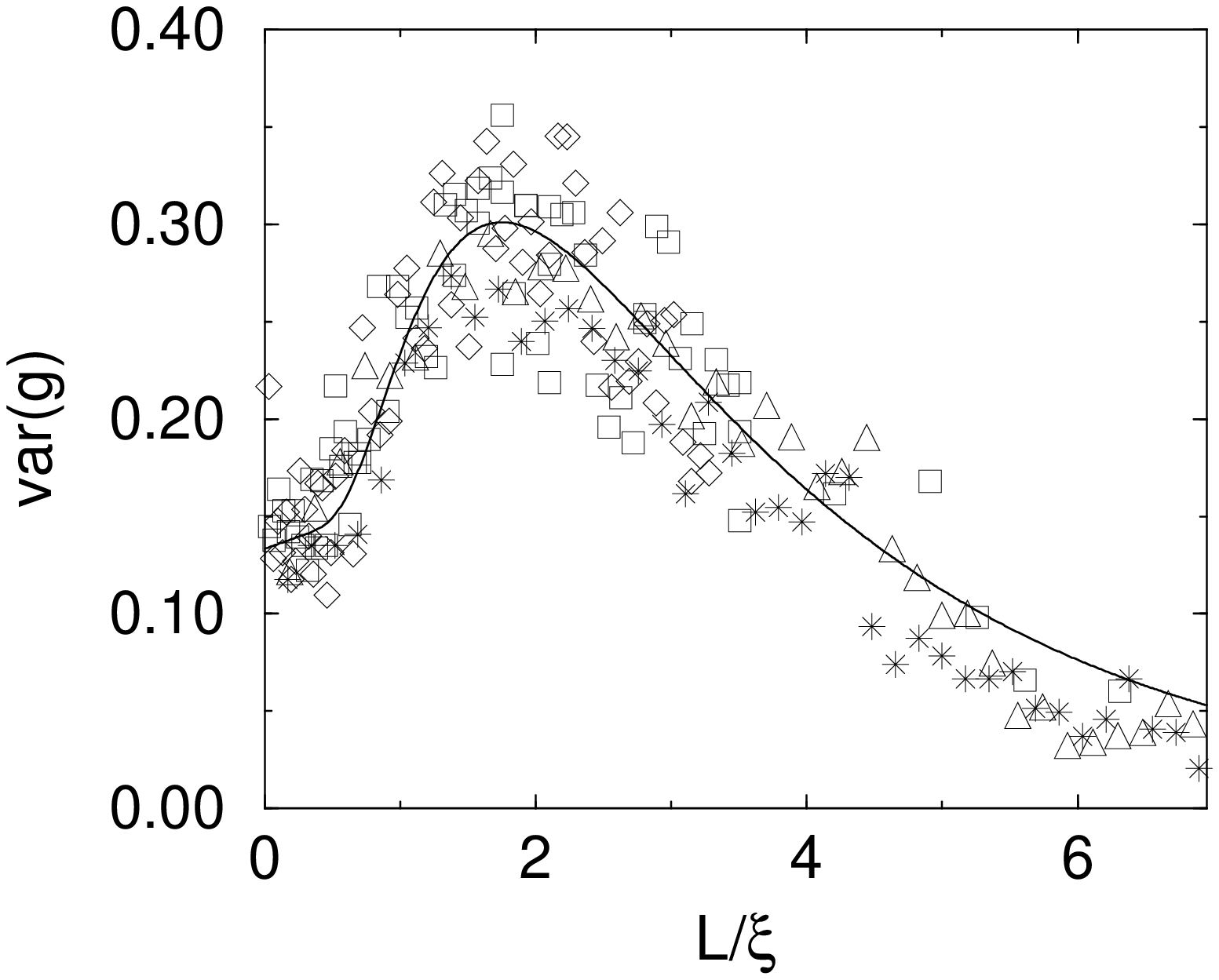,width=120mm,angle=0}}
%\centerline{\epsfxsize=120mm\epsfbox{tube.eps}}  
\vspace{3mm}
\caption{The same as Fig.~\ref{cond_gse_num}, but for the conductance
variance.} 
\label{var_gse_num} 
\end{figure}

Let us discuss the behavior of Eqs.~(\ref{mmz_e2.87}) in the limits of
short ($L\ll\xi$) and long ($L\gg\xi$) wire.
Condition $L\ll\xi$ corresponds to the metallic (weak localization)
region, $\langle g\rangle\gg 1$. We find in this case the following
perturbative (in $L/\xi$) expansion for $\langle g\rangle$, $\langle
g^2\rangle$ and $\mbox{var}(g)$:
\bea
\langle g\rangle(L) &=& \frac{2\xi}{L} - \frac{2}{3} +
             \frac{2}{45}\frac{L}{\xi} + 
\frac{4}{945}\left(\frac{L}{\xi}\right)^2 + O\left(\frac{L}{\xi}\right)^3,
                                                         \no\\
\langle g^2\rangle(L) &=&
\left(\frac{2\xi}{L}\right)^2 - \frac{8}{3}\frac{\xi}{L} +
                               \frac{52}{45} 
 - \frac{136}{945}\frac{L}{\xi} + O\left(\frac{L}{\xi}\right)^2,  \no\\
{\rm and} \quad {\rm var}(g(L)) &=& 
\frac{8}{15} - \frac{32}{315}\frac{L}{\xi} + O\left(\frac{L}{\xi}\right)^2.
\label{mmz_e3.21}
\eea
The perturbative results for the symplectic case are related to those
for the orthogonal class via the symmetry relations
\cite{JuOp,OpJu,Weg}
$\langle g^n\rangle^{Sp}(L)=(-1/2)^n\langle g^n\rangle^O(-L/2)$
 and have the form
\bea
\langle g\rangle(L) &=& \frac{2\xi}{L} + \frac{1}{3} +
             \frac{1}{90}\frac{L}{\xi}
-\frac{1}{1890}\left(\frac{L}{\xi}\right)^2 + O\left(\frac{L}{\xi}\right)^3 ,
                                                          \no\\
\langle g^2\rangle(L) &=&
\left(\frac{2\xi}{L}\right)^2 + \frac{4}{3}\frac{\xi}{L} +
                               \frac{13}{45} +
  \frac{17}{945}\frac{L}{\xi} + O\left(\frac{L}{\xi}\right)^2, \no\\
{\rm and} \quad {\rm var}(g(L)) &=&  
\frac{2}{15} 
+\frac{4}{315}\frac{L}{\xi} + O\left(\frac{L}{\xi}\right)^2. 
\label{mmz_e3.22}
\eea
Finally, for unitary symmetry we have 
\bea
\langle g\rangle(L) &=& \frac{2\xi}{L} - 
             \frac{2}{45}\frac{L}{\xi} + O\left(\frac{L}{\xi}\right)^3,
                                                         \no\\
\langle g^2\rangle(L) &=&
\left(\frac{2\xi}{L}\right)^2 +
                               \frac{4}{45} +
                                O\left(\frac{L}{\xi}\right)^2,  \no\\
{\rm and} \quad {\rm var}(g(L)) &=&   
\frac{4}{15} + O\left(\frac{L}{\xi}\right)^2.
\label{mmz_e3.23}
\eea
The expressions for $\langle g\rangle$ start from the Ohm's law term
$2\xi/L=4\pi\nu A D/L$ (we remind that $g$ is measured in units of
$e^2/h=e^2/2\pi\hbar$ and includes factor $2$ due to the spin, while
$\nu$ is the density of states per spin projection). The other terms
constitute the weak-localization corrections. The leading terms in
$\mbox{var}(g)$  are well-known values of the universal conductance
fluctuations in the case of quasi-1D geometry.

The opposite condition $L\gg\xi$ defines the region of strong
localization. In this case, the asymptotic behavior of $\langle
g^n\rangle$, $n=1,2$,  is as follows:
\bea
{\rm GOE:} && \langle g^n\rangle
=2^{-3/2-n}\pi^{7/2}(\xi/L)^{3/2}e^{-L/2\xi} \no\\
{\rm GUE:} && \langle g^n\rangle
=2^{3-n}\pi^{3/2}(\xi/L)^{3/2}e^{-L/4\xi} \no\\
{\rm GSE:} && \langle g^n\rangle
=2^{15/2-n}\pi^{3/2}(\xi/L)^{3/2}e^{-L/8\xi} 
\label{e6.4}
\eea
Let us recall that $\xi$ is defined here as $\xi=2\pi\nu AD$
independently of the symmetry. The formulas (\ref{e6.4}) demonstrate
therefore well-known dependence of the localization length on the
symmetry of the ensemble, $L_{loc}\propto\beta$. Let us stress,
however, that in the GUE$'$ case (broken time-reversal and spin-rotation
symmetries), the results for which can be obtained via the relation
(\ref{e6.3}) the localization length is the same as in GSE
\cite{eflar,m94}. This is because the transition GSE$\to$GUE$'$ not only
changes $\beta$ from 4 to 2, but also breaks the Kramers degeneracy,
increasing by factor of 2 the number of coupled channels. These two
effects compensate each other. More generally, localization length is
proportional to $\beta/s$, where $s$ is the degeneracy factor. 

\paragraph{DMPK equations.}
As has been already mentioned, calculation of higher moments $\langle
g^n\rangle$ with $n>2$ and thus of the whole distribution function
${\cal P}(g)$ has not been achieved in the supersymmetry approach
because of technical difficulties (necessity to increase the size of
the $Q$-matrix with increasing $n$). The conductance distribution in
the localized regime $L\gg\xi$ can be approximately calculated from
the Dorokhov-Mello-Pereyra-Kumar (DMPK) approach. Within this
approach, pioneered by Dorokhov \cite{dorokhov82} and developed by
Mello, Pereyra and Kumar \cite{mpk}, one derives a Fokker-Planck
(diffusion) equation for the distribution  of transmission
eigenvalues $T_n$ (the eigenvalues of the transmission-matrix product
$tt^\dagger$). This equation is conveniently written in terms of the
distribution function $P(\lambda_1,\lambda_2,\ldots;L)$, where
$\lambda_i=(1-T_i)/T_i$ and has the form
\bea
&& l{\partial P\over\partial L}=
{2\over\gamma}\sum_{n=1}^N{\partial\over\partial\lambda_n}
\lambda_n(1+\lambda_n)J{\partial\over\partial\lambda_n}J^{-1}P\ ,   
\label{e6.5} \\
&& J=\prod_{i=1}^N\prod_{j=i+1}^N |\lambda_i-\lambda_j|^\beta\ ,
\no
\eea
where $l$ is the mean free path, $N$ is the number of transverse
modes, and $\gamma=\beta N+2-\beta$. Overview of the results obtained
within the DMPK approach can be found in the review article of
Beenakker \cite{been-rev}. It has been shown recently \cite{bf} that in the
limit $N\gg 1$ the DMPK approach is equivalent to the supersymmetric
$\sigma$-model considered above. Let us note that the DMPK approach is
restricted to calculation of transport properties of the quasi-1D
conductor (which can be expressed through the transmission eigenvalues
$T_i$), while the supersymmetry method allows to study all sorts of
quantities which can be expressed through the Green's functions
(e.g. statistics of levels, eigenfunction amplitudes, local density of
states etc, -- see other sections of this article).

In the localized regime $L\gg Nl$ the distribution function
$P(x_1,\ldots,x_N;L)$ with $\lambda_i=\sinh^2 x_i$ (so that
$T_i=1/\cosh^2 x_i$) takes the form
\cite{dorokhov82,dorokhov83,pichard91} 
\be
\label{e6.6}
P(x_1,\ldots,x_l)=\left({\gamma l\over 2\pi
L}\right)^{N/2}\prod_{n=1}^N\exp 
\left[-{\gamma l\over 2L}\left(x_n-{L\over\xi_n}\right)^2\right] \ ,\
1\ll x_1\ll x_2\ll\ldots\ll x_n,
\ee
where $\xi_n=\gamma l/(1+\beta n-\beta)$ are the inverse Lyapunov
exponents. In the limit $L\to\infty$ all $x_n$ tend to non-random
values $L/\xi_n$, which is a manifestation of the Oseledec theorem
 (matrix generalization of the ``law of large numbers'')
\cite{oseledec,crisanti,pichard81}. For $L/Nl\gg 1$ the conductance
\be
\label{e6.7}
g=G/(e^2/h)=2\sum_{n=1}^N T_n
\ee
is dominated by $x_1$, e.g. $g\simeq 2/\cosh^2 x_1\simeq 8 e^{-2x_1}$,
which implies the Gaussian distribution of $\ln g$,
\be
\label{e6.8}
P(\ln g)\simeq\left({\gamma l\over 8\pi L}\right)^{1/2}
\exp\left[-{\gamma l\over 8L} 
\left(\ln g+{2L\over \gamma l}\right)^2\right]\ ,
\ee
with the average $\langle \ln g\rangle=-{2L\over \gamma l}$ and the
variance ${\rm var}(g)={4L\over \gamma l}=-2\langle \ln
g\rangle$. We have already encountered the same log-normal
distribution in Sec.~\ref{s3.2.3}, when we calculated the distribution
of the product of the wave function intensities in two points located
close to the opposite edges of the sample. The result (\ref{e6.8}) is
fully confirmed by numerical simulations as illustrated in
Fig.~\ref{log_cond_num}.

\begin{figure}
\centerline{\epsfig{file=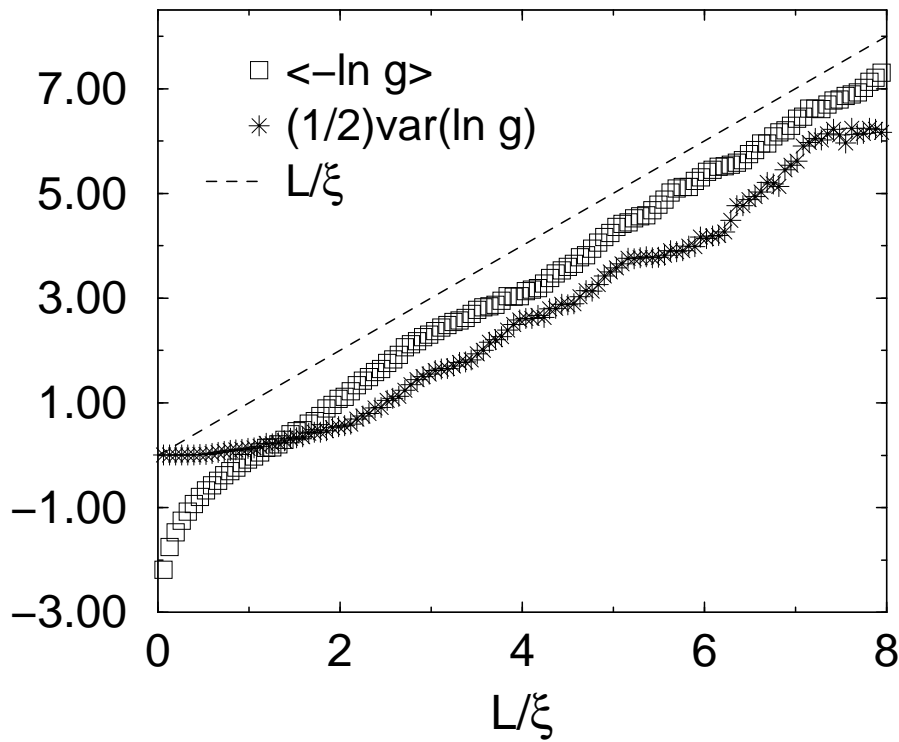,width=120mm,angle=0}}
%\centerline{\epsfxsize=120mm\epsfbox{tube.eps}}  
\vspace{3mm}
\caption{The average and the variance of the conductance logarithm as
a function of the sample length (for the symplectic symmetry class).
The box size $N$ and the number of channels $N_{\rm ch}$ are equal to
10. Each data point corresponds to an average over 100 realizations of
disorder. The dashed line corresponds to the formula (\ref{e6.8}).} 
\label{log_cond_num} 
\end{figure}

The log-normal form (\ref{e6.8}) of the conductance distribution holds
for $g<1$ only (since one transmission eigenvalue cannot produce a
conductance larger than unity); for $g>1$ the distribution $P(g)$
decays very fast. The moments $\langle g^n\rangle$ with $n\ge 1$ are
thus determined by the probability to have $g\sim 1$, which yields
(with exponential accuracy)
\be
\label{e6.9}
\langle g^n\rangle\sim\exp\left(-{L\over 2\gamma l}\right)\,
\ee
in full agreement with the asymptotics of the first two moments,
Eqs.~(\ref{e6.4}), found from the supersymmetric
approach. [Eqs.~(\ref{e6.4}) contain also preexponential factors, for
derivation of which the approximation (\ref{e6.6}), (\ref{e6.8}) of
the solution of the DMPK equations is not sufficient.]

Finally, we note that in the strictly 1D case Abrikosov
\cite{abrikosov81} derived the exact result for the conductance
distribution function $P(g)$ for arbitrary value of $L/l$. In the
limit $L/l\gg 1$ it approaches the log-normal distribution (also
found by Melnikov \cite{melnikov80a,melnikov80b}), which can be obtained
from Eq.(\ref{e6.8}) by setting $N=1$ (i.e. $\gamma=2$). In fact, 
essentially equivalent results were obtained much earlier in the
context of the classical wave propagation, see
Refs.\cite{gertsenshtein,gazaryan,papanicolaou}.

\section{Statistics of wave intensity in optics}
\label{c7}

\setcounter{equation}{0}

As was already mentioned in the introduction, our statistical
considerations are applicable not only to properties of eigenfunctions
and energy levels of quantum particles (electrons), but also to those
of intensities and eigenfrequencies of classical waves. 
This is related to similarity of the stationary
Schr\"odinger equation and the wave equation. Propagation of the
classical field $\psi_\omega$ with frequency $\omega$ in an
inhomogeneous medium is described by the wave equation
\be
\label{e7.1}
[\nabla^2+k_0^2(1+\mu({\bf r}))]\psi_\omega({\bf r})=0\ ,
\ee
supplemented by appropriate sources. Here $k_0=\omega/c$, with $c$
the speed of propagation in the average medium, and $\mu({\bf r})$
describes fluctuations of the refraction index. The field
$\psi_\omega$ can describe a component of the electromagnetic or
acoustic wave. The impurity diagrammatic technique
\cite{frish,pingsheng1,pingsheng2,shapiro90} and the $\sigma$-model
approach \cite{sjohn83,elattari} can be developed in full analogy with
the case of the Schr\"odinger equation in random potential.

Let us consider an open system with a permanently radiating source
(for example, a point-like source would correspond to addition of the
term $\propto\delta({\bf r}-{\bf r_0})$ in the r.h.s. of
Eq.~(\ref{e7.1}). The problem of fluctuations of the wave intensity
$\psi^2_\omega({\bf r})$ in such a situation has a very long
history. Almost a century ago Rayleigh proposed a distribution
which bears his name:
\be
P_o(\tilde{I})= \exp(-\tilde{I}) \ ,
\label{mps_rayleigh}
\ee
where $\tilde{I}$ is the intensity normalized to its average
value, $\tilde{I}=I/\av{I}$. A simple statistical argument leading to
Eq.~(\ref{mps_rayleigh}) is based on representing $\psi_\omega({\bf
r})$ as a sum of many random contributions (plane waves with random
amplitudes and phases). This is essentially the same argument that was
used by Berry to describe fluctuations of eigenfunctions $\psi_i$ in
chaotic billiards and which leads to the RMT statistics of
$|\psi_i^2({\bf r})|$ (see Section~\ref{c3}). Let us mention, however,
a difference between the two cases (emphasized by Pnini and Shapiro
\cite{pnini96}). In the case of an open system, $\psi_\omega$ is a sum
of traveling waves, while for the closed system $\psi_i$ is
represented as a sum of standing waves. As a result, the Rayleigh
statistics has the same form as the statistics of eigenfunction
amplitudes in a closed system with broken time-reversal invariance
(unitary class), where the eigenfunctions are complex. 

Diagrammatic derivation of Eq.~(\ref{mps_rayleigh}) is very simple 
\cite{shapiro86} (see below); for the case of a smooth randomness 
an essentially equivalent derivation \cite{dashen79} using 
path-integral arguments was given. 
However, similarly to distributions of other quantities
studied above (eigenfunction amplitude, local DOS, relaxation time
etc.), distribution of optical intensities show deviations from the
Rayleigh law, which will be studied below. More
specifically, we will consider fluctuations of the intensity $I({\bf
r}, {\bf r_0})$ at a point ${\bf r}$ induced by a point-like source
at ${\bf r_0}$, with the both points ${\bf r}$ and ${\bf r_0}$ located
in the bulk of the sample. We will assume quasi-1D geometry of the
sample with length $L$ much larger than the transverse dimension $W$
(see Fig.~\ref{tube}). 

\begin{figure}
\centerline{\epsfig{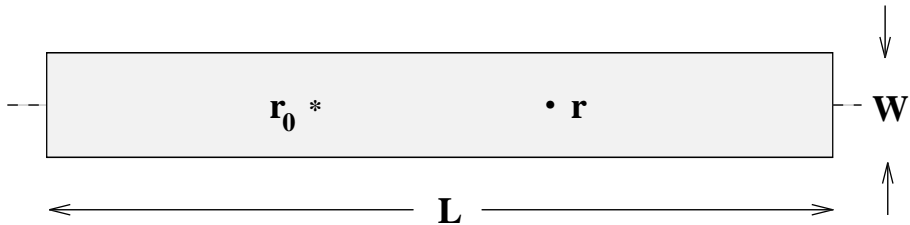}}
%\centerline{\epsfxsize=120mm\epsfbox{tube.eps}}  
\vspace{3mm}
\caption{Geometry of the problem. Points ${\bf r_0}=(x_0,y_0,z_0)$ and
${\bf r}=(x,y,z)$ are the positions of the source and of the
observation point (detector) respectively. From \cite{mps}.}
\label{tube} 
\end{figure}

Let us note that there was a considerable activity recently in
studying  statistics of the transmission coefficients $T_{ab}$
of a disordered waveguide \cite{shnerb,kkbb,nvr,kk,been1}. 
In this formulation of the problem, a source
and a detector of the radiation are located outside the sample. The
source produces a plane wave injected in an incoming channel $a$, and
the intensity in an outgoing channel $b$ is measured. The transmission
coefficients are related to the transmission matrix $t$ (already
mentioned in Section~\ref{c6}) as $T_{ab}=|t_{ab}|^2$. One can also
define the transmittance $T_a=\sum_a T_{ab}$ summed over the outgoing
channels. Finally, the total transmittance $T=\sum_{a,b} T_{ab}$ is
the optical analogue of the conductance $g$ (fluctuations of which
were studied in the Sec.~\ref{c6}). Combining the diagrammatic
approach with the results on distribution of transmission eigenvalues 
in the metallic regime (following from the DMPK equations),
Nieuwenhuizen and van Rossum \cite{nvr} calculated the distribution
functions $P(T_a)$ and $P(T_{ab})$ for $g\gg 1$. The results have the
following form:
\bea
&& P(s_a)=\int_{-i\infty}^\infty{dx\over 2\pi i} \exp[xs_a-\Phi_{con}(x)]
\ ;\label{e7.2}\\ 
&& P(s_{ab})=\int_0^\infty{dv\over v}\int_{-i\infty}^\infty{dx\over
2\pi i} \exp\left[-{s_{ab}\over v}+xv-\Phi_{con}(x)\right]\ ;
\label{e7.3}\\
&&\Phi_{con}(x)=g\ln^2\left(\sqrt{1+x/g}+\sqrt{x/g}\right)\ ,
\label{e7.4}
\eea
where $s_a=T_a/\langle T_a\rangle$ and 
$s_{ab}=T_{ab}/\langle T_{ab}\rangle$. For large $g\gg 1$ (i.e. in the
``metallic'' regime), the distribution $P(s_a)$, Eq.~(\ref{e7.2}), has
a Gaussian form,
\be
P(s_a)\simeq\sqrt{{3g\over 4\pi}}\exp\left[-{3g\over
4}(s_a-1)^2\right]\ ,  
\label{e7.5}
\ee
with log-normal tails at small $s_a\ll 1$ and an exponential asymptotics
at large $s_a\gg 1$ of the form $P(s_a)\sim\exp(-gs_a)$. The distribution
 of the channel-to-channel transmission coefficients $s_{ab}$
is a close relative of the distribution of the point-to-point
transmitted intensity $I({\bf r},{\bf r_0})$ which was discussed
above and will be studied in detail below. For not too large $s_{ab}$
it has the Rayleigh form, $P(s_{ab})\sim e^{-s_{ab}}$, with the
leading perturbative correction given by
\be
P(s_{ab})\simeq e^{-s_{ab}}\left[1+{1\over
3g}(s_{ab}^2-4s_{ab}+2)\right]\ ,\qquad s_{ab}\ll\sqrt{g}
\label{e7.6}
\ee
(this form of the correction to the Rayleigh law for the intensity
distribution was found in a number of papers, see
Refs.~\cite{zavorotnyi,gochelashvili,dashen79,shnerb}). In the intermediate
region $\sqrt{g}<s_{ab}<g$ Eq.(\ref{e7.3}) yields
\be
\label{e7.7}
P(s_{ab})\simeq \exp\left[-s_{ab}\left(1+{1\over
3g}s_{ab}+\ldots\right)\right] 
\ee
(correction of this type was also obtained earlier,
Refs.~\cite{gochelashvili,dashen84}). 
For large $s_{ab}>g$ the distribution acquires a stretched-exponential
form
\be
\label{e7.8}
P(s_{ab})\sim\exp\left(-2\sqrt{gs_{ab}}\right).
\ee
We will return to Eq.~(\ref{e7.8}) below, when comparing the results
for the distribution of $I({\bf r_0},{\bf r})$ with $P(s_{ab})$. 
In the localized regime both distributions $P(s_a)$ and $P(s_{ab})$
are determined by the single (largest) transmission eigenvalue and
acquire the same log-normal form as the conductance distribution, 
Eq.~(\ref{e6.8}). We note also that van Langen, Brouwer and Beenakker
\cite{been1} were able to calculate $P(s_a)$ and $P(s_{ab})$ in the case of the
unitary symmetry, $\beta=2$, in the whole range of the parameter
$L/\xi$ (from weak to strong localization). The distributions
$P(s_{a})$ and $P(s_{ab})$ were studied experimentally by Garcia and
Genack \cite{garcia91,genack93} and by Stoychev and Genack
\cite{stoychev97}; their findings are in good agreement with the
theoretical results. 

Now we return to the statistics of the intensity $I(\r,\ro)$. 
The field at the  point
$\r$ is given by the (retarded) Green's function
$G_R (\r,\ro)$ and the radiation intensity
is $I(\r,\ro)= |G_R (\r,\ro)|^2$.
The average intensity $\av{I(\r,\ro)}$
is given diagrammatically by a diffuson 
$T({\bf{r_1}},{\bf{r_2}})$, attached to two external
vertices. The vertices are short-range objects and can 
be approximated by a $\delta$-function times 
$(\ell/4\pi )$, so that
$\av{I(\r,\ro)}=
(\ell/4\pi )^2 T({\bf{r}},\ro)$.
For the quasi one-dimensional geometry, the expression for  the
diffuson reads 
\be
T({\bf r},{\bf r_o})=\left(\frac{4\pi}{\ell}\right)^2
\frac{3}{4\pi}\frac{\left[z_{<}(L-z_{>})\right]}
{A\ell L}
%z_{<}(L-z_{>})
\label{mps_diffuson}
\ee
where $\ell$ is the elastic mean free path, $A$
is the cross-section of the tube, 
$z$-axis is directed  along the sample,
$z_{<}=\mbox{min} (z,z_o)$ 
and $z_{>}=\mbox{max} (z,z_o)$. We assume that 
$|z-z_0|\gg W$. 

The intensity distribution $P(I)$ is obtained, in the diagrammatic
approach, by calculating the moments
$\av{I^n}$ of the intensity. In the leading approximation
\cite{shapiro86}, one should draw $n$ retarded and $n$ advanced
Green's functions and insert ladders between pairs
$\left\{G_R,G_A\right\}$ in all possible ways. This leads
to $\av{I^n}=n!\av{I}^n$ and, thus, to Eq.~(\ref{mps_rayleigh}).

\begin{figure}
\centerline{\epsfig{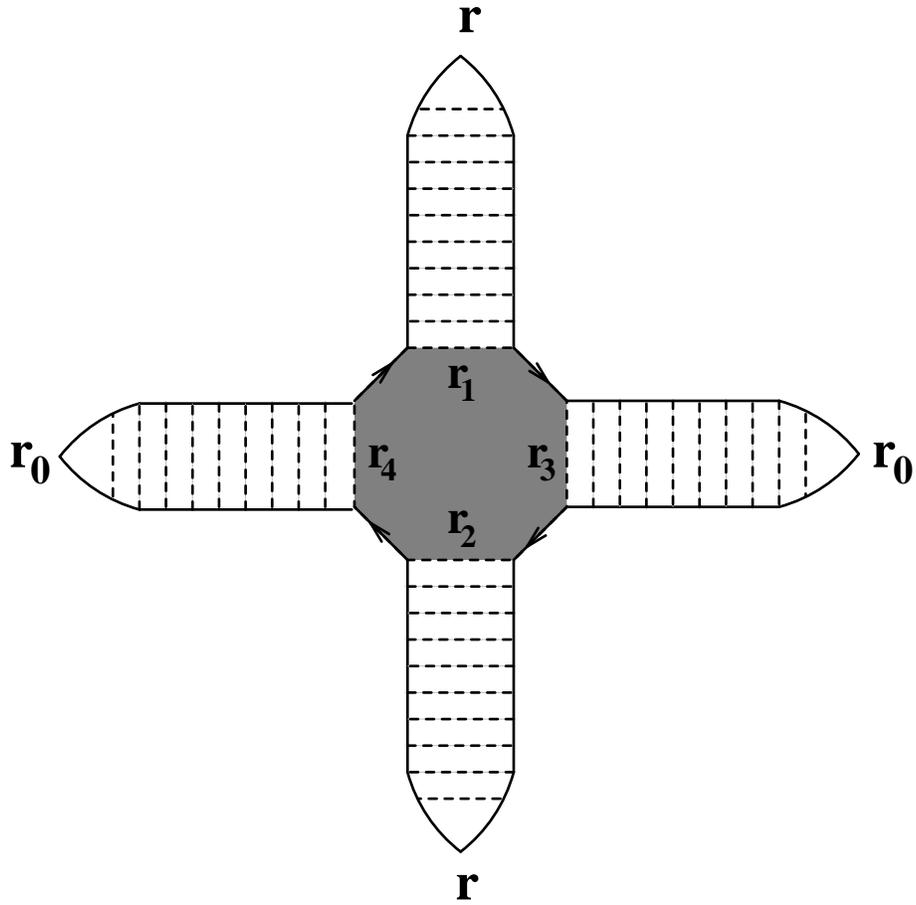}}
%\centerline{\epsfxsize=100mm\epsfbox{hbox1.eps}}  
\vspace{3mm}
\caption{Diagram for a pair of interacting diffusons. The external
vertices contribute the factor $(\ell/4\pi)^4$. The shaded region
denotes the internal interaction vertex, see Eq.~(\ref{mps_C2}). 
From \cite{mps}.}
\label{hbox} 
\end{figure}

Corrections to the Rayleigh result come from diagrams with
intersecting ladders, which describe interaction between
diffusons. The leading correction is due to
pairwise interactions. The diagram in Fig.~\ref{hbox} 
represents a pair of ``colliding'' diffusons. The algebraic 
expression for this diagram is 
\bea
 C(\r,\ro) &=& 2 \left(\frac{\ell}{4\pi}\right)^4
\int\left(\prod_{\imath=1}^4 d^3{\bf{r_\imath}}\right) \nonumber\\  
&\times & T(\r,{\bf{r_1}}) T(\r,{\bf{r_2}})
T({\bf{r_3}},\ro) T({\bf{r_4}},\ro)  \nonumber\\
&\times & \left\{
\left( \frac{\ell^5}{48\pi k^2_o} \right)
\int d^3{\bf\rho} \left[ 
(\nabla_1 +\nabla_2)\cdot (\nabla_3 +\nabla_4)\right. \right.\nonumber
\\ 
& + & 2 \left. \left.
(\nabla_1\cdot\nabla_2) + 2 (\nabla_3 \cdot\nabla_4)\right]
\prod_{\imath=1}^4 \delta({\bf\rho}-{\bf{r_\imath}})
\right\} \ ,
\label{mps_C2}
\eea
where  $k_o$ is the wave number and 
$\nabla_{\imath}$ acts on ${\bf{r_\imath}}$.
The factor $\left(\ell/4\pi\right)^4$ comes from the 
4 external vertices of the diagram, the $T$'s represent
the two incoming and two outgoing diffusons and
the expression in the curly brackets corresponds to 
the internal (interaction) vertex (``Hikami box'')
\cite{hikami81,glk80}.
Finally, the factor 2 accounts for the two possibilities
of inserting a pair of ladders between the outgoing
Green's functions. Integrating by parts and employing
the quasi one-dimensional geometry of the problem, we obtain
(for $z_o<z$):
\be
 C(z,z_o) \simeq 2 \av{I(z,z_o)}^2
\left( 1 + \frac{4}{3\gamma}\right) \ ,
\label{mps_Cgamma}
\ee
where $\av{I(z,z_o)}={3\over 4\pi}{z_0(L-z)\over A\ell L} $ is the average
intensity, 
\be
\gamma = 2g {L^3\over L^2(3z+z_o) - 2Lz(z+z_o) + 2z^2_o (z-z_o)} \gg 1
\ ,
\label{mps_gamma}
\ee
and $g=k^2_o\ell A/3\pi L\gg 1$ is the dimensionless
conductance of the tube.
For simplicity, we will assume that the source and the detector are
located relatively close to each other, so that  $|z-z_o|\ll L$, in
which case Eq.~(\ref{mps_gamma}) reduces to $\gamma= gL^2/2z(L-z)$. (All the
results are found to be qualitatively the same in the generic
situation $z_0\sim z-z_0 \sim L-z\sim L$.)

In order to calculate $\av{I^n}$ one has to
compute a combinatorial factor which counts
the number $N_i$ of diagrams with $i$ pairs of
interacting diffusons. This number is 
$N_i=(n!)^2/[{2^{2i}}{i!}{(n-2i)!}]\simeq
(n!/i!)(n/2)^{2i}$.
For not too large $n\ll\gamma^{1/2}$ it is sufficient to keep the
$i=1$ contribution, yielding the leading perturbative correction \cite{mps},
\be
\label{e7.9}
\langle \tilde{I}^n\rangle=n!\left[1+{2\over 3\gamma}n(n-1)\right]\ ,
\ee
or in terms of the distribution function,
\be
\label{e7.10}
P(\tilde{I})=e^{-\tilde{I}}\left[1+{2\over
3\gamma}(\tilde{I}^2-4\tilde{I}+2)\right]\ ,\qquad \tilde{I}\ll\gamma^{1/2}.
\ee
For larger $n$ (or, equivalently, $\tilde{I}$) we have to sum up the
series over $i$, yielding 
\be
\frac{\av{I^n}}{\av{I}^n}=
n!\sum_{\imath=0}^{[n/2]} \frac{1}{i!}
\left(\frac{2n^2}{3\gamma}\right)^i
\simeq n!\exp(2n^2/3\gamma)  \  .
\label{mps_moments2}
\ee
Although $i$ cannot exceed $n/2$, the sum in
Eq.~(\ref{mps_moments2}) can be extended to $\infty$, if the
value of $n$ is restricted by the condition $n\ll\gamma$. 
Eq.~(\ref{mps_moments2}) represents the leading exponential correction to
the Rayleigh distribution. Let us discuss now effect of higher order
``interactions'' of diffusons. 
Diagrams with 3 intersecting diffusons will contribute
a correction of $n^3/\gamma^2$ in the exponent of
Eq.~(\ref{mps_moments2}), which is small compared to the leading correction
in the whole region $n\ll\gamma$, but becomes larger than unity 
for $n\gtrsim\gamma^{2/3}$. Likewise, diagrams with 4 intersecting
diffusons produce a $n^4/\gamma^3$ correction, etc.
Restoring the distribution $P(I)$, we find \cite{mps}
\be
P(\tilde{I})\simeq\exp\left\{-\tilde{I}+\frac{2}{3\gamma}\tilde{I}^2
+O\left({\tilde{I}^3\over\gamma^2}\right)+\ldots\right\} \ ,
\label{mps_intermediate}
\ee
Eqs.~(\ref{e7.9}), (\ref{e7.10}), (\ref{mps_intermediate}) are
analogous to those found in the case of transmission coefficient
statistics, (\ref{e7.6}), (\ref{e7.7}), with the only difference that
the parameter $g$ is now replaced by $\gamma/2$ (if we consider the
limit $z_0\to 0$, $z\to L$, then $\gamma\to 2g$, so that both results
are consistent). As has been already mentioned, deviations of this form
from the Rayleigh distribution of intensities were found earlier by
various authors
\cite{zavorotnyi,gochelashvili,dashen79,dashen84,shnerb}; a value of the
parameter governing the strength of deviations (here $1/\gamma$)
depends, however, on the geometry of the problem. 
It should be realized that Eq.~(\ref{mps_intermediate})
is applicable only for $\tilde{I}\ll \gamma\sim g$ and, thus, does not
determine the far asymptotics of $P(I)$. The latter is
unaccessible by the perturbative diagram technique and 
is handled below by the supersymmetry method.

 For technical simplicity, we will
assume now that the time reversal symmetry is broken by some
magnetooptical effects (unitary ensemble).
The moments of the 
intensity at point $\r$ due to the source at
$\ro$ are given by
\be
\av{I^n}= \left(-\frac{k_0^2}{16\pi^2}\right)^n 
\int [DQ] 
\left(Q_{12,bb} (z)\right)^n
\left(Q_{21,bb} (z_0)\right)^n
e^{-S[Q]},
\label{mps_superI}
\ee
 where $S[Q]$ is the zero-frequency \mbox{$\sigma$-model}
action, 
\be
S[Q]=-\frac{\pi\nu D}{4}\int d^3\r\, {\rm Str}
(\nabla Q)^2\ , 
\label{mps_sigma}
\ee 
which reduces in the quasi-1D geometry to
$$
S[Q]=-(gL/8) \int dz\, {\rm Str}(dQ/dz)^2.
$$
Assuming again that the two points
$\r$ and $\ro$ are sufficiently close to each other, $|z-z_0|\ll L$
 and taking into account slow variation of the $Q$-field
along the sample, we can replace the product 
$Q_{12,bb}(z) Q_{21,bb}(z_0)$ 
by 
$Q_{12,bb}(z)Q_{21,bb}(z)$. 
We get then the following result for the distribution of the
dimensionless intensity $y= (16\pi^2/k_0^2)I$:
\be
P(y)=\int dQ \delta( y+ Q_{12,bb} Q_{21,bb})Y(Q) \ ,
\label{mps_dist1}
\ee
where $Y(Q)$ is a function of a single supermatrix $Q$ defined 
by Eq.~(\ref{anom_1}). 
Using the fact that the function $Y(Q)$ depends only on the 
eigenvalues $1\le\lambda_1<\infty$, 
$-1\le\lambda_2\le 1$, we find
\bea
P(y)&=&\left(\frac{d}{dy} +y\frac{d^2}{dy^2}\right)
\int d\lambda_1 d\lambda_2 \nonumber \\
&\times& \left(\frac{\lambda_1 +\lambda_2}{\lambda_1 -\lambda_2}
\right) Y(\lambda_1 ,\lambda_2 )
\delta(y + 1-\lambda^2_1) \ .
\label{mps_dist2}
\eea
The function $Y(\lambda_1,\lambda_2)$ can be evaluated at $g\gg 1$ via
the saddle-point method (see \cite{m96} and Section~\ref{s4.3}) with the
result
\be
Y(\lambda_1 ,\lambda_2) 
\simeq\exp\left\{-\frac{\gamma}{2}
\left[\theta^{2}_1 +\theta^{2}_2 \right]
\right\} \ .
\label{mps_SP}
\ee
where $\lambda_1\equiv\cosh\theta_1$, $\lambda_2\equiv\cos\theta_2$
($0\leq\theta_1<\infty$, $0\leq\theta_2\leq\pi$).
In fact, the dependence of $Y$ on $\theta_2$ is
not important, within the exponential accuracy,
because it simply gives a prefactor after the
integration in Eq.~(\ref{mps_dist2}). Therefore, 
the distribution function $P(y)$ 
is given by 
\be
P(y)\sim Y(\lambda_1=\sqrt{1+y} ,\lambda_2=1)
\sim
\exp\left(-\gamma\theta^{2}_1 /2 \right) \ , 
\label{mps_dist3}
\ee 
where $\theta_1 =\ln (\sqrt{1+y}+\sqrt{y})$.
Finally, after normalizing $y$ to its average value
$\av{y}=2/\gamma$, we obtain \cite{mps}:
\be
P(\tilde{I})\simeq\exp\left\{-\frac{\gamma}{2} 
\left[\ln^2
\left(\sqrt{1+2\tilde{I}/\gamma} +\sqrt{2\tilde{I}/\gamma}\right)
\right]\right\} \ .
\label{mps_final}
\ee
For $\tilde{I}\ll \gamma$, Eq.~(\ref{mps_final}) reproduces the
perturbative expansion (\ref{mps_intermediate}), while for
$\tilde{I}\gg\gamma$ it implies the log-normal asymptotic behavior
of the distribution $P(\tilde{I})$:
\be
\ln P(\tilde{I}) \simeq -(\gamma/8)\ln^2(8\tilde{I}/\gamma)\ .
\label{mps_LN}
\ee

 The log-normal ``tail'' (\ref{mps_LN}) should be contrasted
with the stretched-exponential  asymptotic behavior of the distribution
of transmission coefficients, Eq.~(\ref{e7.8}). As was found in
\cite{mps}, these two results match each other in the following way. 
When the points $z$ and $z_0$ approach the sample edges, $z_0=L-z\ll
L$, an intermediate regime of a stretched-exponential behavior emerges:
\begin{equation}
\ln P(\tilde{I})\simeq\left\{
\begin{array}{ll}
-\tilde{I}+{1\over 3g}\tilde{I}^2+\ldots\ , & \ \ \tilde{I}\ll g \\
-2\sqrt{g\tilde{I}}\ , & \ \ g\ll \tilde{I}\ll g\left({L\over
z_0}\right)^2\\ 
-{gL\over 8z_0}\ln^2\left[16\left({z_0\over L}\right)^2
{\tilde{I}\over g}\right]\ , & \ \
\tilde{I}\gg g\left({L\over z_0}\right)^2 \ .
\label{mps_strexp}
\end{array} \right.
\end{equation}
Thus, when the source and the detector move toward the sample edges,
the range of validity of the stretched-exponential behavior becomes
broader, while the log-normal ``tail'' gets pushed further away. In
contrast, when the source and the detector are located deep in the
bulk, $z_0\sim L-z\sim L$, the stretched-exponential regime
disappears, and the Rayleigh distribution crosses over directly to the
log-normal one at $\tilde{I}\sim g$. 

Let us now describe the physical mechanisms standing behind these
different forms of $P(\tilde{I})$ \cite{mps}. The Green's function
$G^R({\bf r_0},{\bf r})$ can be expanded in eigenfunctions of a
non-Hermitean (due to open boundaries) ``Hamiltonian'' as 
$$G^R({\bf r_0},{\bf r})=\sum_i\psi_i^*({\bf r_0})\psi_i({\bf r})
(k_0^2-E_i+i\gamma_i)^{-1}.$$
 Since the level widths $\gamma_i$ are
typically of order of the Thouless energy $E_c\sim D/L^2$, there is
typically $\sim g$ levels contributing appreciably to the sum. 
In view of the random phases of the wave functions, this  
leads to a Gaussian
distribution of $G^R({\bf r_0},{\bf r})$ with zero mean, and thus
to the Rayleigh distribution of $I({\bf r_0},{\bf r})=|G^R({\bf
r_0},{\bf r})|^2$, with the moments $\langle\tilde{I}^n\rangle=n!$. 
The stretched-exponential behavior results from such disorder
realizations where one of the states $\psi_i$ has large amplitudes in
the both points ${\bf r_0}$ and ${\bf r}$. Considering both 
$\psi_i({\bf r_0})$ and $\psi_i({\bf r})$ as independent random
variables with Gaussian distribution and taking into account
that only one (out of $g$) term contributes in this case
to the sum for $G^R$, we find
$\langle\tilde{I}^n\rangle\sim n!n!/g^n$, corresponding to the above
stretched-exponential form of $P(\tilde{I})$. Finally, the log-normal
asymptotic behavior corresponds to those disorder realizations 
where $G^R$ is dominated by an anomalously localized state, which has
an atypically small width $\gamma_i$ (the same mechanism determines
the log-normal asymptotics of the distribution of local density of
states, see Section~\ref{s4.3}).

\section{Statistics of energy levels and eigenfunctions in a ballistic
system with surface scattering}
\label{c8}

\setcounter{equation}{0}

In the preceding part of this article we considered 
statistical properties of spectra of disordered diffusive systems.
Using the supersymmetric $\sigma$-model approach, we were able
 to demonstrate the relevance of the random matrix
theory (RMT) and to calculate deviations from its predictions both for
the level and eigenfunction 
statistics.  Generalization of these results to the case of a chaotic
{\it ballistic} system (i.e. quantum billiard) has become a topic of
great research interest.  For ballistic disordered systems the
$\sigma$-model has been proposed \cite{mk3}, with the Liouville operator
replacing the diffusion operator in the action.  It has also been
conjectured that the same $\sigma$-model in the limit of vanishing
disorder describes statistical properties of spectra of individual
classically chaotic system.  This conjecture was further developed in
\cite{agam,aasa1,aasa2} 
where the $\sigma$-model was obtained by means of
energy averaging, and the Liouville operator was replaced by its
regularization --- the Perron-Frobenius operator.  

However, straightforward application of the results of
Refs.~\cite{km,aa,fm95a,bm97} to the case of an individual chaotic system is
complicated by the fact that the eigenvalues of the Perron-Frobenius
operator are unknown, while its eigenfunctions are extremely singular.
For this reason the $\sigma$-model approach has so far failed to provide
explicit results for any particular ballistic system.
  
In this section, we consider a ballistic system with
surface disorder leading to diffusive scattering of a particle in each
collision with the boundary. This models behavior of a quantum particle
in a box with a rough boundary which is irregular on the scale of the
wave length.  Since the particle loses memory of its direction of motion
after a single collision, this model describes a limit of an ``extremely
chaotic'' ballistic system, with typical relaxation time being of
order of the flight time. (This should be contrasted with the case of a
relatively slight distortion of an integrable billiard
\cite{borgonovoi,frahm97a,frahm97b}.) 
One might naively think that all results for such a model could be
obtained by setting $l\approx L$ in a system with bulk disorder. In fact,
the level statistics in a system with bulk disorder and arbitrary
relation between mean free path $l$ and system size $L$ was studied in
\cite{algef93,algef95,agfish96}.  However, the results presented below 
show that systems with bulk and surface disorder are
not equivalent. 

To simplify the calculations, 
we will assume a circular geometry of the billiard.  A similar
problem was studied numerically in Ref. \cite{louis96,louis97} for a square
geometry. We consider only the case of unitary symmetry (broken
time-reversal invariance); generalization to the orthogonal case is
straightforward. We follow Ref.~\cite{bmm2} in the presentation below.
The level statistics for the same problem
was independently studied in Ref. \cite{DEK}.  Very recently, the same
approach was used \cite{samokhin99} to calculate the persistent
current in a ring with diffusive scattering.

Our starting point is the
sigma-model for ballistic disordered systems \cite{mk3,mk2}. The effective
action for this model has the form 
\begin{eqnarray} 
\label{surf_model1}
F[g({\bf r}, {\bf n})] =  \frac{\pi\nu}{4} \int d{\bf r} {\rm Str}
\left[ i\omega \Lambda \langle g({\bf r}) \rangle -
\frac{1}{2\tau({\bf r})} 
\langle g({\bf r}) \rangle^2 \right. 
 -  \left. 2v_F \langle \Lambda U^{-1} {\bf n} \nabla U
\rangle \right].  
\end{eqnarray}
Here a $8\times 8$ supermatrix field $g$ is defined on the energy
shell of the phase space, i.e. $g$  depends on the
coordinate ${\bf r}$ and direction of the momentum ${\bf n}$.
The momentum dependence of the field $g$ distinguishes the ballistic
$\sigma$-model from the diffusive case where the supermatrix field $Q$
depends on ${\bf r}$ only.  The
angular brackets denote averaging over ${\bf n}$: $\langle {\cal O}
({\bf n}) \rangle = \int d{\bf n} {\cal O}({\bf n})$ with the
normalization $\int d{\bf n} = 1$. Like in the case of the diffusive
$\sigma$-model, the matrix $g$ is constrained by
the condition $g^2({\bf r}, {\bf n}) = 1$, and can be represented as
$g = U \Lambda U^{-1}$, with $\Lambda = {\rm diag}
(1,1,1,1,-1,-1,-1,-1)$.  Since we are interested in the clean limit with no
disorder in the bulk, the second term in the action (\ref{surf_model1})
containing the elastic mean free time $\tau$ is zero everywhere except
at the boundary where it modifies the boundary condition (see below).

As was explained in Section \ref{c2},
the statistical properties of energy levels  
are governed by the structure of the action
in the vicinity of the homogeneous configuration of the $g$-field,
$g({\bf r}, {\bf n})=\Lambda$. Writing $U = 1 - W/2+\ldots$, we find
the action in the leading order in $W$,
\begin{equation} 
\label{surf_model2}
  F_0[W] = -\frac{\pi\nu}{4} \int d{\bf r} d{\bf n} {\rm Str}\, \left[
    W_{21} \left( \hat K - i\omega \right) W_{12} \right],
\end{equation}
where the indices $1,2$ refer to the ``advanced-retarded'' decomposition
of $W$, and $\hat K$ is the Liouville operator, 
$\hat K \equiv v_F {\bf n} \nabla$.
This ``linearized'' action has the same form as that of a diffusive
system, with the diffusion operator being replaced by the Liouville
operator. This enables us to use the results derived for the diffusive
case by substituting the eigenvalues and eigenfunctions of the operator
$\hat K$ for those of the diffusion operator.

The operator $\hat K$ should be supplemented by a boundary condition,
which depends on the form of the surface roughness.  As a model
approximation we consider purely diffuse scattering
\cite{fuchs,abrikos} for 
which the distribution function $\varphi({\bf r},{\bf n})$ of the
outgoing particles is constant and is fixed by flux
conservation\footnote{Exact form of the boundary condition depends on
the underlying microscopic model. In particular, the diffuse
scattering can be modelled by surrounding the cavity by a 
disordered layer with a bulk mean free path $l$ and a thickness
$d\gg l$. The corresponding boundary condition \cite{chandra,morfesh}
differs from Eq.~(\ref{e8.1}) by a parameterless function of order
unity. For a review of the boundary conditions
corresponding to various microscopic realizations of the rough surface
see \cite{okulov79}.}:
\begin{equation}
\label{e8.1}
\varphi({\bf r}, {\bf n}) = \pi \int_{({\bf N}{\bf n'}) > 0}
\left( {\bf N} {\bf n'} \right) \varphi ({\bf r}, {\bf n'})
d{\bf n'}, \ \ \ \left ({\bf N} {\bf n} \right) < 0.
\end{equation}
Here the point ${\bf r}$ lies at the surface, and ${\bf N}$ is an
outward normal to the surface. This boundary condition should be
satisfied by the eigenfunctions of $\hat K$.

The eigenvalues $\lambda$ of the operator $\hat K$ corresponding to 
angular momentum $l$ obey the equation
\begin{equation} 
\label{surf_values}
\tilde J_l(\xi) \equiv -1 + \frac{1}{2} \int_0^{\pi} d\theta \sin\theta
\exp \left[ 2il\theta + 2 \xi \sin\theta \right]  = 0,
\end{equation}
where $\xi \equiv R\lambda/v_F$, and $R$ is the radius of the circle. For
each value of $l=0,\pm1,\pm2,\ldots$ Eq.(\ref{surf_values}) has a set of
solutions $\xi_{lk}$ with $\xi_{lk}=\xi_{-l,k}=\xi^*_{l,-k}$, which
can be labeled with $k=0,\pm1,\pm2,\ldots$ (even $l$) or
$k=\pm1/2,\pm3/2,\ldots$ (odd $l$). For $l=k=0$ we have $\xi_{00}=0$,
corresponding to the zero mode $\varphi ({\bf r},
{\bf n})=\mbox{const}$. All other eigenvalues have positive real part
$\mbox{Re}\,\xi_{lk} > 0$ and govern the relaxation of the
corresponding classical system to the homogeneous distribution in the
phase space.  

\begin{figure}
\centerline{\epsfxsize=120mm\epsfbox{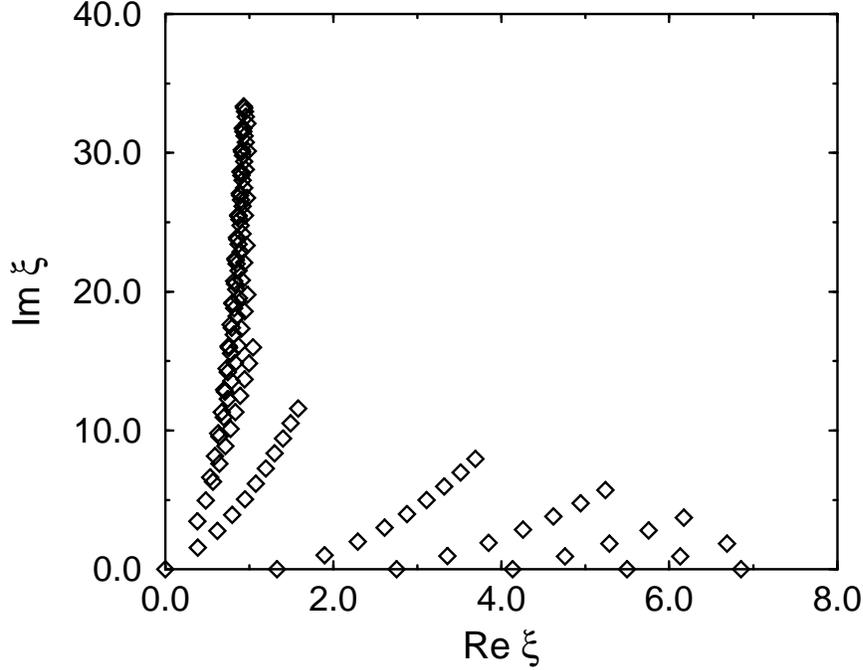}}
\caption{First $11\times 11$  ($0 \le k,l < 11$)
eigenvalues of the Liouville operator $\hat K$ in
units of $v_F/R$, as given by Eq. (\protect\ref{surf_values}). From
Ref.~\cite{bmm2}.}  
\label{perron}
\end{figure}

The asymptotic form of the solutions of Eq.~(\ref{surf_values}) for large
$\vert k \vert$ and/or $\vert l \vert$ can be obtained by using the
saddle-point method,
\begin{eqnarray} 
\label{surf_valas}
\xi_{kl} \approx \left\{ \begin{array}{lr} 0.66 l + 0.14 \ln l + 0.55 \pi i
k, & 0 \le k \ll l \\ 
(\ln k)/4 + \pi i (k+1/8), & 0 \le  l \ll k
\end{array} \right. .
\end{eqnarray}
Note that for $k=0$ all eigenvalues are real, while for high values of
$k$ they lie close to the imaginary axis and do not depend on $l$ (see
Fig.~\ref{perron}).

\paragraph{Level statistics, low frequencies.} 

As was explained in Section~\ref{s2.2} [see Eq.~(\ref{e2.16})], 
in the range of relatively low frequencies (which for
our problem means $\omega \ll v_F/R$, see below) the level correlation
function  $R(s=\omega/\Delta)$  has the form 
\begin{equation} 
\label{surf_lowen}
R(s)-1 = \delta(s) - {\sin^2 \pi s\over (\pi s)^2} +
{\cal A} \left ({R\Delta\over \pi v_F}\right)^2 \sin^2\pi s.
\end{equation}
The first two terms correspond to the
zero-mode approximation and are given by RMT, while the last one
is the non-universal correction to the RMT results. The information
about the operator $\hat K$ enters through the dimensionless constant
${\cal A}=\sum'\xi_{kl}^{-2}$, where the prime indicates that the eigenvalue
$\xi_{00}=0$ is excluded.  The value of ${\cal A}$, as well as the
high-frequency behavior of $R(s)$ (see below), can be extracted from
the spectral function \cite{ashk} 
\begin{equation} 
\label{surf_sum1}
S (\omega) = \sum_{l} S_l (\omega);\ \ \ S_l(\omega) \equiv \sum_k 
\left( \lambda_{kl} - i\omega \right)^{-2}.
\end{equation}
According to the Cauchy theorem, $S_l$ can be
represented as an integral in the complex plane,
$$S_l (\omega) = \left( \frac{R}{v_F} \right)^2 \frac{1}{2\pi i} \oint_C
\frac{1}{(z - i\omega R/v_F)^2} \frac{\tilde J_l' (z)}{\tilde J_l (z)}
dz,$$
where the contour $C$ encloses all zeroes of the function $\tilde
J_l (z)$. Evaluating the residue at $z=i\omega R/v_F$, we find
\begin{equation}
\label{surf_sl}
S_l (\omega) = - (R/v_F)^2 \left. \frac{d^2}{d z^2} \right\vert_{z =
  i\omega R/v_F} \ln \tilde J_l (z) .
\end{equation}
Considering the limit $\omega \to 0$ and subtracting the contribution
of $\lambda_{00}=0$, we get 
\begin{equation} 
\label{surf_s0}
  {\cal A} = -19/27 - 175 \pi^2/1152 + 64/(9\pi^2) \approx -1.48.
\end{equation}
In contrast to the diffusive case, this constant is negative: the level
repulsion is enhanced with respect to result for RMT.  
Eq.(\ref{surf_lowen})
is valid as long as the correction is small compared to the RMT result,
i.e. provided $\omega$ is below the inverse time of flight, $v_F/R$.

\paragraph{Level statistics, high frequencies.}  

In the range $\omega\gg\Delta$
the connected part of the level correlation function 
$R(\omega)-1$ can be decomposed into the smooth
Altshuler-Shklovskii (AS) part $R_{\rm AS} (\omega) =
(\Delta^2/ 2\pi^2) {\rm Re}\,S(\omega)$ \cite{ashk} and the part
$R_{\rm osc}$ which oscillates on the scale of the level
spacing (see Sec.~\ref{s2.2}).  
Evaluating the asymptotic behavior of $S_l(\omega)$ from
Eq.(\ref{surf_sl}), we find in the high-frequency regime when $\omega \gg
v_F/R$:  
\begin{equation} 
\label{surf_highen0}
R_{\rm AS} (\omega) = 
\left( \frac{\Delta R}{v_F} \right)^2 \left(
\frac{v_F}{2\pi\omega R} \right)^{1/2} \cos \left( 4 \frac{\omega R}{v_F} -
\frac{\pi}{4} \right).
\end{equation} 

The oscillating part of the level correlation function
$R_{\rm osc}(s)$ for frequencies $\omega \gg
\Delta$ is given by Eq.~(\ref{e2.23}) with the spectral determinant
$$D(s) = s^{-2}\prod_{kl\ne(00)}  (1- is\Delta/\lambda_{kl} )^{-1} 
(1 + is\Delta/\lambda_{kl} )^{-1}.$$ 
Since $\Delta^{-2}\partial^2\ln D(s)/\partial
s^2=-2\mbox{Re}\,S(\omega)$, we can restore $D(s)$ from
Eqs.(\ref{surf_sum1}), (\ref{surf_sl}) up to a factor of the form $\exp(c_1 +
c_2 s)$, with $c_1$ and $c_2$ being arbitrary constants. These
constants are fixed by the requirement that Eq.(\ref{e2.23}) in the
range $\Delta\ll\omega\ll v_F/R$ should reproduce the low-frequency
behavior (\ref{surf_lowen}).  As a result, we obtain
\begin{equation} 
\label{spdet1}
D(s) = \left( \frac{\pi}{2} \right)^6 \frac{1}{N} \prod_l
\frac{1}{\tilde J_l (i s N^{-1/2}) \tilde J_l (-i s N^{-1/2})}.
\end{equation}
Here $N =(v_F/R\Delta)^2 = (p_F R/2)^2$ is the number of electrons
below the Fermi level. For high frequencies $\omega \gg v_F/R$ this
yields the following expression for the oscillating part of the level
correlation function: 
\begin{equation} 
\label{surf_highen2}
R_{\rm osc} (\omega) = \frac{\pi^4}{128} 
\left( \frac{\Delta R}{v_F} \right)^2
\cos \left( \frac{2\pi \omega}{\Delta} \right).
\end{equation} 
It is remarkable that the amplitude of the oscillating part does not
depend on frequency. This is in contrast to the diffusive case, where
in the AS regime ($\omega$ above the Thouless energy) the oscillating
part $R_{\rm osc} (\omega)$ is exponentially small, see Eq.~(\ref{e2.26}).

\paragraph{The level number variance.} 

The smooth part of the level
correlation function can be best illustrated by plotting the variance
of the number of levels in an energy interval of width $E = s\Delta$, 
\begin{equation} 
\label{surf_lnv}
\Sigma_2 (s) = \int_{-s}^s \left(s - \vert \tilde{s} \vert
\right) R(\tilde{s}) d\tilde{s},  
\end{equation}
A direct calculation gives for $ s \ll N^{1/2}$
\begin{equation} 
  \pi^2 \Sigma_2 (s) 
  =  1 + \gamma + \ln (2\pi s) +
    {\cal A} s^2/(2N) 
\label{surf_lnv1}
\end{equation}
and for $s \gg N^{1/2}$
\begin{eqnarray}
  \pi^2 \Sigma_2 (s) &=& 1 + \gamma + \ln \frac{16 N^{1/2}}{\pi^2}
  \nonumber \\ 
  &&  - \frac{\pi^2}{16}  \left( \frac{2N^{1/2}}{\pi s}
  \right)^{1/2} \cos \left( \frac{4 s}{N^{1/2}} - \frac{\pi}{4}
  \right). 
\label{surf_lnv2}
\end{eqnarray}
Here $\gamma \approx 0.577$ is Euler's constant, and ${\cal A}$ is defined by
Eq.~(\ref{surf_s0}). The first three terms at the rhs of
Eq.~(\ref{surf_lnv1}) 
represent the RMT contribution (curve 1 in Fig.~\ref{surf}).

As seen from Fig.~\ref{surf}, the two 
asymptotics (\ref{surf_lnv1}) and (\ref{surf_lnv2})
perfectly match in the intermediate regime, $s\sim N^{1/2}$. Taken
together, they provide a complete description of $\Sigma_2(s)$.
According to Eq.(\ref{surf_lnv2}), the level number variance saturates at
the value $\Sigma_2^{(0)} = \pi^{-2} (1 + \gamma + \ln
(16N^{1/2}/\pi^2))$, in contrast to the behavior found for diffusive
systems \cite{ashk} or ballistic systems with weak bulk disorder
\cite{algef93,agfish96}. The saturation occurs at energies $s \sim
N^{1/2}$, or in 
conventional units $E \sim v_F/R$. This saturation of $\Sigma_2 (s)$,
as well as its oscillations on the scale set by short periodic orbits,
is expected for a generic chaotic billiard \cite{berry1,berry2}. It is
also in 
good agreement with the results for $\Sigma_2 (s)$ found numerically
for a tight-binding model with moderately strong disorder on boundary
sites \cite{louis96,louis97}. 

\begin{figure}
{\epsfxsize=120mm\centerline{\epsfbox{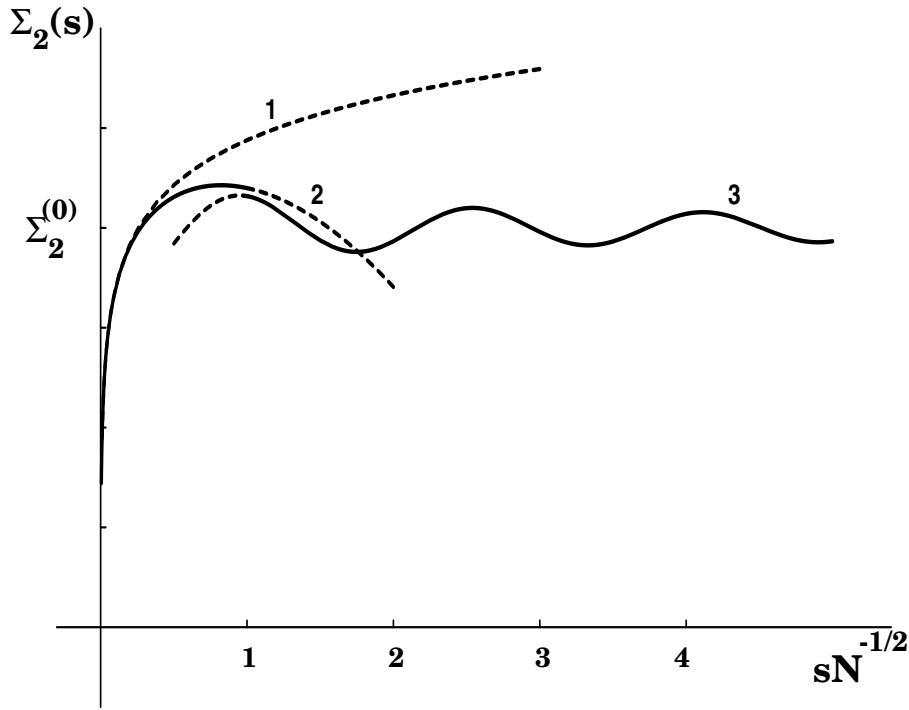}}}
\caption{Level number variance $\Sigma_2 (E)$ as a function of
energy; $s = E/\Delta$. Curve 1 shows the RMT result, while curves 2
and 3 correspond to asymptotic regimes of low (\protect\ref{surf_lnv1}) and
high (\protect\ref{surf_lnv2}) frequencies. The saturation value
$\Sigma_2^{(0)}$ is given in the text. From Ref.~\cite{bmm2}.} 
\label{surf}
\end{figure}

\paragraph {Eigenfunction statistics.} 

Now we turn to correlations of the
amplitudes of an eigenfunction in two different points defined by
Eq.~(\ref{corrrev_e1}).  As was discussed in Sec.~\ref{s3.3.3}, these
correlations are governed by the ballistic propagator 
$\Pi_B({\bf r_1}, {\bf r_2})$ (see Eq.~(\ref{corrrev_ball1}) and the
text preceding it). Direct calculation gives:
\begin{eqnarray} 
\Pi_B ({\bf r_1}, {\bf r_2}) & = & \Pi_1 ({\bf r_1}, {\bf r_2}) +
\Pi_2 ({\bf r_1}, {\bf r_2}), 
\end{eqnarray}
\begin{eqnarray} 
\label{surf_green}
\Pi_1 ({\bf r_1}, {\bf r_2}) &=& \Pi_B^{(0)}({\bf r_1} - {\bf r_2})-
V^{-1}\int d{\bf r'_1}\Pi_B^{(0)}({\bf r'_1} - {\bf r_2}) \\ 
&&\hspace{-2cm}
-V^{-1}\int d{\bf r'_2}\Pi_B^{(0)}({\bf r_1} - {\bf r'_2})
+V^{-2}\int d{\bf r'_1}d{\bf r'_2}
\Pi_B^{(0)}({\bf r'_1} - {\bf r'_2}); \nonumber\\
\Pi_2 ({\bf r_1}, {\bf r_2}) & = & \frac{1}{4\pi p_FR}
\sum_{k=1}^{\infty} \frac{4k^2 - 1}{4k^2} \left( \frac{r_1r_2}{R^2}
\right)^k \cos k \left( \theta_1 - \theta_2 \right) \nonumber
\end{eqnarray}
where $\Pi_B^{(0)}({\bf r}) = 1/(\pi p_F|{\bf r}|)$, and
$(r,\theta)$ are the polar coordinates. This formula has
a clear interpretation. The function $\Pi_B$ can be represented as a sum
over all paths leading from ${\bf r_1}$ to ${\bf r_2}$, with
possible surface scattering in between. In particular, 
$\Pi_1$ corresponds to direct trajectories from ${\bf r_1}$ to
${\bf r_2}$ with no reflection from the surface, while  the
contribution $\Pi_2$ is due to the surface scattering. The first term
in the numerator $4k^2-1$ comes from trajectories with only one
surface reflection, while the second sums up contributions from
multiple reflections.

Let us summarize the main results of this section.
We have used the ballistic $\sigma$-model approach to
study statistical properties of levels and eigenfunctions in a billiard
with diffusive surface scattering, which exemplifies a ballistic system
in the regime of strong chaos. It was found that the level repulsion
and the spectral rigidity are enhanced compared to RMT. In particular,
the level number variance saturates at the scale of the inverse time of
flight, in agreement with Berry's prediction for a generic chaotic system
\cite{berry1,berry2}.  As another manifestation of the strong spectral rigidity,
the oscillating part of the level correlation function does not vanish at
large level separation. 
We calculated also the ballistic analog of the diffusion propagator
in this model, which governs correlations of eigenfunction
amplitudes in different spatial points. 

While we focused our attention on the statistics of
levels and wave functions in a closed ballistic sample, the surface
nature of scattering will also modify statistical properties of the
transport characteristics for an open system. In this connection,
we mention  the recent papers \cite{leadbeater98,sanchez99}
where  quantum localization
and fluctuations of the transmission coefficients and
of the conductance were studied for a quasi-1D wire with a rough
surface. A number of important differences compared to the case of 
bulk disorder was found. 

Note that the motion in a quasi-1D wire with surface scattering
is closely related to the PRBM ensemble of Sec.~\ref{s5.4}. According
to (\ref{e8.1}), the probability density for a particle to leave the
surface after a scattering event with an angle $\theta\ll 1$ with respect
to the surface is ${\cal P}(\theta)\sim\theta$. Since the distance of
the ballistic flight is $r\sim \theta^{-1}$ for small $\theta$, this
yields ${\cal P}(r)\sim r^{-3}$ for $r$ larger than the transverse
size of the wire. Thus,  we get a power-law ``tail''
at large $r$ of the form (\ref{prbm_3}) with $\alpha=3/2$. According
to Sec.~\ref{s5.3}, this is precisely the marginal value separating
the regions of conventional $K_0^{-1}(q,0)\propto q^2$ and
unconventional $K_0^{-1}(q,0)\propto q^{2\alpha-1}$ behavior of the
diffusion propagator $K_0(q,\omega)$. It is clear that at $\alpha=3/2$
the propagator acquires a logarithmic correction, $K_0^{-1}(q,0)\propto
q^2\ln q$, see Ref.~\cite{leadbeater98} for details.

\section{Electron-electron interaction in disordered mesoscopic
systems}
\label{c9}

\setcounter{equation}{0}

In the preceding sections we have considered statistical properties of
energy levels and eigenfunctions of a single particle in a disorder
potential. However, if an electronic system is considered,
the Coulomb interaction between the electrons has to be taken into
account. The influence of the
electron-electron interaction on transport properties of
disordered systems has been intensively studied during the last two
decades, in particular in connection with such phenomena as weak
localization \cite{alar} and universal conductance fluctuations
\cite{lsf}. The 
electron-electron interaction sets the length scale $l_\phi$ (phase
breaking length) below which the electron wave function preserves its
coherence. Also, interplay of the interaction and disorder leads to a
singular (at Fermi energy or at zero temperature)
correction to the density of states and to the conductivity
\cite{alar}.

More recently,
another kind of problems has attracted the research interest: 
to what extent does the electron-electron interaction
influence the properties of the electron spectrum in a disordered dot?
This ineterest is largely motivated by 
a recent progress in nanotechnology which allows to observe experimentally
discrete electronic levels in semiconductor quantum dots
\cite{sivan94,tarucha96} and in small metallic grains \cite{ralph95,agam97}.
In fact, there are two types of the quantum dot spectra studied experimentally
via measuring their I--V characteristics:
(i) excitation spectrum, when excited levels are probed in a dot with
given number of electrons by increasing the source-drain voltage, and
(ii) addition spectrum, when electrons are added one by one by
changing the gate voltage. 

As concerns the excitation spectrum, it
was found in \cite{sia,blanter} that the quasiparticle
levels with energies below the 
Thouless energy $E_c$ (counted from the Fermi energy)
have a width smaller than the mean level spacing and thus form a
well-defined discrete spectrum.
To show this, let us calculate the r.m.s. value of the matrix element
$V_{ijkl}$ of the screened Coulomb interaction which describes a
decay of the quasiparticle state $|i\rangle$ to a three-particle (more
precisely, two particles $+$ one hole) state $|jkl\rangle$. 
Using Eqs.~(\ref{corrrev_states}), (\ref{corrrev_stateso}) for a
diffusive dot, one finds \cite{blanter} 
$\langle|V_{ijkl}|^2\rangle\sim(\Delta/g)^2$. On the other hand, the
density of states of the three-particle states is $\nu_3(E)\sim
E^2/\Delta^3$, so that the Golden Rule width of the one-particle
states is $\Gamma(E)=2\pi\langle |V|^2\rangle\nu_3(E)\sim E^2/\Delta
g^2$. Comparing $\Gamma(E)$ with $\Delta$, we get the above
threshold $E\sim g\Delta = E_c$. The number of the discrete excited
levels is thus of order of the intrinsic dimensionless conductance $g$
of the dot (not to be confused with the tunneling conductance in the
Coulomb blockade regime determined by the contacts). These findings
are in agreement with experiment \cite{sivan94}. 

The fine structure of these excited single particle levels was studied
theoretically
in \cite{agkl,mf97,silvestrov97} (see also \cite{mejia98,leyronas99})
where the Hamiltonian of the many-body
interacting problem was considered as a tight-binding model in the Fock
space, with matrix elements of the Coulomb interaction playing a role
of the hopping terms. It was found that only for the energy above
$E_*\sim\sqrt{g}\Delta$ is the Golden Rule applicable and the levels
have the regular Lorentzian shape. For smaller energies the
quasiparticle excitation consist essentially of a single peak with a
small admixture of other many-particle exact eigenstates. This
corresponds to the Anderson localization in the Fock space. 

Though the properties of the single-particle excitations discussed
above are of most interest, one can also discuss statistical properties
of exact many-body levels. For excitation energy $E\gg\Delta$ a
typical state consists of $\sim\sqrt{E/\Delta}\gg 1$ quasiparticles,
each of them having an energy $\sim\sqrt{E\Delta}$. Comparing the
density of those states to which any of such states is coupled with
the matrix element of the interaction, one estimates the border for
the complete delocalization of many-body states as $E_{ch}\sim\Delta
g^{2/3}$ \cite{jacquod97,mf97} (see also \cite{silvestrov98}). For
energies $E\ll E_{ch}$ the many-body level statistics is Poissonnian,
while for the energies above $E_{ch}$ it should acquire the
Wigner-Dyson form. Numerical studies of statistical properties of
spectra of many-body fermionic systems 
\cite{berkovits94,berkovits96,weinmann97,berkovits98,pascaud98}
indeed show such a crossover from the Poisson to RMT statistics with
increasing excitation energy.

Statistical properties of the addition spectrum have been studied
by several experimental groups recently
\cite{sivan,simmel,marcus1,marcus99a,marcus99b}. We will address the
theoretical aspects of this problem
(following Refs. \cite{bmm1,m98}) in Sec.~\ref{s9.1}. 

%Sec.~\ref{s9.2} deals with a related problem
%of the quantum correction and mesoscopic fluctuations of the
%polarizability of a mesoscopic sample. 

Let us briefly mention another topic
which has attracted a great deal of research
interest recently. This is the problem of localization of two
interacting particles raised by Shepelyansky \cite{shepel94} (see also
an early paper by Dorokhov \cite{dorokhov90}). 
It was suggested in \cite{shepel94} that the
effective localization length $\xi_2$
of two interacting particles in a 1D
disordered sample can be much larger than the localization length
$\xi_1$ of a non-interacting particle. This paper stimulated a considerable
analytical and numerical activity. In the quasi-1D case, the problem
can be mapped onto an ensemble of random banded matrices with strongly
fluctuating diagonal elements, which was studied numerically in
\cite{jacquod95} and analytically (via the supersymmetric
$\sigma$-model approach) in \cite{fm95b,frahm95}. The results
confirmed the original conjecture of Shepelyansky,
$\xi_2\propto\xi_1^2$. In the case of a strictly 1D system, the
situation is somewhat more complicated, since the corresponding random
matrix ensemble was found \cite{ps}
to be of the power-law random banded matrix
type (see Sec.~\ref{s5.4}) in the critical regime $\alpha=1$. Still,
the localization length enhancement was predicted,
$\xi_2\propto\xi_1^\mu$, but with a non-universal exponent $\mu$
($1<\mu<2$) depending on the interaction strength. Generalization of
the Shepelyansky's idea to the vicinity of the Anderson transition
point in higher-dimensional systems was proposed by Imry
\cite{imry95}. A detailed review of recent activity in this
direction can be found in \cite{gmw-rev}.

\subsection{Coulomb blockade: fluctuations in the addition spectra of
quantum dots}
\label{s9.1}

The Coulomb blockade effect shows up the most distinctly in the
addition spectrum experiment, where one measures the conductance of an
almost isolated quantum dot as a function of the gate voltage. Detailed
description of the experimental setup can be found e.g. in
\cite{dotrev,marcus97a}. At low temperature $T$ one observes a sequence
of almost equidistant peaks separated by ``valleys'' where the
conductance is very small. 
The positions of the peaks are determined
by the condition that the energies of the dot with $N$ and $N+1$
electrons are equal. We will assume below that the following
inequality is met: $\Gamma\ll kT \ll \Delta$, where
$\Gamma=\Gamma_l+\Gamma_r$ is the sum of the tunneling rates to the
leads. The peak width is then set by the temperature, while the
height is given by \cite{beenakker91} 
\be
\label{e9.1}
g_{\rm max}={e^2\over h} {\pi\over 2kT} {\Gamma_l \Gamma_r\over
\Gamma_l+\Gamma_r}. 
\ee
The peak heights show strong fluctuations induced by RMT-like
fluctuations of eigenfunction amplitudes, as predicted in
\cite{jsa,prigefii} and observed experimentally in
\cite{chang96,folk96}. 
 In the valleys between the peaks, it is
energetically costly (of order of the charging energy $e^2/C$, where
$C$ is the dot capacitance) to add an electron to (or to remove from)
the dot, and the conductance is determined by virtual processes
(so-called elastic cotunneling \cite{averin90}) and is strongly
suppressed. 

The issue of the statistics of the peak spacings was addressed for the
first time in \cite{sivan}. Basing on numerical data for very small
systems, the authors of \cite{sivan} concluded that the
r.m.s. deviation of the peak spacing $S_N$ is proportional to the
charging energy $e^2/C$, with a coefficient $\approx 0.15$. We will
show below (our consideration will closely follow
Refs.~\cite{bmm1,m98}) that the fluctuations are in fact much smaller,
of order of the mean level spacing $\Delta$. 
Let us note that in an analogous problem for classical
particles the fluctuation magnitude $\mbox{rms}(S_N)$
would indeed be proportional to the mean value 
$\langle S_N\rangle$ \cite{morris,koulakov}. The physical reason for
smaller 
fluctuations in the quantum case is in the delocalized nature of the
electronic wave functions, which are spread roughly uniformly over the
system. These theoretical conclusions were confirmed recently by
thorough experimental studies, as discussed in the end of the
section. 

The simplest theoretical model which may be used to study distribution
of the peak spacings is as follows. One considers a dot as a fixed size
diffusive mesoscopic sample and assumes that changing a gate voltage
by an amount $\delta V_g$ simply reduces to a uniform change of the
potential inside the dot by a constant $\gamma\, \delta V_g$, 
with certain numerical coefficient $\gamma$ (``lever arm''). Such a
model was used for numerical simulations of the addition spectra in
Refs.~\cite{sivan,prus}. We start by considering the statistics of
peak spacings within this model \cite{bmm1}; we will later return
to the approximations involved and relax some of them. 
We will neglect the spin degree of
freedom of electrons first; inclusion of the spin will be also
discussed in the end of the section.

\begin{figure}
{\epsfxsize=120mm\centerline{\epsfbox{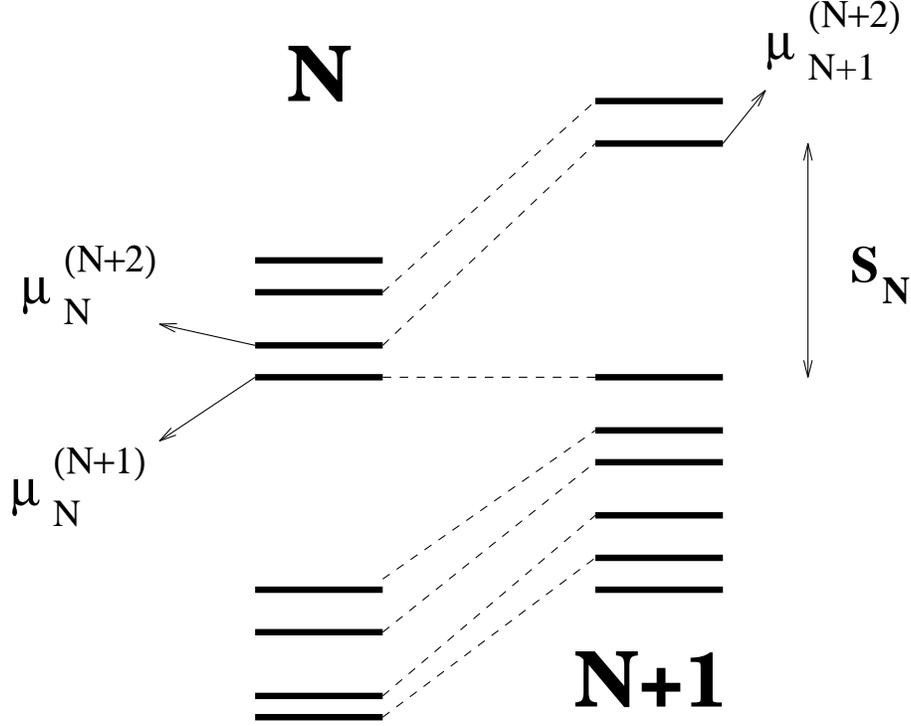}}}
\caption{Energy levels before and after the electron $\#(N+1)$ has
been added to the dot. From Ref.~\cite{bmm1}.} 
\label{coul}
\end{figure}

The distance between the two consecutive conductance peaks is given by
(Fig.~\ref{coul})
\begin{eqnarray}
S_N & = & ({\cal E}_{N+2}-{\cal E}_{N+1}) - 
 ({\cal E}_{N+1}-{\cal E}_{N}) \nonumber \\
    & = & \mu_{N+1}^{N+2} - \mu_N^{N+1}, \label{corrrev_dot1}
\end{eqnarray}
where ${\cal E}_N$ is the ground state of a sample with $N$ electrons. In the
second line of Eq.(\ref{corrrev_dot1}) we rewrote $S_N$ in terms of the
Hartree-Fock single electron energy levels, with $\mu_i^j$ denoting
the energy of the state $\#j$ in the dot containing $i$ electrons. It is
convenient to decompose $S_N$ in the following way
\begin{eqnarray}
S_N & = & (\mu_{N+1}^{N+2}-\mu_{N}^{N+2}) +
(\mu_{N}^{N+2}-\mu_{N}^{N+1}) \nonumber \\
& \equiv & E_1 + E_2 
\label{corrrev_dot2}
\end{eqnarray}
The quantity $E_2$ is the distance between the two levels of the same 
one-particle (Hartree-Fock) Hamiltonian $\hat{H}_N$ (describing a dot
with $N$ electrons) and is expected to obey RMT; in particular
$\langle E_2\rangle=\Delta$ and $\mbox{r.m.s.}(E_2)=a\Delta$ with a
numerical coefficient $a$ of order of unity [$a=0.52$ ($0.42$) for the
orthogonal (resp. unitary) ensemble]. On the other hand, $E_1$ is a
shift of the level $\#(N+2)$ due to the change of the Hamiltonian
$\hat{H}_N\to\hat{H}_{N+1}$ accompanying addition of the electron
$\#(N+1)$ to the system. 

The effective interaction $U({\bf r},{\bf r'})$ between the electrons
$\#(N+1)$ and $\#(N+2)$ can be found from the RPA-type equation,
\be
\label{coul_rpa}
U({\bf r},{\bf r'})=U_0({\bf r}-{\bf r'})-\int d{\bf r_1}d{\bf r_2}
U_0({\bf r}-{\bf r_1})P_0({\bf r_1},{\bf r_2})U({\bf r_2},{\bf r'})\,
\ee
where 
$U_0(r)=e^2/\epsilon r$, $\epsilon$ is the dielectric constant, and
\be
\label{coul_polariz}
P_0({\bf r_1},{\bf r_2})=\nu[\delta({\bf r_1}-{\bf r_2})-V^{-1}]
\ee
is the polarization operator. Its solution has the form \cite{bmm1}
\begin{equation} 
\label{coulmeso_RPAres}
U({\bf r}, {\bf r'}) = \bar U + \delta U ({\bf r}) + \delta 
U ({\bf r'}) + U_{\kappa} ({\bf r}, {\bf r'}).
\end{equation}
Here $\bar{U}\equiv e^2/C$ is a constant (charging energy), 
$\delta U({\bf r})$ is the change of
the self-consistent potential due to addition of one electron
(i.e. difference in the self-consistent potential in the dots 
with $N$ and $N+1$ electrons), and $U_\kappa({\bf r})$ is the 
screened Coulomb interaction.  In particular, in the experimentally
most relevant 2D case (which we will consider below) 
and assuming a circular form of the dot with
radius $R$, we have
\begin{equation}
\delta U({\bf r}) = - {e^2\over 2\epsilon\kappa_s R}(R^2-r^2)^{-1/2},
\label{corrrev_dot3a}
\end{equation}
while $U_\kappa$ is given in the Fourier space by
$\tilde{U}_\kappa({\bf q})=2\pi e^2/\epsilon(q+\kappa_s)$ with the
inverse screening length $\kappa_s=2\pi e^2\nu/\epsilon$. 

According to Eq.~(\ref{coulmeso_RPAres}), the term $E_1$ can be
decomposed into the following
three contributions 
\begin{eqnarray}
E_1&=& e^2/C+\int d{\bf r}\left(|\psi_{N+1}^2({\bf r})|+
|\psi_{N+2}^2({\bf r})|\right)\delta U({\bf r}) \nonumber\\
&&  
+\int d{\bf r}d{\bf r'}|\psi_{N+1}^2({\bf r})||\psi_{N+2}^2({\bf r'})|
U_{\kappa}(|{\bf r}-{\bf r'}|) \nonumber \\
&=&E_1^{(0)}+E_1^{(1)}+E_1^{(2)}
\label{corrrev_dot3}
\end{eqnarray}
The first term in Eq.~(\ref{corrrev_dot3}) (the
charging energy) determines the average value $\langle E_1\rangle$ and
thus the average peak spacing $\langle S_N\rangle$ (since $e^2/C\gg
\Delta$ for a large dot with $N\gg 1$). This is the only contribution
to $E_1$ kept by the so-called constant interaction model,
which in addition neglects fluctuations of the capacitance $C$. 
Consequently, fluctuations of $S_N$ within the constant interaction model
are determined solely by fluctuations of the single-particle level
spacing $E_2$ and thus follow RMT: $\mbox{r.m.s.}(S_N)=a\Delta$. 

The term $E_1$ in Eq.~(\ref{corrrev_dot2}) is however an additional source of
fluctuations and is thus responsible for the enhancement of
fluctuations in comparison with RMT. In principle, all the three terms 
$E_1^{(0)}$, $E_1^{(1)}$, and $E_1^{(2)}$ in Eq.~(\ref{corrrev_dot3})
contribute to this enhancement. Fluctuations of the first one,
$E_1^{(0)}=e^2/C$ are due to the fact that the capacitance is slightly
different from its purely geometric value because of a finite value of
the screening length. The corresponding correction to $C$ can be
expressed in terms of the polarization operator $P({\bf r},{\bf r'})$.
The latter is a fluctuating quantity (because of
fluctuations of the eigenfunctions in the Fermi sea) and contains
a random part $P_r({\bf r},{\bf r'})$ leading to the following
expression for the random part of the charging energy:
\begin{equation}
(e^2/C)_r=2\int d{\bf r} d{\bf r'}\delta U({\bf r})P_r({\bf r},{\bf
r'}) \delta U({\bf r'}).
\label{corrrev_dot4}
\end{equation}
Evaluating the fluctuations of the polarization operator, 
we find \cite{bm3}
\begin{eqnarray}
\mbox{var}(E_1^{(0)}) &=& {48\over\beta}\nu^2\ln g\left[{1\over V}\int
d{\bf r_1} d{\bf r_2}\delta U({\bf r_1})\Pi({\bf r_1},{\bf
r_2}) \delta U({\bf r_2})\right]^2 \nonumber \\
& \propto & {1\over \beta} \ln g \left({\Delta\over g}\right)^2
\label{corrrev_dot5}
\end{eqnarray}

Now we consider fluctuations of the last term, $E_1^{(2)}$, in
Eq.(\ref{corrrev_dot3}). Using Eqs.~(\ref{corrrev_fin1}),
(\ref{corrrev_fin1o}) for the correlations of
eigenfunction amplitudes in two remote points, the variance of
$E_1^{(2)}$ is found to be
\begin{eqnarray} 
\mbox{var}(E_1^{(2)}) & = & {4\over \beta^2 V^4}\int
d{\bf r_1}d{\bf r'_1}d{\bf r_2}d{\bf r'_2}U_\kappa(|{\bf r_1}-{\bf r'_1}|)
U_\kappa(|{\bf r_2}-{\bf r'_2}|)\Pi({\bf r_1},{\bf r_2}) \Pi({\bf
r'_1},{\bf r'_2}) \nonumber\\ 
&\approx &{4\Delta^2\over \beta^2 V^2}\int d{\bf r_1}d{\bf r_2}
\Pi^2({\bf r_1},{\bf r_2}) \nonumber \\
&\propto & {1\over \beta^2}\left({\Delta\over g}\right)^2.
\label{corrrev_dot6}
\end{eqnarray}
Finally, fluctuations of the term $E_1^{(1)}$ can be also evaluated
with a help of Eqs.~(\ref{corrrev_fin1}), (\ref{corrrev_fin1o}), yielding
\begin{eqnarray}
\mbox{var}(E_1^{(1)})&= &{4\over \beta V^2}\int d{\bf r_1}d{\bf r_2} 
\delta U({\bf r_1})\Pi({\bf r_1},{\bf
r_2}) \delta U({\bf r_2})
\nonumber \\
&\propto &{1\over\beta}{\Delta^2\over g}.
\label{corrrev_dot7}
\end{eqnarray}

It is seen that for $g\gg 1$
all the contributions Eqs.~(\ref{corrrev_dot5})--(\ref{corrrev_dot7})
are parametrically small compared to the RMT fluctuations (which are
$\sim\Delta$). Fluctuations of the term $E_1^{(1)}$ related to the
change $\delta U({\bf r})$ of the self-consistent potential represent
parametrically leading contribution 
 to the enhancement of the  peak spacing fluctuations with respect to
RMT. 

Let us now discuss approximations made in the course of the above
derivation:
\begin{enumerate}

\item[i)] The dot was supposed to be diffusive in the
calculation. For a ballistic dot one should replace $\Pi({\bf r},{\bf
r'})$ by its ballistic counterpart. This would
mean that the parameter $g$ is replaced by $\sim
N^{1/2}\sim L/\lambda_F$, where $N$ 
is the number of electrons in the dot and $L$ the characteristic
linear dimension. The numerical coefficient would
depend, however, on ``how strongly chaotic'' is the dot. The role of the
eigenfunctions fluctuations and correlations (``scars'') in
enhancement of the peak spacing fluctuations was studied 
in \cite{stopa} via numerical
simulations of a dot with $N\approx 100$ electrons. 

\item[ii)] It was assumed that changing the gate voltage results in a
spatially uniform change of the potential in the sample. This has led us to
the expression Eq.(\ref{corrrev_dot3a}) for the change of the self-consistent
potential $\delta U({\bf r})$ accompanying the addition of one
electron to the 
dot. This result would correspond to a gate located far enough from
the sample. In a more realistic situation, when the gate is relatively
narrow and located close to the sample, the potential change $\delta
U({\bf r})$ (as well as the additional electron density) will be
concentrated on the side of the dot facing the gate. The change of
the potential $\delta U({\bf r})$ corresponds then to a slight
deformation  of the dot with adding each electron
to it. To estimate $\delta U({\bf r})$ in this case, we can consider
a model problem of a point-like charged object (modeling the gate) located a
distance $d$ from the edge of the dot. Assuming that the dot size is
larger than $d$, we can approximate the dot [while calculating $\delta
U({\bf r})$] by a half plane. In one period of the Coulomb blockade
oscillations the gate charge changes by $e$ (electron charge). The
charge distribution induced in the dot is 
\be
\label{e9.2}
\delta\rho(x,y)=-{e\over \pi^2}\sqrt{{d\over
x}}{1\over\sqrt{(d+x)^2+y^2}}\ ,
\ee
where the closest to the gate point of the dot is chosen as the
coordinate origin and the $x$ axis is directed along the dot edge.
The corresponding change of the potential is 
$\delta U({\bf r})=(e^2\nu/\epsilon)^{-1}\delta\rho({\bf r})$. Substituting
this into the first line of Eq.~(\ref{corrrev_dot7}), we come to the
same result 
$$\mbox{r.m.s.}(E_1^{(1)})\propto \Delta/\sqrt{\beta g},$$ 
in the diffusive limit $d>l$, and to
$$\mbox{r.m.s.}(E_1^{(1)})\sim\Delta/\sqrt{\beta k_F d}$$
in the case $d<l$.

Effect of the quantum dot deformation on the peak spacing fluctuations
has been recently considered in Ref.~\cite{vallejos}. The authors of 
\cite{vallejos}
characterized the strength of the deformation by a phenomenological
dimensionless parameter $x$ and assumed that it can be large
($x\gtrsim1$), strongly affecting the spacing distribution.
However, the above estimates indicate that for a typical geometry this
parameter is much less than unity, $x\sim 1/\sqrt{g}$ or $x\sim
1/\sqrt{k_F d}$.

\item[iii)] It was assumed that the dot energy and the measured gate
voltage are related through a constant (or smoothly varying)
coefficient $\gamma$. This ``lever arm'' $\gamma$ depends, however on
the dot-gate capacitance, which is also a fluctuating quantity. If the
gate size and the distance to the gate is of the  order of the
size of the dot, these fluctuations should be of the same order as
fluctuations of the dot self-capacitance 
and thus lead to additional fluctuations parametrically
 small compared to $\Delta$, see Eq.~(\ref{corrrev_dot5}).

\item[iv)] The calculation was done within the random phase
approximation, which assumes that the ratio of the interparticle
Coulomb energy to the kinetic energy is small, 
$r_s\equiv {\sqrt{2}e^2/\epsilon
\hbar v_F}\ll 1$ ($v_F$ is the Fermi velocity). 
However, most of the experimental realizations of semiconductor
quantum dots correspond to $r_s\approx 1$. Since this value is still 
considerably lower than the Wigner crystallization threshold, the
calculations should be still valid, up to a numerical factor
$\alpha(r_s)$ [depending on $r_s$ only and such that $\alpha(r_s\ll
1)=1$]. 

\item[v)] We considered the model of spinless electrons up to now. 
Let us briefly discuss the role of the spin degree of freedom. 
Within the constant interaction model, it would lead to a bimodal
distribution \cite{prus} of peak spacings 
\begin{equation}
{\cal P}(S_N)={1\over 2}\left[\delta(S_N-e^2/C)+
{1\over 2\Delta}P_{WD}\left({S_N-e^2/C\over
2\Delta}\right) \right],
\label{corrrev_dot8}
\end{equation}
where $P_{WD}(s)$ is the Wigner-Dyson distribution and $\Delta$ denotes
the level spacing in the absence of spin degeneracy. 
The value of the
coefficient $a$ in the relation $\mbox{r.m.s.}(S_N)=a\Delta$ is then
increased (compared to the spinless case)  and is equal to 1.24 (1.16)
for the 
orthogonal (resp. unitary) ensemble. Taking into account fluctuations
of eigenfunctions (and thus of $E_1$) 
however modifies the form of the distribution. The value of 
the term $E_1^{(2)}$ representing the interaction between two
electrons is larger in the case when $\psi_{N+2}$ and $\psi_{N+1}$
correspond to two spin-degenerate states (i.e. have the same spatial
dependence of the wave function), since
\begin{eqnarray}
\label{corrrev_dot9}
&&\!\!\!\left\langle\int d{\bf r}d{\bf r'}|\psi_{i}^2({\bf r})|
|\psi_{i}^2({\bf r'})|
U_{\kappa}(|{\bf r}-{\bf r'}|)\right\rangle-
\left\langle\int d{\bf r}d{\bf r'}|\psi_{i}^2({\bf r})|
|\psi_{j}^2({\bf r'})|
U_{\kappa}(|{\bf r}-{\bf r'}|)\right\rangle \nonumber \\
&&\hspace{1cm}
={2\over\beta V^2}\int d{\bf r}d{\bf r'}k_d(|{\bf r}-{\bf r'}|)
U_{\kappa}(|{\bf r}-{\bf r'}|)\sim \Delta
\end{eqnarray}
for $r_s\sim 1$ (the coefficient depends on $r_s$, see
\cite{bmm1}). Therefore, filling a state $\psi_{i\uparrow}$ pushes
up the level $\psi_{i\downarrow}$ (with respect to other eigenstates) 
by an amount of order of
$\Delta$. This removes a bimodal structure of the distribution of peak
spacings and slightly modifies the value of the coefficient $a$. 

\end{enumerate}

Basing on the above analysis, we can make the following general
statement. Imagine that we fix $r_s\sim 1$ (i.e. fix the electron
density and thus the Fermi wave length) and the system geometry, and then
start to increase the linear dimension $L$ of the system. Then, while
the average value of the peak spacing $S_N$ scales as $\langle
S_N\rangle\approx e^2/C\propto 1/L$, its fluctuations will scale
differently: $\mbox{r.m.s.}(S_N)\sim\Delta\propto 1/L^2$. This
conclusion is also corroborated by diagrammatic calculations of
Ref.~\cite{berkovits97}.

As was mentioned in the beginning of this section,
this result should be contrasted with that for classical
particles \cite{morris,koulakov}, where
the fluctuations are proportional to the mean value 
$\langle S_N\rangle$. The parametrically  smaller
fluctuations in the quantum case are due to the delocalized nature of the
electronic wave functions, which are spread roughly uniformly over the
system. 

The above prediction was confirmed by recent experiments
\cite{marcus1,marcus2,marcus99b}, where a thorough study of the peak spacing
spacing statistics was carried out. It was found that the
low-temperature value of $\mbox{r.m.s.}(S_N)$, as well the typical
temperature scale for its change are approximately given by the mean
level spacing $\Delta$ (while in units of $E_c$ the magnitude of
fluctuations was as small as 1--4\%). 

Several recent papers studied statistical properties of the peak
spacing numerically. Stopa \cite{stopa} used the density functional
theory and found  the fluctuations of the addition
energies to be approximately $0.7\Delta$, in agreement
with the above results. 
Disappearance of the bimodal character of the peak
spacing distribution with increasing $r_s$ was observed recently in
numerical simulations by Berkovits \cite{berkovits98a} via exact
diagonalization. However, the system size and the number of particles
in the exact diagonalization studies are very small, so that it is
difficult to draw any quantitative conclusions concerning the
dots with large number of electrons from the results of
\cite{berkovits98a}.

\section{Summary and outlook}
\label{c10}

\setcounter{equation}{0}

In this article we have reviewed the recent progress in the study of
statistical properties of disordered electronic systems. We have
discussed  statistics of energy levels and eigenfunction
amplitudes, as well as of several related quantities (local density of
states, escape time, conductance, etc.) In most of the article  the
supersymmetric $\sigma$-model approach was used, as a unique and
powerful tool  
allowing one to calculate various distribution and correlation
functions. Within this approach, we have employed a number of
complementary methods of treating the $\sigma$-model: exact solution
(in particular, transfer-matrix method in 1D), perturbation theory,
renormalization group, saddle-point method. The main emphasis has been
put onto system-specific deviations from the universal predictions of
the random matrix theory. The results presented constitute a detailed and,
in many respects, complete description of fluctuations of spectra and wave
functions of disordered systems. 

Still, there is a number of directions in this field in which a more
complete understanding is needed, so that the corresponding research
remains active at present. Let us point out some of them.

\begin{itemize}

\item Statistical properties of energy levels and wave functions of
ballistic systems whose classical counterparts are chaotic require
more theoretical studies. It remains  to be understood what are the
conditions of applicability of the ballistic $\sigma$-model derived in
\cite{mk3,aasa1,aasa2}. Apparently, a certain amount of disorder
present in a system (or, in other words, some ensemble averaging)
is needed to justify the derivation in these
papers, but the precise conditions have not been quantified yet
(see Ref.~\cite{zirnbauer99} where this problem is discussed in the context
of correlations of quantum maps).
Also, a discrepancy between the results of the $\sigma$-model and of
the semiclassical approach in treating repetitions of periodic orbits
\cite{argaman93,bogkeat} is to be resolved. It remains to be seen
whether the 
ballistic $\sigma$-model approach can be developed to predict the
weak-localization corrections (recent attempts
\cite{aleiner96,smith98} did not lead to any definite conclusions),
the asymptotic ``tails'' of the distributions (like those discussed in
Section \ref{c4} for diffusive systems) etc.

\item Interplay of disorder and electron-electron interaction in
mesoscopic systems continues attracting  considerable research interest. 
In Sec.~\ref{c9} of this article 
we discussed only the issue of fluctuations in an (almost)
closed system. In the experiment, coupling of the dot to the outside
world is controlled by the gate(s) and may be varied. There arise a
rich variety of  regimes depending on the coupling strength,
magnetic field, and the strength of the Coulomb interaction (parameter
$r_s$). Fluctuations in open quantum dots have been studied recently,
both theoretically \cite{aleiner96a,aleiner98,kaminski98}
and experimentally
\cite{cronenwett97,cronenwett98,marcus99a,marcus99b}, but the issue is
not fully understood yet. In Ref.~\cite{zhitenev97} it was
found that application of a strong magnetic field affects dramatically
the addition spectrum of a quantum dot, leading to strong fluctuations
and to bunching of the Coulomb blockade peaks. A consistent
explanation of these experimental results is still missing.

Very recently, several groups
\cite{levit99,bonci99,walker99a,walker99b,cohen99} 
simultaneously performed self-consistent
Hartree-Fock calculations of addition spectra of quantum dots.
Though the method allows one to study considerably larger systems as
compared to the exact 
diagonalization method, the obtained results were not
sufficient to extract unambiguously a scaling dependence of the spacing
distribution function on the parameters of the problem.
A drawback of the Hartree-Fock method with the bare Coulomb
potential is that the exchange interaction does not get screened,
whereas the screening was crucially important for the 
theoretical consideration in Section \ref{s9.1}. 
More work in this direction may be expected in the nearest future.

\item The physics of the Anderson metal-insulator transition, including
fluctuations at the critical point, remains an actively studied
field. Parameters of modern computers allow one to evaluate   
numerically critical indices and various distribution
functions with a high accuracy. It would be
very interesting to study
numerically the critical point in higher dimensions (see, in this
respect, recent paper \cite{zhar98}) as well as in an effectively
infinite-dimensional tree-like sparse random matrix model and in a
long-range 1D model (power-law random banded matrix
ensemble considered in Sec.~\ref{s5.4}). 
It remains to be seen what is the status of the conjecture
of Ref.~\cite{ckl} relating the spectral compressibility to the
eigenfunction multifractality. Also, the form of the conductance
distribution at the critical point remains an open
problem. Ref.~\cite{akl} predicted a log-normal form of the
distribution function at $g\gg \langle g\rangle \sim 1$, similarly to
the distribution functions of local density of states and of
relaxation times. However, in contrast to the latter two quantities,
an anomalously large conductance cannot be explained by a single
anomalously localized state (since a single state cannot produce the
conductance larger than 1). It seems much more probable that the
conductance distribution falls off much more fast, in a Gaussian (or
similar) fashion at large $g$, but no corresponding calculation has been
available so far. Technically, the problem is that calculating higher
moments of the conductance requires increasing of the size of the
$Q$-matrix in the $\sigma$-model, which complicates tremendously a
non-perturbative treatment of the distribution function. For recent
numerical studies of the problem see
Refs.~\cite{slevin,wang99,markos99}. 

\end{itemize}

\newpage

\vspace*{5cm}

{\LARGE \bf Acknowledgments}

\vspace{1cm}

My own work reviewed in this paper was done in collaboration
with the late A.G.~Aronov, Ya.M.~Blanter,
F.-M.~Dittes, Y.V.~Fyodorov, V.E.~Kravtsov, A.~M\"uller-Groeling,
B.A.~Muzykantskii, R.~Pnini, T.~Seligman, B.~Shapiro, M.R.~Zirnbauer. 
It is my pleasure to thank all of them.
I am also indebted to many other colleagues,
in particular,  O.~Agam, B.L.~Altshuler, R.~Berkovits, S.~Fishman,
K.~Frahm, Y.~Gefen, D.E.~Khmelnitskii, I.V.~Lerner, L.S.~Levitov,
C.~Marcus, K.~Muttalib, D.G.~Polyakov, P.G.~Silvestrov,
I.E.~Smolyarenko, A.M.~Tsvelik, P.~W\"olfle, and I.Kh.~Zharekeshev for
numerous stimulating discussions of the topics addressed in the review.
This work was supported by SFB195 der Deutschen
Forschungsgemeinschaft.

\appendix

\noindent
\section{Abbreviations}

\vspace{0.5cm}

\begin{tabular}{lp{0.7\linewidth}}

AKL    & Altshuler, Kravtsov, and Lerner, Ref. \cite{akl}\\
ALS    & anomalously localized state\\
%% CE     & canonical ensemble\\
DMPK   & Dorokhov-Mello-Pereira-Kumar (equations)\\
DOS    & density of states\\
%% GCE    & grand canonical ensemble\\
GOE    & Gaussian orthogonal ensemble\\
GSE    & Gaussian symplectic ensemble\\
GUE    & Gaussian unitary ensemble\\
IPR    & inverse participation ratio\\
LDOS   & local density of states\\
LN     & logarithmically-normal\\
PRBM   & power-law random banded matrix\\
RBM    & random banded matrix\\
RG     & renormalization group\\
RPA    & random phase approximation\\
RMT    & random matrix theory\\
WD     & Wigner-Dyson (level statistics)\\
1D, 2D, 3D & one-dimensional, two-dimensional, three-dimensional

\end{tabular}

\end{document}